%% file: PhDThesis_MW.tex
\RequirePackage[l2tabu]{nag}		% Warns for incorrect (obsolete) LaTeX usage
\documentclass[a4paper,11pt,leqno,openbib]{memoir} %add 'draft' to turn draft option on (see below)
\usepackage{datetime}
\usepackage{ifpdf}
\ifpdf
\fi
\ifdraftdoc 
	\usepackage{draftwatermark}				%Sets watermarks up.
	\SetWatermarkScale{0.3}
	\SetWatermarkText{\textbf Draft: \today}
\fi
\newsubfloat{figure}
\newsubfloat{table}
\settrimmedsize{297mm}{210mm}{*}
\settypeblocksize{634pt}{448.13pt}{*} 
\setulmargins{4cm}{*}{*} 
\setlrmargins{*}{*}{1.5} 
\setmarginnotes{17pt}{51pt}{\onelineskip} 
\setheadfoot{\onelineskip}{2\onelineskip} 
\setheaderspaces{*}{2\onelineskip}{*} 
\checkandfixthelayout
\usepackage{fouriernc}
\usepackage[T1]{fontenc}
\OnehalfSpacing 
\setsecnumdepth{subsection} 
\maxsecnumdepth{subsubsection}
\usepackage{calc,soul,fourier}
\makeatletter 
\newlength\dlf@normtxtw 
\newsavebox{\feline@chapter} 
\newcommand\feline@chapter@marker[1][4cm]{%
	\sbox\feline@chapter{% 
		\resizebox{!}{#1}{\fboxsep=1pt%
			\colorbox{black}{\color{white}\thechapter}% 
		}}%
		\rotatebox{90}{% 
			\resizebox{%
				\heightof{\usebox{\feline@chapter}}+\depthof{\usebox{\feline@chapter}}}% 
			{!}{\scshape\so\@chapapp}}\quad%
		\raisebox{\depthof{\usebox{\feline@chapter}}}{\usebox{\feline@chapter}}%
} 
\newcommand\feline@chm[1][4cm]{%
	\sbox\feline@chapter{\feline@chapter@marker[#1]}% 
	\makebox[0pt][c]{% aka \rlap
		\makebox[1cm][r]{\usebox\feline@chapter}%
	}}
\makechapterstyle{daleifmodif}{

	\renewcommand\printchapternum{\null\hfill\feline@chm[2.5cm]\par}

} 
\makeatother 

\chapterstyle{daleifmodif}
\makepagestyle{myvf} 
\makeoddfoot{myvf}{}{\thepage}{}
\makeevenfoot{myvf}{}{\thepage}{} 
\makeheadrule{myvf}{\textwidth}{\normalrulethickness} 
\makeevenhead{myvf}{\tiny\textsc{\leftmark}}{}{} 
\makeoddhead{myvf}{}{}{\tiny\textsc{\rightmark}}
\newcommand{\clearemptydoublepage}{\newpage{\thispagestyle{empty}\cleardoublepage}}
\makeindex
\usepackage{import}

\usepackage{color}

\usepackage[normalem]{ulem}

\usepackage{lipsum}					%Needed to create dummy text
\usepackage{amsfonts} 					%Calls Amer. Math. Soc. (AMS) fonts
\usepackage[centertags]{amsmath}			%Writes maths centred down
\usepackage{stmaryrd}					%New AMS symbols
\usepackage{amssymb}					%Calls AMS symbols
\usepackage{amsthm}					%Calls AMS theorem environment
\usepackage{newlfont}					%Helpful package for fonts and symbols
\usepackage{layouts}					%Layout diagrams
\usepackage{graphicx}					%Calls figure environment
\usepackage{longtable,rotating}			%Long tab environments including rotation. 
\usepackage[utf8]{inputenc}			%Needed to encode non-english characters 
\usepackage{dsfont}									
\usepackage{colortbl}					%Makes coloured tables
\usepackage{wasysym}					%More math symbols
\usepackage{mathrsfs}					%Even more math symbols
\usepackage{float}						%Helps to place figures, tables, etc. 
\usepackage{verbatim}					%Permits pre-formated text insertion
\usepackage{upgreek}					%Calls other kind of greek alphabet
\usepackage{latexsym}					%Extra symbols
\usepackage[square,numbers,
		     sort&compress]{natbib}		%Calls bibliography commands 
\usepackage{url}						%Supports url commands
\usepackage{etex}						%eTeXÕs extended support for counters
\usepackage[spanish,main=english,german,french]{babel}		%For languages characters and hyphenation
\usepackage{color}                    				%Creates coloured text and background
\usepackage[colorlinks=true,
		     allcolors=black]{hyperref}              %Creates hyperlinks in cross references
\usepackage{memhfixc}					%Must be used on memoir document 
\usepackage{enumerate}					%For enumeration counter
\usepackage{footnote}					%For footnotes
\usepackage{microtype}					%Makes pdf look better.
\usepackage{rotfloat}					%For rotating and float environments as tables, 
\usepackage{alltt}						%LaTeX commands are not disabled in 
\usepackage[version=0.96]{pgf}			%PGF/TikZ is a tandem of languages for producing vector graphics from a 
\usepackage{tikz}						%geometric/algebraic description.
\usetikzlibrary{arrows,shapes,snakes,
		       automata,backgrounds,
		       petri,topaths}				%To use diverse features from tikz		
\usepackage[bb=boondox]{mathalfa}

\usepackage{bbm} %to make bbm symbols
\usepackage[makeroom]{cancel}

\newcommand{\pgftextcircled}[1]{                                                                    %Defines encircled text
    \setbox0=\hbox{#1}%
    \dimen0\wd0%
    \divide\dimen0 by 2%
    \begin{tikzpicture}[baseline=(a.base)]%
        \useasboundingbox (-\the\dimen0,0pt) rectangle (\the\dimen0,1pt);
        \node[circle,draw,outer sep=0pt,inner sep=0.1ex] (a) {#1};
    \end{tikzpicture}
}
\newcommand{\blackged}{\hfill$\blacksquare$}
\newcommand{\whiteged}{\hfill$\square$}
\newcounter{proofcount}
\renewenvironment{proof}[1][\proofname.]{\par
 \ifnum \theproofcount>0 \pushQED{\whiteged} \else \pushQED{\blackged} \fi%
 \refstepcounter{proofcount}
 \normalfont 
 \trivlist
 \item[\hskip\labelsep
       \itshape
   {\textbf\em #1}]\ignorespaces
}{%
 \addtocounter{proofcount}{-1}
 \popQED\endtrivlist
}
\let\oldsqrt\sqrt
\def\sqrt{\mathpalette\DHLhksqrt}
\def\DHLhksqrt#1#2{%
\setbox0=\hbox{$#1\oldsqrt{#2\,}$}\dimen0=\ht0
\advance\dimen0-0.2\ht0
\setbox2=\hbox{\vrule height\ht0 depth -\dimen0}%
{\box0\lower0.4pt\box2}}
\newcommand{\mycaption}[2][\@empty]{
	\captionnamefont{\scshape} 
	\changecaptionwidth
	\captionwidth{0.9\linewidth}
	\captiondelim{.\:} 
	\indentcaption{0.75cm}
	\captionstyle[\centering]{}
	\setlength{\belowcaptionskip}{10pt}
	\ifx \@empty#1 \caption{#2}\else \caption[#1]{#2}
}
\newcommand{\mysubcaption}[2][\@empty]{
	\subcaptionsize{\small}
	\hangsubcaption
	\subcaptionlabelfont{\rmfamily}
	\sidecapstyle{\raggedright}
	\setlength{\belowcaptionskip}{10pt}
	\ifx \@empty#1 \subcaption{#2}\else \subcaption[#1]{#2}
}
\usepackage{lettrine}
\newcommand{\initial}[1]{%
	\lettrine[lines=3,lhang=0.33,nindent=0em]{
		\color{black}
     		{\textsc{#1}}}{}}
\theoremstyle{plain}

\theoremstyle{plain}

\theoremstyle{plain}
\theoremstyle{definition}

\theoremstyle{plain}

\theoremstyle{plain}

\theoremstyle{plain}

\numberwithin{equation}{section}

\begin{document}
\input{notations}
\input{env}
\frontmatter

\input{frontmatter/title}
\clearemptydoublepage
\addtocontents{toc}{\par\nobreak \mbox{}\hfill{\textbf Page}\par\nobreak}
\pagenumbering{arabic}
\setcounter{page}{3}
\input{frontmatter/abstract}
\clearemptydoublepage
\addcontentsline{toc}{chapter}{List of Contents}
\renewcommand{\contentsname}{List of Contents}
\maxtocdepth{subsection}
\tableofcontents*
\clearemptydoublepage
\listoftables
\addtocontents{lot}{\par\nobreak\textbf{{\scshape Table} \hfill Page}\par\nobreak}
\clearemptydoublepage
\listoffigures
\addtocontents{lof}{\par\nobreak\textbf{{\scshape Figure} \hfill Page}\par\nobreak}
\clearemptydoublepage

\input{frontmatter/acknowledgement}
\clearemptydoublepage
%
\input{frontmatter/declaration}
\clearemptydoublepage
\savepagenumber
\mainmatter
\restorepagenumber
\import{chapter01/}{chap_intro.tex}
\clearemptydoublepage
\import{chapter02/}{chap_prelim.tex}
\clearemptydoublepage
\import{chapter03/}{chap_entropyvector.tex}

\clearemptydoublepage
\import{chapter04/}{chap_inner.tex}
\clearemptydoublepage
\import{chapter05/}{chap_nonshan.tex}
\clearemptydoublepage
\import{chapter06/}{chap_quantclass.tex}
\clearemptydoublepage
\import{chapter07/}{thermoaxioms.tex}
\clearemptydoublepage
\import{chapter08/}{thermomacro.tex}
\clearemptydoublepage
\import{chapter09/}{conclusion.tex}
\clearemptydoublepage
%
%
%BIBLIOGRAPHY
\bibliographystyle{siam}

\end{document}

%% file: notations.tex
\renewcommand{\labelitemi}{$\bullet$}

\renewcommand{\qedsymbol}{$\blacksquare$}

\newcommand{\indep}{\rotatebox[origin=c]{90}{$\models$}}
\newcommand{\ot}{\otimes}
\newcommand{\eps}{\epsilon}
\newcommand{\sth}{ \ \mathrm{s.t.} \ }
\newcommand{\trans}{\mathrm{T}}
\newcommand{\tr}[1]{\operatorname{tr}\left( #1 \right)}
\newcommand{\partr}[2]{\operatorname{tr}_{#2}\left( #1 \right)}
\newcommand{\trdist}[2]{\operatorname{D}\left( #1 | #2 \right)}
\newcommand{\lin}[1]{\mathcal{L}\left( #1 \right)} %linear operators...
\newcommand{\linH}[1]{\mathcal{L}_\mathrm{H}\left( #1 \right)}

%Convex cones etc.
\newcommand{\cone}{C}
\newcommand{\polycone}{P}
\newcommand{\convhull}[1]{\mathrm{conv}\left\{ #1 \right\}}
\newcommand{\conichull}[1]{\mathrm{cone}\left\{ #1 \right\}}
\newcommand{\polyhedron}{P}
\newcommand{\pro}[1]{\pi(#1)}
\newcommand{\proO}{\pi}
\newcommand{\idmat}{\openone}
\newcommand{\card}[1]{\left| #1 \right|}
\newcommand{\hrep}{\mathcal{H}}
\newcommand{\vrep}{\mathcal{V}}
\newcommand{\rpos}[1]{\mathbb{R}_{\geq 0}^{#1}}
\newcommand{\colvec}[3]{\begin{pmatrix} #1 \\ #2 \\ #3 \end{pmatrix}}
\newcommand{\threemat}[9]{\begin{pmatrix} #1 & #2 & #3 \\ #4 & #5 & #6 \\ #7 & #8 & #9 \end{pmatrix}}
\newcommand{\twomat}[4]{\begin{pmatrix} #1 & #2 \\ #3 & #4 \end{pmatrix}}

%Quantum 
\newcommand{\ket}[1]{| #1 \rangle}
\newcommand{\bra}[1]{\langle #1 |}
\newcommand{\braket}[2]{\langle #1|#2\rangle}
\newcommand{\ketbra}[2]{|#1\rangle\!\langle#2|}
\newcommand{\proj}[1]{\ketbra{#1}{#1}}
\newcommand{\identity}{\openone}
\newcommand{\id}{\mathbbm{1}}
\newcommand{\cptp}{\mathcal{E}}

%Entropies
\newcommand{\Hmin}[1]{H_{\min}\left( #1 \right)}
\newcommand{\Hzero}[1]{H_{0}\left( #1 \right)}
\newcommand{\Hmax}[1]{H_{\max}\left( #1 \right)}
\newcommand{\HH}[1]{H_{\mathrm{H}}\left( #1 \right)}
\newcommand{\Hmineps}[2]{H^{#1}_{\min}\left( #2 \right)}
\newcommand{\Hzeroeps}[2]{H^{#1}_{0}\left( #2 \right)}
\newcommand{\HzeroepsO}[1]{H^{#1}_{0}}
\newcommand{\HminepsO}[1]{H^{#1}_{\min}}
\newcommand{\HHeps}[2]{H^{#1}_{\mathrm{H}}\left( #2 \right)}
\newcommand{\HHepsO}[1]{H^{#1}_{\mathrm{H}}}
\newcommand{\epsball}[2]{\mathcal{B}^{#1}\left( #2 \right)}
\newcommand{\epsballsub}[2]{\mathcal{B}^{#1}_\leq \left( #2 \right)}
\newcommand{\maxEV}{\lambda_{\max}}
\newcommand{\minEV}{\lambda_{\min}}
\newcommand{\ithEV}[1]{\lambda_{{i}}\left( #1 \right)}
\newcommand{\ithEVk}[2]{\lambda_{{#2}}\left( #1 \right)}
\newcommand{\ithEVO}{\lambda_{{i}}}
\newcommand{\rank}{\operatorname{rank}}

%Matrices
\newcommand{\mshan}[1]{M^{#1}_\mathrm{SH}}

%Entropy cones
\newcommand{\marset}[1]{\Gamma^{*}_{\mathcal{M}}\left( #1 \right)}
\newcommand{\marcone}[1]{\overline{\Gamma^{*}_{\mathcal{M}}}\left( #1 \right)}
\newcommand{\marconeO}{\overline{\Gamma^{*}_{\mathcal{M}}}}
\newcommand{\enset}[1]{\Gamma^{*}\left( #1 \right)}
\newcommand{\ensetk}[1]{\Gamma^{*}_{ #1 }}
\newcommand{\encone}[1]{\overline{\Gamma^{*}}\left( #1 \right)}
\newcommand{\enconek}[1]{\overline{\Gamma^{*}_{ #1 }}}
\newcommand{\enconekO}{\overline{\Gamma^{*}}}

\newcommand{\outmarcone}[1]{\Gamma_{\mathcal{M}}\left( #1 \right)}
\newcommand{\outmarconeO}{\Gamma_{\mathcal{M}}}
\newcommand{\outcone}[1]{\Gamma \left( #1 \right)}
\newcommand{\outconek}[1]{\Gamma_{#1}}
\newcommand{\outconekO}{\Gamma}

\newcommand{\inmarcone}[1]{\Gamma^{\mathrm{I}}_{\mathcal{M}}\left( #1 \right)}
\newcommand{\incone}[1]{\Gamma^{\mathrm{I}}\left( #1 \right)}
\newcommand{\inconek}[1]{\Gamma^{\mathrm{I}}_{#1}}

\newcommand{\dist}[1]{\mathcal{P}\left( #1 \right)}
\newcommand{\distk}[1]{\mathcal{P}_{ #1 }}
\newcommand{\mardist}[1]{\mathcal{P}_{\mathcal{M}}\left( #1 \right)}
\newcommand{\mardistO}{\mathcal{P}_{\mathcal{M}}}

%Causal Structures
\newcommand{\bellsc}{\mathrm{B}}
\newcommand{\inst}{\mathrm{IC}}
\newcommand{\bil}{\mathrm{BI}}
\newcommand{\tri}{\mathrm{C}_3}
\newcommand{\pn}{\mathrm{P}_n}
\newcommand{\pfour}{\mathrm{P}_4}
\newcommand{\pfive}{\mathrm{P}_5}
\newcommand{\psix}{\mathrm{P}_6}
\newcommand{\smone}{\mathrm{IC}}
\newcommand{\smtwo}{\mathrm{B}}
\newcommand{\smthree}{\mathrm{S}}
\newcommand{\smcaus}{C}
\newcommand{\caus}{C}

%Superscripts for causal structures
\newcommand{\cCQQ}{\mathrm{CQQ}}
\newcommand{\cCCQ}{\mathrm{CCQ}}
\newcommand{\cC}{\mathrm{C}}
\newcommand{\qQ}{\mathrm{Q}}
\newcommand{\gG}{\mathrm{G}}
\newcommand{\iI}{\mathrm{I}}

\newcommand{\cE}{\mathcal{E}}
\newcommand{\cF}{\mathcal{F}}
\newcommand{\cH}{\mathcal{H}} %used for Hilbert spaces for instance
\newcommand{\cI}{\mathcal{I}}
\newcommand{\cM}{\mathcal{M}} %Used to denote the marginal scenario
\newcommand{\cP}{\mathcal{P}}
\newcommand{\cS}{\mathcal{S}} %Used for density operators
\newcommand{\cX}{\mathcal{X}} %Used for feasible sets
\newcommand{\cY}{\mathcal{Y}} %Used for feasible sets

\newcommand{\bH}{\textbf{H}} %entropyvector
\newcommand{\zerovec}{\textbf{0}} %zerovector

%Thermo
\newcommand{\maj}{\prec_\mathrm{M}}
\newcommand{\pmaj}{\prec_\mathrm{p}}
\newcommand{\tmaj}{\prec_{\mathrm{T}}}
\newcommand{\cZ}{\mathcal{Z}} %partition function

\newcommand{\nuparrow}{\cancel\uparrow}

\newcommand*{\defeq}{\mathrel{\vcenter{\baselineskip0.5ex \lineskiplimit0pt
                     \hbox{\scriptsize.}\hbox{\scriptsize.}}}%
                     =}

\newcommand{\MW}[1]{{\color{purple}[#1]}}
%=========================================================

\newcommand{\curl}[1]{\mathrm{Curl}(#1)}
\newcommand\todown{\searrow}
\def\v{\mathrm{v}}
\newcommand\hu{\hat{u}}
\newcommand\invf{\mathcal{F}^{-1}}
\newcommand\dom{\mathcal{O}}
\def\fou{\mathcal{F}}
\def\lf{\dot{S}}
\def\r3{\mathbb{R}^3}
\def\rd{\mathbb{R}^d}
\def\rp{\mathbb{R}^{+}}
\newcommand\myeq{\stackrel{\mathclap{\normalfont\small\mbox{def}}}{=}}
\def\E{\mathbb{E}}
\def\itri{(I - \Delta)}
\def\tp{\tilde{\mathbb{P}}}
\def\hp{\hat{\mathbb{P}}}
\def\tf{\tilde{\mathcal{F}}}
\def\hf{\hat{\mathcal{F}}}
\def\tom{\tilde{\Omega}}
\def\ho{\hat{\Omega}}
\def\rH{\mathrm{H}}
\def\rV{\mathrm{V}}
\def\rD{\mathrm{D}}
\def\rK{\mathrm{K}}
\def\T{\mathbb{T}}
\def\lipzc{2}
\newcommand{\tunk}{{\tilde{u}}_{{n}_{k}}}
\newcommand{\tu}{\tilde{u}}
\newcommand{\bcal}{\mathcal{B}}
\newcommand{\ccal}{\mathcal{C}}
\newcommand{\ecal}{\mathcal{E}}
\newcommand{\fcal}{\mathcal{F}}
\newcommand{\lcal}{\mathcal{L}}
\newcommand{\ocal}{\mathcal{O}}
\newcommand{\vcal}{\mathcal{V}}
\newcommand{\zcal}{\mathcal{Z}}
\newcommand{\kcal}{\mathcal{K}}
\newcommand{\unk}{{u}_{{n}_{k}}}
\newcommand{\taun}{{\tau}_{n}}
\newcommand{\fmath}{\mathbb{F}}
\newcommand{\smath}{\mathbb{S}}
\newcommand{\norm}[3]{\vert  #1 {\vert }_{#2}^{#3}} %%26/12/2013
\newcommand{\Norm}[3]{\Vert  #1 {\Vert }_{#2}^{#3}} %%26/12/2013
\newcommand{\lb}{\langle}
\newcommand{\rb}{\rangle}
\newcommand{\Xn}{{X}_{n}}
\newcommand{\Mn}{{M}_{n}}
\newcommand{\tMn}{{\tilde{M}}_{n}}
\newcommand{\Pn}{{P}_{n}}
\newcommand{\tPn}{{\tilde{P}}_{n}}
\newcommand{\Jn}[1]{{J}^{n}_{#1}}
\newcommand{\p}{\mathbb{P}}
\newcommand\un{u_n(t)}
\newcommand\uns{u_n(s)}
\newcommand\tus{\tilde{u}(s)}
\newcommand\tut{\tilde{u}(t)}
\newcommand\tuns{\tilde{u}_n(s)}
\newcommand\tusi{\tilde{u}(\sigma)}
\newcommand\tunsi{\tilde{u}_n(\sigma)}
\newcommand\tun{{\tilde{u}_n}}
\newcommand{\embed}{\hookrightarrow }
\newcommand{\normh}[3]{|  #1 {|}_{#2}^{#3}} %%31/10/2014
\newcommand{\Normh}[3]{\Bigl|  #1 {\Bigr| }_{#2}^{#3}} %%31/10/2014
\newcommand{\ilsk}[3]{{\langle #1 , #2 \rangle}_{#3}}
\newcommand{\Ilsk}[3]{\Bigl( #1 , #2 {\Bigr)}_{#3}}
\newcommand{\dual}[3]{{\lb #1 , #2 \rb}_{#3}}
\newcommand{\Dual}[3]{{\Bigl< #1  , #2 \Bigr>}_{#3}}
\newcommand{\dirilsk}[3]{{\bigl( \! \bigl( #1 , #2 \bigr) \! \bigr)}_{#3}}
\newcommand{\ddual}[4]{{}_{#1}\lb #2 ,#3 {\rb }_{#4}}
\newcommand{\dDual}[4]{{}_{#1}{\Bigl< #2 ,#3 \Bigr>}_{#4}}
\newcommand{\tOmega}{\tilde{\Omega}}
\newcommand{\tomega}{\tilde{\omega}}
\newcommand{\tfcal}{\tilde{\fcal}}
\newcommand{\tft}{\tilde{\ft}}
\newcommand{\te}{\tilde{\mathbb{E}}}
\newcommand{\tW}{\tilde{W}}
\newcommand{\ttW}{\tilde{\tilde{W }}}
\def\st{{t \wedge \tau_R^n}}
%\DeclareSymbolFont{bbold}{U}{bbold}{m}{n}
%\DeclareSymbolFontAlphabet{\mathbbold}{bbold}
\newcommand\ind{\mathbb{1}}
\newcommand{\bX}{x}
\newcommand{\D}{\mathrm{d}}

\newcommand{\then}{\Rightarrow}
\newcommand\Eb{\mathbb{E}}
\newcommand\N{\mathbb{N}}
\newcommand\Z{\mathbb{Z}}
\newcommand\Q{\mathbb{Q}}
\newcommand\ps{{(\Omega, \mathcal{F}, \mathbb{P})}}
\newcommand\R{\mathbb{R}}
\newcommand\Pb{\mathbb{P}}
\newcommand\s{$\sigma$\textrm{-field} }
\newcommand\notsubset{\subset\hspace{-3.5mm}{/}\hspace{1.5mm}}
\newcommand\1{u_1}
\newcommand\2{u_2}
\newcommand\xt{X_T}
\def\del{\dot{\Delta}}
\def\lap{\nabla}
\def\invf{\mathcal{F}^{-1}}
\def\dj{\delta_j}
\def\apo{(I + \delta B)^{-1}}
\def\T{\mathbb{T}}
\def\A{\mathrm{A}}
\def\mcal{\mathcal{M}}
\def\divv{\mathrm{div}}
\def\rE{\mathrm{E}}
\def\e{\mathbb{E}}

%% file: env.tex
\newtheorem{theorem}{Theorem}[section]

\newtheorem{acknowledgement}[section]{Acknowledgement}
\newtheorem{alg}[theorem]{Algorithm}
\newtheorem{axiom}[theorem]{Axiom}
\newtheorem{case}[theorem]{Case}
\newtheorem{claim}[theorem]{Claim}
\newtheorem{conclusion}[theorem]{Conclusion}
\newtheorem{condition}[theorem]{Condition}
\newtheorem{conjecture}[theorem]{Conjecture}
\newtheorem{corollary}[theorem]{Corollary}
\newtheorem{criterion}[theorem]{Criterion}
\newtheorem{defn}[theorem]{Definition}
\newtheorem*{ass}{Assumptions}

%\declaretheorem{axiom}
\newtheorem*{axiom_modified}{Axiom 7.3.4'}

\newtheorem{example1}[theorem]{Example}
\newtheorem{exercise1}[theorem]{Exercise}
\newtheorem{lemma}[theorem]{Lemma}
\newtheorem{notation}[theorem]{Notation}
\newtheorem{problem}[theorem]{Open Problem}
\newtheorem{proposition}[theorem]{Proposition}
\newtheorem{remark1}[theorem]{Remark}
\newtheorem{solution}[theorem]{Solution}
\newtheorem{summary}[theorem]{Summary}
\newtheorem{digression1}[theorem]{Digression}

%% The definitions below are modified from my SCBST lectures I replaced [Theorem] by [theorem]
\newtheorem{Definition}[theorem]{Definition}
\newtheorem{Remark}[theorem]{Remark}
\newtheorem{Proposition}[theorem]{Proposition}
\newtheorem{Lemma}[theorem]{Lemma}
\newtheorem{Corollary}[theorem]{Corollary}
\newtheorem{Exercise}[theorem]{Exercise}%[section]
\newtheorem{Example}[theorem]{Example}
\newtheorem{comments}{Comments}
\renewcommand{\thecomments}{}

\newtheorem{rem}{Remark}
\renewcommand{\therem}{}
\newtheorem{note}{Notation}
\renewcommand{\thenote}{}

\newtheorem{example0}{Example}
\renewcommand{\theexample0}{}
\newcommand\U{u^{\nu}}

\newtheorem{motivation}[theorem]{Motivation}

\newenvironment{definition}{
\begin{defn}
	\normalfont}{
\end{defn}
}

\newenvironment{algorithm}{
\begin{alg}
	\normalfont}{
\end{alg}
}

\newenvironment{remark}{
\begin{remark1}
	\normalfont}{
\end{remark1}
}

\newenvironment{example}{
\begin{example1}
	\normalfont}{
\end{example1}
}

\newenvironment{exercise}{
\begin{exercise1}
	\normalfont}{
\end{exercise1}
}

\newenvironment{digression}{
\begin{digression1}
	\normalfont}{
\end{digression1}
}

\newtheoremstyle{AppALem}{1}{1}
  {\itshape}{0pt}{\bfseries}{.}{ }
   {\thmname{Lemma }\thmnumber{A.{#2}}{\thmnote{}}}
   \theoremstyle{AppALem}\newtheorem{lemmaA}{Lemma}

\newtheoremstyle{AppBLem}{1}{1}
  {\itshape}{0pt}{\bfseries}{.}{ }
   {\thmname{Lemma }\thmnumber{B.{#2}}{\thmnote{}}}
   \theoremstyle{AppBLem}\newtheorem{lemmaB}{Lemma}
   
   \newtheoremstyle{AppCLem}{1}{1}
  {\itshape}{0pt}{\bfseries}{.}{ }
   {\thmname{Lemma }\thmnumber{C.{#2}}{\thmnote{}}}
   \theoremstyle{AppCLem}\newtheorem{lemmaC}{Lemma}
   
   \newtheoremstyle{AppDLem}{1}{1}
  {\itshape}{0pt}{\bfseries}{.}{ }
   {\thmname{Lemma }\thmnumber{D.{#2}}{\thmnote{}}}
   \theoremstyle{AppDLem}\newtheorem{lemmaD}{Lemma}

\newtheoremstyle{AppBRem}{1}{1}
  {\itshape}{0pt}{\bfseries}{.}{ }
   {\thmname{Remark }\thmnumber{B.{#2}}{\thmnote{}}}
   \theoremstyle{AppBRem}\newtheorem{remarkB}{Remark}

\newtheoremstyle{AppBCor}{1}{1}
  {\itshape}{0pt}{\bfseries}{.}{ }
   {\thmname{Corollary }\thmnumber{B.{#2}}{\thmnote{}}}
   \theoremstyle{AppBCor}\newtheorem{corB}[lemmaB]{Lemma}

\newtheoremstyle{AppBThm}{1}{1}
  {\itshape}{0pt}{\bfseries}{.}{ }
   {\thmname{Theorem }\thmnumber{B.{#2}}{\thmnote{}}}
   \theoremstyle{AppBThm}\newtheorem{theoremB}{Theorem}
   
\newtheoremstyle{AppCThm}{1}{1}
  {\itshape}{0pt}{\bfseries}{.}{ }
   {\thmname{Theorem }\thmnumber{C.{#2}}{\thmnote{}}}
   \theoremstyle{AppCThm}\newtheorem{theoremC}{Theorem}

\newcommand\coma[1]{{\color{red} #1}}
\newcommand\dela[2]{{\color{green}\sout{#1}#2}}
\newcommand\repa[2]{{\color{green}\sout{#1}}{\color{blue}{#2}}}
\newcommand\adda[1]{{\color{blue}#1}}
\newcommand\think[1]{}
\definecolor{darkred}{rgb}{0.9,0.1,0.1}%%%COLOR FOR SIDE COMMENT
\newcommand{\hcomment}[1]{\marginpar{\raggedright\scriptsize{\textcolor{darkred}{#1}}}}
\definecolor{darkblue}{rgb}{0.1,0.1,0.9}%%%COLOR FOR SIDE COMMENT
\newcommand{\hcommentg}[1]{\marginpar{\raggedright\scriptsize{\textcolor{darkblue}{#1}}}}

%% file: frontmatter/title.tex
\begin{titlingpage}
\begin{SingleSpace}
\calccentering{\unitlength} 
\begin{adjustwidth*}{\unitlength}{-\unitlength}
\vspace*{13mm}
\begin{center}
{\HUGE Quantum Causal Structure and Quantum Thermodynamics}\\[4mm]
\vspace{31mm}
{\large By}\\
\vspace{9mm}
{\large\textsc{Mirjam Sarah Weilenmann}}\\
\vspace{31mm}
{\large
\textsc{Doctor of Philosophy}}\\
\vspace{21mm}
{\large
\textsc{University of York}\\
\textsc{Mathematics}
}\\
\vspace{21mm}
\vspace{9mm}
{\large\textsc{October 2017}}
\vspace{12mm}
\end{center}
\end{adjustwidth*}
\end{SingleSpace}
\end{titlingpage}

%% file: frontmatter/abstract.tex
%UoY:

%The abstract shall follow the title page. It shall provide a synopsis of the thesis, stating the nature and scope of work undertaken and the contribution made to knowledge in the subject treated. It shall appear on its own on a single page and shall not exceed 300 words in length. The abstract of the thesis may, after the award of the degree, be published by the University in any manner approved by the Senate, and for this purpose, the copyright of the abstract shall be deemed to be vested in the University.

\chapter*{Abstract}
\addcontentsline{toc}{chapter}{Abstract}
\begin{SingleSpace}
\initial{T}his thesis reports progress in two domains, namely causal structures and microscopic thermodynamics, both of which are highly pertinent in the development of quantum technologies.
Causal structures fundamentally influence the development of protocols for quantum cryptography and microscopic thermodynamics is crucial for the design of quantum computers.

The first part is dedicated to the analysis of causal structure, which encodes the relationship between observed variables, in general restricting the set of possible correlations between them. Our considerations rely on a recent entropy vector method, which we first review.
We then develop new techniques for deriving entropic constraints to differentiate between causal structures. We provide sufficient conditions for entropy vectors to be realisable within a causal structure and derive new, improved necessary conditions in terms of so-called non-Shannon inequalities. 
We also report that for a family of causal structures, including the bipartite Bell scenario and the bilocal causal structure, entropy vectors are unable to distinguish between classical and quantum causes, in spite of the existence of quantum correlations that are not classically reproducible. Hence, further development is needed in order to understand cause from a quantum perspective.

In the second part we explore an axiomatic framework for modelling error-tolerant processes in microscopic thermodynamics. Our axiomatisation allows for the accommodation of finite precision levels, which is crucial for describing experiments in the microscopic regime. Moreover, it is general enough to permit the consideration of different error types. The framework leads to the emergence of manageable quantities that give insights into the feasibility and expenditure of processes, which for adiabatic processes are shown to be smooth entropy measures. 
Our framework also leads to thermodynamic behaviour at the macroscopic scale, meaning that for thermodynamic equilibrium states a unique function provides necessary and sufficient conditions for state transformations, like in the traditional second law.

\end{SingleSpace}
\clearpage

%% file: frontmatter/acknowledgement.tex
\chapter*{Acknowledgements}
\addcontentsline{toc}{chapter}{Acknowledgements}
\pagenumbering{arabic}
\setcounter{page}{11}
\begin{SingleSpace}

This thesis has been written under the supervision of Roger Colbeck, to whom I am grateful for his support and guidance. I have benefited greatly from his ideas and his method of pursuing research through many discussions. His individual, clear view on different topics, his attention to detail and his strive for punchy wordings have influenced my work presented in this thesis. I also particularly thank Roger for his seemingly unlimited patience and the freedom he granted me to develop my own intellectual pursuits.

I also thank the other members (and former members) of the quantum information and foundations research group, Peter Brown, Paul Busch, Thomas Cope, Spiros Kechrimparis, Sammy Ragy, Oliver Reardon-Smith, Tony Sudbery, Stefan Weigert and Victoria Wright, for many conversations and their contribution to the enjoyable working atmosphere. I would like to especially thank Stefan Weigert and Sabine Bieli for their hospitality upon my arrival in the UK and Paul Busch for his support regarding housing issues, which facilitated my life outside of research.

My research has greatly benefited from discussions and advice from my collaborators, Lea Krämer, Philippe Faist and Renato Renner, who all shared ideas and insights with me that advanced my view and understanding. My work also benefited from conversations with other inspiring researchers, in particular from conversations with Elie Wolfe, who shared his enthusiasm and insights into causal structures with me, and with Lídia del Rio, whose advice regarding several aspects of research and life I greatly appreciated. I would furthermore like to thank Tony Short and Sam Braunstein for investing their time as my examiners.

In the last three years, many people made my life outside science enjoyable, in particular my family and my friends back in Switzerland, the conversations and exchange with whom were always a great pleasure and support. I also met many wonderful people here in York. I would like to particularly thank my office mates for tolerating my swearing at the computer whenever he wasn't doing what he was meant to.

Finally, I would like to thank Peter Brown for taking his time to read through this thesis and for his supportive feedback, including the exhilarant penguin drawings.

\end{SingleSpace}
\clearpage

%% file: frontmatter/declaration.tex
%
%
% UoB guidelines:
%
% Author's declaration
%
% I declare that the work in this dissertation was carried out in accordance
% with the requirements of the University's Regulations and Code of Practice
% for Research Degree Programmes and that it has not been submitted for any
% other academic award. Except where indicated by specific reference in the
% text, the work is the candidate's own work. Work done in collaboration with,
% or with the assistance of, others, is indicated as such. Any views expressed
% in the dissertation are those of the author.
%
% SIGNED: .............................................................
% DATE:..........................
%
\chapter*{Author's declaration}
\addcontentsline{toc}{chapter}{Author's Declaration}
\begin{SingleSpace}
\begin{quote}

%When submitting the thesis, the author shall draw attention to any material contained in it that has been presented before including the full references for any papers published or under review. It should conrm that the work in the thesis is your own, and has not been submitted for examination at this or any other institution for another award. A basic declaration could read as:

\initial{I} declare that this thesis is a presentation of original work and I am the sole author. This work has been carried out under the supervision of Dr.\ Roger Colbeck and has not previously been presented for an award at this, or any other, University. Chapters~\ref{chap:entropy_vec},~\ref{chap:inner},~\ref{chap:nonshan} and~\ref{chap:classquant} are largely based on the published work that is listed below. Chapters~\ref{chap:microthermo} and~\ref{chap:thermomacro} have been researched in collaboration with Dr.\ Lea Kr\"amer, Dr.\ Philippe Faist and Prof.\ Renato Renner. All sources are acknowledged and listed in the Bibliography. \\

%If you have included any published work within your thesis, this needs to be indicated in this section, with full references, as does any collaborative work that you may have undertaken with the names of your colleagues.

%\initial{I} declare that the work in this thesis was carried out in accordance with the requirements of the University's regulations for Research Degree Programmes and I am the sole author. This work was carried out under the supervision of Prof. Zdzis$\l$aw Brze\'zniak, and has not previously been presented for an award at this, or any other, University. All sources are acknowledged and have been listed in the Bibliography.\\

\bigskip

\bigskip

\bigskip

\bigskip

\bigskip

\noindent{\large\textbf{List of publications and preprints}}\\
\begin{itemize}
\item[(1)~] M.\ Weilenmann and R.\ Colbeck, \textit{Inability of the entropy vector method to certify nonclassicality in linelike causal structures,} Phys.\ Rev.\ A 94, 042112 (2016).

\bigskip

\item[(2)~] M.\ Weilenmann and R.\ Colbeck, \textit{Non-Shannon inequalities in the entropy vector approach to causal structures,} arXiv:1605.02078 (2016), (submitted to journal).

\bigskip

\item[(3)~]  M.\ Weilenmann and R.\ Colbeck, \textit{Analysing causal structures with entropy,} arXiv:1709.08988 (2017), Proc.\ Roy.\ Soc.\ A 473, 2207 (2017).

\end{itemize}
%\vspace{1.5cm}
%\noindent
%\hspace{-0.75cm}\textsc{SIGNED: .................................................... DATE: ..........................................}
\end{quote}
\end{SingleSpace}
\clearpage

%% file: chapter01/chap_intro.tex
\let\textcircled=\pgftextcircled
\chapter{Introduction and Synopsis}
\label{chap:intro}

\initial{T}echnological progress is driven by our prevailing desire for comfort and prosperity. Starting with the development of tools in prehistoric times, technological progress has since led to inventions like the television for our entertainment and has enabled projects as ambitious as expeditions into space. Influenced by our needs, technology often imitates our own abilities. To perform physical work we build engines and for tedious computations we rely on computers, with both quickly outperforming our own bodies and brains. 
Technological development is strongly inspired by the nature surrounding us and by our understanding thereof. Among the theories that explain nature, quantum theory stands out due to its accurate experimental predictions and explanatory power. It is in the interest of technological progress to push the application of quantum effects forward, with the aim to discover and exploit significant advantages over current classical technologies.

Quantum effects are employed in cryptographic protocols that rely on Bell's theorem~\cite{Bell1964}, examples include key distribution~\cite{Bennett1984, Ekert1991} and the expansion and amplification of private randomness~\cite{Colbeck2009, Colbeck2012}. Conceptually, these developments have moved the prerequisites of security proofs from computational hardness assumptions (as made for instance when implementing RSA~\cite{RSA}) to assuming the validity of quantum theory. They have also led us to scrutinize the assumptions that are implicit in our considerations. In cryptographic protocols relying on Bell's theorem the crucial assumptions are most conveniently encoded in an underlying causal structure~\cite{Wood2012}. 

Quantum effects also yield computational advantages. For example Shor's factoring algorithm~\cite{Shor1999} allows for the factorisation of primes in polynomial time, which is a significant speedup compared to the known classical algorithms.\footnote{Note that this an issue for the security of RSA mentioned before, which is based on the complexity of this same problem.}  However, exploiting such effects for computation brings with it the challenge of building the necessary quantum devices. Miniaturization of computer chips raises questions regarding the heat generation in computations, theoretical limits for which have been given in terms of Landauer's principle~\cite{Landauer1961}.  
Pragmatic questions regarding computational devices, such as how to best cool microscopic systems, are currently under active investigation~\cite{Linden2010, Brunner2014b}. To what extent realistic heat engines are optimised with quantum effects has not been conclusively answered~\cite{Brunner2012, Skrzypczyk2014, Roulet2017}.

\bigskip

This thesis is concerned with both the analysis of causal structures and microscopic thermodynamics. It is composed of two parts, the first of which compares classical and quantum correlations generated in causal structures beyond the Bell scenario, whereas the second presents an error-tolerant approach to quantum thermodynamics. 
Both topics are examined with information-theoretic concepts: causal structures are analysed in terms of observed correlations and their entropies, thermodynamics is approached with quantum resource theories. In order to facilitate the understanding of this thesis, the relevant concepts and tools are introduced in Chapter~\ref{chap:prelim}, which an expert-reader may prefer to skip. In the following we outline the results presented in the subsequent chapters.

\bigskip

\subsection*{Part~I -- Entropic analysis of causal structures}

For centuries, progress in physics was aided by our intuitive understanding of the world we experience. However, since the emergence of quantum theory with its numerous exotic features, our
intuition has been challenged.
A concept that requires reconsideration in light of quantum theory is causality, which is the topic of Part~I of this thesis.
We address the specific question of how to compare correlations that can be obtained with classical resources to those generated from quantum systems. We analyse the observational differences between quantum phenomena and their classical counterparts in causal structures beyond the Bell scenario~\cite{Bell1964}, an analysis which is facilitated with the use of entropy measures. Such considerations may shed light into the difference of classical and quantum cause from a pragmatic perspective and complement recent work that generalises the classical concepts of cause and causal influence to the quantum realm~\cite{Costa2015, Allen2016}.

\smallskip

\noindent
{\bf Chapter~\ref{chap:entropy_vec} -- Outline of the entropic techniques for analysing causal structures.}
In this chapter, we motivate the use of entropy measures for analysing causal structures and review several entropic (and non-entropic) techniques, bringing different contributions together in a coherent way. The main approach, which we focus on, is the so-called entropy vector method and its fine-graining through post-selection, where probability distributions are analysed through the lens of entropy vectors. These methods allow for the derivation of necessary conditions for correlations to be realisable within a causal structure. 
We prove several fundamental statements, some of which have been implicitly assumed in the literature without formal proof. We also establish connections between different contributions to the literature, relating work that has not been previously united. If not indicated otherwise, statements, proofs and examples are our own. The chapter is based on~\cite{review}.

\smallskip

\noindent
{\bf Chapter~\ref{chap:inner} -- Inner approximations to the entropy cones of causal structures.}
The outer approximations to the set of entropy vectors realisable within a causal structure, introduced in Chapter~\ref{chap:entropy_vec}, are supplemented with corresponding inner approximations. These mainly serve as a means for proving results about entropy vectors in subsequent chapters. In addition, they provide a relatively efficient criterion for showing that the entropy vector method is insufficient for certifying incompatibility of specific probability distributions with certain causal structures. In such cases alternative techniques (for instance those reviewed in Chapter~\ref{chap:entropy_vec}) must be considered instead, which may often lead to problems of computational feasibility. 
This is the first time, inner approximations to the sets of entropy vectors realisable in a causal structure are considered. The chapter is an elaborated version of a part of~\cite{nonshan_short}. 

\smallskip

\noindent
{\bf Chapter~\ref{chap:nonshan} -- Exploring the gap between inner and outer approximations with non-Shannon inequalities.} In this chapter we improve on current entropic techniques by taking so-called non-Shannon inequalities into account. These allow us to derive new, stronger necessary conditions for probability distributions to be realisable within a fixed (classical) causal structure, which are suitable as certificates for the incompatibility of distributions with the supposed causal structure as used in cryptographic protocols.
Our results apply to the entropy vector method as well as to the fine-grained post-selection technique. They improve on the distinction of different classical causal structures and may lead to a better distinction between classical and quantum causes. 

We illustrate our methods on the example of the so-called triangle causal structure, for which we derive numerous new entropy inequalities including infinite families in the classical case and several inequalities for so-called hybrid scenarios. We conjecture that the number of linear inequalities needed to fully characterise the set of realisable entropy vectors in this scenario is infinite.\footnote{We are aware that this contradicts a claim from the literature~\cite{Chaves2014,Chaves2015}, which we however disprove in Section~\ref{sec:insuff}.} Furthermore, we derive a new probability distribution that is not reproducible in the classical triangle scenario but that remains compatible entropically. We supplement this finding with more general searches for incompatible entropy vectors.
The main parts of this chapter are available in~\cite{nonshan_short}.

\smallskip

\noindent
{\bf Chapter~\ref{chap:classquant} -- Entropic distinction between classical and quantum causal structures.} In this chapter we analyse the capability of the entropy vector method to distinguish classical and quantum (and more general non-signalling) resources under the assumption of a fixed causal structure. Our main result of this chapter is that for a large class of causal structures, which generalise the Bell scenario, the entropy vector method is unable to make such a distinction. Furthermore, we show that, along with several other causal structures, the same holds for the so-called bilocal causal structure, which is relevant in the context of entanglement swapping. These results point to a stark limitation of the entropy vector method: no function of the entropies of the observed variables can distinguish classical and quantum causes in these scenarios. It remains an interesting open problem whether the entropy vector method is ever able to make such a distinction or whether it merely expresses causal restrictions independent of the underlying resources. The relation to the post-selection technique that partially salvages the approach is also discussed. The content of this chapter is published as~\cite{linelike}.

\bigskip

\subsection*{Part~II -- Error-tolerant approach to microscopic thermodynamics}

Quantum theory and thermodynamics are two conceptually different theories that successfully complement each other. Quantum theory describes systems at the microscopic scale, whilst thermodynamics is our principal theory for describing large scale systems in (or close to) thermodynamic equilibrium. A connection of the micro and the macro-regime can be established by extending traditional, classical thermodynamics to incorporate microscopic classical and quantum effects.
Thermodynamics at the microscopic scale is often analysed in terms of quantum resource theories~\cite{Janzing2000_cost, Horodecki2003, Horodecki2011}, considerations which have conceptually changed our understanding of phenomena as deep as the second law. From a modern viewpoint this is not merely understood as a statistical statement that holds for large systems with high probability~\cite{Jarzynski1997, Crooks1999_fluct}, but it has been generalised to a family of second laws that remain valid at the microscopic scale~\cite{Brandao2015b}. 

Contrary to the systems considered in phenomenological thermodynamics, the behaviour of microscopic systems is notably affected by small perturbations. Hence, to synchronise theoretical considerations with real-world applications, quantum resource theories should leave the idealised regime behind, accounting for experimental imprecision and errors.

\smallskip

\noindent
{\bf Chapter~\ref{chap:microthermo} -- Axiomatic framework for error-tolerant resource theories.}
We present an axiomatic framework for error-tolerant resource theories, which follows up on a tradition of axiomatic frameworks for classical thermodynamics~\cite{Caratheodory1909, Giles1964, Lieb1998, Lieb1999, Lieb2001, Lieb2013, Lieb2014}. Contrary to previous frameworks, our approach is particularly suitable for describing systems on the microscopic scale. It is novel in that it does not rely on any continuous features such as the existence of a scaling operation~\cite{Lieb1998, Lieb1999, Lieb2001, Lieb2013, Lieb2014}, moreover it explicitly accounts for errors and experimental imprecision. Relying on our framework, we deduce necessary conditions and sufficient conditions for error-tolerant state transformations in terms of real valued functions, for which we derive useful properties. Our axiomatisation is partly inspired by~\cite{Lieb2013, Lieb2014}.

We illustrate all our elaborations with the example of the resource theory of adiabatic processes~\cite{Lieb1998, Lieb1999, Lieb2001, Lieb2013, Lieb2014, Horodecki2003, Horodecki2003b, Weilenmann2015}, where we consider several ways to quantify errors, all of which comply with our axioms. This allows us to recover well-known entropic quantities, a smooth min-entropy and a smooth max-entropy, which we find to be relevant for characterising adiabatic processes in the microscopic regime.
The elaborations in this chapter are unpublished and there is concurrent work that introduces errors to specific quantum resource theories~\cite{Hanson2017, VanderMeer2017, Horodecki2017}. Our work in this chapter is, however, more general in the sense that its axiomatic nature allows for the treatment of different resource theories and different types of errors. Furthermore, the  necessary conditions and the sufficient conditions we derive have not been previously considered. 

\smallskip

\noindent
{\bf Chapter~\ref{chap:thermomacro} -- Macroscopic thermodynamics from microscopic considerations.} 
In this chapter we consider the macroscopic behaviour that emerges from the error-tolerant resource theories. We introduce thermodynamic equilibrium states, for which we show that an asymptotic equipartition property is obeyed~\cite{TCR}. Furthermore, one single quantity specifies necessary and sufficient conditions for state transformations between thermodynamic equilibrium states, a behaviour that we recognise from the second law and axiomatisations of macroscopic thermodynamics~\cite{Lieb1998, Lieb1999, Lieb2001, Lieb2013}. For adiabatic processes, we recover the von Neumann entropy and the Boltzmann entropy. The work presented in this chapter is unpublished. 

\bigskip

\noindent
We complete this thesis with some concluding remarks and point to several open problems in Chapter~\ref{chap:conclusion}.

%% file: chapter02/chap_prelim.tex
\let\textcircled=\pgftextcircled
\chapter{Preliminaries}
\label{chap:prelim}

\initial{T}his chapter presents the background and the main mathematical tools employed in the remainder of this thesis. Section~\ref{sec:convex_geo} introduces convex geometry with a focus on polyhedral computation and the basics of linear and semidefinite programming. Section~\ref{sec:entropy_meas} is concerned with entropy measures and their properties.
The causal structures introduced in Section~\ref{sec:causal_intro} are the central topic of Chapters~\ref{chap:entropy_vec},~\ref{chap:inner},~\ref{chap:nonshan} and ~\ref{chap:classquant}. Finally, quantum resource theories, discussed in Section~\ref{sec:resource}, are the main prerequisite for Chapters~\ref{chap:microthermo} and~\ref{chap:thermomacro}. In each section, we refer to literature that introduces the topics at hand more broadly.

\bigskip

\subsubsection*{Brief overview on notational conventions}
Throughout this thesis, we shall often make use of the usual quantum formalism, where states are represented by density operators, i.e., trace one positive semidefinite operators on a Hilbert space $\cH$, $\rho \in \cS(\cH)$. Whenever we allow for sub-normalised states, i.e., positive semidefinite operators with $ 0 \leq \tr{\rho} \leq 1$, we will use the notation $\rho \in \cS_{\leq}(\cH)$. Positivity of an operator $\rho$ is denoted $\rho \geq 0$, $\rho \geq \sigma$ means that $\rho-\sigma \geq 0$. Hermitian conjugation is denoted  as $\rho^{\dagger}$.   Elements of $\cH$, are denoted in the usual bra-ket notation, $\ket{\psi} \in \cH$, the elements of the dual space, $\cH^{*}$, are denoted by $\bra{\phi} \in \cH^{*}$. Transformations of a quantum system can be expressed as channels, $\cE: \cS(\cH_X) \rightarrow \cS(\cH_Y)$, where $\cE$ is a completely positive trace-preserving map (CPTP map) that maps a state $\rho \in \cS(\cH_X)$ to a state $ \cE(\rho) \in \cS(\cH_Y)$. It will also sometimes be convenient to consider trace non-increasing quantum operations instead. To describe measurements, we will rely on the formalism of positive operator valued measures (POVMs), $\left\{ F^{x} \right\}_{x}$, i.e., sets of positive semidefinite Hermitian operators that sum to the identity operator.~\footnote{This definition is sufficient for this thesis, as we will be concerned with measurements that have discrete outcomes exclusively.}

We shall furthermore consider bits (and qubits), where we use addition modulo $2$, which is denoted $\oplus$. In this context we will also use the logic gates $\operatorname{AND}$ and $\operatorname{OR}$. We shall sometimes consider matrices, $M$, and denote their transposition with respect to the basis they are represented in as $M^{\trans}$. Logarithms are taken with respect to base $2$ throughout this thesis, denoted $\log_2$. The natural numbers $\mathbb{N}$ are chosen not to include $0$.

\section{Convex geometry and optimisation} 
\label{sec:convex_geo}
An introduction to polyhedral computation is given in~\cite{FukudaPolyhedral}. Further information on convex sets, with a focus on their application in optimisation problems, can be found in~\cite{Grotschel, Boyd}. In the following we introduce the basic terminology.

\subsection{Convex and polyhedral cones}
We restrict our considerations to vector spaces $V=\mathbb{R}^{n}$ over the ordered field $\mathbbm{R}$.

\begin{definition}
A \emph{convex cone} is a set $\cone \subseteq V$ such that for any $h_1$, $h_2 \in \cone$ and any $\theta_1$, $\theta_2 \geq 0$, 
\begin{equation}
\theta_1 h_1 + \theta_2 h_2 \in \cone \ .
\end{equation}
\end{definition}

Of particular interest to us are polyhedral cones. We remark here, that there are several conventions on how to define convex polyhedra and convex polytopes. Nonetheless, we shall stick to the following convention throughout this thesis.
\begin{definition}
A \emph{polyhedral cone} is a polyhedron, i.e., a set $\polycone \subseteq V$ that can be written as the solution set of a finite number of linear inequalities 
\begin{equation}
\polycone=\left\{ h \in V \mid A h \leq b \text{ and } C h =d \right\} \ ,
\end{equation} 
where $A \in \mathrm{R}^{m \times n}$, $b \in \mathrm{R}^{m}$, $C \in \mathrm{R}^{k \times n}$ and $d \in \mathrm{R}^{k}$, and a cone, i.e., for any $h \in \polycone$ and any $\theta \geq 0$, $\theta h \in \polycone$.\footnote{The equalities could also be expressed as inequalities here, this is, however, not common in the convex optimisation literature.}
A bounded polyhedron is called a \emph{polytope}.
\end{definition}
A polyhedron is hence given by the intersection of a set of half-spaces $\left\{ h \in V \mid A h \leq b \right\}$ and hyperplanes $\left\{ h \in V \mid C h = d \right\}$ and is by construction convex.
A polyhedral cone can be concisely written as 
\begin{equation}
\polycone=\left\{ h \in V \mid A h \leq 0 \right\} \ ,
\end{equation} 
for some matrix $A$. The description of a polyhedron (or polyhedral cone) in terms of inequalities is called its half-space representation, or \emph{$\hrep$-representation}.
Another representation of a polyhedron (or polyhedral cone) is given in terms of its vertices and extremal rays. For this we need a little more terminology.
\begin{definition}
The \emph{convex hull} of a set of points $S=\left\{s_1, s_2, \ldots, s_k \right\} \subseteq V$ is defined as
\begin{equation}
\convhull{S}\defeq \left\{\sum_{i=1}^{k} \theta_i s_i \ \middle| \ \theta_i \geq 0 \ \forall i \text{ and } \sum_{i=1}^{k} \theta_i=1  \right\} \ .
\end{equation}
The \emph{conic hull} of a set $T=\left\{t_1, t_2, \ldots, t_l \right\}\subseteq V$ is defined as
\begin{equation}
\conichull{T}\defeq \left\{\sum_{i=1}^{l} \theta_i t_i \ \middle| \ \theta_i \geq 0 \ \forall i \right\} \ .
\end{equation}
\end{definition}

\begin{theorem}[Minkowski-Weyl]
For a set $\polyhedron \subseteq V$, the following two statements are equivalent.
\begin{enumerate}[(1)]
\item $\polyhedron$ is a polyhedron.
\item There exist finite sets $S$, $T \subseteq V$, such that 
\begin{equation}
\polyhedron=\convhull{S}+\conichull{T} \ ,
\end{equation}
where $+$ denotes the usual (Minkowski-)sum of two sets.
\end{enumerate}
\end{theorem}
The latter statement also means that $\polyhedron$ is finitely generated, it defines a polyhedron's vertex or \emph{$\vrep$-representation}. For a polyhedral cone $S$ can be taken to be empty (or $S=\left\{\zerovec \right\} \subseteq V$) in its $\vrep$-representation, which corresponds to $b=\zerovec$ in the hyperplane description.

The conversion between the two representations, called vertex ($\hrep$ to $\vrep$) or facet-enumeration ($\vrep$ to $\hrep$) respectively, is a demanding task, for which no general algorithm is known whose runtime is polynomial in input and output size and in the dimension of the polyhedron~\cite{Avis1997}, meaning that all known algorithms have super-polynomial worst-case running-time. There are two main types of algorithms, so-called graph traversal algorithms and incremental algorithms~\cite{Avis1997}. While the former encounter issues when dealing with degeneracies, the latter are problematic for high-dimensional problems, where intermediate results may become too large~\cite{Avis1997, FukudaPolyhedral}. The computational work of this thesis has been carried out with PORTA~\cite{Christof}, an implementation of an incremental algorithm.

\subsection{Projections of convex cones: Fourier-Motzkin elimination} \label{sec:fm_elim}
Several classes of mappings are known to preserve convexity (see Chapter 2.3 of~\cite{Boyd} for details). One example that is of interest to us are the affine transformations.
\begin{definition}
An \emph{affine transformation} is a mapping 
\begin{align*}
 f: \mathbb{R}^{n} &\rightarrow \mathbb{R}^{m}, \\
x &\mapsto f(x)=Ax+b,
\end{align*}
where $A \in \mathbb{R}^{m \times n}$, $b \in \mathbb{R}^{m}$.
\end{definition}
The image of a convex set $S$ under an affine function $f$, denoted $f(S)$ is convex. Note that linear transformations are affine with $b=\zerovec$. We are interested in a subset of the affine transformations, namely \emph{projections}, $\proO$, which are by definition idempotent, $\proO \cdot \proO = \proO$.

In general, affine transformations do not map convex cones to convex cones. Take for instance the polyhedral cone $\polycone= \left\{h \in \mathbb{R}^{2} \mid (1,-1) h = 0 \right\} \subseteq \mathbb{R}^{2}$, which is a line through $\zerovec \in \mathbb{R}^{2}$, and take the affine mapping $f(h)=h+b$, where $b=(1,0)^{\trans}$. Then $f(\polycone)= \left\{h \in \mathbb{R}^{2} \mid (1,-1) h = 1 \right\}$ is not a cone. We can, on the other hand, straightforwardly show that all linear transformations map convex cones to convex cones and polyhedral cones to polyhedral cones. In particular, this holds for linear projections.
\begin{lemma}\label{lemma:polycones}
Let $\polycone \subseteq \mathbb{R}^{n}$ be a convex or polyhedral cone and let $f:\mathbb{R}^{n} \rightarrow \mathbb{R}^{m}$ be a linear transformation. Then $f(\polycone) \subseteq \mathbb{R}^{m}$ is a convex or polyhedral cone respectively.
\end{lemma}
\begin{proof}
A linear transformation $f:\mathbb{R}^{n} \rightarrow \mathbb{R}^{m}$ can be written as $f(x)=Ax$, where $A \in R^{m \times n}$. Take any two vectors $x_1$, $x_2 \in f(\polycone)$, then there exist $h_1$, $h_2 \in \polycone$ such that $x_1=A h_1$ and $x_2=A h_2$.\footnote{Note that the choice of $h_1$ and $h_2$ may not be unique.} Now since $\polycone$ is a convex cone, for any $\theta_1$, $\theta_2 \geq 0$, also $\theta_1 h_1 + \theta_2 h_2 \in \polycone$ and by linearity $f(\theta_1 h_1 + \theta_2 h_2)= \theta_1 x_1 + \theta_2 x_2 \in f(\polycone)$.

Now let $\polycone$ furthermore be polyhedral. From the $\vrep$-description of $\polycone$ it follows that any $h \in \polycone$ can be written as $h=\sum_{i=1}^{k} \lambda_i v_i$, where $\lambda_i \geq 0$ and $v_i \in \mathbb{R}^{n}$ for all $i=1, \ldots, k$ with $k$ finite. Now as for any $x \in f(\polycone)$ there exists an element $h\in \polycone$, such that $x=A h$, by linearity, $x=\sum_{i=1}^{k} \lambda_i Av_i$, i.e., any $x \in f(\polycone)$ can be written as a conic combination of the $k$ vectors $Av_i \in f(\polycone)$.\footnote{Note that a smaller set of vectors may be sufficient to construct any vector $x \in f(\polycone)$ in this way.} 
\end{proof}

An orthogonal projection of a convex cone to the first $k$ out of its $n$ components (which we consider w.l.o.g. here), is achieved with a linear transformation $\proO$ that can be expressed as an $n \times n$-matrix with a $k \times k$-identity matrix as its first block and zeros everywhere else. For a polyhedral cone in its $\vrep$-representation, $\polycone=\conichull{h_1,\ldots, h_k}$, such projection is given as
\begin{equation}
\polycone=\conichull{\pro{h_1},\ldots, \pro{h_k}} \ ,
\end{equation}
where some of the vectors $\pro{h_i}$ may be redundant.\footnote{This follows by linearity, similar to the proof of Lemma~\ref{lemma:polycones}.}
Redundancy of the resulting vectors can be efficiently checked with a linear program.

In its $\hrep$-representation, $\polycone=\left\{ h \mid Ah \leq 0 \right\}$, a standard algorithm that achieves the orthogonal projection of $\polycone$ is Fourier-Motzkin elimination~\cite{Williams1986}. It eliminates the $1$-st to $k$-th components of $h$ from the system of inequalities. The basic procedure for eliminating one of these components, $h_i$, is the following:
\begin{enumerate}[(1)]
\item Partition the inequalities $Ah \leq 0$ into three sets: $H_{+}$ contains all inequalities where the variable $h_i$ has a positive coefficient, i.e.,  all inequalities $j$ where $A_{j i} \geq 0$, $H_{-}$ contains all inequalities where $h_i$ has a negative coefficient and $H_{0}$ is the set of inequalities without $h_i$.
\item If either $H_{+}$ or $H_{-}$ are empty, the inequalities in the repective other set can be disregarded. If $H_{+}$ and $H_{-}$ are both non-empty, rearrange all $\card{H_+}$ inequalities in $H_{+}$ to have the form \begin{equation}{h_i \leq f_j(h_1, \ldots, h_{i-1}, h_{i+1}, \ldots, h_n)} \end{equation} 
and rewrite all $\card{H_{-}}$ inequalities in $H_{-}$ such that \begin{equation}{h_i \geq g_k(h_1, \ldots, h_{i-1}, h_{i+1}, \ldots, h_n)} \ ,\end{equation} 
where $f_j$ and $g_k$ are real vectors encoding the coefficients of the $j$-th inequality of $H_{+}$ and the $k$-th inequality of $H_{-}$ (with $1 \leq j \leq \card{H_{+}}$ and $1 \leq k \leq \card{H_{-}}$); $\card{\cdot}$ denotes the cardinality.
Then combine each inequality in $H_{+}$ with each inequality in $H_{-}$ to an inequality 
\begin{equation}\label{eq:newfm}
{g_k(h_1, \ldots, h_{i-1}, h_{i+1}, \ldots, h_n) \leq f_j(h_1, \ldots, h_{i-1}, h_{i+1}, \ldots, h_n)} \ .
 \end{equation}
\item The union of the inequalities $H_{0}$ and the $\card{H_{+}}\cdot \card{H_{-}}$ newly  generated inequalities \eqref{eq:newfm} characterise the projection of $\polycone$. Some of these inequalities may be redundant, which can be efficiently checked with a linear program.
\end{enumerate}
The procedure can be repeated with each variable to be eliminated. In the following we give a small example for illustration.

\begin{example}
Consider the convex polyhedral cone of Figure~\ref{fig:convexcone}(a), with $\hrep$-representation 
\begin{equation}\polycone= \left\{ \colvec{x}{y}{z} \in \mathbb{R}^{3} \ \middle| \ \threemat{-2}{0}{1}{0}{-2}{1}{0}{0}{-1} \colvec{x}{y}{z} \leq \colvec{0}{0}{0} \right\} \ .
\end{equation} Its $\vrep$-representation can be written as
\begin{equation}\label{eq:hullexample}
\polycone=\conichull{ \colvec{1}{0}{0} , \colvec{0}{1}{0} , \colvec{1}{1}{2} } \ .
\end{equation}

The projection of $\polycone$ onto the $xy$-plane, displayed in of Figure~\ref{fig:convexcone}(b), can be found by applying the projection matrix, 
\begin{equation}
\Pi_{xy}=\begin{pmatrix} 1 & 0 & 0 \\ 0 & 1 & 0  \end{pmatrix} \ ,
\end{equation}
to each of the three vectors in \eqref{eq:hullexample} and then taking the conic hull of the three resulting vectors, which yields
\begin{equation}\polycone_{xy}= \conichull{ \colvec{1}{0}{} , \colvec{0}{1}{} , \colvec{1}{1}{} }= \rpos{2} \ .
\end{equation}
The same can be obtained in the $\hrep$-representation by means of a Fourier-Motzkin elimination.
\begin{enumerate}[(1)]
\item The three sets of constraints are $H_{+}=\left\{ z-2x \leq 0 , \  z-2y \leq 0  \right\}$, $H_{-}=\left\{-z \leq 0 \right\}$ and $H_{0}=\{ \}$.
\item The inequalities of $H_{+}$ and $H_{-}$ are rearranged and combined to yield the two inequalities
$0 \leq 2x $ and $0 \leq 2y $.
\item The resulting cone can be written as 
\begin{equation}
\polycone_{xy}=\left\{ \colvec{x}{y}{} \in \mathbb{R}^{2} \ \middle| \ \twomat{-2}{0}{0}{-2} \colvec{x}{y}{} \leq \colvec{0}{0}{} \right\}= \rpos{2}  \ .
\end{equation}
\end{enumerate}

Similarly, the projection to the $yz$-plane shown in Figure~\ref{fig:convexcone}(c) can be obtained as the conic hull of the projection of a (non-zero) vector on each extremal ray, which yields  
\begin{equation}
\polycone_{yz}= \conichull{ \colvec{0}{0}{} , \colvec{1}{0}{} , \colvec{1}{2}{} } \subsetneq \rpos{2} \ .
\end{equation}
The Fourier-Motzkin elimination proceeds in this case as follows.
\begin{enumerate}[(1)]
\item The three sets of constraints are $H_{+}=\{ \}$, $H_{-}=\left\{z-2x \leq 0 \right\}$ and $H_{0}=\left\{  z-2y \leq 0, \ -z \leq 0   \right\}$.
\item As $H_{+}=\{ \}$, there are no new constraints generated here.
\item The resulting cone is constrained by the inequalities $H_{0}$ and can be written as \begin{equation}\polycone_{yz}=\left\{ \colvec{y}{z}{} \in \mathbb{R}^{2} \ \middle| \ \twomat{-2}{1}{0}{-1} \colvec{y}{z}{} \leq \colvec{0}{0}{} \right\}  \ .
\end{equation}
\end{enumerate}
\label{example:fm_ex}
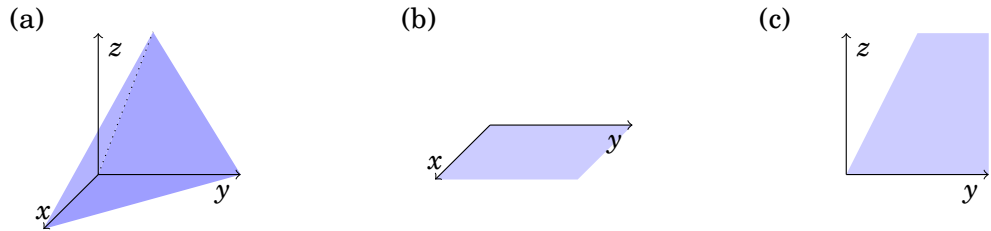
\begin{figure}
\centering 
\resizebox{0.9\columnwidth}{!}{%
\begin{tikzpicture} [scale=0.9]
\node (A) at (-1,2.2,0) {(a)};
\node (B) at (4.5,2.2,0) {(b)};
\node (C) at (9.5,2.2,0) {(c)};

\node (AA) at (13.5,0,0) {};

\fill[blue!38!white] (0,0,0) -- (2,0,0) -- (0,0,2);
\fill[blue!35!white] (0,0,0) -- (2,0,0) -- (1.25,2.5,1.25);
\fill[blue!30!white] (0,0,0) -- (0,0,2) -- (1.25,2.5,1.25);
\draw[dotted] (0,0,0) -- (1.25,2.5,1.25);
\draw[->] (0,0,0) -- (2,0,0) node[anchor=north east]{$y$};
\draw[->] (0,0,0) -- (0,2,0) node[anchor=north west]{$z$};
\draw[->] (0,0,0) -- (0,0,2) node[anchor=south]{$x$};

\fill[blue!20!white] (5.5,0.7,0) -- (7.5,0.7,0) -- (7.5,0.7,2) -- (5.5,0.7,2);
\draw[->] (5.5,0.7,0) -- (7.5,0.7,0) node[anchor=north east]{$y$};
\draw[->] (5.5,0.7,0) -- (5.5,0.7,2) node[anchor=south]{$x$};

\fill[blue!20!white] (10.5,0,0) -- (12.5,0,0) -- (12.5,2,0) -- (11.5,2,0);
\draw[->] (10.5,0,0) -- (12.5,0,0) node[anchor=north east]{$y$};
\draw[->] (10.5,0,0) -- (10.5,2,0) node[anchor=north west]{$z$};

\end{tikzpicture}
}%
\caption[Projections of a polyhedral cone]{Illustration of Example~\ref{example:fm_ex}. For the polyhedral cone of (a), we display its projections to the $xy$-plane and to the $yz$-plane in (b) and (c) respectively. Although not shown in the graphics, the extremal rays of the above cones extend to infinity. }
\label{fig:convexcone}
\end{figure}

\end{example}

Performing Fourier-Motzkin-elimination results in computational problems due to the algorithm's complexity. Applied to a system of $n_0$ inequalities it can yield up to $\left(\frac{n_0}{2}\right)^{2}$ inequalities in the first elimination step. Iterating the procedure over $n$ steps can produce up to $4 \left(\frac{n_0}{4}\right)^{2^{n}}$ inequalities. This doubly exponential growth of the algorithm in the worst case is the main reason for its inefficiency. The elimination algorithm can be adapted by implementing a few rules to remove some of the many redundant inequalities produced in each step. They are collectively known as \u{C}ernikov rules~\cite{Cernikov1960,Chernikov1965} and comprehensively explained in~\cite{Bastrakov2015}. Nevertheless, the number of necessary inequalities can still grow exponentially~\cite{Monniaux2010}. An implementation of the Fourier-Motzkin elimination algorithm (including some \u{C}ernikov rules) is available as part of the PORTA software~\cite{Christof}.

\subsection{Convex optimisation}

Convex optimisation is concerned with minimising or maximising a convex function over a convex set. Two special cases are linear and semidefinite programming, which are recurrent in this thesis and introduced here. A detailed exhibition of convex optimisation is available in the book~\cite{Boyd}, semidefinite programming is also comprehensively introduced in~\cite{Watrous2009}.

\subsubsection{Linear Programming}

A linear program (LP) is concerned with the minimisation of an affine function over a polyhedron. A possible formulation of a primal program and its dual is given in the following, where $x$, $c \in \mathbb{R}^{n}$, $y$, $b \in \mathbb{R}^{m}$ and $A \in \mathbb{R}^{n \times m}$.

\bigskip
 
\noindent
\begin{minipage}{.5\textwidth}
\centering
\textbf{Primal program}
\begin{alignat*}{2}
    &\text{minimise}   \quad   && c^{\trans}x  \\
    &\text{subject to} \quad &&Ax \geq b \\
    &\quad &&x \geq 0 \\
\end{alignat*}
\end{minipage}%
\begin{minipage}{.5\textwidth}
\centering
\textbf{Dual program}
\begin{alignat*}{2}
    &\text{maximise}   \quad   && b^{\trans} y  \\
    &\text{subject to} \quad &&A^{\trans} y \leq c \\
     &\quad &&y \geq 0 \\
\end{alignat*}
\end{minipage}

The \emph{feasible sets} for primal and dual program are $\cX=\left\{ x \geq 0 \mid A x \geq b \right\} $ and $\cY=\left\{ y \geq 0 \mid A^{\trans} y \leq c \right\} $ and their \emph{optimal feasible solutions} are
 $\alpha=\inf_{x \in \cX} c^{\trans}x$ and  $\beta=\sup_{y \in \cY} b^{\trans}y$. \emph{Weak duality} says that the optimal feasible solution to the primal (minimisation) problem is always larger or equal to its dual (maximisation) problem, i.e.,  $\alpha \geq \beta$. Whenever the optimal solution to the primal program equals that of the dual, we call them \emph{strongly dual}. In the case of LPs strong duality only fails if both programs are infeasible~\cite{Boyd}.
 
In this thesis, LPs will mainly be used as efficient tools to prove redundancy of inequalities and to confirm that certain vectors can be written as convex or conic combinations of others. For instance, redundancy of the $i$-th inequality in a system $A x \geq b$, can be checked by minimising that inequality subject to all other inequalities. If there is a solution $x$ such that the optimal feasible solution is strictly smaller than the corresponding $b_i$, then the inequality is irredundant, otherwise redundant. We perform corresponding computations in Mathematica.

\subsubsection{Semidefinite Programming}
A semidefinite program (SDP) is an optimisation of an objective function over the convex cone of positive semidefinite matrices on a Hilbert space subject to additional linear constraints. For the exposition of semidefinite programming we follow the treatment of Ref.~\cite{Watrous2009, Dupuis2013_DH}. 
Let $B$, $X \in \linH{\mathbb{C}^{n}}$, $C$, $Y \in \linH{\mathbb{C}^{m}}$, where $\linH{\mathbb{C}^{n}}$ denotes the Hermitian operators on $\mathbb{C}^{n}$ and let the map $\Phi: \linH{\mathbb{C}^{n}} \rightarrow \linH{\mathbb{C}^{m}}$ be linear and hermiticity-preserving. A possible phrasing of a primal program and its dual is given in the following.

\bigskip

\noindent
\begin{minipage}{.5\textwidth}
\centering
\textbf{Primal program}
\begin{alignat*}{2}
    &\text{minimise}   \quad   && \tr{C^{\dagger}X} \\
    &\text{subject to} \quad && \Phi(X) \geq B \\
    &\quad &&X \geq 0 \\
\end{alignat*}
\end{minipage}%
\begin{minipage}{.5\textwidth}
\centering
\textbf{Dual program}
\begin{alignat*}{2}
    &\text{maximise}   \quad   && \tr{B^{\dagger}Y}  \\
    &\text{subject to} \quad && \Phi^{\dagger}(Y) \leq C \\
     &\quad &&Y \geq 0 \\
\end{alignat*}
\end{minipage}
We point out the analogy to the LP of the previous section here: vectors are replaced by Hermitian operators and the dot product in the objective function is replaced by the Hilbert-Schmidt inner product. The feasible sets for primal and dual program are $\cX=\left\{ X \geq 0 \mid \Phi(X) \geq B \right\} $ and $\cY=\left\{ Y \geq 0 \mid \Phi^{\dagger}(Y) \leq C \right\} $ and their optimal feasible solutions are $\alpha=\inf_{X \in \cX} \tr{C^{\dagger}X}$ and  $\beta=\sup_{Y \in \cY} \tr{B^{\dagger}Y}$. They are weakly dual, i.e., $\alpha \geq \beta$. Contrary to LPs, SDPs are not always strongly-dual, even if $\alpha$ and $\beta$ are both finite. 
Sufficient conditions for strong-duality are given by the \emph{Slater conditions}:
\begin{enumerate}[(1)]
\item If $\beta$ is finite and there exists an operator $X > 0$ such that $\Phi(X) > B$ (i.e., the primal program is \emph{strictly feasible}), then $\alpha=\beta$ and there exists $Y \in \cY$ such that $\beta=\tr{B^\dagger Y}$.

\item If $\alpha$ is finite and there exists an operator $Y > 0$ such that $\Phi^{\dagger}(Y) < C$ (the dual is strictly feasible), then $\alpha=\beta$ and there exists $X \in \cX$ such that $\alpha=\tr{C^\dagger X}$.
\end{enumerate}
To identify optimal solutions we shall make use of the \emph{complementary slackness conditions} for the optimal feasible solutions $X$ and $Y$~\cite{Boyd, Dupuis2013_DH},
\begin{align}
(B-\Phi(X))Y &=0  \\
(C-\Phi^{\dagger}(Y))X &= 0.
\end{align}

\section{Information-theoretic entropy measures} \label{sec:entropy_meas}
In information theory, entropy was first introduced by Shannon~\cite{Shannon1948} to characterise the information content of data and has since been ubiquitous within the literature (a trend that is continued within this thesis). A variety of different entropy measures have been introduced to help characterise different information processing tasks, examples are data compression~\cite{Shannon1948, Koenig2009IEEE_OpMeaning}, randomness extraction~\cite{Bennett1995,Koenig2009IEEE_OpMeaning}, state merging~\cite{Horodecki2005,Berta2009arXiv}, decoupling~\cite{Dupuis2010, Dupuis2014} and tasks related to hypothesis testing~\cite{MDSFT, Wilde2014, Mosonyi2015}. A good reference that provides a thorough exposition of most of these entropy measures is~\cite{TomamichelBook}. In the following, we briefly introduce some of them, namely the Shannon and the von Neumann entropy, which are relevant throughout this thesis, as well as the R\'enyi entropies, which shall be mentioned again in Chapter~\ref{chap:entropy_vec}. Smooth entropies will crucially reappear in Chapters~\ref{chap:microthermo} and~\ref{chap:thermomacro}.

\subsection{Shannon and von Neumann entropy}

\begin{definition}
For a discrete random variable $X$ taking values $x \in \mathcal{X}$ with probability distribution $P_X$ its \emph{Shannon entropy} is defined as
\begin{equation} 
H(X) \defeq - \sum_{x \in \mathcal{X}} P_X(x) \log_2
P_X(x)\, ,
\end{equation}
where $0 \log_2 (0)$ is taken to be $0$ (note that $\lim_{p \rightarrow 0^{+}} p \log_2 p=0$).
\end{definition}

The Shannon entropy describes the information content of data or the uncertainty one has about it. It is sometimes described in terms of a random variable $I_X\defeq-\log_2P_X$ called the \emph{surprisal} or \emph{self-information} of $X$. This name expresses that the lower the probability of an event $X=x$, the larger the surprise about this event when sampling from $X$. For instance, if $X$ is a uniform binary random variable with $x \in \left\{ 0, 1\right\}$ then the surprisal of the event $X=0$ is $1$, whereas for a random variable that deterministically outputs $0$, the event $X=0$ has a surprisal of $0$. The entropy is the expected value of $I_X$, i.e., the average surprisal.

Similarly, the conditional entropy quantifies the uncertainty one has about data that can be modelled as a random variable $X$, if one has access to a random variable $Y$ which is jointly distributed with $X$. More specifically, it is the average over all $x$ of the uncertainties $H(X|Y=y)$ .

\begin{definition}\label{def:shan_ent}
The \emph{conditional entropy} of two jointly distributed discrete random variables $X$ and $Y$ taking values $x \in \mathcal{X}$ and $y \in \mathcal{Y}$, respectively with joint distribution $P_{XY}$, is
\begin{equation}
H(X|Y)\defeq - \sum_{y \in \mathcal{Y}} P_Y(y) \sum_{x \in \mathcal{X}} P_{X|Y}(x,y) \log_2P_{X|Y}(x,y) \ ,
\end{equation}
where $P_{X|Y}$ denotes the conditional distribution. 
\end{definition}

The conditional entropy can be written as the difference of two unconditional entropies $H(X|Y)= H(XY)-H(Y)$. Further entropic quantities of interest that can be written as linear combinations of entropies are the \emph{mutual information} of two jointly distributed random variables $X$ and $Y$,
\begin{equation}
I(X:Y)\defeq H(X)+H(Y)-H(XY) \ , \label{eq:mutinfo}
\end{equation}
and its conditional version, the \emph{conditional mutual information} between two jointly distributed random variables $X$ and $Y$ given a third, $Z$,
\begin{equation}
I(X:Y|Z)\defeq H(XZ)+H(YZ)-H(Z)-H(XYZ) \ . \label{eq:condmutinfo}
\end{equation}
The mutual information is a measure for the dependence between the variables $X$ and $Y$ as it specifies the amount by which the uncertainty about one decreases if one learns the other, $I(X:Y)= H(X)-H(X|Y)$. If their mutual information is zero, two random variables are deemed independent. We will consider independencies as statements about mutual and conditional mutual information more in-depth in Section~\ref{sec:causal_intro}.

In subsequent chapters we shall also revisit the \textit{interaction information}~\cite{McGill1954} of three jointly distributed discrete random variables $X$, $Y$ and $Z$,\footnote{There are two different definitions of the interaction information in the literature, one convention is adopted here, whereas the other would consider $I(X : Y : Z)= I(X : Y | Z)-I(X : Y )$, the negative of what we call the interaction information, instead.}
\begin{equation}
I(X : Y : Z)\defeq I(X : Y )-I(X : Y | Z) \ .
\end{equation}
It is symmetric under the exchange of the three variables. Intuitively, it expresses how much the third variable reduces the mutual information between the other two. 
Of further relevance for our considerations is the \textit{Ingleton quantity}~\cite{Ingleton} of four jointly distributed discrete random variables $W$, $X$, $Y$ and $Z$,
\begin{equation} \label{eq:ingleton}
I_\mathrm{ING}(W,X ; Y,Z)\defeq I(W:X|Y) + I(W:X|Z) + I(Y:Z) - I(W:X) \ .
\end{equation}

\bigskip

The von Neumann entropy is the generalisation of the Shannon entropy to the quantum realm.
\begin{definition}
The \emph{von Neumann entropy} of a density operator $\rho_{X} \in \cS(\cH_X)$ is defined as
\begin{equation} \label{eq:von_neumann}
H(X)\defeq -\tr{\rho_X \log_2 \left( \rho_X \right)}\, .
\end{equation}
\end{definition}
We use the same symbol for the von Neumann entropy as for the Shannon entropy (cf.\ Definition~\ref{def:shan_ent}).\footnote{If one reinterprets random variables as diagonal states then the Shannon entropy becomes a restriction of the von Neumann entropy to said states, which justifies this notation.} This terminology will prove particularly convenient in Chapters~\ref{chap:entropy_vec},~\ref{chap:nonshan} and~\ref{chap:classquant}.

We define the \emph{conditional von Neumann entropy} as
\begin{equation} \label{eq:cond_vn}
H(X|Y)\defeq H(XY)-H(Y).
\end{equation} 
Mutual- and conditional mutual information are defined in analogy with equations~\eqref{eq:mutinfo} and~\eqref{eq:condmutinfo}, but where $H$ denotes the von Neumann entropy instead of the Shannon entropy.

\subsection{R\'enyi entropies} \label{sec:renyi}

While in the asymptotic regime of infinitely many independent repetitions of an information processing task the Shannon and the von Neumann entropy are the characteristic quantities, a variety of different measures are relevant in the \textit{single shot} regime of only few repetitions of the task. An important family of entropy measures are the quantum $\alpha$-R\'{e}nyi entropies~\cite{Renyi1960_MeasOfEntrAndInf}. The classical versions of these quantities are recovered from the following definition when representing random variables as diagonal states. 

\begin{definition}
For a quantum state $\rho_X \in \cS(\cH_X)$, the \emph{$\alpha$-R\'{e}nyi} entropy is
\begin{equation}\label{eq:renyi}
H_{\alpha}(X)\defeq\frac{1}{1-\alpha} \log_2 \tr{\rho_X^{\alpha}} \ ,
\end{equation}
for $\alpha\in(0,\infty)\setminus\{1\}$, the cases $\alpha=0,1,\infty$ are defined via the relevant limits. 
\end{definition}

These quantities may essentially be split into two main types, the \emph{min-entropies} for $\alpha > 1$ and the \emph{max-entropies} for $\alpha < 1$, families of quantities all members of which take similar values~\cite{Renner2004ISIT}. Shannon and von Neumann entropy are recovered via the limits $\alpha \rightarrow 1$, i.e., $H_1(X)=H(X)$. There are also several generalisations of~\eqref{eq:renyi} to families of conditional $\alpha$-R\'{e}nyi entropies. The most common family are the so-called sandwiched R\'{e}nyi entropies, $H_{\alpha}(X|Y)$, introduced in~\cite{MDSFT}, which, 
like several other generalisations~\cite{Petz,FL,Beigi} obey the data processing property,
\begin{equation} \label{eq:cond_rel}
H_\alpha(X | YZ) \leq  H_\alpha(X | Y) \ .
\end{equation}
Alternatively one could define conditional entropy as 
\begin{equation} \label{eq:cond_rel_2}
\tilde{H}_{\alpha}(X|Y)\defeq H_{\alpha}(XY)-H_{\alpha}(Y) \ ,
\end{equation} analogously to~\eqref{eq:cond_vn}. Such definition would, however, not obey~\eqref{eq:cond_rel}. For details on (conditional) R\'{e}nyi entropies we refer to~\cite{TomamichelBook}.

\subsection{Smooth entropy measures}

Whenever errors in information processing tasks are accounted for, the relevant quantities are the \textit{smooth} versions of the corresponding entropy measures. Errors are usually quantified in terms of a deviation from a target state and as such quantified by distance measures. 
A relevant distance measure in this context is the generalised trace distance~\cite{TomamichelBook}.

\begin{definition} \label{def:trdist}
The \emph{generalised trace distance} of two (sub-normalised) states $\rho$, $\sigma \in \cS_{\leq} \left(\cH \right)$ is 
\begin{equation}
\trdist{\rho}{\sigma} \defeq \frac{1}{2} \| \rho -\sigma \|_1 + \frac{1}{2} | \tr{\rho-\sigma} | ,
\end{equation} 
where for a bounded operator $\tau$, $\| \tau \|_1 \defeq \sum_{i=1} s_i(\tau)$ is the $1$-Schatten norm, i.e., the sum of the singular values of $s_i(\tau)$.
\end{definition}
Note that for states that have the same trace, e.g., if they are both normalised, the above coincides with the usual trace distance.
Throughout this thesis, $\epsball{\eps}{\rho} \defeq \left\{ \rho' \in \cS(\cH)  \mid \trdist{\rho}{\rho'} \leq \eps \right\}$ shall denote the set of all states that are $\eps$-close to the state $\rho \in \cS(\cH)$, measured in trace distance. Sometimes we shall include sub-normalised states in which case $\epsballsub{\eps}{\rho} \defeq \left\{ \rho' \in \cS_{\leq}(\cH)  \mid \trdist{\rho}{\rho'} \leq \eps \right\}$.
We will rely on the following instance of the min-entropy, which is obtained from the family of R\'{e}nyi entropies as $\alpha \rightarrow \infty$.
\begin{definition} \label{def:smooth_min}
The \emph{smooth min-entropy} is defined as
\begin{equation} 
\Hmineps{\eps}{\rho} \defeq \max_{\rho' \in \epsballsub{\eps}{\rho}} \Hmin{\rho'}
\end{equation}
with $\Hmin{\rho'} \defeq -\log_2 \maxEV \left(\rho'\right)$, where $\maxEV \left(\rho'\right)$ denotes the largest eigenvalue of $\rho'$.
\end{definition}

As max-entropy we consider the generalised entropy measure introduced in~\cite{Dupuis2013_DH}, given in the following. 

\begin{definition}
The \emph{smooth max-entropy}, $\HHeps{1-\eps}{\rho}$, is defined in terms of the following semidefinite program with optimal solution $2^{\HHeps{1-\eps}{\rho}}$,

\bigskip

\noindent
\begin{minipage}{.5\textwidth}
\centering
\textbf{Primal program}
\begin{align*} 
    \min \quad &\frac{1}{1-\eps} \tr{Q}\\
    \sth \quad &\tr{Q \rho}\geq 1-\eps \\
    \quad \quad &Q \leq \mathbb{I} \ \\
   %\quad Q \geq 0 \ 
\end{align*}
\end{minipage}%
\begin{minipage}{.5\textwidth}
\centering
\textbf{Dual program}
\begin{align*}
    \max \quad &\alpha - \frac{\tr{X}}{1-\eps} \\
    \sth \quad &\alpha \rho \leq \mathbb{I} + X \\
   \quad X \geq 0 \\
   \quad \alpha \geq 0 
\end{align*}
\end{minipage}
\end{definition}
Note that for the above program with $\eps>0$ and a state $\rho \in \cS(\cH)$ the primal program is bounded (as $Q \leq \mathbb{1}$) and the dual is strictly feasible with $X= \alpha \mathbbm{1}$. The complementary slackness conditions are 
\begin{align}
(\alpha \rho -X) Q &=Q \ , \\
\tr{ Q \rho} X &= (1-\eps) X \ , \\
 Q  X &=  X \ .
\end{align} 
$\HHepsO{1-\eps}$ has several useful properties. First of all, it is monotonically increasing in $\eps$, like $\HminepsO{\eps}$~\cite{Dupuis2013_DH}. Moreover, it relates nicely to the original min and max-entropies~\cite{Dupuis2013_DH}. For $\eps \rightarrow 1$, $\HHepsO{1-\eps}$ converges to the min-entropy
\begin{equation}
\Hmin{\rho}=\lim_{\eps \rightarrow 1} \HHeps{1-\eps}{\rho} \  .
\end{equation}
It furthermore relates to the max-entropy defined as the R\'{e}nyi entropy for $\alpha \rightarrow 0$, $ \Hzero{\rho}\defeq \log_2\left( \rank \rho \right)$
in the sense that $\HHeps{1}{\rho}=\Hzero{\rho}$, which can be seen from the feasibility of the primal and dual programs with $Q= \Pi_\rho$, $\alpha= \frac{1}{\minEV\left(\rho\right)}$ and $X=\frac{1}{\minEV\left(\rho\right)} \rho -\Pi_\rho$, where $\Pi_\rho$ is the projector onto the support of $\rho$ and $\minEV\left(\rho\right)$ is the smallest eigenvalue of $\rho$.
The following lemma establishes a relation to the smooth version of this max-entropy, defined as
\begin{equation} 
\Hzeroeps{\eps}{\rho} \defeq \max_{\rho' \in \epsballsub{\eps}{\rho} } \Hzero{\rho'} \ . 
\end{equation}

\begin{lemma} \label{lemma:max_entropies}
For any $0 < \eps <1$ the smooth entropies $\HzeroepsO{\eps}$ and $\HHepsO{1-\eps}$ are related as
\begin{equation}
\log_2\left(2^{ \Hzeroeps{\eps}{\rho}} - 1 \right) \leq \HHeps{1-\eps}{\rho} + \log_2\left(1-\eps\right) \leq \Hzeroeps{\eps}{\rho} \ .
\end{equation}
\end{lemma}

\begin{proof} 
Let $\rho \in \cS(\cH)$ with $\dim{\left(\cH \right)}=n$ and eigenvalues $\left\{\ithEV{\rho}\right\}_{i=1}^{n}$. Now consider these eigenvalues in decreasing order and let $r=\max \left\{ i \mid \sum_{i}^{n} \ithEV{\rho} \geq \eps \right\}$ and let $\eps_1= \sum_{r+1}^{n} \ithEV{\rho}$, let $h(\eps)=\ithEVk{\rho}{r+1}$ and $\eps_2 =\eps-\eps_1$, as is also illustrated in Figure~\ref{fig:optimal_max}.
\begin{figure}
\centering
\includegraphics[width=0.6\textwidth]{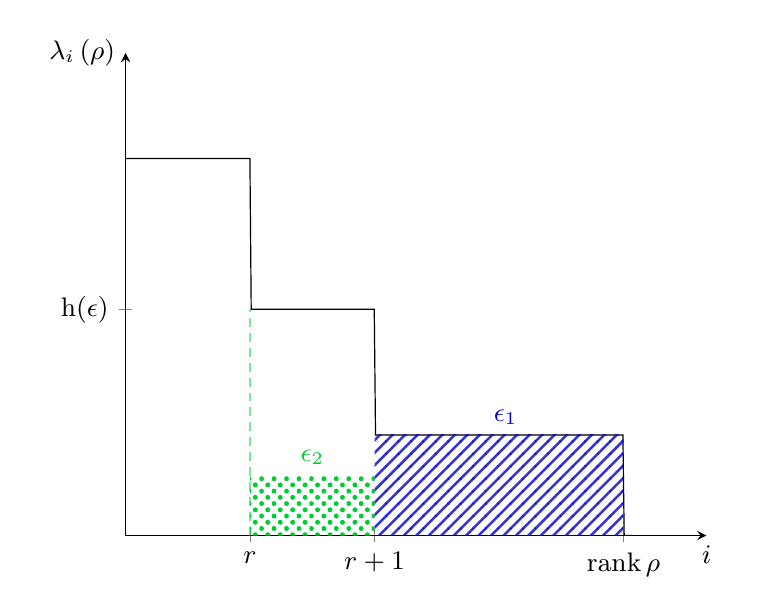}
\caption[Illustration of parameters introduced to prove Lemma~\ref{lemma:max_entropies}]{Spectrum of $\rho$ displayed as a step function. This illustrates how the parameters $r$, $h(\eps)$, $\eps_1$ and $\eps_2$ are specified.}
\label{fig:optimal_max}
\end{figure}

We first show that 
\begin{equation}
\HHeps{1-\eps}{\rho} + \log_2 \left(1-\eps \right)= \log_2 \left( r + 1 - \frac{\eps_2}{h(\eps)} \right) \ .
\end{equation}
To this end, let us consider the primal program used to define $\HHepsO{1-\eps}$ with
\begin{equation}
 Q = \sum_{i=1}^{r+1} \ket{i} \bra{i} - \frac{\eps_2}{h(\eps)} \ket{r+1}\bra{r+1} \ .
\end{equation}
First, note that as $\frac{\eps}{h(\eps)} \leq 1$, $0 \leq Q \leq 1$ and that
$\frac{1}{1-\eps} \tr{Q} = \frac{1}{1-\eps} \left( r + 1 - \frac{\eps_2}{h(\eps)} \right) $ and $ \tr{Q\rho} = 1 - \eps $.

To confirm that this choice of $Q$ leads to the optimal solution, we consider the dual program with variables
\begin{align}
  \alpha &= \frac{1}{h(\eps)} \\
    X &= \sum_{i=1}^{r} \left(\frac{\ithEV{\rho}}{h(\eps)} - 1 \right) \ket{i} \bra{i}  \  ,
\end{align} 
which are both positive, obey $\alpha \rho \leq \mathbb{1} + X$ and yield the same value as the primal program,
\begin{align}
    \alpha - \frac{\tr{X}}{1-\eps} 
    &= \frac{1}{h(\eps)} - \frac{1}{1-\eps} \left(\frac{1}{h(\eps)} \left(\sum_{i=1}^{r} \ithEV{\rho} \right) - r\right) \\
    &= \frac{1}{h(\eps)} - \frac{1}{1-\eps} \left(\frac{1}{h(\eps)} \left(1-\eps_1-h(\eps)\right) - r\right) \\
    &= \frac{1}{1-\eps} \left( 1-\frac{\eps_2}{h(\eps)}+r \right) \ .
\end{align}
Hence, we have found the optimal solution and it directly follows that
\begin{equation} 
\HHeps{1-\eps}{\rho} + \log_2(1-\eps)   = \log_2 \tr{Q}    = \log_2 \left( r + 1 -\frac{\eps_2}{h(\eps)} \right) \ .
\end{equation}
    
Now, since $ \left( r+1 \right) h(\eps)\leq 1 $, we have
\begin{align}
\log_2 \left(1-\frac{\eps_2}{h(\eps)}+ r \right)
 &= \log_2 \left((1+r) \left(1-\frac{\eps_2}{(r+1) h(\eps) }\right) \right) \\
 &\leq \Hzeroeps{\eps}{\rho} + \log_2 (1-\eps_2) \\
 &\leq \Hzeroeps{\eps}{\rho} \ ,
\end{align}
where in the first step we use that $\Hzeroeps{\eps}{\rho}= \log_2 \left(1+r \right)$.
On the other hand, $\eps_2 \leq h(\eps)$ implies
\begin{align}
 \log_2 \left(1-\frac{\eps_2}{h(\eps)}+r \right)
 \geq \log_2 r 
 = \log_2 \left(2^{\Hzeroeps{\eps}{\rho}}-1 \right) \ .
\end{align}
\end{proof}

\section{Causal structures} \label{sec:causal_intro}
Causal relations among a set of variables impose mathematical restrictions on their possible joint distribution, which can be conveniently represented with a causal structure. 

\begin{definition}
A \emph{causal structure}, $C$, is a set of variables arranged in a directed acyclic graph~(DAG), in which a subset of the nodes is assigned as observed.
\end{definition}

The directed edges of the graph are intended to represent causation, perhaps by propagation of some influence, and cycles are excluded to avoid the well-known paradoxes associated with causal loops.  We will interpret causal structures in different ways depending on the supposed physics of whatever is mediating the causal influence.\footnote{Note that there are other ways to define causal structures, as mentioned in Section~\ref{sec:other_stuff}.} In all of these situations, only the observed variable itself and no additional unobserved system can be transmitted via a link between two observed variables. This understanding of the causal links encodes a Markov condition. Note that in situations where one might want to describe a notion of future instead of direct causation this interpretation of the graph is not convenient~\cite{Colbeck2013,Colbeck2016}.

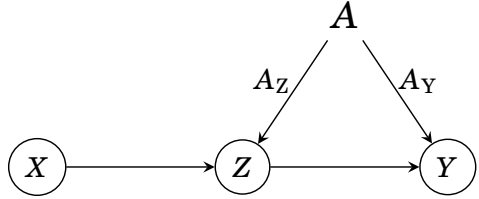
\begin{figure}
\centering 
\resizebox{0.4\columnwidth}{!}{%
\begin{tikzpicture}
\node[draw=black,circle,scale=0.75] (1) at (-2,-0.5) {$X$};
\node[draw=black,circle,scale=0.75] (2) at (-0,-0.5) {$Z$};
\node[draw=black,circle,scale=0.75] (3) at (2,-0.5) {$Y$};
\node (4) at (1,1) {$A$};

\draw [->,>=stealth] (1)--(2);
\draw [->,>=stealth] (2)--(3);
\draw [->,>=stealth] (4)--(2) node [above,pos=0.8,yshift=+1ex] {$\scriptstyle A_\mathrm{Z}$};
\draw [->,>=stealth] (4)--(3) node [above,pos=0.8,yshift=+1ex] {$\scriptstyle A_\mathrm{Y}$};

\end{tikzpicture}
}%
\caption[Pearl's instrumental scenario]{Pearl's instrumental scenario. The nodes $X$, $Y$ and $Z$ are observed, $A$ is unobserved. In the classical case this can be understood in the following way: A random variable $X$ and an unobserved $A$ are used to generate another random variable $Z$. Then $Y$ is generated from $A$ and the observed output of node $Z$. In particular, no other information can be forwarded from $X$ through the node $Z$ to $Y$.
In the quantum case, the source $A$ shares a quantum system $\rho_A\in\cS(\cH_A)$, where $\cH_A\cong\cH_{A_{\mathrm{Z}}}\otimes\cH_{A_{\mathrm{Y}}}$. The subsystem $A_{\mathrm{Z}}$ is measured to produce $Z$ and likewise for $Y$. The subsystems $A_{\mathrm{Z}}$ and $A_{\mathrm{Y}}$ are both considered to be parents of $Z$ (and $Y$).}
\label{fig:instrumental}
\end{figure}

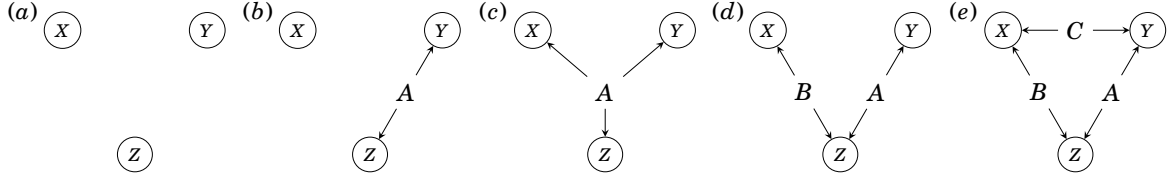
\begin{figure}
\centering
\resizebox{0.98 \columnwidth}{!}{%
\begin{tikzpicture}[scale=0.55]
\node (a)  at (-9.6,2.5) {$(a)$};
\node[draw=black,circle,scale=0.75] (X1) at (-8.5,2) {$X$};
\node[draw=black,circle,scale=0.75] (Y1) at (-4.5,2) {$Y$};
\node[draw=black,circle,scale=0.75] (Z1) at (-6.5,-1.46) {$Z$};

\node (b) at (-3.1,2.5) {$(b)$};
\node[draw=black,circle,scale=0.75] (X) at (-2,2) {$X$};
\node[draw=black,circle,scale=0.75] (Y) at (2,2) {$Y$};
\node[draw=black,circle,scale=0.75] (Z) at (0,-1.46) {$Z$};
\node (A) at (1,0.28) {$A$};

\node (c)  at (3.4,2.5) {$(c)$};
\node[draw=black,circle,scale=0.75] (X2) at (4.5,2) {$X$};
\node[draw=black,circle,scale=0.75] (Y2) at (8.5,2) {$Y$};
\node[draw=black,circle,scale=0.75] (Z2) at (6.5,-1.46) {$Z$};
\node (A2) at (6.5,0.28) {$A$};

\node (d)  at (9.9,2.5) {$(d)$};
\node[draw=black,circle,scale=0.75] (X3) at (11,2) {$X$};
\node[draw=black,circle,scale=0.75] (Y3) at (15.0,2) {$Y$};
\node[draw=black,circle,scale=0.75] (Z3) at (13,-1.46) {$Z$};
\node (A3) at (14,0.28) {$A$};
\node (B3) at (12,0.28) {$B$};

\node (e)  at (16.4,2.5) {$(e)$};
\node[draw=black,circle,scale=0.75] (X4) at (17.5,2) {$X$};
\node[draw=black,circle,scale=0.75] (Y4) at (21.5,2) {$Y$};
\node[draw=black,circle,scale=0.75] (Z4) at (19.5,-1.46) {$Z$};
\node (A4) at (20.5,0.28) {$A$};
\node (B4) at (18.5,0.28) {$B$};
\node (C4) at (19.5,2) {$C$};

\draw [->,>=stealth] (A)--(Y);
\draw [->,>=stealth] (A)--(Z);

\draw [->,>=stealth] (A2)--(Y2);
\draw [->,>=stealth] (A2)--(Z2);
\draw [->,>=stealth] (A2)--(X2);

\draw [->,>=stealth] (A3)--(Y3);
\draw [->,>=stealth] (A3)--(Z3);
\draw [->,>=stealth] (B3)--(X3);
\draw [->,>=stealth] (B3)--(Z3);

\draw [->,>=stealth] (A4)--(Y4);
\draw [->,>=stealth] (A4)--(Z4);
\draw [->,>=stealth] (B4)--(X4);
\draw [->,>=stealth] (B4)--(Z4);
\draw [->,>=stealth] (C4)--(X4);
\draw [->,>=stealth] (C4)--(Y4);
\end{tikzpicture}
}%
\caption[Causal structures with three observed variables]{Assuming no direct causal influences among the three observed
  variables $X$, $Y$ and $Z$, the above are the only possible causal
  structures (up to relabelling). $A$, $B$ and $C$ correspond to unobserved variables.}
\label{fig:3variables}
\end{figure}

The smallest causal structure that leads to interesting insights and one of the most thoroughly analysed ones is Pearl's instrumental causal structure, $\inst$~\cite{Pearl1995}. It is displayed in Figure~\ref{fig:instrumental}. 
Another example that has recently gained a lot of attention is the triangle causal structure, $\tri$, of Figure~\ref{fig:3variables}(e)~\cite{Branciard2012, Fritz2012, Steudel2015, Chaves2014, Henson2014, Chaves2015}. It represents a situation where three parties make observations, $X$, $Y$ and $Z$ respectively, on systems, $A$, $B$ and $C$, that are shared between two parties each. This may for instance be realised in a communication protocol where three parties aim to obtain (correlated) data without ever having interacted as a group, however, having previously shared systems pairwise. Excluding direct causal influences among the observed variables $X$, $Y$ and $Z$, there are only five distinct causal structures with three observed nodes (cf.\ Figure~\ref{fig:3variables}).
The causal structures $\inst$ and $\tri$ will be revisited several times within Part~I of this thesis.

\subsection{Classical causal structures as Bayesian networks}\label{sec:bayesian}

\begin{definition}
A \emph{classical causal structure}, $C^{\cC}$, is a causal structure in which each node of the DAG has an associated random variable.
\end{definition}

It is common to use the same label for the node and its associated random variable, which are all assumed to be discrete. Thus, a classical causal structure $C^{\cC}$ with $n$ nodes $X_1, X_2, \ldots, X_n$ has $n$ associated random variables $X_i$ each taking values in an alphabet $\cX_i$. The DAG encodes constraints on their joint distribution $P_{X_1 X_2 \cdots X_n} \in \distk{n}$, where $\distk{n}$ denotes the set of all joint distributions of $n$ discrete random variables. We will often rely on a concise notation for statements about such distributions. Whenever we omit the values of the involved random variables, this means that the given relation holds for any of their values, e.g. the statement $P_{X_1 X_2}=P_{X_1}P_{X_2}$ means that for all $x_1 \in \cX_1$ and for  all $x_2 \in \cX_2$, $P_{X_1X_2}(x_1, x_2)=P_{X_1}(x_1) P_{X_2}(x_2)$. A central notion for analysing joint distributions of random variables is independence.

\begin{definition}
Let $X_S$, $X_T$, $X_U$ be three subsets of a set of jointly distributed random variables $\left\{X_1, X_2, \ldots, X_n \right\}$. Then $X_S$ and $X_T$ are said to be \emph{conditionally independent} given $X_U$ if and only if their joint distribution $P_\mathrm{X_S X_T X_U}$ can be written as
\begin{equation}
P_\mathrm{X_S X_T X_U}= P_\mathrm{X_S|X_U} P_\mathrm{X_T|X_U} P_\mathrm{X_U} \ .
\end{equation}
Conditional independence of $X_S$ and $X_T$ given $X_U$ is denoted as $X_S \indep X_T | X_U$.
Two sets $X_S$ and $X_T$ are said to be \emph{(marginally) independent} if 
\begin{equation}
P_\mathrm{X_S X_T}= P_\mathrm{X_S} P_\mathrm{X_T} \, ,
\end{equation}
concisely written $X_S \indep X_T$. \footnote{Note that this can also be considered as conditional independence conditioned on the empty set.}
\end{definition}

Classical causal structures are interpreted as Bayesian networks~\cite{Spirtes2000,Pearl2009}. 
Here, the standard terminology and the tools that are needed to understand the subsequent chapters are introduced. In a DAG that represents a Bayesian network a set of nodes, $X$, has \emph{ancestors}, $X^{\downarrow}$, all nodes from which a directed path points to at least one node in $X$, and \emph{descendants}, $X^{\uparrow}$, all nodes which can be reached from a node in $X$ along a directed path. The direct ancestors, i.e., the nodes from which there is a direct arrow to a node in $X$, are called $X$'s \emph{parents}, $X^{\downarrow_{1}}$, the direct descendants are called its \emph{children}, $X^{\uparrow_{1}}$. We also introduce a symbol for $X$'s \emph{non-descendants} $X^{\nuparrow}$, which are all nodes except for the ones in $X$ and all of their descendants.

\begin{definition}\label{def:compat}
Let $C^{\cC}$ be a classical causal structure with nodes $\left\{X_1,~X_2,~\ldots~,~X_n \right\}$. A probability distribution $P_\mathrm{X_1 X_2 \ldots X_n} \in \distk{n}$ is \emph{(Markov) compatible} with $C^{\cC}$ if it can be decomposed as 
\begin{equation}
P_\mathrm{X_1 X_2 \ldots X_n}= \prod_{i} P_\mathrm{X_{i}|X^{\downarrow_{1}}_{i}} \ .
\end{equation}
\end{definition}

The compatibility constraint encodes all conditional independencies of the random variables in the causal structure $C^{\cC}$. Nonetheless, whether a particular set of variables is conditionally independent of another is more easily read from the DAG, as explained in the following.

\begin{definition}
 Let $X$, $Y$ and $Z$ be three pairwise disjoint sets of nodes in a DAG $G$. The sets $X$ and $Y$ are said to be directionally separated or \emph{d-separated} by $Z$, if $Z$ blocks any path from any node in $X$ to any node in $Y$. 
A path is \emph{blocked} by $Z$, if the path contains one of the following: $i \rightarrow z \rightarrow j$ or  $i \leftarrow z \rightarrow j$ for some nodes $i$, $j$ and a node $z \in Z$ in that path, or if the path contains $i \rightarrow k \leftarrow j$, where $k \notin Z$. 
\end{definition}

The d-separation of the nodes in a causal structure is directly related to the conditional independence of its variables. The following proposition corresponds to Theorem~1.2.5
from~\cite{Pearl2009}, previously introduced in~\cite{Verma1988, Meek1995}. It justifies the application of d-separation as a means to identify independent variables.

\begin{proposition}[Verma \& Pearl] \label{prop:dseparation} Let $C^{\cC}$ be
  a classical causal structure and let $X$, $Y$ and $Z$
  be pairwise disjoint subsets of nodes in $C^{\cC}$.  If a
  probability distribution $P$ is compatible with $C^{\cC}$, then the
  d-separation of $X$ and $Y$ by $Z$ implies the conditional
  independence $X \indep Y | Z$.  Conversely, if for every
  distribution $P$ compatible with $C^{\cC}$ the conditional
  independence $X \indep Y | Z$ holds, then $X$ is d-separated from
  $Y$ by $Z$.
\end{proposition}

The compatibility of probability distributions with a classical causal
structure is
conveniently determined with the following proposition, which has also
been called the parental or local Markov condition before
(Theorem~1.2.7 in~\cite{Pearl2009}).
\begin{proposition}[Pearl] \label{prop:localmark}
Let $C^{\cC}$ be a classical causal structure. A probability distribution $P$ is compatible with $C^{\cC}$ if and only if every variable in $C^{\cC}$ is independent of its non-descendants, conditioned on its parents.
\end{proposition}
Hence, to establish whether a probability distribution is compatible
with a certain classical causal structure, it is sufficient to check that every variable $X$ is independent of its non-descendants $X^{\nuparrow}$ given its parents $X^{\downarrow_{1}}$, concisely written 
$X \indep X^{\nuparrow} | X^{\downarrow_{1}}$, 
i.e., it suffices to check one constraint per variable.  In particular, explicitly checking for all possible sets of nodes
whether they obey the independence relations implied by
d-separation is unnecessary. Each relevant constraint can be conveniently expressed in terms of the conditional mutual information~\eqref{eq:condmutinfo}
as\footnote{To see this, note that the relative entropy
  $D(P \| Q) \defeq  \sum_x P_X(x)\log_2 \left(P_X(x)/Q_X(x)\right)$ satisfies
  $D(P\| Q)=0\ \Leftrightarrow \ P = Q$, and that
  $I(X:X^{\nuparrow} | X^{\downarrow_{1}})=D(P_\mathrm{X X^{\nuparrow}
    X^{\downarrow_{1}}} \| P_\mathrm{X | X^{\downarrow_{1}}}
  P_\mathrm{X^{\nuparrow} | X^{\downarrow_{1}}} P_\mathrm{
    X^{\downarrow_{1}}})$.}
\begin{equation} \label{eq:indepentr}
I(X:X^{\nuparrow}| X^{\downarrow_{1}})=0 \ .
\end{equation}

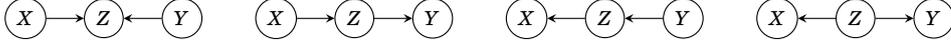
\begin{figure}
\centering 
\resizebox{0.8\columnwidth}{!}{%
\begin{tikzpicture}[scale=0.8]
%top left
\node[draw=black,circle,scale=0.7] (1) at (-3.25,0.5) {$X$};
\node[draw=black,circle,scale=0.7] (2) at (-2,0.5) {$Z$};
\node[draw=black,circle,scale=0.7] (3) at (-0.75,0.5) {$Y$};

%top right
\node[draw=black,circle,scale=0.7] (4) at (0.75,0.5) {$X$};
\node[draw=black,circle,scale=0.7] (5) at (2,0.5) {$Z$};
\node[draw=black,circle,scale=0.7] (6) at (3.25,0.5) {$Y$};

%bottom left
\node[draw=black,circle,scale=0.7] (7) at (4.75,0.5) {$X$};
\node[draw=black,circle,scale=0.7] (8) at (6,0.5) {$Z$};
\node[draw=black,circle,scale=0.7] (9) at (7.25,0.5) {$Y$};

%bottom right
\node[draw=black,circle,scale=0.7] (10) at (8.75,0.5) {$X$};
\node[draw=black,circle,scale=0.7] (11) at (10,0.5) {$Z$};
\node[draw=black,circle,scale=0.7] (12) at (11.25,0.5) {$Y$};
\draw [->,>=stealth] (1)--(2);
\draw [->,>=stealth] (3)--(2);
\draw [->,>=stealth] (4)--(5);
\draw [->,>=stealth] (5)--(6);
\draw [->,>=stealth] (9)--(8);
\draw [->,>=stealth] (8)--(7);
\draw [->,>=stealth] (11)--(10);
\draw [->,>=stealth] (11)--(12);
\end{tikzpicture}
}%
\caption[Insufficiency of conditional independencies for characterising causal links]{While in the left causal structure $X \protect\indep Y$, the other three networks share the conditional independence relation $X \protect\indep Y | Z$. This illustrates that the conditional independencies are not sufficient for characterising the causal links among a set of random variables.}
\label{fig:causex}
\end{figure}

While these conditional independence relations capture some features of
the causal structure, they are insufficient to completely capture the
causal relations between variables, as is illustrated with Figure~\ref{fig:causex}. 
In this case, the probability distributions themselves are unable to
capture the difference between the considered causal structures: correlations
are insufficient to determine causal links between random variables.  
External interventions allow for the exploration of causal links beyond the conditional independencies~\cite{Pearl2009}. However, we do not consider these here. 

Let $C^{\cC}$ be a classical causal structure involving $n$ random variables $\left\{X_1,~X_2,~\ldots~,~X_n \right\}$. 
The restricted set of distributions that are compatible with the causal structure $C^{\cC}$ is
\begin{equation}
\dist{C^{\cC}} \defeq  \left\{P \in \distk{n} \ \middle| \ P=\prod_{i=1}^{n}P_{\mathrm{X_i | X_{i}^{\downarrow_{1}}}} \right\} \ .
\end{equation}

\begin{example}[Distributions compatible with the instrumental scenario]
The classical instrumental scenario of Figure~\ref{fig:instrumental} allows for any four variable distribution in the set
\begin{equation}
\dist{\inst^{\cC}}=\left\{ P_{\mathrm{AXYZ}} \in \distk{4} \ \middle| \ P_{\mathrm{AXYZ}}= P_{\mathrm{Y | AZ}} P_{\mathrm{Z | AX}} P_{\mathrm{X}} P_{\mathrm{A}} \right\} \ .
\end{equation}
\end{example}

Many interesting scenarios, for instance the instrumental scenario, involve unobserved variables that are suspected to cause some of the correlations between the variables we observe, graphically the distinction of unobserved nodes from the observed ones is made by omitting the circle around their label (cf.\ Figure~\ref{fig:instrumental}). The unobserved variables may impose constraints on the possible joint distributions of the observed variables, a well-known example being a Bell inequality~\cite{Bell1964}\footnote{For a detailed discussion of the significance of Bell inequality violations on classical causal structures see~\cite{Wood2012}.}.

For a classical causal structure, $C^\cC$ on $n$ random variables $\left\{X_1,~X_2,~\ldots~,~X_n \right\}$, the restriction to the set of observed variables is called its \emph{marginal scenario}, denoted $\cM$. Here, we assume w.l.o.g.\ that the first $k\leq n$ variables are observed and the remaining $n-k$ are not. We are thus interested in the correlations among the first $k$ variables that can be obtained as the marginal of some distribution over all $n$ variables. 
Without any causal restrictions the set of all probability distributions of the $k$ observed variables is
\begin{equation}
\mardistO\defeq  \left\{P \in \distk{k} \ \middle| \ P=\sum_{X_{k+1},\ldots,X_n} P_\mathrm{X_1 X_2 \ldots X_n} \right\} \ ,
\end{equation}
 i.e., $\mardistO = \distk{k}$. For a classical causal structure, $C^{\cC}$, on the set of variables $\left\{X_1,~ X_2,~\ldots~,~X_n \right\}$, marginalising all distributions $P \in \dist{C^{\cC}}$ over the $n-k$ unobserved variables leads to the set 
 \begin{equation}\mardist{C^{\cC}} \defeq  \left\{ P \in \distk{k} \ \middle| \ P=\sum_{X_{k+1},\ldots,X_n} \prod_{i=1}^{n} P_\mathrm{X_i | X_{i}^{\downarrow_{1}}} \right\} \ .
 \end{equation}
In contrast to the unrestricted case, this set of distributions can in general not be recovered by considering a causal structure that involves only $k$ observed random variables, as can be seen with the following example.

\begin{example}[Observed distributions in the triangle and related scenarios.]
The five causal structures displayed in Figure~\ref{fig:3variables} illustrate that the unobserved variables crucially influence the possible observed distributions. The conditional independencies among the observed variables in each example are listed in Table~\ref{table:3variablescenarios}.
\begin{table}\footnotesize
\centering
\begingroup
\def\inlinedisplayeqn#1{\vspace*{0.5ex}$\displaystyle #1$\vspace*{0.5ex}}
\begin{tabular}{|m{3.0cm}|m{6.3cm}|m{4.3cm}|}
    \hline
     \hspace{0pt}\textbf{Causal Structure}  &\hspace{0pt}\textbf{Compatible Distributions} &\hspace{0pt}\textbf{Observed Independence} \\ 
		\hline
    \hspace{0pt} (a) & \vspace{4pt}\hspace{0pt}\inlinedisplayeqn{P_{XYZ}=P_{X}P_{Y}P_{Z}} & \hspace{0pt}${I(X:YZ)=0} , $ ${I(Y:XZ)=0}, $ ${I(Z : XY)=0}$ \\ 
 		\hline
    \hspace{0pt} (b) & \vspace{4pt}\hspace{0pt}\inlinedisplayeqn{P_{XYZ}=\sum_{A}P_{X}P_{Y\mid A}P_{Z\mid A}P_{A}} &
 $I(X:YZ)=0$ \\ 
 		\hline
    \hspace{0pt} (c) &  \vspace{4pt}\hspace{0pt}\inlinedisplayeqn{P_{XYZ}=\sum_{A}P_{X\mid A}P_{Y\mid A}P_{Z\mid A}P_{A}} &
 None \\ 
 		\hline
    \hspace{0pt} (d) & \vspace{4pt}\hspace{0pt}\inlinedisplayeqn{P_{XYZ}=\sum_{A,B}P_{X\mid B}P_{Y\mid A}P_{Z\mid AB}P_{A}P_{B}} &
 $I(X:Y)=0$ \\ 
 		\hline
    \hspace{0pt} (e) & \vspace{4pt}\hspace{0pt}\inlinedisplayeqn{P_{XYZ}=\! \! \! \!\sum_{A,B,C}\! \! \! \! \! P_{X\mid AC}P_{Y\mid AC}P_{Z\mid AB}P_{A}P_{B}P_{C}} &
None \\ 
	\hline		
		\hline
    \end{tabular}
\endgroup
\caption[Compatible distributions for three variable causal structures]{Compatible distributions and conditional independence relations for the three variable causal structures of Figure~\ref{fig:3variables}.}
\label{table:3variablescenarios}
\end{table}
While causal structures (a), (b) and (d) each exhibit different independencies, this is not the case for examples (c) and (e).
Nonetheless, while the causal structure of Figure~\ref{fig:3variables}(c) does not impose any restrictions on the compatible $P_{XYZ}$, the distributions that are compatible with the classical triangle causal structure, $\tri^\cC$, are
\begin{equation}
\mardist{\tri^{\cC}}= \left\{ P_{\mathrm{XYZ}} \in \distk{3} \ \middle| \ P_{\mathrm{XYZ}}= \sum_{A, B, C} P_{\mathrm{X | BC}} P_{\mathrm{Y | AC}} P_{\mathrm{Z| AB}} P_{\mathrm{A}} P_{\mathrm{B}} P_{\mathrm{C}} \right\} \ .
\end{equation}
For instance, perfectly correlated bits
$X$, $Y$ and $Z$, i.e., those with joint distribution
\begin{equation} \label{eq:perfcor}
P_{XYZ}(x,y,z)= 
\begin{cases}
\frac{1}{2} &x=y=z\\
0 &\textrm{otherwise},
\end{cases}
\end{equation}
are not achievable in $\tri^\cC$~\cite{Steudel2015}\footnote{Note that this is even the case if nodes $A$, $B$ and $C$ share quantum or non-signalling resources~\cite{Henson2014}.}.
The set $\mardist{\tri^\cC}$ is furthermore not convex, which can be seen by considering the perfect correlations~\eqref{eq:perfcor} (which are not in $\mardist{\tri^{\cC}}$) as a convex combination of the distribution where $X$, $Y$ and $Z$ are always $0$ and the distribution where $X$, $Y$ and $Z$ are always $1$ (both of which are in $\mardist{\tri^\cC}$).
\end{example}

\subsection{Causal structures involving quantum and more general resources}\label{sec:quantum_ns_causal}

There are situations, when we observe correlations that have not been generated with classical resources, but instead rely on systems from a more general theory. Of particular relevance are quantum systems, that have been shown to lead to stronger correlations than their classical counterpart by Bell~\cite{Bell1964}. The concept of a generalised causal structure was introduced in~\cite{Henson2014}, the idea being to have one framework in which classical, quantum and more general systems, for instance non-local boxes~\cite{Tsirelson1993,Popescu1994FP}, can be shared by unobserved nodes and where theory independent features of networks and corresponding bounds on our observations may be identified. 

\begin{definition}
A \emph{generalised causal structure} $C^{\gG}$ is a causal structure which for each observed node has an associated random variable and for each unobserved node has a corresponding non-signalling resource allowed by a generalised probabilistic theory.
In a \emph{quantum causal structure}, $C^{\qQ}$, the non-signalling resources are restricted to be quantum systems.
\end{definition}

Generalised probabilistic theories may be conveniently described in the operational-probabilistic framework~\cite{Chiribella2010}. Circuit elements correspond to so-called tests that are connected by wires, which represent propagating systems. In general, such a test has an input system, and two outputs: an output system and an outcome. In case of a trivial input system we are talking about a preparation-test and in case of a trivial output system of an observation-test. In the causal structure framework there is a test associated to each node. However, each such test has only one output: for unobserved nodes this is a general resource state, for observed nodes it is a random variable. Furthermore, resource states do not allow for signalling from the future to the past, i.e., we are considering so-called causal operational-probabilistic theories. 
A distribution $P$ over the observed nodes of a generalised causal structure $C^{\gG}$ is \emph{compatible with $C^{\gG}$} if there exists a causal operational-probabilistic theory, a resource for each unobserved edge in that theory and a test for each node that allow for the generation of $P$. We denote the set of all compatible distributions $\mardist{C^{\gG}}$.

In the arguably most relevant case of a quantum causal structure, $C^{\qQ}$, unobserved systems can be mathematically described in the usual density operator formalism. A framework that allows for analysing quantum causal structures was introduced in~\cite{Chaves2015}. For unity of description with $C^{\gG}$~\cite{Henson2014}, our account of quantum causal structures deviates slightly from this approach.~\footnote{In~\cite{Chaves2015} nodes correspond to quantum systems. All outgoing edges of a node together define a completely positive trace preserving (CPTP) map with output states corresponding to the joint state associated with its children. Similarly, the CPTP map associated to the input edges of a node must map the states of the parent nodes to the node in question. In~\cite{Henson2014}, on the other hand, edges correspond to states whereas the transformations occur at the nodes. We employ this latter approach as the quantum system associated with an unobserved node in~\cite{Chaves2015} is split into subsystems in such a way that the latter can be conveniently considered to label its outgoing edges, which will be useful for the considerations in the following chapters. However, both viewpoints are equally valid.}
For a quantum causal structure $C^{\qQ}$, nodes without input edges
correspond to the preparation of a quantum state described by a density operator on a Hilbert space, e.g., $\rho_A \in \cS(\cH_A)$ for a node $A$, where for observed nodes this state is required to be classical. 
For each directed edge in the graph there is a corresponding subsystem with Hilbert space labelled by the edge's input and output nodes. For instance, if $Y$ and $Z$ are the only children of $A$ then there are associated spaces $\cH_{A_Y}$ and $\cH_{A_Z}$ such that $\cH_A=\cH_{A_Y}\otimes\cH_{A_Z}$. For convenience of exposition, edges and their associated systems share the same label.~\footnote{Note that in the classical case these subsystems may all be taken to be copies of the system itself.} At an unobserved node, a CPTP map from the joint state of all its input edges to the joint state of its output edges is performed. A node is always labelled by its output state. For an observed node the latter is classical. Hence, it corresponds to a random variable that represents the output statistics obtained in a measurement by applying a POVM to the input states.\footnote{
Preparation and measurement can also be seen as CPTP maps with classical input and output systems respectively, thus allowing for a unified description.} If all input edges are classical this can be interpreted as a stochastic map between random variables.
A distribution, $P$, over the observed nodes of a causal structure $C^{\qQ}$ is \emph{compatible with $C^{\qQ}$} if there exists a quantum state labelling each unobserved node (with subsystems for each unobserved edge) and transformations, i.e., preparations and CPTP maps for each unobserved node as well as POVMs for each observed node, that allow for the generation of $P$ by means of the Born rule. We denote the set of all compatible distributions $\mardist{ C^{\qQ}}$.

\begin{example}[Compatible distributions for the quantum instrumental scenario]
For the quantum instrumental scenario of Figure~\ref{fig:instrumental}, the set of compatible distributions is,
\begin{equation}
\mardist{ \inst^{\qQ} }=\left\{ P_{XYZ}  \in \cP_3 \ \middle| \ P_{XYZ}=\tr{(E_X^{Z} \otimes  F_Z^{Y}) \rho_{A}} P_X \right\} \ .
 \end{equation}  
 A state  $\rho_A \in \cS(\cH_{A_Z}\otimes \cH_{A_Y})$ is prepared. Depending on the random variable $X$, a POVM  $\left\{E_X^{Z} \right\}_Z$ on $\cH_{A_Z}$ is applied to generate the output distribution of the observed variable $Z$. Depending on the latter, another POVM $\left\{ F_Z^{Y}\right\}_Y$ is applied to generate the distribution of $Y$.
\end{example}

Generalised causal structures not only encompass the quantum case, but classical causal structures can also be viewed as a special case thereof~\cite{Henson2014, Fritz2015}. In a classical causal structure, $C^{\cC}$, the edges of the DAG represent the propagation of classical information, and, at a node with incoming edges, the random variable can be generated by applying an arbitrary function to its parents.  We are hence implicitly assuming that all the information about the parents is transmitted to its children (otherwise the set of allowed functions would be restricted). This does not pose a problem since classical information can be copied.  In the general case, $C^{\gG}$ (or $C^{\qQ}$), on the other hand, the no-cloning theorem means that the children of a node cannot (in general) all have access to the same information as is present at that node.
There is no notion of a joint state of all nodes in the causal structure and it is hence not clear how one could condition on an unobserved system. Moreover, in generalised probabilistic theories there is no consensus on the representation of states and their dynamics (unlike the density operator formalism in quantum mechanics).  To circumvent these issues, the classical notion of d-separation has been reformulated to derive  conditional independence relations among observed variables that hold in any causal operational probabilistic theory.~\cite{Henson2014}. 

\begin{proposition}[Henson, Lal \& Pusey] \label{prop:dseparationgDAG} Let $C^{\gG}$ be a generalised causal structure  and let $X$, $Y$ and $Z$ be pairwise disjoint subsets of observed nodes in  $C^{\gG}$.  If a probability distribution $P$ is compatible with $C^{\gG}$, then the d-separation of $X$ and $Y$ by $Z$ implies the conditional independence $X \indep Y | Z$.  Conversely, if for every distribution $P$ compatible with $C^{\gG}$ the conditional independence $X \indep Y | Z$ holds, then $X$ is d-separated from $Y$ by $Z$ in $C^{\gG}$.
\end{proposition}

Ref.~\cite{Henson2014} provides sufficient conditions for identifying causal structures, $C$, for which the only restrictions on the compatible distributions over the observed variables for $C^{\cC}$ are
those that follow from the d-separation of these variables. Since, by
Proposition~\ref{prop:dseparationgDAG}, these conditions also hold in
$C^\qQ$ and $C^\gG$, this implies $\mardist{C^\cC}=\mardist{C^\qQ}=\mardist{C^\gG}$. 
For causal structures with up to six nodes, there are $21$ examples (and over $10 000$ adaptations thereof) where such equivalence does not hold and where further relations among the observed variables have to be taken into account~\cite{Henson2014, Pienaar2016}.  $\inst$ and $\tri$ are two such examples.
\footnote{Note that $\tri$ is the only example from Figure~\ref{fig:3variables} where this is the case. That compatible classical and quantum distributions are the same, the cases we usually care about, can be straightforwardly checked: In structures (a), (b) and (c) all joint distributions are allowed for the variables that share a common cause in the classical case. Hence, quantum systems do not enable any stronger correlations. This can also be seen as for any quantum state $\rho_A$ shared at $A$ and measured later the correlations can be classically reproduced if $A$ sends out the same classical output statistics to the parties directly.
In structure (d) no non-classical quantum correlations exist either~\cite{Fritz2012}. This is also
fairly intuitive: the quantum measurements performed at $X$ and $Y$ could be equivalently performed at the sources $B$ and $A$ respectively, such that these sources distribute cq-states of the form
$\sum_{x} P_X(x) \ketbra{x}{x} \otimes \rho_{B_Z}^{x}$ and $\sum_{y}
P_Y(y) \ketbra{y}{y} \otimes \rho_{A_Z}^{y}$ instead.  The same
correlations can be achieved classically by taking random variables
$B=X$ and $A=Y$ (these being distributed according to $P_X$ and
$P_Y$).  Since $\rho_{B_Z}^{x}$ and $\rho_{A_Z}^{y}$ are functions of
$X$ and $Y$, the statistics formed by measuring such states can be
computed classically via a probabilistic function (this function could
be made deterministic by taking $B=(X,W)$, where $W$ is distributed
appropriately).
}

\subsubsection{Other ways to explore quantum and generalised causal structures} \label{sec:other_stuff}
The approach to quantum and generalised causal structures above is based on adaptations of the theory of Bayesian networks to the respective settings and on retaining the features that remain valid, for instance the relation between d-separation and independence for observed variables~\cite{Henson2014}. Other approaches to generalise classical networks to the quantum realm have been pursued in Ref.~\cite{Leifer2013}, where conditional quantum states, analogous to conditional probability distributions, were introduced.

Recent articles have also proposed generalisations of Reichenbach's principle~\cite{Reichenbach1956} to the quantum realm~\cite{Pienaar2015, Costa2015, Allen2016}. Whilst in Ref.~\cite{Pienaar2015} a new graph separation rule, q-separation, was introduced, \cite{Costa2015, Allen2016} rely on a formulation of quantum networks in terms of quantum channels and their Choi states.

A  very active research area is the exploration \emph{indefinite causal structures}~\cite{Hardy2005, Hardy2007, Hardy2009}. There are several approaches to this, for instance the process matrix formalism~\cite{Oreshkov2012},  which has lead to the derivation of so called causal inequalities and the identification of signalling correlations that are achievable in this framework, but not with any predefined causal structure~\cite{Oreshkov2012,Baumeler2014}. 
Another framework for describing such scenarios is the theory of quantum combs~\cite{Chiribella2013}, illustrated by a quantum switch, a quantum bit controlling the circuit structure in a quantum computation. 
A recent framework of so-called causal boxes with the aim to model cryptographic protocols is also available~\cite{Portmann2017}.

\section{Quantum resource theories for thermodynamics}\label{sec:resource}

Resource theories deal with questions like ``Is a particular task possible with the resources at hand?'' or, ``Which resources do we need to complete a certain task and in what quantity?''. Such an agent based view is naturally taken in information theory where the aim is often to complete communication tasks as efficiently as possible. In thermodynamics, such perspective is less traditional, as we are used to describe systems without reference to an observer. However, macroscopic thermodynamics can also be phrased as a resource theory, a perspective which led Lieb and Yngvason to the derivation of the second law from a small set of axioms~\cite{Lieb1998,Lieb1999,Lieb2001,Lieb2013,Lieb2014}.

Usually, a resource theory is defined in terms of a class of \emph{free operations} on some state space, which are deemed cheap or free in the sense that they are easy to perform. The free operations impose a structure on the state space by inducing a partial ordering of the states: For two states $\rho$ and $\sigma$, 
\begin{equation}
\rho \prec \sigma
\end{equation}
if and only if there is a free operation transforming $\rho$ to $\sigma$.
Mathematically, $\prec$ is usually taken to be a preorder, that is, it is reflexive and transitive — it is always possible to obtain a resource $\rho$ from itself, i.e., the identity operation is free, and if $\rho$ can  be  transformed  into $\sigma$ and $\sigma$ into $\tau$, then there is a free operation transforming $\rho$ to $\tau$, i.e., free operations can be composed sequentially.
In macroscopic thermodynamics, these free operations are adiabatic processes, defined as those operations on a system that leave the environment unchanged, except for a weight having risen or fallen in the process~\cite{Lieb1998,Lieb1999,Lieb2001,Lieb2013,Lieb2014}.

With respect to $\prec$, states can be assigned values. Intuitively, a state $\rho$ is more valuable than a state $\sigma$ if it allows for the generation of a larger set of states and the completion of more tasks than $\sigma$. Hence, if the two states can be ordered with $\prec$, then $\rho$ is more valuable than $\sigma$ if and only if $\rho \prec \sigma$ but $\sigma \not\prec \rho$\footnote{In general, there are incomparable states, for which neither $\rho \prec \sigma$ nor $\sigma \prec \rho$, their value assignment has to be treated with care.}. This is quantified in terms of  \emph{monotones}, functions from the state space to $\mathbb{R}$ that are monotonic with respect to the ordering. Quantifying resources is an essential part of a resource theoretic framework and facilitates the comparison of different resources. It often allows for the derivation of practically relevant conditions for state transformations~\cite{Lieb2013,Lieb2014,Kraemer2016}. For the resource theory of macroscopic thermodynamics these quantities correspond to the Boltzmann entropy and appropriate min- and max-entropies~\cite{Weilenmann2015}.

In the microscopic regime, \emph{quantum resource theories} provide a natural means to model (thermodynamic) processes. The free operations are usually a restricted class of CPTP maps, acting on the set of all quantum states. They yield bounds for work extraction and more generally give us laws that determine which transformations are possible, given fine control of an experimental setup. 
In the following we specify the resource theory of adiabatic operations, that will be of interest for our considerations in Chapters~\ref{chap:microthermo} and~\ref{chap:thermomacro}. For details on resource theories and further examples we refer to the vast literature~\cite{Janzing2000_cost,Horodecki2011,Renes2014, Horodecki2003, Horodecki2003b, Browne2003, Braunstein2005, Plenio2005, Brandao2008, Gour2008, Gour2009, Horodecki2009, Vaccaro2011, Skotiniotis2012, Toloui2012, Brandao2013_resource, Barnett2013, Gour2013, Marvian2013, Marvian2014, Marvian2014a, Veitch2014, Grudka2014, Rivas2014, Baumgratz2014a, Levi2014, Halpern2014c, Brandao2015, DelRio2015, Lostaglio2015, Lostaglio2015b, Korzekwa2016, Sparaciari2016, Kraemer2016, Chitambar2016, Chitambar2016a, Marvian2016, Winter2016a, Napoli2016, Streltsov2016, Gour2016, Theurer2017}.

\subsection{The resource theory of adiabatic processes} \label{sec:noisy}

The idea of considering a (microscopic) resource theory of adiabatic processes originates from Lieb and Yngvason's framework for macroscopic thermodynamics mentioned above. The interaction with an environment and a weight that leaves no trace in the environmental system except for a change in the weight translates to the following operations~\cite{Weilenmann2015}. 

\begin{definition}
The resource theory of (microscopic) \emph{adiabatic processes} is defined by the set of free operations which map a state $\rho \in \cS(\cH_S)$ to another state
\begin{equation}
\sigma=\partr{U (\rho \otimes \tau) U^{\dagger}}{A} \  , 
\end{equation}
where $\tau$ is a sharp state on some Hilbert space $\cH_A$, meaning a state for which all its non-zero eigenvalues take the same value, $U$ is an arbitrary unitary and the partial trace is taken over the subsystem $A$, where it is further required that $\partr{U (\rho \otimes \tau) U^{\dagger}}{S}=\tau$. 
\end{definition}

The application of an arbitrary unitary is enabled by explicitly modelling a weight system~\cite{Allahverdyan2004, Aberg2014} and the condition $\partr{U (\rho \otimes \tau) U^{\dagger}}{S}=\tau$ is due to the requirement that the operations leave no trace on the environmental system after the interaction; for details regarding this correspondence we refer to~\cite{Weilenmann2015}.
These operations introduce a partial ordering of all possible states, which can be expressed in terms of majorisation.

\begin{definition}
Let $\rho$, $\sigma \in \cS(\cH)$ with Hilbert space dimension $\dim{(\cH)}=n$ be two states with spectra $\left\{\ithEV{\rho} \right\}_i$ and $\left\{\ithEV{\sigma} \right\}_i$ respectively, ordered such that $\ithEVk{\rho}{1} \geq \ithEVk{\rho}{2} \geq \ldots \geq \ithEVk{\rho}{n}$ and $\ithEVk{\sigma}{1} \geq \ithEVk{\sigma}{2} \geq \ldots \geq \ithEVk{\sigma}{n}$. Then $\rho$ \emph{majorises} $\sigma$, denoted $\rho \maj \sigma$, if for all $1 \leq k \leq n$,
\begin{equation}
\sum_{i=1}^{k} \ithEV{\rho} \geq \sum_{i=1}^{k} \ithEV{\sigma} \ .
\end{equation}
\end{definition}
Mathematically, $\maj$ is a \emph{preorder}, meaning it is reflexive, and transitive. The following proposition specifies its relation to adiabatic processes, which has been proven in~\cite{Weilenmann2015} based on results regarding the resource theory of noisy operations~\cite{Horodecki2003,Horodecki2003b,Gour2013}.

\begin{proposition} 
For $\rho$, $\sigma \in \cS(\cH)$ the following are equivalent.
\begin{enumerate}[(1)]
\item $\rho$ can be transformed into $\sigma$ by an adiabatic operation.
\item $\rho \maj \sigma$.
\end{enumerate}
\end{proposition}
We remark here that there are pairs of states, $\rho$, $\sigma \in \cS(\cH)$, where neither can be converted into the other by an adiabatic operation, i.e., there exist states $\rho$ and $\sigma$, for which neither $\rho \maj \sigma$, nor $\sigma \maj \rho$.

An equivalent formulation of majorisation is given in terms of \emph{step functions}. Instead of comparing sums of eigenvalues of the respective states, one can use that for $\rho$, $\sigma \in \cS(\cH)$, $ \rho \maj \sigma$ if and only if
\begin{equation}
 \int_0^x f_\rho(x') \ dx \geq \int_0^x f_\sigma(x') \ dx'  
\end{equation}
for all $0 \leq x < n$, where $n=\dim(\cH)$ is the dimension of the system and $f_\rho(x)$ and $f_\sigma(x)$ are the step functions defined as 
\begin{equation}
 f_\rho (x) = \begin{cases}
 \ithEV{\rho}  &\text{for} \quad i-1 \leq x < i,\\
0 &\text{otherwise} .
\end{cases}
\end{equation}

The monotonic functions with respect to adiabatic operations are numerous, they include the R\'{e}nyi entropies introduced in Section~\ref{sec:renyi}~\cite{Karamata, Gour2013}.

%=========================================================

%% file: chapter03/chap_entropyvector.tex
\let\textcircled=\pgftextcircled
\part{Entropic analysis of causal structures}

\chapter{Outline of the entropic techniques for analysing causal structures}
\label{chap:entropy_vec}

\initial{I}n this Chapter we introduce the entropy vector approach to causal structures, which provides the necessary background and motivates the considerations of Chapters~\ref{chap:inner},~\ref{chap:nonshan} and~\ref{chap:classquant}, which are concerned with novel contributions.
This chapter, on the other hand, reviews the state of the art of current entropic techniques, illustrated with numerous examples that are intended to make the topic easily accessible. With our exposition, we fill a few gaps on the way, proving several facts about entropy cones some of which have been implicitly assumed in the literature but not rigorously proven. Wherever there is no other specification, the examples presented and the proofs conducted are our own.  We also establish connections between different contributions to the field that have not been previously demonstrated.

We start with a short motivation why entropic techniques are a convenient means to analyse causal structures in Section~\ref{sec:causalstructures}, before introducing the entropy vector approach and its fine-grained application in post-selected causal structures in Sections~\ref{sec:entropicappr} and~\ref{sec:post-selection}. In Section~\ref{sec:further_techniques}, we give an overview on alternative computational techniques and provide some information on their relation to the entropy vector method. 

\section{Motivation for the entropic analysis}\label{sec:causalstructures}

Deciding whether a causal explanation that involves unobserved variables is compatible with observed statistical data is a central scientific task. In the context of quantum information theory and quantum cryptography we are usually interested in the \emph{in}compatibility of statistical data with a particular classical causal structure~\cite{Ekert1991,Mayers1998,Barrett2005b,Acin2006,Colbeck2009, Colbeck2011,Pironio2010,Vazirani2014,Miller2014}, an idea that lies behind Bell's theorem~\cite{Bell1964} (see also~\cite{Wood2012}).
Since classical causal structures are mathematically rather well-understood, it is in principle known how to check whether observed correlations are compatible with a given causal structure by means of quantifier elimination techniques~\cite{Geiger1999}. However, compatibility tests for networks that involve unobserved systems are only computationally feasible for small cases~\cite{Garcia2005, Lee2015}. Explicit complexity results are available for the Bell scenario, based on so-called correlation and cut polytopes~\cite{Pitowsky1991, Avis2004}, where it was shown that testing whether certain correlations are achievable in the Bell scenario with two parties and $m$-dichotomic observables each is NP-complete~\cite{Avis2004}. Quantum and generalised causal structures are less explored in this respect.

Finding good heuristics to help identify correlations that are (in)compatible with a causal structure is an important endeavour that is currently actively researched~\cite{Chaves2012, Chaves2013, Fritz2013, Chaves2014, Chaves2014b, Henson2014, Steudel2015, Chaves2015, Rosset2015, Chaves2015a, Pienaar2015, Chaves2016, Pienaar2016, Wolfe2016, Kela2017, Navascues2017, Miklin2017}. 
Entropic considerations~\cite{Chaves2012, Chaves2013, Fritz2013, Chaves2014, Chaves2014b, Henson2014,  Steudel2015, Chaves2015, Chaves2016, Pienaar2016, Kela2017, Miklin2017} have several advantages. First of all, entropic restrictions on possible correlations in a causal structure are independent of the dimension of the involved random variables. Hence, considerations based on entropy measures enable the derivation of constraints that are valid for arbitrarily large alphabet sizes of all involved observed and unobserved systems. Secondly, the entropic techniques detailed below map the set of compatible observed distributions for a classical causal structure, which is often not convex, to a convex set of so-called entropy vectors, and, in addition, they conveniently represent independencies by linear relations between entropies instead of polynomial constraints (see also Section~\ref{sec:bayesian}).  
These properties make entropy a useful means for distinguishing different causal structures in many situations.

\section{Entropy vector approach}\label{sec:entropicappr}
Characterising the joint distributions of a set of random variables or alternatively considering a multi-party quantum state in terms of its entropy (and the entropy of its marginals) has a tradition in information theory dating back to Shannon~\cite{Shannon1948, Yeung1997, Pippenger2003}.
However, only recently has this approach been extended to account for causal structure~\cite{Chaves2012, Fritz2013}.  In Sections~\ref{sec:classicalcone} and \ref{sec:causal} respectively, we review this approach with and without imposing causal constraints. All our considerations are concerned with discrete random variables, for extensions of the approach to continuous random variables (and their limitations) we refer the reader to~\cite{Chan2003, Fritz2013}.

\subsection{Entropy cones without causal restrictions} \label{sec:classicalcone}
The entropy cone for a set of $n$ random variables was
introduced in~\cite{Yeung1997}. For a set of $n\geq 2$ jointly distributed random variables,
$\Omega\defeq\left\{X_{1},~X_{2},~\ldots~,~X_{n}\right\}$, we denote their
probability distribution as $P \in \mathcal{P}_{n}$, where $\cP_{n}$
is the set of all probability mass functions of $n$ jointly
distributed random variables. The Shannon entropy maps any subset of
$\Omega$ to a non-negative real value:
\begin{align*}
\begin{split}
 H: \mathscr{P}(\Omega) \rightarrow \left[0, \infty \right), \\
X_S \mapsto H(X_S),
\end{split}
\end{align*}
where $\mathscr{P}(\Omega)$ denotes the power set of $\Omega$, and $H(\{\})=0$.
The entropy of the joint distribution of the random variables $\Omega$
and that of all its marginals can be expressed as components of
a vector in $\mathbbm{R}_{\geq 0}^{2^n-1}$, ordered in the following as\footnote{Since the empty set always has zero entropy, it is omitted from the entropy vector.}
$$\left(H(X_1),~H(X_2),~\ldots~,~H(X_n), H(X_1X_2),~H(X_1X_3),~\ldots~,~H(X_1 X_2~\ldots~X_n) \right) \ .$$
We use $\bH(P) \in \mathbbm{R}^{2^{n}-1}_{\geq 0}$ to denote the
vector corresponding to a particular distribution $P \in
\mathcal{P}_{n}$.  The set of all such vectors is
\begin{equation}
\ensetk{n}  \defeq  \left\{ v \in \mathbbm{R}_{\geq 0}^{2^{n}-1} \ \middle| \ \exists P \in \distk{n} \sth  v=\bH(P) \right\} \ .
\end{equation}
Its closure $\enconek{n}$ includes vectors $v$ for which
there exists a sequence of distributions $P_k\in\distk{n}$ such that $\bH(P_k)$ tends to $v$ as $k\rightarrow \infty$.
It is known that the \emph{entropy cone} $\enconek{n}$ is a convex cone for any
$n\in \mathbbm{N}$~\cite{Zhang1997}. 
As such, its boundary may be characterised in terms of (potentially infinitely many) linear constraints. Because $\enconek{n}$ is difficult to characterise, we will in the following consider various approximations.

\subsubsection{Outer approximation: the Shannon cone} 
The standard outer approximation to $\enconek{n}$ is the polyhedral cone constrained by the \emph{Shannon inequalities} listed in the following, which are obeyed by any entropy vector of a set of jointly distributed random variables:\footnote{Note that sometimes $H(\{\})=0$ is included as an extra constraint, but we keep this implicit here.}
\begin{itemize}
\item Monotonicity: For all $X_T$, $X_S \subseteq \Omega$,
\begin{equation}\label{eq:monotonicity}
{H(X_S \setminus X_T) \leq H(X_S)} \ .
\end{equation}
\item Submodularity: For all $X_S$, $X_T \subseteq \Omega$,
\begin{equation} \label{eq:submodularity}
{H(X_S \cap X_T) +  H(X_{S} \cup X_{T})  \leq  H(X_{S}) + H(X_{T})} \ .
\end{equation}
\end{itemize}
$X_S \setminus X_T$ denotes the relative complement of $X_T$ in $X_S$, $X_S \cap X_T$ denotes the intersection of $X_S$ and $X_T$ and $X_{S} \cup X_{T}$ denotes their union.
We remark here that with the convention $H(\{ \})=0$ the above inequalities also imply that entropy is positive. The monotonicity constraints correspond to positivity of conditional entropy, $H(X_S \cap X_T|X_S \setminus X_T)\geq 0$, and submodularity is equivalent to the
positivity of the conditional mutual information, $I(X_S \setminus X_T :X_T \setminus X_S |X_S \cap X_T) \geq 0$. The monotonicity and submodularity constraints can all be generated from a minimal set of $n + n (n-1) 2^{n-3}$ inequalities~\cite{Yeung1997}: for the monotonicity constraints it is sufficient to consider the $n$ constraints with $X_S=X_i$, for some $X_i\in\Omega$ and $X_S \cup X_T =\Omega$; for the submodularity constraints it is sufficient to consider those with $X_S \setminus X_T=X_i$ and $X_T \setminus X_S=X_j$ with $i<j$ and where $X_U\defeq X_S \cap X_T$ is any subset of $\Omega$ not containing $X_i$ or $X_j$, i.e., submodularity constraints of the form $I(X_i :X_j |X_U) \geq 0$.

These $n+n(n-1)2^{n-3}$ independent Shannon inequalities can be expressed in terms of a ${(n+n(n-1)2^{n-3})\times(2^{n}-1)}$ dimensional matrix, which we call $\mshan{n}$, such that for any $v\in\ensetk{n}$ the conditions $\mshan{n} \cdot v\geq 0$ hold\footnote{This  condition is to be interpreted as the requirement that each component of $\mshan{n} \cdot v$ is non-negative.}.  Hence, a violation of $\mshan{n} \cdot v\geq 0$ by a vector for $v\in\mathbb{R}_{\geq0}^{2^{n}-1}$ certifies that there is no distribution $P\in\distk{n}$ such that $v=\bH(P)$. It follows that the \emph{Shannon cone},
\begin{equation}
\outconek{n} \defeq  \left\{v \in \mathbbm{R}_{\geq 0}^{2^{n}-1} \ \middle| \
 \mshan{n} \cdot v \geq 0 \right\} \, ,
\end{equation}
is an outer approximation to the set of achievable entropy vectors $\ensetk{n}$~\cite{Yeung1997}.

\begin{example}\label{example:3shannon}
The three variable Shannon cone is 
$\outconek{3}= \left\{v \in \mathbbm{R}_{\geq 0}^{7} \ \middle| \ \mshan{3} \cdot v \geq 0 \right\}$, where 
\begin{equation}
\mshan{3}= \left( \begin{array}{ccccccc}
0 & 0 & 0 & 0 & 0 & -1 & 1 \\
0 & 0 & 0 & 0 & -1 & 0 & 1 \\
0 & 0 & 0 & -1 & 0 & 0 & 1 \\
1 & 1 & 0 & -1 & 0 & 0 & 0 \\
1 & 0 & 1 & 0 & -1 & 0 & 0 \\
0 & 1 & 1 & 0 & 0 & -1 & 0 \\
-1 & 0 & 0 & 1 & 1 & 0 & -1 \\
0 & -1 & 0 & 1 & 0 & 1 & -1 \\
0 & 0 & -1 & 0 & 1 & 1 & -1 
\end{array} \right) .
\end{equation}
The first three rows are monotonicity constraints, the remaining six ensure submodularity.
\end{example}

\subsubsection{Beyond the Shannon cone}\label{sec:nonShan}
For two variables the Shannon cone coincides with the set of achievable entropy vectors, $\outconek{2}= \ensetk{2}$, while for three random variables this holds only for its closure, i.e.\ $\outconek{3}=\enconek{3}$ but $\ensetk{3}\subsetneq\enconek{3}$~\cite{Han81,Zhang1997}. For $n \geq 4$ further independent linear inequalities are needed to fully characterise $\enconek{n}$, the first of which was discovered in~\cite{Zhang1998}.

\begin{proposition}[Zhang \& Yeung] \label{prop:zhangyeung} For any four discrete random variables $X_1$, $X_2$, $X_3$ and $X_4$ the following inequality holds:
\begin{multline} \label{eq:zy}
- H(X_1) - H(X_2) -\frac{1}{2} H(X_3)  + \frac{3}{2}H(X_1X_2) + \frac{3}{2}H(X_1X_3) 
+ \frac{1}{2}H(X_1X_4) + \frac{3}{2}H(X_2X_3)   \\
+ \frac{1}{2}H(X_2X_4) -\frac{1}{2}H(X_3X_4) - 2 H(X_1X_2X_3) - \frac{1}{2}H(X_1X_2X_4) \geq 0.
\end{multline}
This is in the following abbreviated as $\Diamond_{X_1 X_2 X_3 X_4} \geq 0$.
\end{proposition}

For $n\geq4$ the convex cone $\enconek{n} \subsetneq \outconek{n}$ cannot be characterised by finitely many linear inequalities~\cite{Matus2007}, i.e., it is not polyhedral (in our terminology), and a complete linear characterisation of $\enconek{n}$ is not available. Two infinite families of valid inequalities are given in the following.\footnote{For $s=1$ both inequalities are equivalent to \eqref{eq:zy}.}

\begin{proposition}[Mat\'{u}\u{s}] \label{prop:matusfam}
Let $X_1$, $X_2$, $X_3$ and $X_4$ be random variables and let $s \in \mathbb{N}$. Then the following inequalities hold:
\begin{multline} \label{eq:matus1}
s \left[  I(X_1:X_2|X_3)  + I(X_1:X_2|X_4) + I(X_3:X_4)
 -  I(X_1:X_2)\right] \\
 + I(X_1:X_3|X_2) + \frac{s(s+1)}{2} \left[I(X_2:X_3|X_1) + I(X_1:X_2|X_3) \right] \geq  0,
\end{multline}
\begin{multline}
s \left[ I(X_2:X_3|X_1) + 2 I(X_1:X_2|X_3) +I(X_1:X_2|X_4) + I(X_3:X_4) 
 - I(X_1:X_2) \right] \label{eq:matus2} \\
+ I(X_1:X_3|X_2)  
+ \frac{s(s-1)}{2} \left[ I(X_2:X_4|X_1) 
+ I(X_1:X_2|X_4) \right] \geq 0.
\end{multline}
\end{proposition}

Several additional entropy inequalities have been discovered~\cite{Zhang1997, Zhang1998, Makarychev2002, Dougherty2006}. Recently, systematic searches for new inequalities for $n=4$ have been conducted~\cite{Xu2008, Dougherty2011}, which recover most of the previously known inequalities; in particular the inequality from Proposition~\ref{prop:zhangyeung} is re-derived and shown to be implied by tighter ones~\cite{Dougherty2011}. The systematic search in~\cite{Dougherty2011} is based on considering additional random variables that obey certain constraints and then deriving four variable inequalities from the known identities for five or more random variables (see also~\cite{Zhang1998, Matus2007}), an idea that is captured by a so-called copy lemma~\cite{Zhang1998, Dougherty2011, Kaced2013}.
In Ref.~\cite{Dougherty2011}, rules to generate families of inequalities have been suggested, in the style of techniques introduced in~\cite{Matus2007} for the derivation of Proposition~\ref{prop:matusfam}.

For more than four variables, only few additional inequalities are known~\cite{Zhang1998, Makarychev2002}. 
Curiously, to our knowledge, all known irredundant four variable non-Shannon inequalities (i.e., the ones found in~\cite{Zhang1998, Makarychev2002, Dougherty2006, Matus2007, Xu2008, Dougherty2011} that are not yet superseded by tighter ones) can be written as a positive linear combination of the \emph{Ingleton quantity}, $I_{\mathrm{ING}}(X_1,X_2;X_3,X_4)$, and conditional mutual information terms~\cite{Dougherty2011}.

\subsubsection{Inner approximations}\label{sec:inner_general}
Such approximations can be defined in terms of \emph{linear rank inequalities}, which are inequalities that are always satisfied by the ranks of the subspaces of a vector space. Analogously to the entropy vectors, these ranks can be represented as vectors, which are called \emph{linearly representable}. It is known that all linear entropy inequalities are also valid for linearly representable vectors but not the other way around~\cite{Hammer2000} (consider Example~\ref{example:rank_viol} for an entropy vector that violates a linear rank inequality). Hence, the linearly representable vectors, $\inconek{n}$, give an inner approximation to $\enconek{n}$.
 
For $n=4$, this inner approximation is the polyhedral cone constrained by the Shannon inequalities and the six permutations of the \emph{Ingleton inequality}~\cite{Ingleton}, 
\begin{equation} \label{eq:ingletoninequ}
{I_{\mathrm{ING}}(X_1, X_2 ; X_3, X_4)\geq 0} \ ,
\end{equation}
for random variables $X_1$, $X_2$, $X_3$ and $X_4$, which are concisely written in a matrix $M_\mathrm{I} \in \mathbb{R}^{6 \times 15}$. The region these inequalities constrain is called the \emph{Ingleton cone},
\begin{equation}
\inconek{4} \defeq  \left\{v \in \mathbbm{R}_{\geq 0}^{15} \ \middle| \ \mshan{4} \cdot v \geq 0 \text{ and } M_\mathrm{I} \cdot v \geq 0 \right\} \ .
\end{equation}
The following example shows that this inner approximation is a  strict subset of the entropy cone, $\inconek{4} \subsetneq \enconek{4}$.

\begin{example} \label{example:rank_viol}
Let $X_1$, $X_2$, $X_3$ and $X_4$ be four jointly distributed random variables. Let $X_3$ and $X_4$ be uniform random bits and let $X_1=\operatorname{AND}(\neg X_3, \neg X_4)$ and $X_2=\operatorname{AND}(X_3,X_4)$.
This distribution~\cite{Matus2007} leads to the entropy vector 
\begin{equation}
v  \approx  (0.81,0.81,1,1,1.50,1.50,1.50,1.50,1.50,2,2,2,2,2,2) \ ,
\end{equation}
for which
\begin{equation}
I_{\mathrm{ING}}(X_1, X_2 ; X_3, X_4) \approx -0.12 \ ,
\end{equation}
in violation of the Ingleton inequality.
\end{example}

$\Gamma^{I}_5$ is defined by the Shannon inequalities, several instances of the Ingleton inequality and $24$ additional inequalities and their permutations~\cite{Dougherty2009}. For $6$ or more variables, a complete list of all linear rank inequalities is not known, nor whether such a list of inequalities would be finite. Over a billion valid linear rank inequalities in $6$ variables have so far been found~\cite{Dougherty2014}.

\subsection{Entropy cones for causal structures} \label{sec:causal}
Under the assumption of a causal structure, $C$, the compatible distributions of the observed variables, $\mardist{C^{\cC}}$, $\mardist{C^{\qQ}}$ and $\mardist{C^{\gG}}$ respectively, can all be mapped to sets of compatible entropy vectors, which can be compared. In each case there is a distinct method to compute an outer approximation to the respective set. 
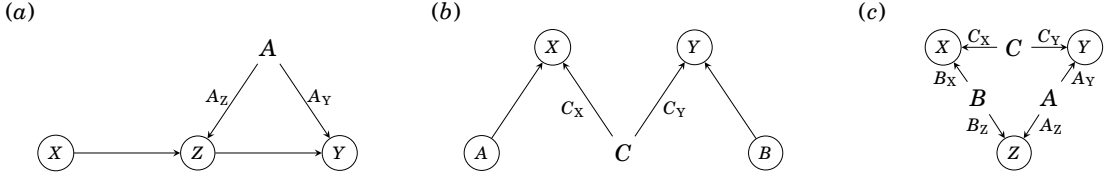
\begin{figure}
\centering 
\resizebox{0.95\columnwidth}{!}{%
\begin{tikzpicture} [scale=1.1]
\node (A) at (-2.5,1.5) {$(a)$};
\node[draw=black,circle,scale=0.75] (1) at (-2,-0.5) {$X$};
\node[draw=black,circle,scale=0.75] (2) at (-0,-0.5) {$Z$};
\node[draw=black,circle,scale=0.75] (3) at (2,-0.5) {$Y$};
\node (4) at (1,1) {$A$};

\draw [->,>=stealth] (1)--(2);
\draw [->,>=stealth] (2)--(3);
\draw [->,>=stealth] (4)--(2) node [above,pos=0.8,yshift=+1ex] {$\scriptstyle A_\mathrm{Z}$};
\draw [->,>=stealth] (4)--(3) node [above,pos=0.8,yshift=+1ex] {$\scriptstyle A_\mathrm{Y}$};

\node (B) at (3.5,1.5) {$(b)$};
\node[draw=black,circle,scale=0.75] (B1) at (4,-0.5) {$A$};
\node(B2) at (6,-0.5) {$C$};
\node[draw=black,circle,scale=0.75] (B3) at (8,-0.5) {$B$};
\node[draw=black,circle,scale=0.75] (B4) at (5,1) {$X$};
\node[draw=black,circle,scale=0.75] (B5) at (7,1) {$Y$};

\draw [->,>=stealth] (B1)--(B4);
\draw [->,>=stealth] (B2)--(B4) node [below,pos=0.8,yshift=-1.5ex] {$\scriptstyle C_\mathrm{X}$};
\draw [->,>=stealth] (B2)--(B5) node [below,pos=0.8,yshift=-1.5ex] {$\scriptstyle C_\mathrm{Y}$};
\draw [->,>=stealth] (B3)--(B5);  

\node (C0) at (9.5,1.5) {$(c)$};
\node[draw=black,circle,scale=0.75] (X) at (10.5,1) {$X$};
\node[draw=black,circle,scale=0.75] (Y) at (12.5,1) {$Y$};
\node[draw=black,circle,scale=0.75] (Z) at (11.5,-0.5) {$Z$};
\node (A) at (12,0.28) {$A$};
\node (B) at (11,0.28) {$B$};
\node (C) at (11.5,1) {$C$};
\node (00) at (13.0,0) { };

\draw [->,>=stealth] (A)--(Y)  node [right,pos=0.15] {$\scriptstyle A_\mathrm{Y}$};
\draw [->,>=stealth] (A)--(Z)  node [right,pos=0.5] {$\scriptstyle A_\mathrm{Z}$};
\draw [->,>=stealth] (B)--(X)  node [left,pos=0.15] {$\scriptstyle B_\mathrm{X}$};
\draw [->,>=stealth] (B)--(Z)  node [left,pos=0.5] {$\scriptstyle B_\mathrm{Z}$};
\draw [->,>=stealth] (C)--(X)  node [above,pos=0.5,yshift=-0.5ex] {$\scriptstyle C_\mathrm{X}$};
\draw [->,>=stealth] (C)--(Y)  node [above,pos=0.5,yshift=-0.5ex] {$\scriptstyle C_\mathrm{Y}$};
\end{tikzpicture}
}%
\caption[Illustration of the instrumental causal structure, the bipartite Bell causal structure and the triangle causal structure]{(a) Pearl's instrumental causal structure, $\inst$, previously introduced in Chapter~\ref{chap:prelim}. Recall that the subsystem  $A_{\mathrm{Z}}$ and $A_{\mathrm{Y}}$ are both considered to be parents of $Z$ (and $Y$).  
 (b) Bipartite Bell scenario, $\bellsc$. The observed variables $A$ and $B$ together with an unobserved system $C$ are used to generate outputs $X$ and $Y$ respectively. In the classical case $C$ can be modelled as a random variable, in the quantum case it is a quantum state on a Hilbert space $\cH_C\cong\cH_{C_{\mathrm{X}}}\otimes\cH_{C_{\mathrm{Y}}}$. (c) Triangle causal structure, $\tri$. Three observed random variables $X$, $Y$ and $Z$ share pairwise common causes $A$, $B$ and $C$, which in the classical case are again modelled by random variables and in the quantum case as states on Hilbert spaces $\cH_A\cong\cH_{A_{\mathrm{Y}}}\otimes\cH_{A_{\mathrm{Z}}}$, $\cH_B\cong\cH_{B_{\mathrm{X}}}\otimes\cH_{B_{\mathrm{Z}}}$ and $\cH_C\cong\cH_{C_{\mathrm{X}}}\otimes\cH_{C_{\mathrm{Y}}}$, respectively.}
\label{fig:generalexamples}
\end{figure}

\subsubsection{Entropy cones for classical causal structures} \label{sec:classical_method}
In the classical case, we aim to characterise (or approximate) the set of entropy vectors of distributions that are compatible with a classical causal structure, $C^\cC$, (or its closure)
\begin{equation}
\marcone{C^{\cC}} \defeq \overline{\left\{ w \in \mathbb{R}^{2^{k}-1}_{\geq 0} \ \middle| \ \exists P \in \mardist{C^{\cC}} \sth w=\bH(P) \right\} } \ .
\end{equation} 
Compared to the unrestricted case introduced in Section~\ref{sec:classicalcone}, the causal structure adds constraints in terms of the encoded conditional independence relations and it requires the consideration of unobserved variables. The first \emph{entropy} inequalities for a marginal scenario were derived in~\cite{Steudel2015}, where certificates for the existence of common ancestors of a subset of the observed random variables of at least a certain size were given. A systematic entropy vector approach was introduced in~\cite{Chaves2012, Fritz2013, Chaves2014b}. Subsequently, outer approximations to the sets of compatible entropy vectors of a variety of causal structures were derived~\cite{Henson2014}. In the following, we outline the systematic approach, providing rigorous proofs of convexity for the relevant sets in Lemma~\ref{lemma:convexity} and Lemma~\ref{lemma:convexity2}, which are important also for the validity of the results of the subsequent chapters.

According to Proposition~\ref{prop:localmark} all entropic restrictions due to the conditional independence relations in a causal structure $C^\cC$ can be captured with a maximum of $n$ independent equalities~\eqref{eq:indepentr}. Their coefficients can be concisely written in terms of a matrix $M_\mathrm{CI}\left(C^{\cC}\right)$, where CI stands for conditional independence.
We define the set
\begin{equation}
\enset{C^{\cC}} \defeq  \left\{ v \in \ensetk{n} \ \middle| \ M_\mathrm{CI}\left(C^{\cC}\right) \cdot v = 0 \right\} \ ,
\end{equation}
which, as the following lemma confirms, is the set of achievable entropy vectors in $C^{\cC}$.  

\begin{lemma} \label{lemma:convexity}
For a causal structure $C^{\cC}$,
\begin{equation}
\enset{C^{\cC}} = \left\{v \in \mathbb{R}_{\geq 0}^{2^{n}-1}
  \ \middle| \ \exists P  =   \prod_{i=1}^{n} P_{\mathrm{X_i |  X_{i}^{\downarrow_{1}}}}\in\cP_n \sth v =\bH(P) \right\} \ .
\end{equation}
Furthermore, its topological closure is the convex cone
\begin{equation}
\encone{C^{\cC}}=\left\{ v \in \enconek{n} \ \middle| \ M_\mathrm{CI}\left(C^{\cC}\right) \cdot v = 0 \right\} \ .
\end{equation}
\end{lemma}
In the following, we will call $\encone{C^\cC}$ the \emph{entropy cone} of $C^\cC$.
We remark here that in spite of the convexity of $\encone{C^{\cC}}$ for any $C^{\cC}$, the set $\mathcal{P}\left(C^{\cC}\right)$ is in general not convex. This alludes to the fact that significant information about the correlations of the random variables is lost via the mapping from $\dist{C^{\cC}}$ to the corresponding entropy cone $\encone{C^{\cC}}$. Nevertheless, the convexity of $\encone{C^{\cC}}$ is one of the aspects that makes entropic considerations (computationally) worthwhile.

\begin{proof}[Proof of Lemma~\ref{lemma:convexity}]
For the causal structure $C^{\cC}$, let the set of compatible entropy vectors be denoted $E \left(C^{\cC}\right)$, i.e., 
$E \left(C^{\cC}\right)\defeq  \left\{v \in \mathbb{R}_{\geq 0}^{2^{n}-1}  \ \middle| \ \exists P  =   \prod_{i=1}^{n} P_{\mathrm{X_i | X_{i}^{\downarrow_{1}}}} \in \distk{n} \sth v = \bH(P) \right\}.$  
According to Proposition~\ref{prop:localmark} and~\eqref{eq:indepentr},  
${E \left(C^{\cC}\right) = \left\{v \in \ensetk{n} \ \middle| \ \forall i \in \left\{1,~\ldots~,n\right\},~ I(X_i : X_{i}^{\nuparrow} |X_{i}^{\downarrow_{1}})=0  \right\}}$.  Applying the definition of $M_\mathrm{CI}\left(C^{\cC}\right)$ yields the desired equality $\enset{C^{\cC}}=E \left(C^{\cC}\right)$.

Now, let us define $F\left( C^{\cC} \right)\defeq \left\{ v \in \enconek{n}\ \middle| \ M_\mathrm{CI}\left(C^{\cC}\right) \cdot v = 0 \right\} \subseteq \mathbb{R}^{2^{n}-1}$. This is a closed convex set, since $\enconek{n}$ is known to be closed and convex and since restricting the closed convex cone $\enconek{n}$ with linear equalities retains these properties.\footnote{More precisely, as the set of solutions to the matrix equality $M_\mathrm{CI}\left(C^{\cC}\right) \cdot v = 0$ is also closed and convex, $F(C^{\cC})$ is the intersection of two closed convex sets and as such also closed and convex. To see that the intersection is convex, consider two subsets $A$, $B \subseteq \mathbb{R}^{2^{n}-1}$. Let $x$, $y \in A \cap B$. Then, due to the convexity of $A$ and $B$, every convex combination of $x$ and $y$ lies within $A$ and within $B$. Thus, $A \cap B$ is convex.}. From this we conclude that $\encone{C^{\cC}}=\overline{\left\{ v \in \ensetk{n} \ \middle| \ M_\mathrm{CI}\left(C^{\cC}\right) \cdot v = 0 \right\}}$ is convex, as $\encone{C^{\cC}} = F\left(C^{\cC}\right)$.
To see this, note that as $F\left(C^{\cC}\right)$ is closed, any element $w \in F\left(C^{\cC}\right)$, in particular any element on its boundary, can be obtained as the limit of a sequence of elements $\left\{w_k \right\}_k$ for $k \rightarrow \infty$, where the $w_k$ lie in the interior of $F\left(C^{\cC}\right)$ for all $k$. Hence $w \in \encone{C^{\cC}}$.
\end{proof}

An outer approximation to $\encone{C^{\cC}}$ can be defined by supplementing the $n$ variable Shannon constraints with the conditional independence equalities implied by $C^\cC$,
\begin{equation}
\outcone{C^{\cC}} \defeq  \left\{ v \in \outconek{n} \ \middle| \  M_\mathrm{CI}\left(C^{\cC}\right) \cdot v = 0 \right\} \ .
\end{equation}

\begin{example}[Outer approximation to the entropy cone of the classical instrumental causal structure]
According to Proposition~\ref{prop:localmark}, the instrumental scenario (see Figure~\ref{fig:generalexamples}(a)) has at most $4$ independent conditional independence equalities~\eqref{eq:indepentr}. We find that there are only two,
${I(A:X)=0}$ and ${I(Y:X\mid AZ)=0}$. Hence,
\begin{equation}
\encone{\inst^{\cC}}= \left\{ v \in \enconek{4} \ \middle| \ M_\mathrm{CI}\left(\inst^{\cC}\right) \cdot v = 0 \right\}
\end{equation}
 with
$$ M_\mathrm{CI}\left(\inst^{\cC}\right)=\left( \begin{array}{ccccccccccccccc}
-1 & -1 & 0 & 0 & 1 & 0 & 0 & 0 & 0 & 0 & 0 & 0 & 0 & 0 & 0 \\
0 & 0 & 0 & 0 & 0 & 0 & 1 & 0 & 0 & 0 & 0 & -1 & -1 & 0 & 1 \\
\end{array} \right) \ , $$
where the coordinates are ordered as
$(H(A),H(X),H(Y),H(Z),H(AX),H(AY),H(AZ),H(XY), \\ H(XZ),H(YZ),H(AXY),H(AXZ),H(AYZ),H(XYZ),H(AXYZ))$.  
An outer approximation is 
\begin{equation}
\outcone{\inst^{\cC}}=\left\{ v \in \outconek{4} \ \middle| \ M_\mathrm{CI}\left(\inst^{\cC}\right) \cdot v = 0 \right\} \ .
\end{equation}
\end{example}

To derive constraints on the entropy vectors of the observed marginal scenario, $\cM$, a marginalisation has to be performed. In the entropic picture, this is achieved by dropping the coordinates that represent entropies of sets of variables containing at least one unobserved variable from the vectors. This corresponds to a projection $\pi_\mathcal{M}:\mathbb{R}^{2^{n}-1}\rightarrow\mathbb{R}^{2^{k}-1}$ that returns all entropy vectors $w$ of the observed variables for which there exists at least one entropy vector $v$ in the original scenario with matching entropies on the observed variables~\cite{Chaves2014}. 

The projection of the whole set of entropy vectors, $\enset{\caus^\cC}$, could be obtained by discarding the components that are projected away from every single entropy vector $v \in \enset{\caus^\cC}$. For a polyhedral cone such as $\outcone{\caus^\cC}$ and sometimes $\encone{\caus^\cC}$, its projection can be more efficiently determined from its extremal rays or bounding hyperplanes respectively (see Section~\ref{sec:fm_elim} for details), e.g. with the Fourier-Motzkin elimination algorithm that has been previously used in this context~\cite{Chaves2012, Fritz2013, Chaves2014, Chaves2014b}.

Without any causal restrictions, the entropy cone $\enconek{n}$ is projected to the marginal cone
\begin{equation}
\marconeO \defeq \left\{w \in \mathbb{R}^{2^k-1}_{\geq 0} \ \middle| \ \exists v \in \enconek{n} \sth w= \pi_\cM(v) \right\} \ .
\end{equation}
Note that if we marginalise $\enconek{n}$ over $n-k$ variables we recover the entropy cone for $k$ random variables, i.e., $\overline{\Gamma^{*}_{\mathcal{M}}} = \overline{\Gamma^{*}_{k}}$. The same applies to the outer approximations: the $n$ variable Shannon cone $\outconek{n}$ is projected to the $k$ variable Shannon cone with the mapping $\pi_\cM$, i.e.,
\begin{equation}
\outmarconeO \defeq  \left\{ w \in \mathbb{R}^{2^k-1}_{\geq 0} \ \middle| \  \exists v \in \outconek{n} \sth w= \pi_\cM(v)  \right\} = \outconek{k} \ .
\end{equation}
This follows because the $n$ variable Shannon constraints contain the corresponding $k$ variable constraints as a subset, and since any vector in $\Gamma_k$ can be extended to a vector in $\Gamma_n$, for instance by taking $H(X_{k+1})=H(X_{k+2})= \cdots = H(X_n)=0$ and $H(X_S \cup X_T)=H(X_S)$ for any $X_T \subseteq \left\{ X_{k+1},~X_{k+2},~\ldots~,~X_n \right\}$.

The following lemma confirms that, given a classical causal structure $C^{\cC}$, the same elimination procedure applies and yields the entropy cone of its marginal scenario, $\marcone{C^\cC}$, which we call its \emph{marginal cone}.

\begin{lemma} \label{lemma:convexity2}
The set of entropy vectors compatible with the marginal scenario of the classical causal structure $C^{\cC}$, is the projection of $\encone{C^\cC}$ onto $\cM$, i.e.,
\begin{equation} \label{eq:convex2}
\marcone{C^{\cC}}= \left\{w \in \mathbb{R}^{2^k-1}_{\geq 0} \ \middle| \ \exists v \in \encone{C^{\cC}} \sth  w= \pi_\cM(v) \right\} \ .
\end{equation}
Furthermore, $\marcone{C^{\cC}}$ is a convex cone.
\end{lemma}
\begin{proof}[Proof of Lemma~\ref{lemma:convexity2}]
Let $\cF=\left\{w \in \mathbb{R}^{2^k-1}_{\geq 0} \ \middle| \ \exists v \in \enset{C^{\cC}} \sth  w= \pi_\cM(v) \right\}$.
Note that $w\in \cF$ implies that there exists $v\in\enset{C^{\cC}}$ such that
$w=\pi_\mathcal{M}(v)$. By Lemma~\ref{lemma:convexity}, we have
$v=\bH(P)$ for some
$P=\prod_iP_\mathrm{X_i|X_i^{\downarrow_{1}}}$.  If we take
$P'=\sum_{X_{k+1},\ldots,X_n}P$ then $w=\bH(P')$ and hence
$w\in\marset{C^{\cC}}$.

Conversely, $w\in\marset{C^{\cC}}$ implies that there exists
$P'\in\mardist{C^{\cC}}$ such that $w=\bH(P')$ and hence there exists $P\in\dist{C^\cC}$ such that $P'=\sum_{X_{k+1},\ldots,X_n}P$. If we take $v=\bH(P)$, then $w=\pi_\mathcal{M}(v)$ and hence $w\in\cF$. Taking the topological closure of both sets implies~\eqref{eq:convex2}.
Furthermore, $\marcone{C^{\cC}}$ is a convex cone by Lemma~\ref{lemma:polycones}.
\end{proof}

The projection of $\outcone{C^{\cC}}$ to the marginal scenario $\cM$ is an outer approximation to $\marcone{C^{\cC}}$, given as
\begin{equation}
\outmarcone{C^{\cC}}\defeq \left\{w \in \mathbb{R}^{2^k-1}_{\geq 0} \ \middle| \ \exists v \in \outcone{C^{\cC}} \sth  w= \pi_\mathcal{M}(v) \right\} \ , 
\end{equation} which can be computationally obtained via a Fourier-Motzkin elimination algorithm from the inequality description of $\outcone{C^{\cC}}$. It can be written as $\outmarcone{C^{\cC}}=\left\{w \in \outmarconeO \ \middle| \ M_\mathcal{M}\left(C^{\cC}\right) \cdot w \geq 0 \right\}$, where the matrix $M_\cM\left(C^{\cC}\right)$ encodes the inequalities on
$\mathbb{R}^{2^k-1}_{\geq 0}$ that are implied by
$M_\mathrm{CI}\left(C^{\cC}\right)\cdot v=0$ and $M_\mathrm{SH}^n\cdot
v\geq 0$ (except for the $k$-variable Shannon constraints, which are already included in $\outmarconeO$). In the following we give two examples, that will both reappear in subsequent chapters.

\begin{example}[Outer approximation to the marginal cone of
  the instrumental scenario~\cite{Henson2014,Chaves2014b}] \label{example:IC_class_obs}
  For the instrumental scenario the outer approximation to its
  marginal cone is found by projecting $\outcone{\inst^{\cC}}$
  to its three variable scenario. This yields
  \begin{equation}
  \outmarcone{\inst^{\cC}} =\left\{w \in
    \outmarconeO \ \middle| \ M_\mathcal{M}\left(\inst^{\cC}\right) \cdot w
    \geq 0 \right\} \ ,
  \end{equation}
  where
  $M_\mathcal{M}\left(IC^{\cC}\right)=\left( \begin{array}{ccccccc}
                                               -1 & 0 & 1 & 0 & 0 & -1 & 1 \\
\end{array} \right)$ corresponds to the inequality $I(X:YZ) \leq H(Z)$ from~\cite{Henson2014,Chaves2014b}. 
\end{example}

\begin{example}[Outer approximation to the marginal cone of the classical triangle scenario from~\cite{Chaves2014, Chaves2015}] \label{example:TRI_class_obs}
For the triangle scenario of Figure~\ref{fig:generalexamples}(c), the marginal cone, $\marcone{\tri^{\cC}}$, 
is restricted by the outer approximation,
\begin{equation}
\outmarcone{\tri^{\cC}}=\left\{w\in \outmarconeO \ \middle| \ M_\cM\left(\tri^{\cC}\right) \cdot w \geq 0 \right\} \ ,
\end{equation}
which was explicitly computed by Chaves et al.~\cite{Chaves2014,Chaves2015} and which is reiterated here. It is obtained from all six variable Shannon inequalities and the conditional independence equalities~\eqref{eq:indepentr}, which are in this case $I(A:BCX)=0$, $I(X:AYZ\mid BC)=0$ and appropriate permutations. In addition to the Shannon constraints there are the following $7$ additional inequalities (including permutations), forming the coefficient matrix $M_\cM\left(\tri^{\cC}\right)$:\footnote{Since no explicit linear description of $\encone{\tri^{\cC}}$ is
known, it is impossible to directly compute
$\marcone{\tri^{\cC}}$ with a variable
elimination. 
}
\begin{align}
- H(X)-H(Y)-H(Z)+H(XY)+H(XZ) &\geq 0, \label{eq:margindep}\\
-5 H(X)-5H(Y)-5H(Z)+4H(XY)+4H(XZ)+4H(YZ)-2H(XYZ) &\geq 0, \label{eq:margindep2}\\
-3 H(X)-3H(Y)-3H(Z)+2H(XY)+2H(XZ)+3H(YZ)-H(XYZ) &\geq 0 \ . \label{eq:margindep3}
\end{align}
We will return to this example in Chapters~\ref{chap:inner} and~\ref{chap:nonshan}. 
\end{example}

\subsubsection{Compatible entropy vectors for quantum causal structures} \label{sec:quantumlatent}
For a quantum causal structure, $C^{\qQ}$, the set of compatible probability distributions over the observed variables, $\mardist{C^{\qQ}}$, leads to a set of compatible entropy vectors,
\begin{equation}
\marset{C^{\qQ}}
\defeq  \left\{w \in \mathbb{R}^{2^{k}-1}_{\geq 0} \ \middle| \ \exists P \in \mardist{ C^{\qQ} } \sth  w= \bH(P) \right\} \ .
\end{equation}

We shall in the following consider the topological closure of this set to ensure that $\marcone{C^{\cC}} \subseteq \marcone{C^{\qQ}}$. In the quantum case, it is, however, unknown whether $\marcone{C^{\qQ}}$ is a convex cone.

A framework for approximating $\marcone{C^{\qQ}}$ was introduced in~\cite{Chaves2015} and corresponding procedures are outlined in the following.\footnote{Our account of quantum causal structures is (slightly) adapted from~\cite{Chaves2015} to unify the description of quantum and generalised causal structures, as explained in Section~\ref{sec:quantum_ns_causal}.}
For the construction of an outer approximation, $\outmarcone{C^{\qQ}}$, an entropy is associated to each random variable and to each subsystem of a quantum state in $C^{\qQ}$ (equivalently to each edge originating at a quantum node), the von Neumann entropy of the respective system. 
Because of the impossibility of cloning, the observed outcomes and the quantum systems they are generated from do not exist simultaneously. Therefore, there is in general no joint multi-party quantum state for all subsystems in a causal structure\footnote{Attempts to circumvent this have been made~\cite{Leifer2013}.} and we cannot sensibly consider the joint entropy of such states and outcomes.  More concretely, if a system $A$ is measured to produce $Z$, then $\rho_{AZ}$ is not defined and hence neither is $H(AZ)$.
\begin{definition}\label{def:coex}
Two subsystems in a quantum causal structure $C^{\qQ}$ \emph{coexist} if neither one is a quantum ancestor of the other. A set of subsystems that mutually coexist is termed \emph{coexisting}.
\end{definition}
A quantum causal structure may have several maximal coexisting sets. Only within such subsets is there a well defined joint quantum state and joint entropy.

\begin{example}[Coexisting sets for the quantum instrumental scenario] \label{example:coexsets} Consider the quantum version of the instrumental scenario, $\inst^\qQ$, illustrated in Figure~\ref{fig:generalexamples}(a). There are three observed variables as well as two edges originating at unobserved (quantum) nodes, hence $5$ variables to consider. More precisely, the quantum node $A$ has two associated subsystems $A_{\mathrm{Z}}$ and $A_{\mathrm{Y}}$. The correlations seen at the two observed nodes $Z$ and $Y$ are formed by measurement on the respective subsystems $A_{\mathrm{Z}}$ and $A_{\mathrm{Y}}$. The maximal coexisting sets in this causal structure are
$\left\{ A_{\mathrm{Y}},~A_{\mathrm{Z}},~X \right\}$,  $\left\{  A_{\mathrm{Y}},~X,~Z \right\}$ and $\left\{ X,~Y,~Z \right\}$.
\end{example}

Without loss of generality we can assume that any  initial, i.e., parentless quantum states (such as $\rho_A$ above) are pure. This is because any mixed state can be purified, and if the transformations and measurement operators are then taken to act trivially on the purifying systems the same statistics are observed. In the causal structure of Example~\ref{example:coexsets}, this implies that $\rho_A$ can be considered to be pure and thus $H(A_{\mathrm{Y}} A_{\mathrm{Z}})=0$. The Schmidt decomposition then implies that $H(A_{\mathrm{Y}})=H(A_{\mathrm{Z}})$. This is computationally useful as it reduces the number of free parameters in the entropic description of the scenario. Furthermore, by Stinespring's theorem~\cite{Stinespring}, whenever a CPTP map is applied at a node that has at least one quantum child, then one can instead consider an isometry to a larger output system. The additional system that is required for this can be taken to be part of the unobserved quantum output (or one of them in case of several quantum output nodes). Each such case allows for the reduction of the number of variables by one, since the joint entropy of all inputs to such a node must be equal to that of all its outputs.

The von Neumann entropy of a multi-party quantum state is submodular~\cite{Lieb1973} and
obeys the following condition:
\begin{itemize}
\item Weak monotonicity~\cite{Lieb1973}: For
  all $X_S$, $X_T\subseteq \Omega$\footnote{ Recall that $H(\left\{ \right\})=0$.}
 \begin{equation}
 H(X_S\setminus X_T) + H(X_T\setminus X_S) \leq H(X_S)+H(X_T) \, .
 \end{equation}
\end{itemize} 
This is the dual of submodularity in the sense that the two
inequalities can be derived from each other by considering
purifications of the corresponding quantum states~\cite{Araki1970}.
Within the context of causal structures, these relations can always be
applied between variables in the same coexisting set. In addition,
whenever it is impossible for there to be entanglement between the
subsystems $X_S\cap X_T$ and $X_S\setminus X_T$ --- for instance if
these subsystems are in a cq-state --- the monotonicity constraint
$H(X_S\setminus X_T)\leq H(X_S)$ holds.  If it is also impossible for
there to be entanglement between $X_S\cap X_T$ and $X_T\setminus X_S$,
then the monotonicity relation $H(X_T\setminus X_S)\leq H(X_T)$ holds
rendering the weak monotonicity relation stated above redundant.

Altogether, these considerations lead to a set of basic inequalities containing some Shannon and some weak-monotonicity inequalities, which are conveniently expressed in a matrix $M_\mathrm{B}\left(C^{\qQ}\right)$. This way of approximating the set of compatible entropy vectors is inspired by work on the entropy cone for multi-party quantum states~\cite{Pippenger2003}.  Note also that there are no further inequalities for the von Neumann entropy known to date~\cite{Linden2005,Cadney2012,Linden2013,Gross2013} (contrary to the classical case where a variety of non Shannon inequalities are known, see Section~\ref{sec:nonShan}). 

The conditional independence constraints in $C^{\qQ}$ cannot be identified by Proposition~\ref{prop:localmark}, because variables do not coexist with any quantum parents and hence conditioning a variable on a
quantum parent is not meaningful. Nonetheless, among the variables in a coexisting set the conditional independencies that are valid for $C^{\cC}$ also hold in $C^{\qQ}$. This can be seen as follows. First, any constraints that involve only observed variables (which always coexist) hold by Proposition~\ref{prop:dseparationgDAG}. Secondly, for unobserved systems only their classical ancestors and none of their descendants can be part of the same coexisting set. An
unobserved system is hence independent of any subset of the same coexisting set with which it shares no ancestors. Note that each of the subsystems associated with a quantum node is considered to be a parent of all of the node's children (see Figure~\ref{fig:generalexamples}(a) for an example). Thirdly, suppose $X_S$ and $X_T$ are disjoint subsets of a coexisting set, $\Xi$, and that the unobserved system $A_X$ is also in $\Xi$.  Then $I(A_X:X_S|X_T)=0$ if $X_T$ d-separates $A_X$ from $X_S$ (in the full
graph including quantum nodes).\footnote{This follows because any quantum states generated from the classical separating variables may be obtained by first producing random variables from the latter (for which the usual d-separation rules hold) and then using these to generate the quantum states in question (potentially after generating other variables in the network), hence retaining conditional independence.} 
The same considerations can be made for sets of unobserved systems. All independence constraints can be assembled in a matrix $M_\mathrm{QCI}\left(C^{\qQ}\right)$.

Among the variables that do not coexist, some are obtained from others by means of quantum operations. These variables are thus related by data processing inequalities~(DPIs)~\cite{BookNielsenChuang2000}.
\begin{proposition}[DPI] \label{prop:DPI} Let $\rho_\mathrm{X_S X_T} \in
  \mathcal{S}(\mathcal{H}_{X_\mathrm{S}} \otimes
  \mathcal{H}_{X_{\mathrm{T}}})$ and $\mathcal{E}$ be a completely
  positive trace preserving (CPTP) map~\footnote{Note that as the map from a quantum state to the diagonal state with entries equal to the outcome probabilities of a measurement is a CPTP map and hence also obeys the DPI.} on
  $\mathcal{S}(\mathcal{H}_{X_{\mathrm{T}}})$\footnote{Note that in
    general $\mathcal{E}$ can be a map between operators on different
    Hilbert spaces, i.e.\
    $\mathcal{E}:\mathcal{S}(\mathcal{H}'_{X_{\mathrm{T}}})
    \rightarrow \mathcal{S}(\mathcal{H}''_{X_{\mathrm{T}}})$. However,
    as we can consider these operators to act on the same larger
    Hilbert space we can w.l.o.g.\ take $\mathcal{E}$ to be a map on
    this larger space, which we call
    $\mathcal{S}(\mathcal{H}_{X_{\mathrm{T}}})$.} leading to a state
  $\rho'_\mathrm{X_{\mathrm{S}} X_{\mathrm{T}}}$.  Then
\begin{equation}
{I(X_{\mathrm{S}} : X_{\mathrm{T}})_{\rho'_\mathrm{X_S X_T}} \leq I(X_{\mathrm{S}} : X_{\mathrm{T}})_{\rho_\mathrm{X_S X_T}}} \ .
\end{equation}
\end{proposition}
The data processing inequalities
provide an additional set of entropy inequalities, which can be expressed in terms of a matrix inequality $M_\mathrm{DPI}\left(C^{\qQ}\right) \cdot
v \geq 0$.\footnote{There are also DPIs for conditional mutual
  information, e.g., $I(A:B|C)_{\rho'_{ABC}}\leq
  I(A:B|C)_{\rho_{ABC}}$ for
  $\rho'_{ABC}=(\cI\otimes\cE\otimes\cI)(\rho_{ABC})$, but these are
  implied by Proposition~\ref{prop:DPI}, so they need not be treated separately.} 
In general, there are a large number of variables for which data processing inequalities hold. It is thus beneficial to derive rules that specify which of the inequalities are needed. 
First, note that whenever a concatenation of two CPTP maps $\mathcal{E}_1$ and $\mathcal{E}_2$, $\mathcal{E}=\mathcal{E}_2 \circ \mathcal{E}_1$, is applied to a state, then any DPIs for inputs and outputs of $\mathcal{E}$ are implied by the DPIs for $\mathcal{E}_1$ and $\mathcal{E}_2$. This follows from combining the DPIs for input and output states of $\mathcal{E}_1$ and $\mathcal{E}_2$ respectively. Hence, the DPIs for composed maps $\mathcal{E}$ never have to be considered as separate constraints. 
Secondly, whenever  a state $\rho_\mathrm{X_S X_T X_R} \in
  \mathcal{S}(\mathcal{H}_{X_\mathrm{S}} \otimes
  \mathcal{H}_{X_{\mathrm{T}}}\otimes
  \mathcal{H}_{X_{\mathrm{R}}})$ can be decomposed as $\rho_\mathrm{X_S X_T X_R}=\rho_\mathrm{X_S X_T}\otimes \rho_\mathrm{X_R}$ and a CPTP map $\mathcal{E}$ transforms the state on $\mathcal{S}(\mathcal{H}_{X_{\mathrm{S}}})$. Then any DPIs for $\rho_\mathrm{X_S X_T X_R}$ are implied by the DPIs for  $\rho_\mathrm{X_S X_T}$, which follows from  $I(X_{\mathrm{S}}:X_{\mathrm{T}}X_{\mathrm{R}})=I(X_{\mathrm{S}}:X_{\mathrm{T}})$, $I(X_{\mathrm{S}}X_{\mathrm{R}}:X_{\mathrm{T}})=I(X_{\mathrm{S}}:X_{\mathrm{T}})$ and $I(X_{\mathrm{S}}X_{\mathrm{T}}:X_{\mathrm{R}})=0$.
Furthermore, whenever a node has classical and quantum inputs, there
is not only a CPTP map generating its output state but this map can
be extended to a CPTP map that simultaneously retains the classical inputs, which leads to tighter inequalities. 
\begin{lemma}\label{lemma:simplify_DPI_classicalquantum}
Let $Y$ be a node with classical and quantum inputs
$X_\mathrm{C}$ and $X_\mathrm{Q}$ and $\cE$ be a CPTP map that acts at
this node, i.e., $\cE$ is a map from
$\cS(\cH_{X_\mathrm{C}}\otimes\cH_{X_\mathrm{Q}})$ to $\cS(\cH_Y)$. Then
$\cE$ can be extended to a map
$\cE':\cS(\cH_{X_\mathrm{C}}\otimes\cH_{X_\mathrm{Q}})\to\cS(\cH_{X_\mathrm{C}}\otimes\cH_Y)$
such that
$\cE':\rho_{X_\mathrm{C}X_\mathrm{Q}}\mapsto\rho'_{X_\mathrm{C}Y}$
with the property that $\rho'_{X_\mathrm{C}Y}$ is classical on $\cH_{X_\mathrm{C}}$ and $\rho'_{X_\mathrm{C}}=\rho_{X_\mathrm{C}}$.
Furthermore, the DPIs for $\cE'$ imply those for $\cE$.
\end{lemma}

\begin{proof} 
  The first part of the lemma follows because classical information
  can be copied, and hence $\cE'$ can be decomposed into first copying
  $X_\mathrm{C}$, and then performing $\cE$.  (Alternatively, we can
  think of $\cE$ as the concatenation of $\cE'$ with a partial trace;
  this allows us to use the same output state $\rho'$ for both maps in
  the argument below.)

  Suppose
  $\cE:\rho_{X_\mathrm{C}X_\mathrm{Q}}\mapsto\rho'_Y$. The
  second part follows because if
  ${I( X_\mathrm{C}  X_\mathrm{Q}  X_\mathrm{S} :
    X_\mathrm{T})_{\rho}} \geq {I(Y  X_\mathrm{S} :
    X_\mathrm{T})_{\rho'}}$
  is a valid DPI for $\cE$ then
  ${I( X_\mathrm{C}  X_\mathrm{Q}  X_\mathrm{S} :
    X_\mathrm{T})_{\rho} \geq I( X_\mathrm{C}  Y  X_\mathrm{S}
    : X_\mathrm{T})_{\rho'}}$
  is valid for $\cE'$.  The second of these implies the first by the
  strong subadditivity relation
  ${I( X_\mathrm{C}  Y  X_\mathrm{S} : X_\mathrm{T})_{\rho'}} \geq {I(Y
   X_\mathrm{S} : X_\mathrm{T})_{\rho'}}$.
\end{proof}

All the above (in)equalities are necessary conditions for a vector to be an entropy vector compatible with the causal structure $C^{\qQ}$. They constrain a polyhedral cone in $\mathbb{R}_{\geq 0}^{m}$, where $m$ is the total number of coexisting sets of $C^{\qQ}$,
\begin{equation}\outcone{C^{\qQ}}\defeq \left\{v \in \mathbb{R}_{\geq 0}^{m} \ \middle| \
  M_\mathrm{B}\left(C^{\qQ}\right) \cdot v \geq 0, 
M_\mathrm{QCI}\left(C^{\qQ}\right) \cdot v=0,\text{ and } M_\mathrm{DPI}\left(C^{\qQ}\right) \cdot v \geq 0 \right\} \ .
\end{equation}

\begin{example}[Entropy inequalities for the quantum instrumental scenario]\label{example:quantum_instrumental}
  For $\inst^{\qQ}$, the cone ${\outcone{\inst^{\qQ}}=\left\{v \in
      \mathbb{R}_{\geq 0}^{15} \ \middle| \ M_\mathrm{B}\left(\inst^{\qQ}\right)
      \cdot v \geq 0, M_\mathrm{QCI}\left(\inst^{\qQ}\right) \cdot
      v=0 \text{ and } M_\mathrm{DPI}\left(\inst^{\qQ}\right) \cdot v
      \geq 0 \right\}}$ involves the matrix
  $M_\mathrm{B}\left(\inst^{\qQ}\right)$ that features $29$
  (independent) inequalities, listed in Section~\ref{sec:29inequ} (note that the only weak monotonicity
  relations that are not made redundant by other basic inequalities are 
  $H(A_Y|A_ZX)+H(A_Y)\geq 0$, $H(A_Z|A_YX)+H(A_Z)\geq 0$, $H(A_Y \mid
  A_Z)+H(A_Y \mid X) \geq 0$ and ${H(A_Z \mid A_Y)}+H(A_Z \mid X) \geq
  0$). In this case a single independence constraint encodes that $X$ is independent of $A_YA_Z$: 
$$M_\mathrm{QCI}\left(\inst^{\qQ}\right)= \left( \begin{array}{ccccccccccccccc}
0 & 0 & -1 & 0 & 0 & -1 & 0 & 0 & 0 & 0 & 0 & 0 & 1 & 0 & 0 
\end{array} \right)\, .$$
Two data processing inequalities are
required (cf.\ Lemma~\ref{lemma:simplify_DPI_classicalquantum}),
$I(A_Z X:A_Y) \geq I(XZ:A_Y)$ and $I(A_YZ:X) \geq I(YZ:X)$, 
which yield a matrix
$$M_\mathrm{DPI}\left(IC^{\qQ}\right)=\left( \begin{array}{ccccccccccccccc}
0 & 0 & 0 & 0 & 0 & 0 & 0 & 0 & 1 & 0 & -1 & 0 & -1 & 1 & 0 \\
0 & 0 & 0 & 0 & 0 & 0 & 0 & 1 & 0 & 0 & 0 & -1 & 0 & -1 & 1 
\end{array} \right).$$
The above matrices are all expressed in terms of coefficients of 
$(H(A_Y), H(A_Z), H(X), H(Y), H(Z), \\ H(A_YA_Z), H(A_YX), H(A_YZ),
H(A_ZX),H(XY), H(XZ), H(YZ), H(A_YA_ZX), H(A_YXZ),
H( X Y Z))$.  Although the notation suppresses the
different states, there is no ambiguity because, for example, the entropy of
$X$ is the same for all states with subsystem $X$. 
\end{example}

From $\outcone{C^{\qQ}}$, an outer approximation to
$\marcone{ C^{\qQ}}$ can be obtained with a Fourier-Motzkin elimination. This leads to
\begin{equation}\outmarcone{C^{\qQ}}\defeq \left\{w \in \mathbb{R}^{2^k-1}_{\geq 0} \ \middle| \ \exists v \in \outcone{C^{\qQ}} \sth  w= \pi_\cM(v) \right\}\ ,
\end{equation} which can be written as
$\outmarcone{C^{\qQ}}=\left\{w \in \outmarconeO \ \middle| \ M_\cM\left(C^{\qQ}\right) \cdot w \geq 0 \right\}.$
The matrix $M_\cM\left(C^{\qQ}\right)$ encodes all
(in)equalities on $\mathbb{R}^{2^k-1}_{\geq 0}$ implied by
$M_\mathrm{B}\left(C^{\qQ}\right) \cdot v \geq 0$,
$M_\mathrm{QCI}\left(C^{\qQ}\right) \cdot v=0$ and
$M_\mathrm{DPI}\left(C^{\qQ}\right) \cdot v \geq 0$ (except for the
Shannon inequalities, which are already included in
$\outmarconeO$). Note that $\outmarcone{C^{\cC}} \subseteq\outmarcone{C^{\qQ}} \subseteq \outmarconeO$, where the first relation holds because all inequalities relevant for quantum states hold in the classical case as well.\footnote{This can be seen by thinking of a classical source as made up of two or more (perfectly correlated) random variables as its subsystems, which are sent to its children and processed there. The Shannon inequalities hold among all of these variables (and also imply any weak monotonicity constraints). The classical independence relations include the quantum ones but may add constraints that involve conditioning on any of the variables' ancestors. These (in)equalities are tighter than the DPIs, which are hence not explicitly considered in the classical case.}

\begin{example}[Outer approximation to the set of compatible entropy vectors for the quantum triangle scenario~\cite{Chaves2015}]\label{example:quantumtriangle}
The outer approximation to $\marcone{\tri^{\qQ}}$ has been computed in~\cite{Chaves2015} and is given here (see also Figure~\ref{fig:generalexamples}(c)). We will revisit this example in Section~\ref{sec:quantumtriangle}. For the systems involved in $\tri^\qQ$, there are only $124 < {2^{9}-1}$ well-defined entropy values in the quantum scenario, subsets of the maximal coexisting sets,
$\left\{A_{\mathrm{Y}},~A_{\mathrm{Z}},~B_{\mathrm{X}},~B_{\mathrm{Z}},~C_{\mathrm{X}},~C_{\mathrm{Y}} \right\}$, $\left\{A_{\mathrm{Y}},~A_{\mathrm{Z}},~B_{\mathrm{Z}},~C_{\mathrm{Y}},~X \right\}$, $\left\{A_{\mathrm{Z}},~B_{\mathrm{X}},~B_{\mathrm{Z}},~C_{\mathrm{X}},~Y \right\}$, $\left\{ A_{\mathrm{Y}},~B_{\mathrm{X}},~C_{\mathrm{X}},~C_{\mathrm{Y}},~Z \right\}$, $\left\{ A_{\mathrm{Z}},~B_{\mathrm{Z}},~X ,~Y \right\}$, $\left\{ A_{\mathrm{Y}},~C_{\mathrm{Y}},~X,~Z \right\}$, $\left\{ B_{\mathrm{X}},~C_{\mathrm{X}},~Y,~Z \right\}$ and $\left\{ X,~Y,~Z \right\}$.

Within these sets positivity, submodularity and weak-monotonicity hold, as well as monotonicity for any cq-states, forming the matrix $M_\mathrm{B}\left(\tri^{\qQ}\right)$.  
The coexisting sets intuitively represent the different stages of the overall system, starting out in a product state $\rho_{\mathrm{ABC}}=\rho_{\mathrm{A_{Y}A_{Z}}} \otimes \rho_{\mathrm{B_{X}B_{Z}}} \otimes \rho_{\mathrm{C_{X}C_{Y}}}$, where $X$, $Y$ and $Z$ are generated by measuring the respective subsystems in any order. The data processing inequalities connect the systems' entropies before and after the individual measurements, they yield another matrix $M_\mathrm{DPI}\left(\tri^{\qQ}\right)$. Conditional independence relations complement these inequalities with a matrix $M_\mathrm{QCI}\left(\tri^{\qQ}\right)$ that encodes the independence constraints
\begin{equation}\label{eq:indepsources}
I(A_\mathrm{Y} A_\mathrm{Z} : B_\mathrm{X} B_\mathrm{Z} C_\mathrm{X} C_\mathrm{Y})=0
\end{equation}
and permutations of the pairs of variables shared by each source.\footnote{These constraints, together with the DPI and the basic inequalities imply all further independencies in the causal structure. For instance, for three sets of variables $X$, $Y$ and $Z$ the inequality $I(X:YZ)=I(X:Y|Z)+I(X:Z)$ holds. Due to the positivity of the mutual information and the conditional mutual information $I(X:YZ)=0$ also implies $I(X:Z)=0$. Similarly, the DPI allow for the derivation of constraints such as $I(A_\mathrm{Y} A_\mathrm{Z} : B_\mathrm{Z} C_\mathrm{Y} X)=0$.}

The marginal cone $\outmarcone{\tri^{\qQ}}$, is given by the Shannon
inequalities for the jointly distributed $X$, $Y$ and $Z$ and the additional
inequality 
\begin{equation}
I(X:Y) + I(X:Z) \leq H(X) \ ,\label{eq:fritzhenson}
\end{equation} 
including permutations of $X$, $Y$ and $Z$~\cite{Chaves2015}.\footnote{We remark that before running the Fourier-Motzkin elimination algorithm we can use equality constraints to eliminate variables and reduce the dimensionality.  We can also take
$\rho_{\mathrm{A_{Y}A_{Z}}}$, $\rho_{\mathrm{B_{X}B_{Z}}}$ and
$\rho_{\mathrm{C_{X}C_{Y}}}$ to be pure to remove six further
variables.}
This outer approximation does not to coincide with the outer approximation $\outmarcone{\tri^{\cC}}$.
\end{example}

In~\cite{Chaves2015}, the quantum method has also been combined with the approach reviewed in Section~\ref{sec:quantum_conditioning} below, where it has been applied to a scenario related to $\inst$ (see Example~\ref{example:information_causality}).

\subsubsection{Compatible entropy vectors for generalised causal structures} \label{sec:gDAGs}

The object of interest for our entropic considerations is set of compatible entropy vectors obtained from $\mardist{C^{\gG}}$, 
\begin{equation}\marset{C^{\gG}}\defeq  \left\{w \in \mathbb{R}^{2^{k}-1}_{\geq 0} \ \middle| \ \exists P \in \mardist{ C^{\gG}} \sth  w= \bH(P) \right\} \ .
\end{equation}

As in the quantum case, we usually consider the topological closure $\marcone{C^{\gG}}$, to ensure $\marcone{C^{\cC}} \subseteq \marcone{C^{\qQ}} \subseteq \marcone{C^{\gG}}$. Whether $\marcone{C^{\gG}}$ is convex is an open question. Causal structures that allow for unobserved non-signalling resources were introduced in  Section~\ref{sec:quantum_ns_causal} in unison with the quantum framework. 
For both, there is no joint state of all systems associated with nodes of the causal structure and one also has to resort to the coexisting sets here (recall Definition~\ref{def:coex}). However, as opposed to the quantum case there is an additional challenge for defining outer approximations: it is not clear how to define entropy for the unobserved non-signalling systems. Hence, a systematic entropic procedure, in which the unobserved variables are explicitly modelled and then eliminated from the description, is not available for generalised causal structures. The
issue is that we are lacking a generalisation of the Shannon and von Neumann entropy to generalised probabilistic theories that is submodular and for which the conditional entropy can be written as the difference of unconditional entropies~\cite{Short2010a, Barnum2012}.

One possible generalised entropy is the \emph{measurement entropy}, which is positive and obeys some of the submodularity constraints (those with $X_S \cap X_T=\{ \}$)~\cite{Short2010a, Barnum2012}. Using this, Ref.~\cite{Cadney2012a} considered the set of possible entropy vectors for a bipartite state in \emph{box
world}, a generalised probabilistic theory that permits all bipartite correlations that are non-signalling~\cite{Popescu1994FP, Barrett07}. They found no further constraints on the set of possible entropy vectors in this setting (hence, contrary to the quantum case, measurement entropy vectors of separable states in box world can violate monotonicity). Other generalised probabilistic theories and multi-party states have, to our knowledge, not been similarly analysed.

Outer approximations to the sets of compatible entropy vectors for generalised causal structures were derived in~\cite{Henson2014}, based on the observed variables and their independencies. For certain generalised causal structures, Ref.~\cite{Henson2014} also provides a few additional inequalities that do not follow from the observed independencies. An example is the inequality ${I(X:Y)}+{I(X:Z)} \leq {H(X)}$ for the triangle causal structure of Figure~\ref{fig:generalexamples}(c), which had previously been established in the classical case~\cite{Fritz2012} and quantum mechanically (cf.\ Example~\ref{example:quantumtriangle} and~\cite{Chaves2015}).

\section{Entropy vector approach with post-selection} \label{sec:post-selection}
A technique that leads to additional, more fine-grained inequalities
is based on post-selecting on the values of parentless classical variables. 
This technique was pioneered by Braunstein and Caves~\cite{Braunstein1988} and has subsequently been used to systematically derive numerous entropy inequalities~\cite{Chaves2012, Fritz2013, Chaves2013, Pienaar2016, Chaves2016}. 

\subsection{Post-selection in classical causal structures}\label{sec:classical_conditioning}
In the following we denote a random variable $X$ post-selected on the event of another random variable, $Y$, taking a particular value, $Y=y$, as $X_{\mid Y=y}$. The same notation is used for a set of random variables $\Omega=\left\{X_1,~X_2,~\ldots~,~X_n \right\}$, whose joint distribution is conditioned on ${Y=y}$, $\Omega_\mathrm{ \mid Y=y}=\left\{X_{1 \mathrm{ \mid Y=y}},~X_{2 \mathrm{ \mid Y=y}},~\ldots~,~X_{ n \mathrm{ \mid Y=y}} \right\}$. The following proposition can be understood as a generalisation of (a part of) Fine's theorem~\cite{Fine1982, Fine1982a}. 

\begin{lemma}\label{lemma:justify_conditioning}
  Let $C^{\cC}$ be a classical causal structure with a parentless
  observed node $X$ that takes values $X = 1,2,\ldots,n$ and let $P$ be
  a compatible joint distribution over all random variables
  $X \cup X^{\uparrow} \cup X^{\nuparrow}$ in $C^{\cC}$. 
  Then there exists a joint distribution $Q$ over the
  $n \cdot \left|X^{\uparrow}\right| + \left|X^{\nuparrow}\right|$
  random variables
  $\Omega_\mathrm{ \mid X} \defeq  X^{\uparrow}_\mathrm{ \mid X=1} \cup
  X^{\uparrow}_\mathrm{ \mid X=2} \cup \cdots \cup X^{\uparrow}_\mathrm{ \mid
    X=n} \cup X^{\nuparrow}$
  such that
  $Q( X^{\uparrow}_\mathrm{ \mid X=x} X^{\nuparrow})=P( X^{\uparrow}
  X^{\nuparrow} \mid X=x)$ for all
  $x\in\{1,\ldots,n\}$. 
\end{lemma}
\begin{proof} 
  The joint distribution over the random variables
  $X^{\uparrow} \cup X^{\nuparrow}$ in $C^{\cC}$ can be written as
  ${P(X^{\uparrow}  X^{\nuparrow})}= {\sum_{x=1}^{n} P( X^{\uparrow}
    \mid X^{\nuparrow} X=x) P( X^{\nuparrow}) P(X=x)}$.
  Now take
\begin{equation}
 Q( X^{\uparrow}_\mathrm{ \mid X=1}~\cdots~ X^{\uparrow}_\mathrm{\mid X=n}
  X^{\nuparrow})= {\prod_{x=1}^{n} {P( X^{\uparrow}\mid
  X^{\nuparrow}X=x)}P( X^{\nuparrow})} \ .
\end{equation}
  This distribution has marginals
  $Q( X^{\uparrow}_\mathrm{ \mid X=x} X^{\nuparrow})={P( X^{\uparrow} \mid X^{\nuparrow} X=x)}  P( X^{\nuparrow})$, as required. 
\end{proof}

It is perhaps easiest to think about this lemma in terms of a
new causal structure $C_X^{\cC}$ on $\Omega_\mathrm{ \mid X}$ that is related to
the original.  Roughly speaking the new causal structure is formed by
removing $X$ and replacing the descendants of $X$ with several copies
each of which have the same causal relations as in the original causal
structure (with no mixing between copies).  More precisely, if $X$ is
a parentless node in $C^{\cC}$ we can form a \emph{post-selected causal
  structure} (post-selecting on $X$) on $\Omega_\mathrm{ \mid X}$, which we call $C_X^{\cC}$, as
follows:
\begin{enumerate}[(1)]
\item For each pair of nodes $A$ and $B$ that are in
$X^{\nuparrow}$ in $C^{\cC}$, make $A$ a parent of $B$ in $C_X^{\cC}$ iff $A$ is a parent of $B$ in $C^{\cC}$. 
\item For each node $B$ that is in $X^{\nuparrow}$ in $C^{\cC}$ and for each node $A_{\mid X=x}$, make $B$ a parent of $A_{\mid X=x}$ in $C_X^{\cC}$ iff $B$ is a parent of $A$ in $C^{\cC}$. 
\item For each pair of nodes, $A_{\mid X=x}$ and $B_{\mid X=x}$, make $B_{\mid X=x}$ a parent of $A_{\mid X=x}$ in $C_X^{\cC}$ iff $B$ is a parent of $A$ in $C^{\cC}$. There is no mixing between different values of $X=x$. 
\end{enumerate}
See~\ref{fig:instrumental_conditioned} for post-selected versions of the causal structures of Figure~\ref{fig:generalexamples}. Note that the technique is not applicable to the triangle causal structure where there are no observed parentless nodes. This viewpoint on post-selection leads to the following corollary of
Lemma~\ref{lemma:justify_conditioning}, which is an alternative
generalisation of Fine's theorem
\begin{corollary}\label{lemma:altfine}
  Let $C^{\cC}$ be a classical causal structure with a parentless
  observed node $X$ that takes values $X = 1,2,\ldots,n$ and let $P$ be
  a compatible joint distribution over all random variables
  $X \cup X^{\uparrow} \cup X^{\nuparrow}$ in $C^{\cC}$.
  Then there exists a joint distribution
  $Q$ compatible with the post-selected causal structure, $C^\cC_X$, such
  that
  $Q( X^{\uparrow}_\mathrm{ \mid X=x} X^{\nuparrow})=P( X^{\uparrow}
  X^{\nuparrow} \mid X=x)$ for all $x\in\{1,\ldots,n\}$.
\end{corollary}

Since they correspond to distributions in the original scenario, the distributions that are of interest in this new causal structure
are the marginals
$P(X^{\uparrow}_\mathrm{ \mid X=x} X^{\nuparrow} X^{\nuparrow})$ for all
$x$ (and their interrelations).  Any constraints on these distributions derived
in the post-selected scenario are by construction valid for the
distributions compatible with the original causal structure.

\begin{example}[Post-selection in the instrumental
  scenario] \label{example:conditional} Consider the causal structure
  $\inst$ where the parentless variable $X$ takes values $x \in \left\{0,1\right\}$. 
  For any $P$ compatible with $\inst^{\cC}$, there exists a distribution
  $Q$ compatible with the post-selected causal structure of
  Figure~\ref{fig:instrumental_conditioned}(a), $\inst^{\cC}_X$, such that
  ${Q(Z_{\mid X=0} Y_{\mid X=0} A)}={P(Z Y \mid A X=0)}P(A)$ and
  ${Q(Z_{\mid X=1} Y_{\mid X=1} A)}={P(Z Y \mid A X=1)}P(A)$. These
  marginals and their relation are of interest for analysing the original
  scenario.
\begin{figure}
\centering 
\resizebox{0.85\columnwidth}{!}{%
\begin{tikzpicture} [scale=0.6]
\node (1) at (-8,0.25){$(a)$};
\node[draw=black,circle,scale=0.6] (2) at (-6,-0.25) {$Z_{\mid X=0}$};
\node[draw=black,circle,scale=0.6] (2b) at (-6,-1.75) {$Z_{\mid X=1}$};
\node[draw=black,circle,scale=0.6] (3) at (2,-0.25) {$Y_{\mid X=0}$};
\node[draw=black,circle,scale=0.6] (3b) at (2,-1.75) {$Y_{\mid X=1}$};
\node (4) at (-2,-1.0) {$A$};

\draw [->,>=stealth] (2)--(3);
\draw [->,>=stealth] (4)--(2);
\draw [->,>=stealth] (4)--(3);
\draw [->,>=stealth] (2b)--(3b);
\draw [->,>=stealth] (4)--(2b);
\draw [->,>=stealth] (4)--(3b);

\node (M1) at (5,0.25){$(b)$};
\node[draw=black,circle,scale=0.6] (M2) at (7,-0.25) {$X_{\mid A=0}$};
\node[draw=black,circle,scale=0.6] (M2b) at (7,-1.75) {$X_{\mid A=1}$};
\node[draw=black,circle,scale=0.6] (M3) at (15,-0.25) {$Y_{\mid B=0}$};
\node[draw=black,circle,scale=0.6] (M3b) at (15,-1.75) {$Y_{\mid B=1}$};
\node (M4) at (11,-1.0) {$C$};

\draw [->,>=stealth] (M4)--(M2);
\draw [->,>=stealth] (M4)--(M3);
\draw [->,>=stealth] (M4)--(M2b);
\draw [->,>=stealth] (M4)--(M3b);
\end{tikzpicture}
}%
\caption[Post-selected instrumental and post-selected Bell causal structure]{(a) Pearl's instrumental scenario post-selected on binary
  $X$, called $\inst_X$. The causal structure is obtained from the $\inst$ by removing $X$ and replacing $Y$ and $Z$ with copies, each of which have the same causal relations as in the original causal structure. (b)
  Post-selected Bell scenario with binary inputs $A$ and $B$.}
\label{fig:instrumental_conditioned}
\end{figure}
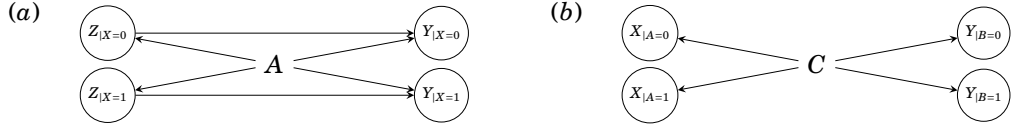
\end{example}

Note that the above reasoning may be applied recursively. Indeed, the causal structure with variables $\Omega_\mathrm{ \mid X}$ may be post-selected on the values of one of its parentless nodes. 
The joint distributions of the nodes $\Omega_{\mid X}$ and the associated causal structure may be analysed in terms of entropies, as illustrated with the following example.

\begin{example}[Entropy inequalities for the post-selected Bell scenario~\cite{Braunstein1988}]\label{example:bell}
  In the bipartite Bell scenario of Figure~\ref{fig:generalexamples}(b) with binary inputs $A$ and $B$,
  Corollary~\ref{lemma:altfine} may be applied first to
  post-select on the values of $A$ and then of $B$. This leads to a
  distribution $Q$ compatible with the post-selected causal structure
  (on $A$ and $B$) shown in Figure~\ref{fig:instrumental_conditioned}(b),
  for which $Q(X_\mathrm{ \mid A=a} Y_\mathrm{ \mid B=b})={P(X Y \mid A=a, B=b)}$ for
  $a, b \in \left\{0,1\right\}$.\footnote{In this case the joint distribution is already known to
    exist by Fine's theorem~\cite{Fine1982, Fine1982a}.}
  Applying the
  entropy vector method to the post-selected causal structure and
  marginalising to vectors,
\begin{multline*}
 \left(H(X_\mathrm{ \mid A=0}),H(X_\mathrm{ \mid A=1}),H(Y_\mathrm{ \mid B=0}),H(Y_\mathrm{ \mid B=1}),  H(X_\mathrm{ \mid A=0}Y_\mathrm{ \mid B=0}), \right. \\ \left. H(X_\mathrm{ \mid A=0}Y_\mathrm{ \mid B=1}),  H(X_\mathrm{ \mid A=1}Y_\mathrm{ \mid B=0}),  H(X_\mathrm{ \mid A=1}Y_\mathrm{ \mid B=1}) \right) \ ,
\end{multline*}
yields the inequality $H(Y_1|X_1)+H(X_1|Y_0)+H(X_0|Y_1)-H(X_0|Y_0) \geq 0$ and its permutations~\cite{Braunstein1988, Chaves2012}.\footnote{Note that whenever the input nodes take more than two values, the latter may be partitioned into two sets, guaranteeing applicability of the inequality. Furthermore, \cite{Chaves2013} showed that these inequalities are sufficient for detecting any behaviour that is not classically reproducible in the Bell scenario where the two parties perform measurements with binary outputs. However, this result is proven indirectly, relying on convex combinations of observed, non-local distributions with other achievable local distributions. Hence, it is not generalisable to non-convex causal structures. Moreover, even for more general causal structures featuring a convex set of achievable distributions it is not clear how such mixing would help certify non-classicality.}
\end{example}

The extension of Fine's theorem to more general Bell scenarios~\cite{Liang2011, Abramsky2011}, i.e., to scenarios involving a number of spacelike separated parties that each choose input values and produce some output random variable (and scenarios that can be reduced to the latter), has been combined with the entropy vector method in~\cite{Chaves2012, Fritz2013}. 
Entropy inequalities that are derived in this way provide novel and non-trivial necessary conditions for the distributions compatible with the original classical causal structure. Ref.~\cite{Fritz2013} introduced this idea and analysed the so-called $n$-cycle scenario, which is of particular interest in the context of non-contextuality and includes the Bell scenario (with binary inputs and outputs) as a special case.\footnote{A full probabilistic characterisation of the $n$-cycle scenario was given in~\cite{Araujo2013}.}
In Ref.~\cite{Chaves2012}, new entropic inequalities for the bilocality scenario, which is relevant for entanglement swapping~\cite{Branciard2010, Branciard2012}, as well as quantum violations of the classical constraints on the 4- and 5-cycle scenarios were derived. 
It was proven in~\cite{Chaves2013} that for the $n$-cycle scenario, the (polynomial number of) entropic inequalities are sufficient for the detection of any non-local distributions (just as the exponential number of inequalities in the probabilistic case~\cite{Araujo2013}). 
In the following we illustrate the method of~\cite{Chaves2012, Fritz2013} with a continuation of Example~\ref{example:conditional}.

\begin{example}[Entropy inequalities for the post-selected instrumental scenario] \label{example:xxx}
  The entropy vector method from Section~\ref{sec:entropicappr} is
  applied to the $5$-variable causal structure of
  Figure~\ref{fig:instrumental_conditioned}(a). The marginalisation is
  performed to retain all marginals that correspond to (post-selected) distributions
  in the original causal structure of Figure~\ref{fig:generalexamples}(a),
  i.e., any marginals of ${P(Y Z \mid X=0)}$ and ${P(Y Z \mid
    X=1)}$.
  Hence, the $5$ variable entropy cone is projected to a cone that
  restricts vectors of the form
  ${(H(Y_\mathrm{ \mid X=0}),~H(Y_\mathrm{ \mid X=1}),~H(Z_\mathrm{ \mid
      X=0}),~H(Z_\mathrm{ \mid X=1}),~H(Y_\mathrm{ \mid X=0}Z_\mathrm{ \mid
      X=0}),~H(Y_\mathrm{ \mid X=1}Z_\mathrm{ \mid X=1}))}$.
  Note that entropies of unobserved marginals such as
  $H(Y_\mathrm{ \mid X=0}Z_\mathrm{ \mid X=1})$ are not included.  With this
  technique, the Shannon constraints for the three variables
  ${(H(Y_\mathrm{ \mid X=0}),~H(Z_\mathrm{ \mid X=0}),~H(Y_\mathrm{ \mid
      X=0}Z_\mathrm{ \mid X=0}))}$
  are recovered (the same holds for $X=1$); no additional constraints
  arise here. (Note that without post-selecting, additional constraints are recovered in this scenario (cf.\ Example~\ref{example:IC_class_obs}).)
  
It is interesting to compare this to the Bell scenario of
Example~\ref{example:bell}. In both causal structures \emph{any} $4$-variable
distributions,
$P_{Z_\mathrm{ \mid X=0}Z_\mathrm{ \mid X=1}Y_\mathrm{ \mid X=0}Y_\mathrm{ \mid X=1}}$
and
$P_{X_\mathrm{\mid A=0}X_\mathrm{ \mid A=1}Y_\mathrm{ \mid B=0}Y_\mathrm{ \mid B=1}}$
respectively, are achievable (the additional causal links in
Figure~\ref{fig:instrumental_conditioned}(b) do not affect the set of
compatible distributions). However, the marginal entropy
vector in the Bell scenario has more components, leading to additional
constraints on the observed variables~\cite{Braunstein1988, Chaves2012}.
\end{example}

In some cases two different causal structures, $C_1$ and $C_2$, can
yield the same set of distributions after marginalising, a fact that
has been further explored in~\cite{Budroni2016b}. When this occurs,
either causal structure can be imposed when identifying the set of
achievable marginal distributions in either scenario. If the
constraints implied by the causal structure $C_1$ are a subset of
those implied by $C_2$, then those of $C_2$ can be used to compute
improved outer approximations on the entropy cone for
$C_1$. Such considerations also yield a criterion
  for indistinguishability of causal structures in certain marginal
  scenarios: if $C_1$ and $C_2$ yield the same set of distributions
  after marginalising then they cannot be distinguished in that
  marginal scenario.
Additionally, valid independence constraints may speed up computations even
if they do not lead to any new relations for the observed variables.\footnote{Note that some care has to be taken when identifying valid constraints for scenarios with causal structure~\cite{Budroni2016b}.}

In situations like Example~\ref{example:xxx}, where no new constraints follow from post-selection, it may be possible to introduce additional input variables to the causal structure, which may be beneficial for instance if the aim is to certify the presence of quantum nodes in a network. The new parentless nodes may then be used to apply Lemma~\ref{lemma:justify_conditioning} and the above entropic techniques. Mathematically, introducing further nodes to a causal structure is always possible. However, this is only interesting if experimentally feasible, e.g.\ if an experimenter has control over certain observed nodes and is able to devise an experiment where he can change their inputs. In the instrumental scenario this may be of interest.

\begin{example}[Variations of the instrumental scenario] 
In the instrumental scenario of Figure~\ref{fig:generalexamples}(a), a measurement on system $A_Z$ is performed depending on $X$ (where in the classical case $A_Z$ can w.l.o.g. be taken to be a copy of the unobserved random variable $A$). Its outcome $Z$ (in the classical case a function of $A$) is used to choose another measurement to be performed on $A_Y$ to generate $Y$ (classically another a copy of $A$). 
It may often be straightforward for an experimenter to choose between
several measurements. In the causal structure this corresponds to
introducing an additional observed input $S$ to the second measurement
(with the values of $S$ corresponding to different measurements on
$A_Y$). Such an adaptation is displayed in Figure~\ref{fig:information_causality_DAG}(a).~\footnote{Note that for ternary $S$ the outer approximation of the post-selected causal structure of Figure~\ref{fig:information_causality_DAG}(d) with Shannon inequalities does not lead to any interesting constraints (as opposed to the structure of Figure~\ref{fig:information_causality_DAG}(e), which is analysed further in Example~\ref{example:information_causality}).}

Alternatively, it may be possible that the first measurement (on $A_Z$) is chosen depending on a combination of different, independent factors, which each correspond to a random variable $X_i$. For two variables $X_1$ and $X_2$ the corresponding causal structure is displayed in Figure~\ref{fig:information_causality_DAG}(b). We will revisit this example in Section~\ref{sec:fourfive}. 
Together, the two adaptations yield the causal structure of Figure~\ref{fig:information_causality_DAG}(c), considered in the context of the physical principle of information causality~\cite{Pawlowski2009} (see also Example~\ref{example:information_causality} below).
\end{example}
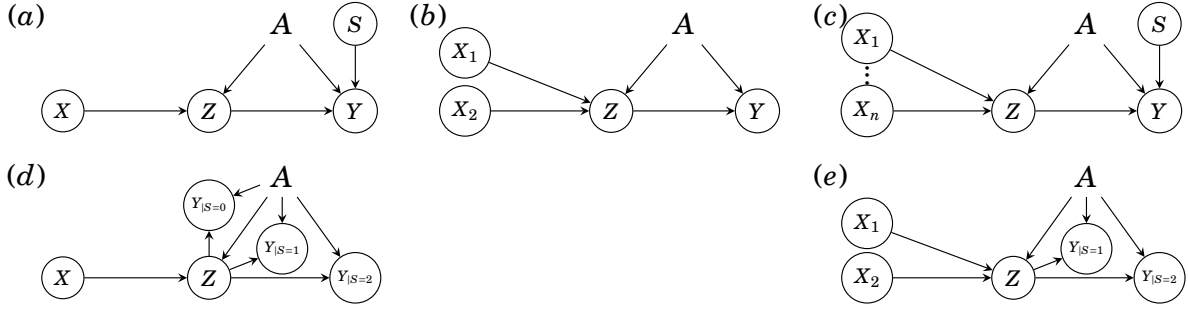
\begin{figure}
\centering 
\resizebox{1\columnwidth}{!}{%
\begin{tikzpicture} [scale=0.9]
\node (B0) at (-13.5,1.3) {$(a)$};
\node[draw=black,circle,scale=0.7] (B1) at (-13,0) {$X$};
\node[draw=black,circle,scale=0.75] (B2) at (-11,0) {$Z$};
\node[draw=black,circle,scale=0.75] (B3) at (-9,0) {$Y$};
\node[draw=black,circle,scale=0.75] (B5) at (-9,1.2) {$S$};
\node (B4) at (-10,1.2) {$A$};
\draw [->,>=stealth] (B1)--(B2);
\draw [->,>=stealth] (B2)--(B3);
\draw [->,>=stealth] (B4)--(B2);
\draw [->,>=stealth] (B4)--(B3);
\draw [->,>=stealth] (B5)--(B3);

\node (A0) at (-8,1.3) {$(b)$};
\node[draw=black,circle,scale=0.7] (A1a) at (-7.5,0.8) {$X_1$};
\node[draw=black,circle,scale=0.7] (A1) at (-7.5,0) {$X_2$};
\node[draw=black,circle,scale=0.75] (A2) at (-5.5,0) {$Z$};
\node[draw=black,circle,scale=0.75] (A3) at (-3.5,0) {$Y$};
\node (A4) at (-4.5,1.2) {$A$};
\draw [->,>=stealth] (A1)--(A2);
\draw [->,>=stealth] (A1a)--(A2);
\draw [->,>=stealth] (A2)--(A3);
\draw [->,>=stealth] (A4)--(A2);
\draw [->,>=stealth] (A4)--(A3);

\node (0) at (-2.5,1.3) {$(c)$};
\node[draw=black,circle,scale=0.7] (1a) at (-2,1) {$X_1$};
\node[draw=black,circle,scale=0.7] (1) at (-2,0) {$X_n$};
\node(1b) at (-2,0.55) {$\cdot$};
\node(1c) at (-2,0.45) {$\cdot$};
\node(1d) at (-2,0.35) {$\cdot$};
\node[draw=black,circle,scale=0.75] (2) at (-0,0) {$Z$};
\node[draw=black,circle,scale=0.75] (3) at (2,0) {$Y$};
\node[draw=black,circle,scale=0.75] (5) at (2,1.2) {$S$};
\node (4) at (1,1.2) {$A$};
\draw [->,>=stealth] (1)--(2);
\draw [->,>=stealth] (1a)--(2);
\draw [->,>=stealth] (2)--(3);
\draw [->,>=stealth] (4)--(2);
\draw [->,>=stealth] (4)--(3);
\draw [->,>=stealth] (5)--(3);

\node (M) at (-2.5,-0.9) {$(e)$};
\node[draw=black,circle,scale=0.7] (M0) at (-2,-1.55) {$X_1$};
\node[draw=black,circle,scale=0.7] (M1) at (-2,-2.3) {$X_2$};
\node[draw=black,circle,scale=0.75] (M2) at (0,-2.3) {$Z$};
\node[draw=black,circle,scale=0.5] (M5) at (1,-1.9) {$Y_{\mid S=1}$};
\node[draw=black,circle,scale=0.5] (M3) at (2,-2.3) {$Y_{\mid S=2}$};
\node (M4) at (1,-0.9) {$A$};
\draw [->,>=stealth] (M0)--(M2);
\draw [->,>=stealth] (M1)--(M2);
\draw [->,>=stealth] (M2)--(M3);
\draw [->,>=stealth] (M2)--(M5);
\draw [->,>=stealth] (M4)--(M2);
\draw [->,>=stealth] (M4)--(M3);
\draw [->,>=stealth] (M4)--(M5);

\node (D0) at (-13.5,-0.9) {$(d)$};
\node[draw=black,circle,scale=0.7] (D1) at (-13,-2.3) {$X$};
\node[draw=black,circle,scale=0.75] (D2) at (-11,-2.3) {$Z$};
\node[draw=black,circle,scale=0.5] (D6) at (-11,-1.3) {$Y_{\mid S=0}$};
\node[draw=black,circle,scale=0.5] (D3) at (-9,-2.3) {$Y_{\mid S=2}$};
\node[draw=black,circle,scale=0.5] (D5) at (-10,-1.9) {$Y_{\mid S=1}$};
\node (D4) at (-10,-0.9) {$A$};
\draw [->,>=stealth] (D1)--(D2);
\draw [->,>=stealth] (D2)--(D3);
\draw [->,>=stealth] (D4)--(D2);
\draw [->,>=stealth] (D4)--(D3);
\draw [->,>=stealth] (D4)--(D5);
\draw [->,>=stealth] (D2)--(D5);
\draw [->,>=stealth] (D4)--(D6);
\draw [->,>=stealth] (D2)--(D6);
\end{tikzpicture}
}%
\caption[Variations of the instrumental scenario and their post-selected versions]{Variations of the instrumental scenario (a), (b) and (c). The causal structure (c) is relevant for the derivation of the information causality inequality. (d) and (e) are the causal structures that are effectively analysed when post-selecting on a ternary $S$ in (a) and on a binary $S$ in (c) respectively.}
\label{fig:information_causality_DAG}
\end{figure}

A second approach that relies on very similar ideas (also justified by
Lemma~\ref{lemma:justify_conditioning}) is taken
in~\cite{Pienaar2016}. For a causal structure $C^{\cC}$ with nodes
$\Omega = X \cup X^{\uparrow} \cup X^{\nuparrow}$, where $X$ is a
parentless node, conditioning the joint distribution over all nodes on
a particular $X=x$ retains the independencies of
$C^{\cC}$.~\footnote{Any conditional independence relation that holds
  for $P(X^{\nuparrow}  X^{\uparrow} \mid X)$ is also valid for
  the $P(X^{\nuparrow}  X^{\uparrow} \mid X=x)$ for all $x$, by
  definition of the conditional distribution.} In particular, the
conditioning does not affect the distribution of the $X^{\nuparrow}$,
i.e., ${P(X^{\nuparrow}\mid X=x)=P(X^{\nuparrow})}$ for all $x$.  The
corresponding entropic constraints can be used to derive entropic
inequalities without the detour over computing large entropy cones,
which may be useful where the latter computations are infeasible. The
constraints that are used in~\cite{Pienaar2016}
are, however, a (diligently but somewhat arbitrarily chosen) subset of
the constraints that would go into the entropic technique detailed
earlier in this section for the full causal structure. Indeed, when
the computations are feasible, applying the full entropy vector method to the
corresponding post-selected causal structure gives a
systematic way to derive constraints, which  are in general
strictly tighter (cf.\ Example~\ref{example:pienaar_structure_analysis}).

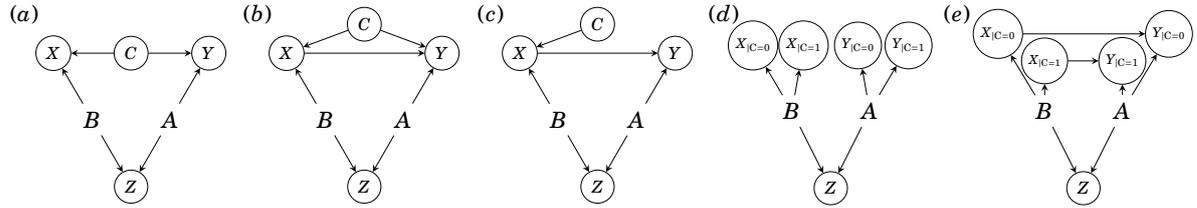
\begin{figure}
\centering
\resizebox{1\columnwidth}{!}{%
\begin{tikzpicture}[scale=0.62]
\node (0) at (-2.75,3) {$(a)$};
\node[draw=black,circle,scale=0.75] (X) at (-2,2) {$X$};
\node[draw=black,circle,scale=0.75] (Y) at (2,2) {$Y$};
\node[draw=black,circle,scale=0.75] (Z) at (0,-1.46) {$Z$};
\node (A) at (1,0.28) {$A$};
\node (B) at (-1,0.28) {$B$};
\node[draw=black,circle,scale=0.75] (C) at (0,2) {$C$};

\draw [->,>=stealth] (A)--(Y); 
\draw [->,>=stealth] (A)--(Z); 
\draw [->,>=stealth] (B)--(X); 
\draw [->,>=stealth] (B)--(Z); 
\draw [->,>=stealth] (C)--(X); 
\draw [->,>=stealth] (C)--(Y); 

\node (M0) at (3.25,3) {$(b)$};
\node[draw=black,circle,scale=0.75] (MX) at (4,2) {$X$};
\node[draw=black,circle,scale=0.75] (MY) at (8,2) {$Y$};
\node[draw=black,circle,scale=0.75] (MZ) at (6,-1.46) {$Z$};
\node (MA) at (7,0.28) {$A$};
\node (MB) at (5,0.28) {$B$};
\node[draw=black,circle,scale=0.75] (MC) at (6,2.75) {$C$};

\draw [->,>=stealth] (MA)--(MY); 
\draw [->,>=stealth] (MA)--(MZ); 
\draw [->,>=stealth] (MB)--(MX); 
\draw [->,>=stealth] (MB)--(MZ); 
\draw [->,>=stealth] (MC)--(MX); 
\draw [->,>=stealth] (MC)--(MY); 
\draw [->,>=stealth] (MX)--(MY);

\node (N0) at (9.25,3) {$(c)$};
\node[draw=black,circle,scale=0.75] (NX) at (10,2) {$X$};
\node[draw=black,circle,scale=0.75] (NY) at (14,2) {$Y$};
\node[draw=black,circle,scale=0.75] (NZ) at (12,-1.46) {$Z$};
\node (NA) at (13,0.28) {$A$};
\node (NB) at (11,0.28) {$B$};
\node[draw=black,circle,scale=0.75] (NC) at (12,2.75) {$C$};
\draw [->,>=stealth] (NA)--(NY); 
\draw [->,>=stealth] (NA)--(NZ); 
\draw [->,>=stealth] (NB)--(NX); 
\draw [->,>=stealth] (NB)--(NZ); 
\draw [->,>=stealth] (NC)--(NX); 
\draw [->,>=stealth] (NX)--(NY);

\node (X0) at (15.25,3) {$(d)$};
\node[draw=black,circle,scale=0.6] (XX0) at (16,2.2) {$X_\mathrm{ |C=0}$};
\node[draw=black,circle,scale=0.6] (XX1) at (17.3,2.2) {$X_\mathrm{ |C=1}$};
\node[draw=black,circle,scale=0.6] (XY0) at (18.7,2.2) {$Y_\mathrm{ |C=0}$};
\node[draw=black,circle,scale=0.6] (XY1) at (20,2.2) {$Y_\mathrm{ |C=1}$};
\node[draw=black,circle,scale=0.75] (XZ) at (18,-1.5) {$Z$};
\node (XA) at (19,0.5) {$A$};
\node (XB) at (17,0.5) {$B$};

\draw [->,>=stealth] (XA)--(XY0); 
\draw [->,>=stealth] (XA)--(XY1);
\draw [->,>=stealth] (XA)--(XZ); 
\draw [->,>=stealth] (XB)--(XX0); 
\draw [->,>=stealth] (XB)--(XX1); 
\draw [->,>=stealth] (XB)--(XZ); 

\node (X0) at (21.25,3) {$(e)$};
\node[draw=black,circle,scale=0.6] (YX0) at (22.3,2.5) {$X_\mathrm{ |C=0}$};
\node[draw=black,circle,scale=0.56] (YX1) at (23.5,1.8) {$X_\mathrm{ |C=1}$};
\node[draw=black,circle,scale=0.6] (YY0) at (25.5,1.8) {$Y_\mathrm{ |C=1}$};
\node[draw=black,circle,scale=0.6] (YY1) at (26.7,2.5) {$Y_\mathrm{ |C=0}$};
\node[draw=black,circle,scale=0.75] (YZ) at (24.5,-1.5) {$Z$};
\node (YA) at (25.5,0.5) {$A$};
\node (YB) at (23.5,0.5) {$B$};

\draw [->,>=stealth] (YA)--(YY0); 
\draw [->,>=stealth] (YA)--(YY1);
\draw [->,>=stealth] (YA)--(YZ); 
\draw [->,>=stealth] (YB)--(YX0); 
\draw [->,>=stealth] (YB)--(YX1); 
\draw [->,>=stealth] (YB)--(YZ); 
\draw [->,>=stealth] (YX0)--(YY1);
\draw [->,>=stealth] (YX1)--(YY0); 

\end{tikzpicture}
}%
\caption[Causal structures from Example~\ref{example:pienaar_structure_analysis}]{Causal structures from Example~\ref{example:pienaar_structure_analysis}. Post-selecting on a binary observed variable $C$ leads to the causal structure (d) in the case of structure (a), whereas both (b) and (c) lead to structure (e). In particular, this shows that the post-selection technique may yield the same results for different causal structures.}
\label{fig:Pienaar_examples}
\end{figure}

So far, the restricted technique has been used in~\cite{Pienaar2016} to derive the entropy inequality 
\begin{equation}
I(X_\mathrm{ \mid C=0}:Z)-I(Y_\mathrm{ \mid C=0}:Z)-I(X_\mathrm{ \mid C=1}:Z)+I(Y_\mathrm{ \mid C=1}:Z)\leq H(Z),\label{eq:pienaar_inequality}
\end{equation}
which is valid for all classical causal structures of Figure~\ref{fig:Pienaar_examples} (previously considered in~\cite{Henson2014}). The inequality was used to certify the existence of classical distributions that respect the conditional independence constraints among the observed variables but that are not achievable in the respective causal structures.~\footnote{These causal structures may thus also allow for quantum correlations that are not classically achievable.}
In the following we look at these three causal structures in more
detail and illustrate the relation between the two techniques.
\begin{example}\label{example:pienaar_structure_analysis}
Applying the post-selection technique for a binary random variable $C$ to the causal structure from Figure~\ref{fig:Pienaar_examples}(a) yields the effective causal structure \ref{fig:Pienaar_examples}(d). The latter can be analysed with the above entropy vector method, which leads to a cone that is characterised by $14$ extremal rays or equivalently in terms of $22$ inequalities, both available in Section~\ref{sec:pienaaranalysis}. 
The inequalities $I(Z:X_\mathrm{ \mid C=1})\geq 0$, $I(Z:Y_\mathrm{ \mid C=0})\geq 0$, $I(X_\mathrm{\mid C=1}:Y_\mathrm{ \mid C=1} \mid Z)\geq 0$ and $H(Z\mid X_\mathrm{ \mid C=0}) \geq {I(X_\mathrm{ \mid C=1} Z:Y_\mathrm{ \mid C=1})}$, which are part of this description, imply \eqref{eq:pienaar_inequality} above.
 
In Figure~\ref{fig:Pienaar_examples}, structures (b) and (c) both lead to the causal structure (e) upon post-selecting on a binary $C$. The latter causal structure turns out to be computationally harder to analyse with the entropy vector method and conventional variable elimination tools on a desktop computer are unable to perform the corresponding marginalisation. Hence, the method outlined in~\cite{Pienaar2016} is a useful alternative here.
\end{example}

\subsection{Post-selection in quantum and general non-signalling causal structures} \label{sec:quantum_conditioning}
 \label{sec:generalised_cone}
 In causal structures with quantum and more general non-signalling
 nodes, Lemma~\ref{lemma:justify_conditioning} is not valid.  For
 instance, Bell's theorem can be recast as the statement that there
 are distributions compatible with the quantum Bell scenario for which
 there is no joint distribution of $X_\mathrm{ \mid A=0}$,
 $X_\mathrm{ \mid A=1}$, $Y_\mathrm{ \mid B=0}$ and $Y_\mathrm{ \mid B=1}$ in the
 post-selected causal structure (on $A$ and $B$) that has the required
 marginals (in the sense of Corollary~\ref{lemma:altfine}).

Nonetheless, the post-selection technique has been generalised to such scenarios~\cite{Chaves2015, Chaves2016}, i.e., it is still possible to post-select on parentless observed (and therefore classical) nodes to take specific values.
In such scenarios the observed variables can be thought of as obtained
from the unobserved resources by means of measurements or tests. If a
descendant of the variable that is post-selected on has quantum or
general non-signalling nodes as parents, then the different instances
of the latter node and of all its descendants do not coexist (even if
they are observed, hence classical). This is because such observed
variables are generated by measuring a quantum or other non-signalling system. Such a system is altered (or destroyed) in a measurement, hence does not allow for the simultaneous generation of different instances of its children due to the impossibility of cloning.
In the quantum case, this is reflected in the identification of the coexisting sets in the post-selected causal structure,\footnote{Different instances of a variable after post-selection have to be seen as alternatives and not as simultaneous descendants of their parent node as the representation of the post-selected causal structure might suggest.} as is illustrated with the following example.

\begin{example}[Information causality scenario in the quantum case~\cite{Chaves2015}]\label{example:information_causality}
Information causality is an information theoretic principle obeyed by classical and quantum physics but not by general probabilistic theories that produce correlations that violate Tsirelson's bound~\cite{Tsirelson1993}, e.g.\ in box world~\cite{Pawlowski2009, Popescu1994FP, Barrett07}. The principle is stated in terms of the optimal performance of two parties in a game and is quantified in terms of an entropy inequality.
The communication scenario used to derive the principle of information causality~\cite{Pawlowski2009} is based on the variation of the instrumental scenario displayed in Figure~\ref{fig:information_causality_DAG}(c). It has been analysed with the entropy vector method in Ref.~\cite{Chaves2015}, an analysis that is presented in the following. 

Conditioning on values of the variable $S$ is possible in the classical and quantum cases. However, whereas in the classical case the variables $Y_\mathrm{ \mid S=s}$ for different $S$ share a joint distribution (according to Lemma~\ref{lemma:justify_conditioning}), they do not coexist in the quantum case.
For bipartite $S$, the coexisting sets are
$\left\{X_1,~X_2,~A_Z,~A_Y\right\}$,
$\left\{X_1,~X_2,~Z,~A_Y\right\}$,
$\left\{X_1,~X_2,~Z,~Y_\mathrm{ \mid S=1}\right\}$ and
$\left\{X_1,~X_2,~Z,~Y_\mathrm{ \mid S=2}\right\}$. The only independence
constraints in the quantum case are that $X_1$, $X_2$ and $\rho_A$ are
mutually independent. Marginalising until only entropies of
$\{X_1, Y_\mathrm{ \mid S=1}\}$, $\{X_2, Y_\mathrm{ \mid S=2}\}$, $\{Z\}$ and
their subsets remain, yields only one non-trivial inequality,
${\sum_{s=1}^{n} I(X_s : Y_\mathrm{ \mid S=s}) \leq H(Z),}$ with
$n=2$.\footnote{Note that in Ref.~\cite{Chaves2015} they derived the more general inequality  $I(X_1 : Y_\mathrm{ \mid S=1})+I(X_2 : Y_\mathrm{ \mid S=2}) \leq
  H(Z)+I(X_1:X_2)$,
  where $X_1$ and $X_2$ are not assumed independent. Furthermore,
  this is also the only inequality found in the classical case when
  restricting to the same marginal scenario~\cite{Chaves2016}.}  The same inequality was previously derived by Pawlowski et al.\ for
general $n$~\cite{Pawlowski2009}, where the choice of marginals
was inspired by the communication task considered. Consequently, Ref.~\cite{Chaves2015} considered other marginal scenarios, namely the entropies of  $\{X_1,X_2,Z,Y_\mathrm{ \mid S=1}\}$, $\{X_1,X_2,Z,Y_\mathrm{ \mid S=2}\}$ and all of their subsets, which led to additional inequalities.
\end{example}

For causal structures allowing for general non-signalling resources, $C^{\gG}$, similar methods have been introduced in~\cite{Chaves2016}. Let $O=X_\mathrm{ O}^{\uparrow} \cup X_\mathrm{ O}^{\nuparrow} \cup X$ be the disjoint union of its observed nodes, where $X_\mathrm{ O}^{\uparrow}$ are the observed descendants and $X_\mathrm{ O}^{\nuparrow}$ the observed non-descendants of $X$. If the variable $X$ takes values $x \in \left\{1,~2,~\ldots,~n \right\}$, this leads to a joint distribution of $X_\mathrm{ O}^{\uparrow} \cup X_\mathrm{ O}^{\nuparrow}$ for each $X=x$, i.e., there is a joint distribution for ${P(X_\mathrm{ O}^{\uparrow} X_\mathrm{ O}^{\nuparrow} \mid X=x)=P(X_\mathrm{ O}^{\uparrow} \mid X_\mathrm{ O}^{\nuparrow}  X=x)P(X_\mathrm{ O}^{\nuparrow})}$ for all $x$, denoted  $P(X_\mathrm{ O { \mid X=x}}^{\uparrow} X_\mathrm{ O}^{\nuparrow})$. 
Because $X$ does not affect the distribution of the independent variables $X_\mathrm{ O}^{\nuparrow}$, the distributions $P(X_\mathrm{ O { \mid X=x}}^{\uparrow} X_\mathrm{ O}^{\nuparrow})$ have coinciding marginals on $X_\mathrm{ O}^{\nuparrow}$, i.e., ${P(X_\mathrm{ O}^{\nuparrow})=\sum_{X_\mathrm{ O {\mid X=x}}^{\uparrow}=s}P(X_\mathrm{ O { \mid X=x}}^{\uparrow} X_\mathrm{ O}^{\nuparrow})}$ for all $x$, where $s$ runs over the alphabet of $X_\mathrm{ O}^{\uparrow}$. This encodes no-signalling constraints.

In terms of entropy, there are $n$ entropy cones, one for each $P(X_\mathrm{ O { \mid X=x}}^{\uparrow} X_\mathrm{ O }^{\nuparrow})$ (which each encode the independencies among the observed variables). According to the above, they are required to coincide on the entropies of $X_\mathrm{ O}^{\nuparrow}$ and on those of all its subsets. These constraints define a convex polyhedral cone that is an outer approximation to the set of all entropy vectors achievable in the causal structure. Whenever the distributions $P(X_\mathrm{ O { \mid X=x}}^{\uparrow} X_\mathrm{ O }^{\nuparrow})$ involve less than three variables and assuming that all constraints implied by the causal structure and no-signalling have been taken into account, this approximation is tight because $\overline{\Gamma^{*}_3}=\Gamma_3$.\footnote{Note that it may not always be obvious how to identify all relevant constraints (see Example~\ref{example:ns_ic} for an illustration).}
An example of the use of this technique can be found in Ref.~\cite{Chaves2016}, which we discuss in the following.

\begin{example}[Information causality scenario in general non-signalling theories]\label{example:ns_ic}
This is related to
Example~\ref{example:information_causality} above and reproduces an analysis from~\cite{Chaves2016}. In this marginal scenario we consider the
Shannon cones for the three sets $\left\{
  X_1, Y_{\mathrm{ \mid S=1}} \right\}$, $\left\{ X_2, Y_{\mathrm{\mid S=2}}
\right\}$ and $\left\{ Z \right\}$ 
as well as the constraints $I(X_1:Y_{\mathrm{ \mid S=1}})\leq H(Z)$ and
$I(X_2:Y_{\mathrm{\mid S=2}})\leq H(Z)$, which are conjectured to hold~\cite{Chaves2016}.
(This conjecture is based on an argument in~\cite{Popescu2014} that covers a special case; we are not aware of a general proof. Nonetheless, for box-world these inequalities are implied by the results from Ref.~\cite{Short2010a}.)
These conditions constrain a polyhedral cone
of vectors
${\left(H(X_1),~H(X_2),~H(Z),~H(Y_{\mathrm{ \mid S=1}}),~H(Y_{\mathrm{ \mid
    S=2}}),~H(X_1 Y_{\mathrm{ \mid S=1}}),~H(X_2 Y_{\mathrm{ \mid S=2}}) \right)}$
with $8$ extremal rays that are all achievable using
PR-boxes~\cite{Tsirelson1993,Popescu1994FP}. Importantly, the
stronger constraint
${I(X_1:Y_{\mathrm{ \mid S=1}})} +I(X_2:Y_{\mathrm{ \mid S=2}})\leq H(Z)$, which
holds in the quantum case (cf.\ Example~\ref{example:information_causality}), does not hold here.
\end{example}

\section{Alternative techniques for analysing causal structures} \label{sec:further_techniques}
Even though the entropy vector method has so far been the most popular approach
to relaxing the problem of characterising the set of probability distributions compatible with a
causal structure, other computational techniques are currently being developed. In the following, we give a brief overview of these methods. 

In this context, we also point out that there are methods that allow
 one to certify that the only restrictions implied by a causal structure are the conditional independence constraints among the observed variables~\cite{Henson2014} as well as procedures to show that the opposite is the case~\cite{Evans2012, Evans2015}. Such methods may (when applicable) indicate whether a causal  structure should be analysed further (corresponding techniques are reviewed in~\cite{Pienaar2016}).

\subsection{Entropy cones relying on R\'{e}nyi and related entropies}\label{sec:renyi_cone}
Entropy vectors may be computed in terms of other entropy measures,
for instance in terms of the $\alpha$-R\'{e}nyi entropies~\cite{Renyi1960_MeasOfEntrAndInf}. One may expect that useful constraints on the compatible distributions can be derived from such entropy vectors.
For $0 < \alpha <1$ and $\alpha > 1$ such cones were analysed
in~\cite{Linden2013a}. In the classical case positivity and
monotonicity are the only linear constraints on the corresponding
entropy vectors for any $\alpha \neq 0, 1$.  For multi-party quantum
states monotonicity does not hold for any $\alpha$, like in the case
of the von Neumann entropy. For $0 < \alpha <1$, there are no
constraints on the allowed entropy vectors except for positivity,
whereas for $\alpha > 1$ there are constraints, but these
are non-linear. The lack of further linear inequalities that generally hold limits the usefulness of
entropy vectors using $\alpha$-R\'{e}nyi entropies for analysing causal structures. To our knowledge it is not known how or whether non-linear inequalities for R\'{e}nyi entropies may be employed for this
task. The case $\alpha=0$ has been considered separately in~\cite{Cadney2012a}, where it was shown that further linear inequalities hold for such entropies. However, only bipartitions of the parties were considered and the generalisation to full entropy vectors is still to be explored.~\footnote{This entropy is furthermore special, in so far as for a quantum state $\rho_\mathrm{X}$,
${H_0(X)= \log_2 \operatorname{rank}\rho_\mathrm{X},}$
which leads to a discrete set of values.}

The above considerations do not mention conditional entropy and hence
could be taken with the definition of Equation~\eqref{eq:cond_rel_2}. Alternatively, one may consider a definition of the R\'{e}nyi conditional entropy, for which Equation~\eqref{eq:cond_rel} holds~\cite{Petz,TCR,MDSFT,FL,Beigi}. 
Doing so means that the conditional R\'{e}nyi entropy cannot be expressed as a difference of unconditional entropies, and so to use entropy vectors we would need to consider the conditional entropies as separate components. Along these lines, one may also think about combining R\'{e}nyi entropies for different values of $\alpha$ and to use appropriate chain rules~\cite{Dupuis2015}. Because of the large increase in the number of variables compared to the number of inequalities it is not clear whether this could yield any useful new conditions.

A second family of entropy measures, related to R\'enyi entropies, are
the Tsallis entropies~\cite{Havrda,Tsallis}, which can be defined by
$H_{T,\alpha}(X):=\frac{1}{1-\alpha}\left(2^{(1-\alpha)H_{\alpha}(X)}-1\right)$.
Little work has been done on these in the context of causal
structures, but some numerical work~\cite{Wajs15} suggests that they
have advantages for detecting non-classicality in the post-selected
Bell scenario (see also~\cite{Rastegin15}).

\subsection{Polynomial inequalities for compatible distributions}
The probabilistic characterisation of causal structures, depends (in
general) on the dimensionality of the observed
variables. Computational hardness results suggest that a full
characterisation is unlikely to be feasible, except in small cases~\cite{Pitowsky1991,Avis2004}. 
Recent progress has been made with the development of procedures to construct polynomial Bell inequalities. A method that resorts to linear programming techniques~\cite{Chaves2015a} has lead to the derivation of new inequalities for the bilocality scenario (as well as a related four-party scenario). 
Another, iterative procedure allows for enlarging networks by adding a party to a network in a particular way.~\footnote{In our terminology adding a party means adding one observed input and one observed output node as well as an unobserved parent for the output, the latter may causally influence one other output random variable in the network.} This allows for the constructions of non-linear inequalities for the latter, enlarged network from inequalities that are valid for the former~\cite{Rosset2015}. 

\subsection{Graph inflations}
A recent approach relies on considering enlarged networks, so called inflations, and inferring causal constraints from those~\cite{Wolfe2016, Navascues2017}. Inflated networks may contain several copies of a variable that each have the same dependencies on ancestors (the latter may also exist in several instances) and which share the same distributions with their originals. Such inflations allow for the derivation of probabilistic inequalities that restrict the set of compatible distributions.\footnote{These ideas also bear some resemblance to the procedures in~\cite{Kela2017}, in the sense that they employ the idea that certain marginal distributions may be obtained from different networks.} 
Inflations allowed the authors of~\cite{Wolfe2016} to refute certain distributions as incompatible with the triangle causal structure from Figure~\ref{fig:instrumental}(c), in particular the so called W-distribution which could neither be proven to be incompatible entropically nor with the covariance matrix approach below.

\subsection{Semidefinite tests relying on covariance matrices}
One may look for mappings of the distribution of a set of observed
variables that encode causal structure beyond considering
entropies. For causal structures with two generations, i.e., one
generation of unobserved variables as ancestors of one generation of
observed nodes, a technique has been found using covariance matrices~\cite{Kela2017}.
Each observed variable is mapped to a vector-valued random variable
and the covariance matrix of the direct sum of these variables 
is considered. Due
to the law of total expectation~\cite{weiss2006course}, this matrix allows for a certain
decomposition depending on the causal structure.
For a particular observed distribution and its covariance matrix, the existence of such a decomposition may be tested via semidefinite programming. The relation of this technique to the entropy vector method is not yet well understood. A partial analysis considering several examples is given in Section X of Ref.~\cite{Kela2017}.

\section{Appendix}

\subsection{Inequalities from Example~\ref{example:quantum_instrumental}} \label{sec:29inequ}
In the following we provide the basic inequalities for the quantum instrumental scenario, $\inst^{\qQ}$, i.e., the constraints making up the matrix $M_B(\inst^{\qQ})$, 
\begin{eqnarray}
I(A_Y:A_Z)&\geq&0 \\
I(A_Y:X)&\geq&0 \\ 
I(A_Z:X)&\geq&0 \\ 
I(A_Y:A_Z|X)&\geq&0 \\ 
I(A_Y:X|A_Z)&\geq&0 \\ 
I(A_Z:X|A_Y)&\geq&0 \\ 
I(A_Y:Z)&\geq&0 \\ 
I(X:Z)&\geq&0 \\
I(X:Z|A_Y)&\geq&0 \\ 
I(A_Y:Z|X)&\geq&0 
\end{eqnarray}
\begin{eqnarray}
I(A_Y:X|Z)&\geq&0 \\
I(X:Y)&\geq&0 \\ 
I(Y:Z)&\geq&0  \\
I(Y:Z|X)&\geq&0 \\ 
I(X:Z|Y)&\geq&0 \\ 
I(X:Y|Z)&\geq&0 \\ 
H(A_Z|X)&\geq&0 \\ 
H(A_Y A_Z|X)&\geq&0 \\ 
H(X|A_Y A_Z)&\geq&0 \\
H(A_Y|XZ)&\geq&0 \\ 
H(X|A_YZ)&\geq&0 \\ 
H(Z|A_YX)&\geq&0 \\
H(X|YZ)&\geq&0 \\
H(Y|XZ)&\geq&0 \\ 
H(Z|XY)&\geq&0 \\ 
H(A_Z|A_Y)+H(A_Z|X) &\geq&0 \\ 
H(A_Y|A_Z)+H(A_Y|X) &\geq&0 \\ 
H(A_Z|A_YX)+H(A_Z) &\geq&0 \\ 
H(A_Y|A_ZX)+H(A_Y) &\geq&0 \ . 
\end{eqnarray}

\pagebreak

Independence constraints and data processing inequalities are provided
in the main text.  If we include these and remove redundant
inequalities we obtain the following set of constraints, which for
convenience we give in matrix form (such that $\encone{\inst^{\qQ}}=\left\{v\in\mathbb{R}^{15}_{\geq
  0} \ \middle| \ M\cdot v\geq 0\right\}$):
  
{
$$M=\left(
\begin{array}{ccccccccccccccc}
 0 & 0 & 0 & 0 & 0 & 0 & 0 & 1 & 0 & 0 & 0 & -1 & 0 & -1 & 1 \\
 0 & 0 & 0 & 0 & 0 & 0 & 0 & 0 & 1 & 0 & -1 & 0 & -1 & 1 & 0 \\
 0 & 0 & -1 & 0 & 0 & -1 & 0 & 0 & 0 & 0 & 0 & 0 & 1 & 0 & 0 \\
 0 & 0 & 0 & 0 & -1 & 0 & 0 & 0 & 0 & 0 & 1 & 1 & 0 & 0 & -1 \\
 0 & 0 & 0 & -1 & 0 & 0 & 0 & 0 & 0 & 1 & 0 & 1 & 0 & 0 & -1 \\
 0 & 0 & -1 & 0 & 0 & 0 & 0 & 0 & 0 & 1 & 1 & 0 & 0 & 0 & -1 \\
 0 & 0 & 0 & 1 & 1 & 0 & 0 & 0 & 0 & 0 & 0 & -1 & 0 & 0 & 0 \\
 0 & 0 & 1 & 0 & 1 & 0 & 0 & 0 & 0 & 0 & -1 & 0 & 0 & 0 & 0 \\
 0 & 0 & 1 & 1 & 0 & 0 & 0 & 0 & 0 & -1 & 0 & 0 & 0 & 0 & 0 \\
 1 & 0 & 0 & 0 & 1 & 0 & 0 & -1 & 0 & 0 & 0 & 0 & 0 & 0 & 0 \\
 1 & 0 & 1 & 0 & 0 & 0 & -1 & 0 & 0 & 0 & 0 & 0 & 0 & 0 & 0 \\
 0 & -1 & 0 & 0 & 0 & 1 & 0 & 0 & 1 & 0 & 0 & 0 & -1 & 0 & 0 \\
 -1 & 0 & 0 & 0 & 0 & 1 & 1 & 0 & 0 & 0 & 0 & 0 & -1 & 0 & 0 \\
 0 & 1 & 1 & 0 & 0 & 0 & 0 & 0 & -1 & 0 & 0 & 0 & 0 & 0 & 0 \\
 0 & 0 & 0 & 0 & 0 & 0 & 0 & 0 & 0 & -1 & 0 & 0 & 0 & 0 & 1 \\
 0 & 0 & 0 & 0 & 0 & 0 & 0 & 0 & 0 & 0 & -1 & 0 & 0 & 0 & 1 \\
 0 & 0 & 0 & 0 & 0 & 0 & -1 & 0 & 0 & 0 & 0 & 0 & 0 & 1 & 0 \\
 0 & 0 & 0 & 0 & 0 & 0 & 0 & -1 & 0 & 0 & 0 & 0 & 0 & 1 & 0 \\
 0 & 0 & 0 & 0 & 0 & 0 & 0 & 0 & 0 & 0 & -1 & 0 & 0 & 1 & 0 \\
 1 & 0 & 0 & 0 & 0 & 0 & 0 & 0 & -1 & 0 & 0 & 0 & 1 & 0 & 0 \\
 0 & 1 & 0 & 0 & 0 & 0 & -1 & 0 & 0 & 0 & 0 & 0 & 1 & 0 & 0 
\end{array}
\right) \ $$
}

\subsection{Entropy inequalities and extremal rays for Example~\ref{example:pienaar_structure_analysis}} \label{sec:pienaaranalysis}
The causal structure of Figure~\ref{fig:Pienaar_examples}(a), previously considered in~\cite{Henson2014, Pienaar2016} is analysed entropically by means of the entropic post-selection technique. 
The outer approximation to the entropy cone of the causal structure of Figure~\ref{fig:Pienaar_examples}(d) is computed and marginalised to vectors \begin{multline*}
\left(H(Z),~H(X_\mathrm{ \mid C=0}),~H(X_\mathrm{ \mid C=1}),~H(Y_\mathrm{ \mid C=0}),~H(Y_\mathrm{ \mid C=1}),~H(X_\mathrm{\mid C=0}Z),~H(X_\mathrm{ \mid C=1}Z),~H(Y_\mathrm{ \mid C=0}Z), \right. \\ \left.~H(Y_\mathrm{\mid C=1}Z),~H(X_\mathrm{ \mid C=0}Y_\mathrm{\mid C=0}),~H(X_\mathrm{ \mid C=1}Y_\mathrm{ \mid C=1}),~H(X_\mathrm{ \mid C=0}Y_\mathrm{ \mid C=0}Z),~H(X_\mathrm{ \mid C=1}Y_\mathrm{\mid C=1}Z) \right).
\end{multline*} 
From this computation, we obtain the following $14$ extremal rays, where each ray is represented by one particular vector on it. (The tip of this pointed polyhedral cone is the zero-vector.)

$$\arraycolsep=1.5pt 
\begin{array}{cccccccccccccccc}
(  1) &&& 0 & 0 & 0 & 0 & 1 & 0 & 0 & 0 & 1 & 0 & 1 & 0 & 1  \\
(  2) &&& 0 & 0 & 0 & 1 & 0 & 0 & 0 & 1 & 0 & 1 & 0 & 1 & 0  \\
(  3) &&& 0 & 0 & 1 & 0 & 0 & 0 & 1 & 0 & 0 & 0 & 1 & 0 & 1  \\
(  4) &&& 0 & 1 & 0 & 0 & 0 & 1 & 0 & 0 & 0 & 1 & 0 & 1 & 0  \\
(  5) &&& 1 & 1 & 1 & 1 & 1 & 2 & 2 & 2 & 2 & 2 & 2 & 2 & 2  \\
(  6) &&& 1 & 0 & 1 & 0 & 1 & 1 & 2 & 1 & 2 & 0 & 2 & 1 & 2  \\
(  7) &&& 1 & 1 & 0 & 1 & 0 & 2 & 1 & 2 & 1 & 2 & 0 & 2 & 1  \\
(  8) &&& 1 & 0 & 0 & 0 & 0 & 1 & 1 & 1 & 1 & 0 & 0 & 1 & 1  \\
(  9) &&& 1 & 0 & 0 & 0 & 1 & 1 & 1 & 1 & 1 & 0 & 1 & 1 & 1  \\
( 10) &&& 1 & 0 & 0 & 1 & 0 & 1 & 1 & 1 & 1 & 1 & 0 & 1 & 1  \\
( 11) &&& 1 & 0 & 1 & 0 & 0 & 1 & 1 & 1 & 1 & 0 & 1 & 1 & 1  \\
( 12) &&& 1 & 1 & 0 & 0 & 0 & 1 & 1 & 1 & 1 & 1 & 0 & 1 & 1  \\
( 13) &&& 1 & 0 & 0 & 1 & 1 & 1 & 1 & 1 & 1 & 1 & 1 & 1 & 1  \\
( 14) &&& 1 & 1 & 1 & 0 & 0 & 1 & 1 & 1 & 1 & 1 & 1 & 1 & 1  \\
\end{array}
$$
The corresponding inequality description is given by the following $2$ equalities and $18$ inequalities (or equivalently $22$ inequalities).
\begin{eqnarray}
\quad H(X_\mathrm{ \mid C=0})+ H(Y_\mathrm{ \mid C=0}) &=& H(X_\mathrm{ \mid C=0}Y_\mathrm{ \mid C=0})\\
\quad H(X_\mathrm{ \mid C=1})+ H(Y_\mathrm{ \mid C=1}) &=& H(X_\mathrm{ \mid C=1}Y_\mathrm{ \mid C=1})\\
\quad H(X_\mathrm{ \mid C=1}Y_\mathrm{ \mid C=1}) &\leq & H(X_\mathrm{ \mid C=1}Y_\mathrm{ \mid C=1} Z)\\
\quad H(X_\mathrm{ \mid C=0}Y_\mathrm{ \mid C=0}) &\leq & H(X_\mathrm{ \mid C=0}Y_\mathrm{ \mid C=0} Z)\\
\quad H(Y_\mathrm{ \mid C=1}Z) &\leq & H(X_\mathrm{ \mid C=1}Y_\mathrm{ \mid C=1} Z)\\
\quad H(Y_\mathrm{ \mid C=0}Z) &\leq & H(X_\mathrm{ \mid C=0}Y_\mathrm{ \mid C=0} Z)\\
\quad H(X_\mathrm{ \mid C=1}Z) &\leq & H(X_\mathrm{ \mid C=1}Y_\mathrm{ \mid C=1} Z)\\
\quad H(X_\mathrm{ \mid C=0}Z) &\leq & H(X_\mathrm{ \mid C=0}Y_\mathrm{ \mid C=0} Z)\\
\quad H(Y_\mathrm{ \mid C=0}Z) &\leq & H(Z) + H(Y_\mathrm{ \mid C=0})\\
\quad H(Y_\mathrm{ \mid C=1}Z) &\leq & H(Z) + H(Y_\mathrm{ \mid C=1}) \\
\quad H(Y_\mathrm{ \mid C=1}) + H(X_\mathrm{ \mid C=1}Z)  &\leq & H(Z) + H(X_\mathrm{ \mid C=1}Y_\mathrm{ \mid C=1})\\
\quad H(Y_\mathrm{ \mid C=0}) + H(X_\mathrm{ \mid C=0}Z)  &\leq & H(Z) + H(X_\mathrm{ \mid C=0}Y_\mathrm{ \mid C=0}) \\
\quad H(Z) + H(X_\mathrm{ \mid C=0}Y_\mathrm{ \mid C=0} Z) &\leq &  H(X_\mathrm{ \mid C=0}Z)+ H(Y_\mathrm{ \mid C=0}Z) \\
\quad H(Z) + H(X_\mathrm{ \mid C=1}Y_\mathrm{ \mid C=1} Z) &\leq &  H(X_\mathrm{ \mid C=1}Z)+ H(Y_\mathrm{ \mid C=1}Z) \\
\quad H(Y_\mathrm{ \mid C=1}) + H(X_\mathrm{ \mid C=1}Z)+ H(X_\mathrm{ \mid C=0}Y_\mathrm{ \mid C=0})  &\leq & H(Y_\mathrm{ \mid C=0}) + H(X_\mathrm{ \mid C=0}Z)+H(X_\mathrm{ \mid C=1}Y_\mathrm{ \mid C=1}Z)\\
\quad H(Y_\mathrm{ \mid C=1}) + H(Y_\mathrm{ \mid C=0}Z)+ H(X_\mathrm{ \mid C=0}Y_\mathrm{ \mid C=0})  &\leq & H(Y_\mathrm{ \mid C=0}) + H(Y_\mathrm{ \mid C=1}Z)+H(X_\mathrm{ \mid C=0}Y_\mathrm{ \mid C=0}Z)
\end{eqnarray}
\begin{eqnarray}
\quad \quad \quad \ H(X_\mathrm{ \mid C=1} Z)+H(Y_\mathrm{ \mid C=1} Z)+H(X_\mathrm{ \mid C=0} Y_\mathrm{ \mid C=0}) &\leq & H(X_\mathrm{ \mid C=0} Z)+H(Y_\mathrm{ \mid C=0} Z)+H(X_\mathrm{ \mid C=1}Y_\mathrm{ \mid C=1}Z)\\
\quad \quad \quad \ H(X_\mathrm{ \mid C=0} Z)+H(Y_\mathrm{ \mid C=0} Z)+H(X_\mathrm{ \mid C=1} Y_\mathrm{ \mid C=1}) &\leq & H(X_\mathrm{ \mid C=1} Z)+H(Y_\mathrm{ \mid C=1} Z) +H(X_\mathrm{ \mid C=0}Y_\mathrm{ \mid C=0}Z)\\
\quad \quad \quad \ H(Y_\mathrm{ \mid C=0})+H(Y_\mathrm{ \mid C=1} Z)+H(X_\mathrm{ \mid C=1}Y_\mathrm{ \mid C=1}) &\leq & H(Y_\mathrm{ \mid C=1})+H(Y_\mathrm{ \mid C=0} Z)+H(X_\mathrm{ \mid C=1}Y_\mathrm{ \mid C=1}Z) \\
\quad \quad \quad H(Y_\mathrm{ \mid C=0})+H(X_\mathrm{ \mid C=0} Z)+H(X_\mathrm{ \mid C=1}Y_\mathrm{ \mid C=1}) &\leq & H(Y_\mathrm{ \mid C=1})+H(X_\mathrm{ \mid C=1} Z)+H(X_\mathrm{ \mid C=0}Y_\mathrm{ \mid C=0}Z) \ . 
\end{eqnarray}

%=========================================================

%% file: chapter04/chap_inner.tex
\let\textcircled=\pgftextcircled
\chapter{Inner approximations to the entropy cones of causal structures}
\label{chap:inner}

\initial{T}he outer approximations to the entropic cones introduced in Chapter~\ref{chap:entropy_vec} have been presented in the literature without further reference to their tightness, except for the case of the Bell scenario~\cite{Chaves2013}. In this chapter, we show that to obtain such results it is sometimes useful to complement the outer approximations with corresponding inner approximations. For certain causal structures we can show that these inner and outer approximations coincide and, hence, we can identify the actual entropy cone with relatively simple means. Inner approximations are furthermore generally relevant for identifying regions of entropy space where the outer approximations discussed in Chapter~\ref{chap:entropy_vec} are provably tight, and regions where there is a gap between inner and outer approximations. The exploration of this gap will be topic of Chapter~\ref{chap:nonshan}.

From the perspective of quantum technologies, it is of interest for device-independent cryptography to be able to identify particular distributions as incompatible with a certain (classical) causal structure. Checking whether their entropy vector lies in an inner approximation can be useful for deciding whether entropy vectors are relevant for the task at hand or whether one should consider a more fine-grained technique such as post-selection or some non-entropic alternative.

In the following we illustrate techniques to identify inner approximations to the entropy cones of causal structures, something that has not been considered in the literature before. In Section~\ref{sec:fourfive} we give inner approximations for causal structures with four and five random variables in terms of linear rank inequalities. In addition, we give heuristic arguments on how to find inner approximations for causal structures with up to five \emph{observed} nodes in a relatively efficient manner. Section~\ref{sec:triangle_inner} is dedicated to the example of the triangle causal structure, where we also briefly outline how to improve such inner approximations (Section~\ref{sec:improveinner}).

\section{Techniques to find inner approximations for causal structures with up to five observed variables} \label{sec:fourfive}
For causal structures $\smcaus$ that involve a total of four or five variables, inner approximations, $\incone{\smcaus^\cC}$, to their entropy cones, $\encone{\smcaus^\cC}$, are obtained by combining $\inconek{4}$ or $\inconek{5}$ respectively with the conditional independence constraints of $\smcaus^{\cC}$. An inner approximation to the corresponding marginal cone, $\inmarcone{\smcaus^\cC}$, is then obtained from $\incone{\smcaus^\cC}$ with a Fourier-Motzkin elimination, like for outer approximations. It is guaranteed that $\inmarcone{\smcaus^\cC}$ is an inner approximation to $\marcone{\smcaus^\cC}$, since it is a projection of an inner approximation $\incone{\smcaus^\cC} \subseteq \encone{\smcaus^{\cC}}$. Hence, inner approximations can be straightforwardly computed for all causal structures with up to five nodes. Examples are the three causal structures of Figure~\ref{fig:HensonStructures}.\footnote{There are an additional $94$ causal structures involving $5$ variables (none for $4$) for which such approximations can be computed and are interesting with regard to Chapter~\ref{chap:classquant}, since they might exhibit a separation between classical and quantum on the level of entropies~\cite{Henson2014}. } In the following we provide an inner approximation for the instrumental scenario, the other two examples are provided in Section~\ref{sec:inner_examples} for completeness.
\begin{figure}
\centering
\resizebox{0.95 \columnwidth}{!}{%
\begin{tikzpicture}[scale=0.85]
\node (a)  at (-10,1) {$(a)$};
\node[draw=black,circle,scale=0.75]  (Z1) at (-4.5,-2) {$Y$};
\node[draw=black,circle,scale=0.75]  (Y1) at (-7,-2) {$Z$};
\node[draw=black,circle,scale=0.75]  (X1) at (-9.5,-2) {$X$};
\node (A1) at (-5.75,0.5) {$A$};
\node (b) at (-3,1) {$(b)$};
\node[draw=black,circle,scale=0.75]  (W) at (-2,-2) {$W$};
\node[draw=black,circle,scale=0.75]  (X) at (-1,0.5) {$X$};
\node (A) at (0,-2) {$A$};
\node[draw=black,circle,scale=0.75]  (Y) at (1,0.5) {$Y$};
\node[draw=black,circle,scale=0.75]  (Z) at (2,-2) {$Z$};
\node (c)  at (3.5,1) {$(c)$};
\node[draw=black,circle,scale=0.75]  (Y2) at (4.5,0.5) {$Y$};
\node[draw=black,circle,scale=0.75]  (Z2) at (6.5,0.5) {$Z$};
\node[draw=black,circle,scale=0.75]  (X2) at (5.5,-0.75) {$X$};
\node (A2) at (4.5,-2) {$A$};
\node (B2) at (6.5,-2) {$B$};
\node (C2) at (7.0,-2) { };
\draw [->,>=stealth] (A2)--(X2);
\draw [->,>=stealth] (B2)--(Z2);
\draw [->,>=stealth] (B2)--(X2);
\draw [->,>=stealth] (A2)--(Y2);
\draw [->,>=stealth] (X2)--(Z2);
\draw [->,>=stealth] (X2)--(Y2);
\draw [->,>=stealth] (A1)--(Y1);
\draw [->,>=stealth] (A1)--(Z1);
\draw [->,>=stealth] (Y1)--(Z1);
\draw [->,>=stealth] (X1)--(Y1);
\draw [->,>=stealth] (W)--(X);
\draw [->,>=stealth] (A)--(X);
\draw [->,>=stealth] (A)--(Y);
\draw [->,>=stealth] (Z)--(Y);
\end{tikzpicture}
}%
\caption[Causal structures with four or five nodes] {Three causal structures, $\smcaus$, for which the outer approximation, $\outmarcone{\smcaus^{\cC}}$ tightly approximates the classical entropy cone $\marcone{\smcaus^{\cC}}$, which also coincides with $\marcone{\smcaus^{\qQ}}$. The observed variables are labelled $W$, $X$, $Y$ and $Z$, the unobserved nodes are called $A$ and $B$.}
\label{fig:HensonStructures}
\end{figure}

\begin{example}[Inner approximation to the entropy cone of the classical instrumental scenario]
\label{example:innerI}
For the classical instrumental scenario, $\inst^\cC$ of Figure~\ref{fig:HensonStructures}(a), we can compute an inner approximation by adding the conditional independence constraints $I(A:X)=0$ and $I(X:Y|AZ)=0$ to the Ingleton cone $\inconek{4}$, as prescribed above. We can, however, also directly prove that $\inmarcone{\inst^{\cC}}=\outmarcone{\inst^{\cC}}$ (and hence also $\inmarcone{\inst^{\cC}}= \marcone{\inst^{\cC}}$) by showing that all Ingleton inequalities are implied by Shannon and conditional independence constraints.
Since $I(A:X)=0$, $I_\mathrm{ ING}\left(A,X;Y,Z \right) \geq 0$ is implied by Shannon and independence constraints and the rewritings of $I_\mathrm{ ING}$ according to \eqref{eq:rewritings1}--\eqref{eq:rewritings4} imply that $I(A:X)=0$ implies all Ingleton inequalities except for $I_\mathrm{ ING}\left(Y,Z;A,X \right) \geq 0$. Now, 
we can rewrite 
\begin{align}
I_\mathrm{ING}\left(Y,Z;A,X \right) &=I(Y:Z|A)+I(Y:X|Z)+I(X:A|Y)-I(X:Y|A) \\
&=I(Y:X|Z)+I(X:A|Y)+I(Y:Z|AX)-I(Y:X|AZ) \ ,
\end{align}
the positivity of which is implied by the Shannon inequalities and the independence constraint $I(X:Y|AZ)=0$.
\end{example}

Implementing all relevant linear rank inequalities of four and five variables (which includes their permutations and the application of the Ingleton inequality to each four variable subset as well as grouping several variables to one (see Section~\ref{sec:inner_general} and Ref.~\cite{Dougherty2009} for details) and then performing a variable elimination may be impractical and computationally too challenging for certain causal structures. For causal structures with strictly more than five random variables an inner approximation cannot be derived from linear rank inequalities in this way at all, since not all linear rank inequalities are known in these cases and since the list of such inequalities may even be infinite (see Section~\ref{sec:inner_general} and Ref.~\cite{Dougherty2014}). 
It is therefore useful to realise that for some causal structures, $\smcaus$, an inner approximation, $\inmarcone{\smcaus^{\cC}}$, can be obtained by intersecting the outer approximation, $\outmarcone{\smcaus^{\cC}}$, with $\inconek{k}$, where $k$ is the number of observed variables (this intersection can practically be computed as long as $k \leq 5$). 
The cones obtained using this intersection are not necessarily inner approximations, and, if they are, have to be proven as such, for example by explicitly constructing distributions that reproduce entropy vectors on each of the extremal rays (which is often straightforward). 
We have computed the respective inner approximations for the first few causal structures listed in Ref.~\cite{Henson2014} (we make the extremal rays as well as distributions recovering entropy vectors on each of them available at~\cite{URL}). In all of these examples the intersection procedure outlined here immediately led to an inner approximation.

In cases where one (or few) of the extremal rays of the cone that is obtained from such an intersection are not straightforwardly confirmed to be reproducible, dropping such rays may still yield a useful inner approximation. However, it may be more suitable in certain cases to take further linear rank inequalities into account in order to obtain a cone of which all extremal rays are achievable. The reason is that it may not always be obvious whether certain rays are reproducible or not, and, if one aims for a useful inner approximation, it is undesirable to drop too many rays.  
To prove certain rays to be unachievable, it is also possible to take non-Shannon inequalities into account (cf.\ Chapter~\ref{chap:nonshan} for a treatment of such inequalities).

\begin{figure}
\centering 
\resizebox{0.4\columnwidth}{!}{%
\begin{tikzpicture} [scale=1.1]
\node[draw=black,circle,scale=0.7] (A1a) at (-7.5,1) {$X_1$};
\node[draw=black,circle,scale=0.7] (A1) at (-7.5,0) {$X_2$};
\node[draw=black,circle,scale=0.75] (A2) at (-5.5,0) {$Z$};
\node[draw=black,circle,scale=0.75] (A3) at (-3.5,0) {$Y$};
\node[draw=gray,circle,scale=0.75] (A5) at (-3.5,1) {\color{gray}$S$};
\node (A4) at (-4.5,1.2) {$A$};
\draw [->,>=stealth] (A1)--(A2);
\draw [->,>=stealth] (A1a)--(A2);
\draw [->,>=stealth] (A2)--(A3);
\draw [->,>=stealth] (A4)--(A2);
\draw [->,>=stealth] (A4)--(A3);
\draw [gray,->,>=stealth] (A5)--(A3);
\draw [->,>=stealth] (A1a)--(A1);
\end{tikzpicture}
}%
\caption[Causal structure underlying the information causality game]{Causal structure underlying the information causality game, $\hat{\inst}$. Alice holds a database, here restricted to two pieces of information $X_1$ and $X_2$. These may not be independent, which is expressed by a potential causal influence from $X_1$ to $X_2$. She is then allowed to send a message $Z$ to Bob, who, takes a guess $Y$. Alice and Bob may have shared some resources before performing the protocol, either some classical randomness, $A$, a quantum system, $\rho_A$, or a resource from a more general theory, which Alice may use in order to choose her message and Bob may use to make his guess. A referee then tells Bob which bit to guess, a request that can also be modelled as a random variable, $S$, with causal link to $Y$, in which case we call the causal structure $\tilde{\inst}$.}
\label{fig:IC_inner_ch4}
\end{figure}
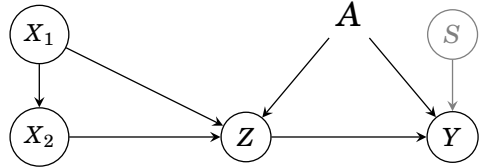

\begin{example} \label{example:IC_inner}
We consider the classical $5$-variable causal structure of Figure~\ref{fig:IC_inner_ch4}, $\hat{\inst}^{\cC}$. We can in principle consider all linear rank inequalities of five random variables combined with all Shannon inequalities and the conditional independence constraints, which would give us an inner approximation, $\inmarcone{\hat{\inst}^{\cC}}$, to the marginal cone, $\marcone{\hat{\inst}^{\cC}}$. This procedure would involve an (impractically) large number of inequalities.

Instead, we consider the outer approximation in terms of Shannon inequalities and conditional independence constraints, $\outmarcone{\hat{\inst}^{\cC}}$, and intersect this cone with the Ingleton cone for the four observed variables, $\inconek{4}$, i.e., we add the Ingleton inequalities~\eqref{eq:ingleton} for the four observed variables to $\outmarcone{\hat{\inst}^{\cC}}$. This is easily achieved computationally, but it does not lead to a restriction of the Shannon outer approximation, which is characterised by $52$ extremal rays.

 Adding the Ingleton inequality for all subsets of four out
    of the five random variables to $\outcone{\hat{\inst}^{\cC}}$ before performing the variable
    elimination, only $46$ extremal rays are recovered. These are straightforwardly confirmed to be   reproducible with entropy vectors in $\hat{\inst}^{\cC}$.

For the $6$ extremal rays of the outer approximation that we have not recovered in this way, we can also show that they are not achievable in $\hat{\inst}^{\cC}$, as they violate the inequalities we obtain when taking non-Shannon inequalities into account in the computation of the outer approximations to $\marcone{\hat{\inst}^{\cC}}$ (cf.\ Chapter~\ref{chap:nonshan} for the treatment of non-Shannon inequalities). Corresponding computational results are presented at the end of this chapter (Section~\ref{sec:technicalities_inner}).
\end{example}
In the following we give a detailed analysis of the inner approximation to the triangle causal structure. 

\section{Inner approximation to the entropy cone of the classical triangle scenario}\label{sec:triangle_inner}
In the following we first derive an inner approximation to the entropy cone of the triangle causal structure in terms of linear rank inequalities and then we outline improvements.

\subsection{Inner approximation by means of linear rank inequalities} \label{sec:inner_tri_lin}
An inner approximation to $\enconek{6}$ in terms of linear rank inequalities is not available (see also Section~\ref{sec:inner_general} and Ref.~\cite{Dougherty2014}). 
Nonetheless, we are able to derive an inner approximation to $\marcone{\tri^{\cC}}$ by relying on Ingleton's inequality. In the following, we apply Ingleton's inequality~\eqref{eq:ingletoninequ} to any subset of four out of the six random variables in $\tri^\cC$ (taking permutations into account).\footnote{Note that these are not all possible instances of the Ingleton inequality in the case of six variables.} We concisely write these additional
inequalities in a matrix $M_\mathrm{I}$ and consider the cone
\begin{equation}
\incone{\tri^{\cC}}=\left\{v \in \outconek{6} \ \middle| \ M_\mathrm{CI}\left(\tri^{\cC}\right) \cdot v = 0, \ M_\mathrm{I} \cdot v \geq 0 \right\}.
\end{equation}
When marginalising this cone we obtain
\begin{equation}
\inmarcone{\tri^{\cC}}= \left\{w \in \outconek{3} \ \middle| \ M_{\mathrm{I},\mathcal{M}}\left(\tri^{\cC}\right) \cdot w \geq 0 \right\},
\end{equation}
where $M_{\mathrm{I},\mathcal{M}}\left(C_3^{\cC}\right)$ contains only
one inequality,\footnote{Inequality~\eqref{eq:intinfo} renders the three Shannon inequalities of the form $I(X:Y|Z)\geq 0$ redundant. $\inmarcone{\tri^{\cC}}$ is thus fully characterised by the six remaining three variable Shannon inequalities 
and \eqref{eq:intinfo}.}
\begin{equation}\label{eq:intinfo}
-I(X:Y:Z) \geq 0.
\end{equation}
This relation can also be analytically derived from the Ingleton
inequality and the conditional independence constraints of
$\tri^{\cC}$, the proof proceeds as follows. There are only
  three instances of the Ingleton inequality that are not implied by
  the conditional independencies and the Shannon inequalities (cf.\
  also Proposition~\ref{prop:ingletonperm}). The independence
  constraint $I(X:Y|C)=0$ and its permutations $I(X:Z|B)=0$ and
  $I(Y:Z|A)=0$ lead to \eqref{eq:intinfo} in all three cases.
  Note also, that the same inner approximation would be obtained by dropping three extremal rays from $\outmarcone{\tri^{\cC}}$ and taking the convex hull of the others.

\begin{proposition} \label{prop:triangleingleton}
$\inmarcone{\tri^{\cC}}$ is a strict inner approximation to the marginal entropy cone of the triangle causal structure, i.e.,
\begin{equation} 
\inmarcone{\tri^{\cC}}\subsetneq \marcone{\tri^{\cC}} \ .
\end{equation}
\end{proposition}
The proof of Proposition~\ref{prop:triangleingleton} is deferred to Section~\ref{sec:triangle_ingleton}. $\inmarcone{\tri^{\cC}}$ provides a certificate for vectors to be realisable as entropy vectors in $C_3^{\cC}$: if a vector $v \in \mathbb{R}^{7}$ obeys all Shannon constraints as well as~\eqref{eq:intinfo}, then it lies in
$\marcone{\tri^{\cC}}$.

It is worth emphasising that not all correlations whose entropy
vectors lie in $\inmarcone{\tri^{\cC}}$
can be realised in $\tri^{\cC}$.  Instead, if ${\bH}(P)\in \inmarcone{\tri^{\cC}}$ then there exists $P'\in\mardist{\tri^{\cC}}$ such that ${\bH}(P')={\bH}(P)$. An example is given by the distribution
\begin{equation}\label{eq:w_cor}
P_{XYZ}(x,y,z)=\left\{\begin{array}{cl}\frac{1}{3}&\text{ }
    (x,y,z)=(1,0,0),(0,1,0),(0,0,1)\\
    0&\text{ otherwise,}\end{array}\right.
\end{equation}
which is not compatible with $\tri^{\cC}$, as shown in~\cite{Wolfe2016}. These correlations are not in $\mardist{\tri^{\cC}}$, but their entropy
vector nevertheless satisfies~\eqref{eq:intinfo}. Our argument
implies that there must be another distribution realisable in $\tri^{\cC}$ with the same entropy vector.\footnote{Another such example is given by the correlations realised in the scenario of Figure~\ref{fig:fritzproof}, which will be considered in detail in Section~\ref{sec:quantcorr}.}
\label{sec:w_correl}

\subsection{Improving inner approximations}\label{sec:improveinner} 
Here, we further approximate the boundary of $\marcone{\tri^\cC}$ from the inside by improving on $\inmarcone{\tri^{\cC}}$. The arguably most straightforward technique is to construct entropy vectors compatible with $\tri^{\cC}$ that lie outside $\inmarcone{\tri^{\cC}}$ and by taking the conic hull of these vectors and a vector on each of the extremal rays of $\inmarcone{\tri^{\cC}}$. Similar techniques have already been considered in the case of $\enconek{4}$, i.e., for four variable entropy vectors without causal restrictions~\cite{Dougherty2011}. 

In the causal context, we can perform random searches, but they have to be set up such that they meet the causal restrictions. Such searches usually yield vectors that lie within $\inmarcone{\tri^{\cC}}$, even when sampling quantum strategies. In the causal context, it thus seems suitable to start with the consideration of simple distributions that can be generated in $\tri^{\cC}$ and that are expected to lead to entropy vectors that lie close to the cone's boundary. Whether a vector lies within $\inmarcone{\tri^{\cC}}$ can then be checked with a linear program (in either the cone's $\hrep$-representation, where we check for violation of any of the inequalities, or in the $\vrep$-representation, where we could also check whether the vector can be written as a conic combination of entropy vectors on the extremal rays of the cone).
Whenever a vector is found that lies outside $\inmarcone{\tri^{\cC}}$, the conic hull of one non-zero vector on every extremal ray and the new vector yields an improved inner approximation to $\marcone{\tri^{\cC}}$.

\begin{example}[Entropy vectors in the classical triangle scenario obtained with functions that map two bits to one]\label{example:twotoone}
We take $A$, $B$ and $C$ to share one uniformly random bit each and we consider the case where bits $X$, $Y$ and $Z$ are generated with deterministic functions from $A$, $B$ and $C$, i.e., the situation where $X=f\left(B, C \right)$, $Y=g\left(A, C \right)$ and $Z=h\left(A, B \right)$, where $f$, $g$ and $h$ are arbitrary deterministic functions from two bits to one. Among the entropy vectors of distributions $P_{XYZ}$ that are generated in this way, three vectors lie outside $\marcone{\tri^{\cC}}$,
\begin{equation}\label{eq:twotoonevector}
\left(\frac{1}{2}+\frac{3}{4} \log_2\left(\frac{4}{3}\right), \ \frac{1}{2}+\frac{3}{4} \log_2\left(\frac{4}{3}\right), \ \frac{1}{2}+\frac{3}{4} \log_2\left(\frac{4}{3}\right), \ \frac{9}{8}+\frac{5}{8} \log_2\left(\frac{8}{5}\right), \ \frac{3}{2}, \ \frac{3}{2}, \ \frac{13}{8}+\frac{3}{8} \log_2\left(\frac{8}{3}\right) \right) \ ,
\end{equation}
with components $\left(H(X), H(Y), H(Z), H(XY), H(XZ), H(YZ), H(XYZ) \right)$, and two permutations thereof that correspond to permutations of $X$, $Y$ and $Z$. 

The new cone can also be converted to its $\hrep$-representation. As our computational tools only allow rational numbers, our vectors~\eqref{eq:twotoonevector} can not be used immediately. However, by rounding appropriately, meaning rounding such that we obtain vectors that can be written as a conic combination of the (unrounded) vectors on the extremal rays -- something we can check with a linear program -- this issue is taken care of and we can derive an inner approximation in $\hrep$-representation (which is slightly relaxed in comparison to the $\vrep$-representation). In the case of the three additional vertices~\eqref{eq:twotoonevector} rounded to a precision of $10^{-6}$ we obtain a set of $26$ inequalities that replace~\eqref{eq:intinfo} in our improved approximation. The details of this are presented in Section~\ref{sec:technicalities_inner_2}.
\end{example}

\smallskip

We remark here that this technique may also be applied to any other causal structure. For less than $6$ observed variables it is often a reasonable strategy to start with a set of linearly representable entropy vectors and to bulge the inner approximation out to obtain better approximations. Since the entropy vector method introduced in Chapter~\ref{chap:entropy_vec} is computationally feasible for causal structures with a small number of nodes, our methods for computing inner approximations are sufficient for the scenarios that are computationally tractable with the entropy vector method. The techniques described here will be relevant for our derivations in the two following chapters.
    
\section{Appendix}  
\subsection{Inner approximations to the marginal cones of the causal structures of Figure~\ref{fig:HensonStructures}}\label{sec:inner_examples} 

In case of the bipartite Bell scenario, $\smtwo$, the causal structure directly implies the following independencies among the four observed  variables\footnote{Note that $A$ and $YB$ do not share any ancestors  (similarly $AX$ and $B$).}:
\begin{align}
I(A:YB)&=0, \label{eq:P4indep}\\
I(AX:B)&=0. \label{eq:P4indep2}
\end{align}
In $\smtwo$ the Shannon inequalities together with~\eqref{eq:P4indep} and \eqref{eq:P4indep2} fully characterise the
set of achievable entropy vectors of the observed nodes, as we argue in the following.

The Shannon inequalities on four variables and Equations~\eqref{eq:P4indep} and~\eqref{eq:P4indep2} are necessary conditions for a vector
$v\in\mathbb{R}^{15}$ to be realisable as an entropy vector compatible with $\smcaus$, i.e., these constraints hold for any $v={\bH}(P_{AXYB})$ where $P_{AXYB}\in \mardist{\smtwo^\cC}$.  They therefore constrain an an outer approximation to $\marcone{\smtwo^\cC}$. In the following we show that this is also an inner approximation in this case. First, we list one vector on each extremal ray of $\marcone{\smtwo^\cC}$, where the components are ordered as
\begin{multline*}
\left( H(A),H(X), H(Y), H(B), H(AX), H(AY), H(AZ), H(XY), \right. \\ 
\left. H(XB), H(YB), H(AXY), 
H(AXB), H(AYB), H(XYB), H(AXYB) \right) \ ,
\end{multline*}
$$\arraycolsep=1.2pt 
\begin{array}{cccccccccccccccc}
(1)  &1 &1 &1 &1 &2 &2 &2 &2 &2 &2 &3 &3 &3 &3 &3 \\
(2)   &0 &1 &1 &1 &1 &1 &1 &2 &2 &2 &2 &2 &2 &2 &2 \\
(3)  &1 &1 &1 &0 &2 &2 &1 &2 &1 &1 &2 &2 &2 &2 &2 \\
(4)  &0 &0 &0 &1 &0 &0 &1 &0 &1 &1 &0 &1 &1 &1 &1 \\
(5)   &0 &0 &1 &0 &0 &1 &0 &1 &0 &1 &1 &0 &1 &1 &1 \\
(6)   &0 &1 &0 &0 &1 &0 &0 &1 &1 &0 &1 &1 &0 &1 &1 \\
(7)   &1 &0 &0 &0 &1 &1 &1 &0 &0 &0 &1 &1 &1 &0 &1 \\
(8)   &0 &0 &1 &1 &0 &1 &1 &1 &1 &1 &1 &1 &1 &1 &1 \\
(9)   &0 &1 &1 &0 &1 &1 &0 &1 &1 &1 &1 &1 &1 &1 &1 \\
(10) &1 &1 &0 &0 &1 &1 &1 &1 &1 &0 &1 &1 &1 &1 &1.
\end{array}
$$
If each of these rays is achievable with a distribution in
$\mardist{\smtwo^\cC}$, then, by convexity of
$\marcone{\smtwo^\cC}$, the outer approximation is also an inner approximation. In other
words, any vector $v$ that obeys the Shannon constraints
and Equations~\eqref{eq:P4indep} and~\eqref{eq:P4indep2} is then the entropy vector of at least one compatible distribution (or the limit of a sequence of such vectors), i.e., $v \in \marcone{\smtwo^\cC}$.  We establish this by taking
$C_1$, $C_2$ and $C_3$ to be uniform random bits and use the following
functions:
\begin{itemize}
\item To recover (1), take $A=C_1$, $X=C_1\oplus C_2$, $Y=C_2 \oplus C_3$ and
  $B=C_3$.
\item To realise (2), let $A=1$ be deterministic and choose $X=C_2$, $Y=C_2\oplus
  C_3$ and $B=C_3$. (3) can be achieved with an analogous strategy, where
  $B=1$ is the deterministic variable.
\item Choose $A=X=Y=1$ and $B=C_3$ to recover (4). (5), (6) and (7) are permutations of this strategy.
\item To obtain entropy vector (8), let $A=X=1$ be deterministic and let $Y=B=C_2$. (9) and (10) are permutations of this.
\end{itemize}

\bigskip

For the Figure~\ref{fig:HensonStructures}(c) we show in the following that $\outmarcone{\smthree^\cC} \cap \outconek{3}=\outmarcone{\smthree^\cC}$ is and inner approximation and hence also its entropy cone $\marcone{\smthree^\cC}$. Its Shannon inequalities and independence constraints lead to an outer approximation that is the conic hull of the following vectors, denoted here as lists of their components, ordered as $(H(X), H(Y), H(Z), H(XY), H(XZ), H(YZ), H(XYZ))$,
$$\arraycolsep=1.5pt 
\begin{array}{cccccccc}
(1)  &1 &1 &2  &2 &2 &2  &2 \\
(2)  &1 &2 &1  &2 &2 &2  &2 \\
(3)  &1 &1 &1  &2 &2 &2  &2 \\
(4)  &1 &2 &2  &2 &2 &2  &2 \\
(5)  &1 &1 &1  &2 &2 &1  &2 \\
(6)  &1 &1 &1  &1 &1 &1  &1 \\
(7)  &1 &0 &1  &1 &1 &1  &1 \\
(8)  &1 &1 &0  &1 &1 &1  &1 \\
(9)  &1 &0 &0  &1 &1 &0  &1 \\
(10) &0 &1 &0  &1 &0 &1  &1 \\
(11) &0 &0 &1  &0 &1 &1  &1 . \\
\end{array}
$$
The following strategies confirm that all of the extremal rays are achievable within the causal structure and, hence, that we have found the associated entropy cone. 
\begin{itemize}
\item The entropy vectors (1) and (2) are recovered by choosing $A$ and $B$ to be uniform bits and $X=A \oplus B$, $Y=B$, $Z=(A, X)$, or $X=A \oplus B$, $Y=(B, X)$, $Z=A$ respectively.
\item (3) is recovered by letting $A$ and $B$ be uniform bits and $X=A \oplus B$, $Y=B$, $Z=A$.
\item The entropy vector (4) is recovered by letting $A$ and $B$ be uniform bits and $X=A \oplus B$, $Y=(B,X)$, $Z=(A,X)$.
\item Let $A$ and $B$ be uniform bits and let $X=A \oplus B$, $Y=B$, $Z=A \oplus X$ to recover (5).
\item $X$ is a uniform bit and $Y=X=Z$ to recover (6).
\item To recover vectors (7) and (8), $A$ or $B$ are taken to be a uniform bit, and $X=A=Z$ or $X=B=Y$ respectively. The remaining variable is deterministic. 
\item Entropy vectors (9)-(11) are obtained by choosing either $X$, $Y$, or $Z$ respectively to be uniform bits and the other two variables to take a value deterministically.
\end{itemize} 

\vfill

\subsection{Technical details regarding Example~\ref{example:IC_inner}}\label{sec:technicalities_inner}  
Computing $\outmarcone{\hat{\inst}^{\cC}}$  
yields a cone with $52$ extremal rays. In the following, we list an entropy vector on each such extremal ray, with components 
\begin{multline*}
(H(X_1), H(X_2), H(Z), H(Y),H(X_1X_2),H(X_1Z),H(X_1Y),H(X_2Z),\\ H(X_2Y),H(ZY),H(X_1X_2Z), H(X_1X_2Y),H(X_1ZY),H(X_2ZY),H(X_1X_2ZY) ).
\end{multline*}
\noindent
\begin{minipage}{.5\textwidth}
{\small
$$\arraycolsep=1.pt 
\begin{array}{cccccccccccccccc}
(1) &1& 1& 1& 1 &  1 &1& 1 &1 &1 &1 &  1 &1& 1& 1 &  1 \\
(2) &1& 1 &1& 0 &  1 &1 &1 &1& 1 &1  & 1& 1& 1 &1 &  1 \\
(3) &1& 0& 1 &1  & 1& 1& 1 &1 &1& 1 &  1& 1& 1& 1 &  1 \\
(4) &0& 1 &1 &1 &  1& 1 &1 &1& 1 &1 &  1 &1 &1 &1 &  1 \\
(5) &1& 1& 0 &0  & 1& 1 &1 &1 &1& 0  & 1 &1& 1& 1  & 1 \\
(6) &1& 0& 1& 0 &  1& 1& 1& 1& 0& 1 &  1& 1& 1 &1  & 1 \\
(7) &0& 1& 1& 0   &1& 1& 0& 1& 1& 1&   1& 1& 1& 1&   1 \\
(8) &0& 0& 1& 1&   0& 1& 1& 1& 1& 1&   1& 1& 1& 1&   1 \\
(9) &1& 0& 0& 0&   1& 1& 1& 0& 0& 0&   1& 1& 1& 0&   1 \\
(10)&0& 1& 0& 0&   1& 0& 0& 1& 1& 0&   1& 1& 0& 1&   1 \\
(11)&0& 0& 1& 0&   0& 1& 0& 1& 0& 1&   1& 0& 1& 1&   1 \\
(12)&0& 0& 0& 1&   0& 0& 1& 0& 1& 1&   0& 1& 1& 1&   1 \\
(13)&1& 1& 1& 1&   2& 2& 2& 2& 2& 1&   2& 2& 2& 2&   2 \\
(14)&1& 1& 1& 1&   1& 2& 2& 2& 2& 2&   2& 2& 2& 2&   2 \\
(15)&1& 1& 1& 1&   1& 2& 1& 2& 1& 2&   2& 1& 2& 2&   2 \\
(16)&1& 0& 1& 1&   1& 2& 1& 1& 1& 2&   2& 1& 2& 2&   2 \\
(17)&0& 1& 1& 1&   1& 1& 1& 2& 1& 2&   2& 1& 2& 2&   2 \\
(18)&1& 0& 1& 1&   1& 2& 2& 1& 1& 2&   2& 2& 2& 2&   2 \\
(19)&0& 1& 1& 1&   1& 1& 1& 2& 2& 2&   2& 2& 2& 2&   2 \\
(20)&1& 1& 1& 0&   2& 2& 1& 2& 1& 1&   2& 2& 2& 2&   2 \\
(21)&0& 1& 1& 2&   1& 1& 2& 2& 2& 2&   2& 2& 2& 2&   2 \\
(22)&1& 0& 1& 2&   1& 2& 2& 1& 2& 2&   2& 2& 2& 2&   2 \\
(23)&1& 1& 1& 2&   1& 2& 2& 2& 2& 2&   2& 2& 2& 2&   2 \\
(24)&1& 1& 2& 1&   2& 2& 2& 2& 2& 2&   2& 2& 2& 2&   2 \\
(25)&1& 1& 2& 1&   2& 3& 2& 3& 2& 3&   3& 3& 3& 3&   3 \\
(26)&1& 1& 1& 2&   2& 2& 3& 2& 3& 2&   3& 3& 3& 3&   3 \\
\end{array}$$
}
\end{minipage}%
\begin{minipage}{.5\textwidth}
{\small
$$\arraycolsep=1.2pt 
\begin{array}{cccccccccccccccc}
(27)&2& 1& 2& 1&   2& 3& 2& 3& 2& 3&   3& 2& 3& 3&   3 \\
(28)&1& 2& 2& 1&   2& 3& 2& 3& 2& 3&   3& 2& 3& 3&   3 \\
(29)&1& 1& 2& 1&   2& 3& 2& 2& 2& 3&   3& 2& 3& 3&   3 \\
(30)&1& 1& 2& 1&   2& 2& 2& 3& 2& 3&   3& 2& 3& 3&   3 \\
(31)&1& 1& 2& 1&   2& 3& 2& 3& 2& 3&   3& 2& 3& 3&   3 \\
(32)&1& 1& 1& 1&   2& 2& 2& 2& 2& 2&   3& 3& 3& 3&   3 \\
(33)&1& 1& 2& 2&   2& 3& 3& 3& 2& 3&   3& 3& 3& 3&   3 \\
(34)&1& 1& 2& 2&   2& 3& 3& 2& 3& 3&   3& 3& 3& 3&   3 \\
(35)&1& 1& 2& 2&   2& 3& 2& 3& 3& 3&   3& 3& 3& 3&   3 \\
(36)&1& 1& 2& 2&   2& 2& 3& 3& 3& 3&   3& 3& 3& 3&   3 \\
(37)&1& 1& 2& 1&   2& 3& 2& 3& 1& 3&   3& 2& 3& 3&   3 \\
(38)&1& 1& 2& 1&   2& 3& 1& 3& 2& 3&   3& 2& 3& 3&   3 \\
(39)&1& 1& 1& 1&   2& 2& 2& 2& 2& 2&   3& 2& 3& 3&   3 \\
(40)&1& 1& 2& 2&   2& 3& 3& 3& 3& 3&   3& 3& 3& 3&   3 \\
(41)&1& 1& 2& 2&   2& 3& 2& 3& 2& 3&   3& 2& 3& 3&   3 \\
(42)&1& 1& 2& 3&   2& 3& 3& 3& 3& 3&   3& 3& 3& 3&   3 \\
(43)&1& 1& 2& 2&   2& 3& 3& 3& 3& 4&   4& 3& 4& 4&   4 \\
(44)&2& 1& 2& 1&   3& 4& 3& 3& 2& 3&   4& 3& 4& 4&   4 \\
(45)&1& 2& 2& 1&   3& 3& 2& 4& 3& 3&   4& 3& 4& 4&   4 \\
(46)&1& 1& 2& 3&   2& 3& 4& 3& 4& 4&   4& 4& 4& 4&   4 \\
(47)&2& 2& 2& 2&   3& 3& 3& 3& 4& 3&   4& 4& 4& 4&   4 \\
(48)&2& 2& 2& 2&   3& 3& 3& 4& 3& 3&   4& 4& 4& 4&   4 \\
(49)&2& 2& 2& 2&   3& 3& 4& 3& 3& 3&   4& 4& 4& 4&   4 \\
(50)&2& 2& 2& 2&   3& 4& 3& 3& 3& 3&   4& 4& 4& 4&   4 \\
(51)&2& 2& 3& 2&   3& 4& 3& 4& 3& 5&   5& 4& 5& 5&   5 \\
(52)&2& 2& 3& 2&   4& 4& 3& 4& 3& 4&   5& 4& 5& 5&   5 .\\
\end{array}
$$
}
\end{minipage}
Taking Ingleton's inequality for any subset of four out of the five random variables into account, we recover a cone characterised by the first $46$ of these extremal rays. We have identified probability distributions compatible with $\hat{\inst}^{\cC}$ that reproduce vertices on each of these rays. Hence their convex hull is an inner approximation to $\marcone{\hat{\inst}^{\cC}}$. For the rays (47)-(52) we can furthermore show that they are all outside $\marcone{\hat{\inst}^{\cC}}$ by resorting to non-Shannon inequalities. 

In the following we list distributions that allow us to recover vectors on the first $46$ of the above extremal rays, which proves that the convex hull of these rays is an inner approximation to $\marcone{\hat{\inst}^{\cC}}$. For this purpose, let $C_1$, $C_2$, $C_3$, $C_4$, $C_5$ and $C_6$ be uniformly random bits. 
\begin{itemize}
\item (1) Let $Y=Z=X_2=X_1=C_1$.
\item (2) Let $Z=X_2=X_1=C_1$ and $Y=1$; 
(3) let $Y=Z=X_1=C_1$ and $X_2=1$; (4) let $Y=Z=X_2=C_1$ and $X_1=1$.
\item (5) Let $X_2=X_1=C_1$ and $Y=Z=1$; (6) let $Z=X_1=C_1$ and $Y=1$, $X_2=1$; (7) let $Z=X_2=C_1$ and $Y=1$, $X_1=1$; (8) let $Y=Z=C_1$ and $X_1=1$, $X_2=1$.
\item (9) Let $X_1=C_1$ and $Y=Z=X_2=1$. (10)-(12) are permutations of this.
\item (13) Let $X_1=C_1$, $X_2=C_2$ and $Y=Z=X_1\oplus X_2$; (14) Let $X_2=X_1=C_1$, let $Y=A=C_2$ and $Z=X_2 \oplus A$.
\item (15) Let $X_2=X_1=C_1$, $A=C_2$, $Z=X_2 \oplus A$ and $Y=Z \oplus A$.
\item (16) Let $X_1=C_1$ and $X_2=1$, let $A=C_2$, $Z=X_1 \oplus A$ and $Y=Z \oplus A$; (17) let $X_1=1$ and $X_2=C_1$, let $A=C_2$, $Z=X_2 \oplus A$ and $Y=Z \oplus A$.
\item (18) Let $X_1=C_1$ and $X_2=1$, let $Y=A=C_2$ and $Z=X_1 \oplus A$; (19) let $X_1=1$ and $X_2=C_1$, let $Y=A=C_2$ and $Z=X_2 \oplus A$; (20) let $X_1=C_1$ and $X_2=C_2$, let $Z=X_1 \oplus X_2$ and $Y=1$.
\item (21) Let $X_1=1$ and $X_2=C_1$, let $A=C_2$, $Z=X_2 \oplus A$ and $Y=(A, Z \oplus A)$; (22) Let $X_1=C_1$ and $X_2=1$, let $A=C_2$, $Z=X_1 \oplus A$ and $Y=(A, Z \oplus A)$.
\item (23) Let $X_1=X_2=C_1$, let $A=C_2$, $Z=X_2 \oplus A$ and $Y=(A, Z \oplus A)$.
\item (24) Let  $X_1=C_1$ and $X_2=C_2$, let $Z=(Z_1,Z_2)=(X_1,X_2)$ and $Y=Z_1 \oplus Z_2$.
\item (25) Let $X_1=C_1$, $X_2=C_2$ and $A=C_3$, let $Z=(X_1\oplus A,X_2 \oplus A)$ and $Y=A$; (26) let $X_1=C_1$, $X_2=C_2$ and $A=C_3$, let $Z=X_1 \oplus X_2 \oplus A$ and $Y=(A, Z \oplus A)$.
\item (27) Let $X_1=(C_1,C_2)$, $X_2=C_1$ and $A=C_3$, let $Z=(Z_1,Z_2)=(C_2\oplus A,X_2 \oplus A)$ and $Y=Z_1 \oplus A$; (28) let $X_1=C_1$, $X_2=(X_1,C_2)$ and $A=C_3$, let $Z=(C_2\oplus A,X_1 \oplus A)$ and $Y=Z_1 \oplus A$.
\item (29) Let $X_1=C_1$, $X_2=C_2$ and $A=C_3$, let $Z=(Z_1,Z_2)=(X_1 \oplus A, X_2)$ and $Y=Z_1 \oplus Z_2 \oplus A$; (30) let $X_1=C_1$, $X_2=C_2$ and $A=C_3$, let $Z=(Z_1,Z_2)=(X_1 , X_2\oplus A)$ and $Y=Z_1 \oplus Z_2 \oplus A$. 
\item (31) Let $X_1=(C_1,C_2)$, $X_2=(C_3,C_4)$ and $A=(C_5,C_6)$, let $Z=(Z_1,Z_2,Z_3,Z_4)=(C_1 \oplus C_5, C_2 \oplus C_6, C_3 \oplus C_6, C_4 \oplus C_5 \oplus C_6)$ and $Y=(Z_1 \oplus Z_3 \oplus C_5 \oplus C_6,Z_2 \oplus Z_4 \oplus C_5)$.\footnote{Note that the entropy values here are double the ones given in the description of the extremal ray.}
\item (32) Let $X_1=C_1$, $X_2=C_2$ and $A=C_3$, let $Z=X_1 \oplus  X_2 \oplus A$ and $Y= A$.
\item (33) Let $X_1=C_1$, $X_2=C_2$ and $A=C_3$, let $Z=(Z_1,Z_2)=(X_1 \oplus A, X_2 \oplus A)$ and $Y=(A, Z_2 \oplus A)$; (34) let $X_1=C_1$, $X_2=C_2$ and $A=C_3$, let $Z=(Z_1,Z_2)=(X_1 \oplus A, X_2)$ and $Y=(Z_1, Z_2 \oplus A)$; (35) let $X_1=C_1$, $X_2=C_2$ and $A=C_3$, let $Z=(Z_1,Z_2)=(X_1 \oplus A, X_2 \oplus A)$ and $Y=(Z_1 \oplus A,  A)$; (36) let $X_1=C_1$, $X_2=C_2$ and $A=C_3$, let $Z=(Z_1,Z_2)=(X_1 , X_2 \oplus A)$ and $Y=(Z_1 \oplus A,  Z_2)$.
\item (37) Let $X_1=C_1$, $X_2=C_2$ and $A=C_3$, let $Z=(Z_1,Z_2)=(X_1 \oplus A , X_2 \oplus A)$ and $Y= Z_2 \oplus A$; (38) let $X_1=C_1$, $X_2=C_2$ and $A=C_3$, let $Z=(Z_1,Z_2)=(X_1 \oplus A , X_2 \oplus A)$ and $Y= Z_1 \oplus A$.
\item (39) Let $X_1=C_1$, $X_2=C_2$ and $A=C_3$, let $Z=X_1 \oplus  X_2 \oplus A$ and $Y=Z \oplus A$.
\item (40) Let $X_1=C_1$, $X_2=C_2$ and $A=C_3$, let $Z=(Z_1,Z_2)=(X_1 \oplus A, X_2 \oplus A)$ and $Y=(Z_1 \oplus Z_2 \oplus A,  A)$; (41) Let $X_1=C_1$, $X_2=C_2$ and $A=C_3$, let $Z=(Z_1,Z_2)=(X_1 \oplus A, X_2 \oplus A)$ and $Y=(Z_1 \oplus  A, Z_2 \oplus A)$; (42) let $X_1=C_1$, $X_2=C_2$ and $A=C_3$, let $Z=(Z_1,Z_2)=(X_1 \oplus A, X_2 \oplus A)$ and $Y=(Z_1 \oplus  A, Z_2 \oplus A, A)$.
\item (43) Let $X_1=C_1$, $X_2=C_2$ and $A=(C_3,C_4)$, let $Z=(Z_1,Z_2)=(X_1 \oplus C_3, X_2 \oplus C_4)$ and $Y=(C_3, Z_1 \oplus Z_2 \oplus C_3 \oplus C_4)$; (44) let $X_1=(C_1,C_2)$, $X_2=C_3$ and $A=C_4$, let $Z=(Z_1,Z_2)=(C_1 \oplus X_2, C_2 \oplus A)$ and $Y=Z_1 \oplus Z_2 \oplus A$; (45) let $X_1=C_1$, $X_2=(C_2, C_3)$ and $A=C_4$, let $Z=(Z_1,Z_2)=(X_1 \oplus C_2, C_3 \oplus A)$ and $Y=Z_1 \oplus Z_2 \oplus A$.
\item (46) Let $X_1=C_1$, $X_2=C_2$ and $A=(C_3,C_4)$, let $Z=(Z_1,Z_2)=(X_1 \oplus C_3, X_2 \oplus C_4)$ and $Y=(C_3,C_4, Z_1 \oplus Z_2 \oplus C_3 \oplus C_4)$.
\end{itemize} 

\subsection{Proof of Proposition~\ref{prop:triangleingleton}} \label{sec:triangle_ingleton}
In the following we prove Proposition~\ref{prop:triangleingleton}.
First, we explicitly calculate the $7$ extremal rays of the
  cone $\inmarcone{\tri^{\cC}}$. This can be
  done computationally and leads to the following extremal rays, which
  are denoted as the components of one particular vector on the
  ray:~\footnote{As usually, we order the components as $(H(X),H(Y),H(Z),H(XY),H(XZ),H(YZ),H(XYZ))$.}
  $$\arraycolsep=1.2pt 
\begin{array}{cccccccc}
(1)  &1 &1 &1 &2 &2 &2 &2 \\
(2)  &0 &0 &1 &0 &1 &1 &1 \\
(3)  &0 &1 &0 &1 &0 &1 &1 \\
(4)  &1 &0 &0 &1 &1 &0 &1 \\
(5)  &0 &1 &1 &1 &1 &1 &1 \\
(6)  &1 &0 &1 &1 &1 &1 &1 \\
(7)  &1 &1 &0 &1 &1 &1 &1 .
\end{array}
$$
These rays can also be analytically shown to be the extremal rays of
$\inmarcone{\tri^{\cC}}$.\footnote{To do
  so, note that in seven dimensions seven inequalities can lead to at
  most seven extremal rays (choosing six of the seven to be
  saturated).  One can then check that each of the claimed rays
  saturates six of the seven inequalities constraining
  $\inmarcone{\tri^{\cC}}$.}  For each
extremal ray we show how to generate a probability distribution
$P\in\mardist{\tri^{\cC}}$ whose entropy
vector $v\in\marcone{\tri^{\cC}}$
lies on the ray. To do so, let $A$, $B$ and $C$ be uniform random
bits.
\begin{itemize}
\item (1): Take $X=C$, $Z=A$ and $Y=A \oplus C$. 
\item (2): Take $Z=A$ and let $X=1$ and $ Y=1$ deterministic. (3) and
  (4) are permutations of this.
\item (5): Choose $Y=A=Z$ and let $X=1$ deterministic. (6) and (7) are permutations of this.
\end{itemize}
In this way, all extremal rays of
$\inmarcone{\tri^{\cC}}$ are achieved
by vectors in $\marcone{\tri^{\cC}}$ and, by
convexity of $\marcone{\tri^{\cC}}$,
we have $\inmarcone{\tri^{\cC}}\subseteq
\marcone{\tri^{\cC}}$.

To show that the inclusion is strict, let $A$, $B$ and $C$ be uniform
random bits. Let $X=\operatorname{AND}(B,C)$,
$Y=\operatorname{AND}(A,C)$ and $Z=\operatorname{OR}(A,B)$. The
marginal distribution $P_\mathrm{XYZ} \in
\mardist{\tri^{\cC}}$ leads to an entropy
vector $v_1=(0.81,~0.81,~0.81,~1.55,~1.5,~1.5,~2.16)$ and an
interaction information of $I(X:Y:Z) \approx 0.04 > 0$ (where all
numeric values are rounded to two decimal places). Hence, $v_1 \in
\marcone{\tri^{\cC}}$ but $v_1
\notin \inmarcone{\tri^{\cC}}$ and therefore
$\inmarcone{\tri^{\cC}}\subsetneq
\marcone{\tri^{\cC}}$. \qed  

\subsection{Inequalities defining the inner approximations in Example~\ref{example:twotoone}}\label{sec:technicalities_inner_2}  
In the following we give the $\hrep$-representation obtained by rounding the vectors \eqref{eq:twotoonevector} of Example~\ref{example:twotoone} to a precision of $10^{-6}$ and then converting the rational vectors from $\vrep$-representation to  $\hrep$-representation. Note that the vectors have been rounded to lie within the cone that is obtained with the original \eqref{eq:twotoonevector}, hence are guaranteed to yield an inner approximation. There are a total of $32$ inequalities, where the first $6$ are the Shannon inequalities that also characterise $\inmarcone{\tri^{\cC}}$ and the following $28$ (including permutation of $X$, $Y$ and $Z$ where applicable) replace inequality~\eqref{eq:intinfo},

{\small
\begin{align}
10451H(X)+  14149 \left(H(Y)+ H(Z)- H(XY) -  H(XZ) \right)-  17847 \left(H(YZ) - H(XYZ) \right) &\leq 0 \\ 
33083 \left(H(X)+  H(Y)- H(XY) \right) +  73761 \left(H(Z)
- H(XZ)- H(YZ)+  H(XYZ)\right) &\leq 0 \\  
237517 \left(H(X)+ H(Y)+ H(Z)\right) - 278195 \left(H(XY) 
+H(XZ)+H(YZ)\right)+ 318873H(XYZ) &\leq 0 \\  
303422 \left(H(X)+ H(Y)+ H(Z)- H(XY) - H(XZ)\right)- 283083 \left(H(YZ)-H(XYZ)\right) &\leq 0 \\ 
344361 \left(H(X)+ H(Y)- H(XY)- H(YZ)+ H(XYZ) \right) + 364700 \left(H(Z)- H(XZ) \right) &\leq 0 \\ 
385300H(X)+ 405639 \left(H(Y)+ H(Z)- H(XY) - H(XZ)-H(YZ)+ H(XYZ)\right) &\leq 0 \\  
959061 \left(H(X)+ H(Y)+ H(Z) \right)- 938722 \left(H(XY)+ H(XZ)+H(YZ) \right)+ 898044H(XYZ) &\leq 0 \\ 
1262483 \left(H(X)+ H(Y)+ H(Z)-H(YZ)\right)-1221805 \left(H(XY)+H(XZ)\right)+1181127H(XYZ) &\leq 0 \\ 
1385039H(X)+1344361 \left(H(Y)+H(Z)-H(XY)-H(XZ)-H(YZ)\right)+1303683H(XYZ) &\leq 0 \ .
\end{align}
}
  
%=========================================================

%% file: chapter05/chap_nonshan.tex
\let\textcircled=\pgftextcircled
\chapter{Exploring the gap between inner and outer approximations with non-Shannon inequalities}
\label{chap:nonshan}

\initial{T}his chapter is dedicated to the exploration of the gap between the inner and the outer approximations to the sets of compatible entropy vectors that were introduced in Chapters~\ref{chap:inner} and~\ref{chap:entropy_vec} respectively. In Section~\ref{sec:nonshan}, we outline a technique to improve on the outer approximations of entropy cones, which leads to the derivation of non-Shannon inequalities for causal structures. In Section~\ref{sec:triangle}, we apply this to improve the entropic characterisation of the classical triangle causal structure, one of the smallest causal structures that features such a gap. In addition to the computational procedure introduced in Section~\ref{sec:nonshan}, we also perform analytic derivations of new inequalities, including infinite families. In Section~\ref{sec:quantum_appl} we analyse the application of non-Shannon inequalities to quantum and so-called hybrid causal structures and in Section~\ref{sec:nonshan_post} we apply such inequalities in combination with the post-selection technique.

\section{Improving outer approximations with non-Shannon inequalities}\label{sec:nonshan}
To improve previous outer approximations to the entropy cone of a classical causal structure, $C^{\cC}$, we consider the techniques introduced in Refs.~\cite{Chaves2012, Fritz2013, Chaves2014b}, previously outlined in Section~\ref{sec:causal}. 
\begin{enumerate}[(1)]
\item Take the Shannon inequalities for the joint distribution of all variables in $C^{\cC}$ (observed and unobserved).
\item Take all conditional independence equalities that are implied by $C^{\cC}$ into account.
\item Eliminate all entropies of unobserved variables (or sets of variables containing such) from the full set of inequalities (e.g.\ by means of a Fourier-Motzkin elimination algorithm~\cite{Williams1986}, see also Section~\ref{sec:fm_elim}).
\end{enumerate}
To improve on this, we realise that the only reason that in most cases the outer approximations $\outmarcone{C^\cC}$ are not tight is the restriction of the approach to Shannon inequalities. If we were to consider the entropy cone of all involved variables $\enconekO$ instead of its Shannon approximation $\outconekO$, we would always recover $\marcone{C^{\cC}}$ with this method.
To improve on the outer approximations we hence adapt the computational procedure above as follows.
\begin{enumerate}[(1)]
\item Take the Shannon inequalities for the joint distribution of all variables in $C^{\cC}$ (observed and unobserved).
\item Add a set of valid and irredundant non-Shannon inequalities.
\item Take all conditional independence equalities that are implied by $C^{\cC}$ into account.
\item Eliminate all entropies of unobserved variables (or sets of variables containing such) from the full set of inequalities.
\end{enumerate}
The procedure leads to new constraints on the entropies of the observed variables in $C^{\cC}$ and as such often yields a new and improved outer approximation to $\marcone{C^{\cC}}$. 

Our results from Section~\ref{sec:fourfive} imply that the consideration of non-Shannon inequalities cannot lead to any further constraints for the causal structures from Figure~\ref{fig:HensonStructures}. For the causal structure $\hat{\inst}^\cC$ from Example~\ref{example:IC_inner}, on the other hand, we find that they do. Furthermore, we find that for the remaining $18$ causal structures listed in~\cite{Henson2014} (independent examples that all involve six variables), non-Shannon inequalities are invariably relevant. We can supplement the inequalities listed in~\cite{Henson2014} with new inequalities, tightening the entropic outer approximations, in all cases.\footnote{The consideration of inequality~\eqref{eq:zy} is sufficient to see this.} 

For scenarios with four (or more) \emph{observed} variables, instances of the non-Shannon inequalities, e.g.\ of inequality~\eqref{eq:zy}, become relevant without reference to the unobserved nodes. Such instances hold whether the unobserved nodes represent classical, quantum or more general non-signalling resources and allow us to tighten the outer approximation of the set of compatible entropy vectors in each case (contrary to when non-Shannon inequalities are applied to unobserved variables, in which case they are only known to hold for classical causal structures). The same reasoning applies to the comparison of different causal structures with the same set of observed variables: while non-Shannon inequalities derived relying on unobserved variables may lead to inequalities that help distinguish the two causal structures, the application of non-Shannon inequalities to the observed variables cannot.
In the following we provide a detailed entropic analysis of one particular example, the triangle scenario. It stands out with respect to the other $17$ examples in the sense that it only involves three observed variables, therefore all applications of non-Shannon inequalities involve the unobserved nodes.

\section{Improved entropic approximation for the classical triangle scenario} \label{sec:triangle} 

Comparing inner and outer approximations to the entropy cone of the classical triangle causal structure (see Example~\ref{example:TRI_class_obs} and Section~\ref{sec:triangle_inner}, as well as Section~\ref{sec:causal_intro}), there are three extremal rays of $\outmarcone{\tri^\cC}$ that are not within our inner approximations, specifically $\bH=(2,3,3,4,4,5,6)$ and permutations (all other extremal rays coincide with those of the inner approximation derived in Section~\ref{sec:triangle_ingleton}). 
In the following we show how to shave off parts of the region between inner and outer approximation with non-Shannon inequalities that prove the additional rays to be inachievable. We do this with the computational method outlined in the previous section alongside a derivation of several infinite families of inequalities.

\subsection{Motivating the search for improved outer approximations to the marginal cone of the triangle causal structure} \label{sec:insuff}
We first prove that the Shannon approximation to the marginal cone of the triangle causal structure is not tight, i.e., that $\marcone{\tri^{\cC}} \subsetneq \outmarcone{\tri^{\cC}}$. This motivates our attempts to derive non-Shannon inequalities to improve current approximations. It also disproves a claim of Refs.~\cite{Chaves2014,Chaves2015}, which together argue that in the marginal scenario there is no entropic separation between the Shannon cone and the classical entropy cone, i.e., that $\outmarcone{\tri^{\cC}}=\marcone{\tri^{\cC}}$.\footnote{The details of this were presented in the Supplementary Information of~\cite{Chaves2015}.} This is of importance here, as the claim of~\cite{Chaves2014,Chaves2015} implies that all non-Shannon inequalities are redundant in this scenario.

\begin{proposition}\label{prop:strictsub}
The marginal entropy cone of the classical triangle scenario is a proper subset of the corresponding marginal Shannon cone, i.e.,
\begin{equation} \marcone{\tri^{\cC}} \subsetneq \outmarcone{\tri^{\cC}} \ . \end{equation}
\end{proposition}

Proposition~\ref{prop:strictsub} implies that better approximations to $\marcone{\tri^{\cC}}$ exist and, hence, the exploration of non-Shannon constraints is of interest in the triangle scenario. Before proceeding with the proof, we note that the outer approximation, $\outmarcone{\tri^{\cC}}$ was explicitly computed by Chaves et al.~\cite{Chaves2014,Chaves2015} and was reiterated as Example~\ref{example:TRI_class_obs}.

\begin{proof} 
We prove this by adding the four valid inequalities
\begin{align}
&\Diamond_{YZAX} \geq 0, \label{eq:ZYindep} \\
&\Diamond_{XZBY} \geq 0, \\
&\Diamond_{XYCZ} \geq 0, \\
&\Diamond_{YZXA} \geq 0,
\end{align}
to $\outcone{\tri^{\cC}}$ and by then performing a Fourier-Motzkin elimination to remove terms involving $A$, $B$ and $C$. The resulting set of inequalities includes the Shannon inequalities
for three jointly distributed random variables as well as the additional inequalities
\begin{align}
-H(X)-H(Y)-H(Z)+H(XY)+H(XZ) &\geq 0, \label{eq:first_inequ} \\
-4H(X)-4H(Y)-4H(Z)+3H(XY)+3H(XZ)+4H(YZ)-2H(XYZ) &\geq 0, \label{eq:zytriangle2}\\
-2H(X)-2H(Y)-2H(Z)+3H(XY)+3H(XZ)+3H(YZ)-4H(XYZ) &\geq 0, \\
-8H(X)-8H(Y)-8H(Z)+7H(XY)+7H(XZ)+7H(YZ)-5H(XYZ) &\geq 0, \label{eq:zytriangle}
\end{align} 
up to permutations of $X$, $Y$ and $Z$, where applicable. All except for inequality~\eqref{eq:first_inequ} (and its permutations) are new as compared to the Shannon approximation from Example~\ref{example:TRI_class_obs}. 
For each new inequality (\eqref{eq:zytriangle2}--\eqref{eq:zytriangle}) we find violations with vectors in the
interior of $\outmarcone{\tri^{\cC}}$, specifically the vector $v=\left( 11, 14, 14, 20, 20, 23, 28 \right)$ or a vector with appropriately permuted $X$, $Y$ and $Z$. Hence, $\marcone{\tri^{\cC}}\subsetneq\outmarcone{\tri^{\cC}}$.
\end{proof}

It is interesting to note in this context, that $\tri^{\cC}$ involves only three observed variables such that without causal restrictions $\outmarconeO$ would be the three variable Shannon cone and non-Shannon inequalities would be irrelevant. Nevertheless, the non-Shannon constraints of the six variable scenario together with the independence constraints lead to new and non-trivial constraints when marginalised, in contrast to many other three variable scenarios. 

\subsection{Improved outer approximations to the entropy cone of the triangle causal structure}\label{sec:newouter}
The insight that $\outmarcone{\tri^{\cC}}$ does not tightly approximate $\marcone{\tri^{\cC}}$ naturally
leads to the search for better approximations that enable a more accurate distinction of the triangle causal structure from other scenarios, such as the one from Figure~\ref{fig:3variables}(c). Such approximations are derived in the following, using non-Shannon inequalities.\footnote{Such inequalities survive the projection to the three variable scenario, despite that they would not be relevant for a three variable marginal scenario, if we were not to suppose any causal constraints.} As there are infinitely many such inequalities, the following reasoning may be applied to reduce their number. All known (and thus far) irredundant four variable non-Shannon entropy inequalities can be written as the sum of the \emph{Ingleton quantity}~\eqref{eq:ingleton} and (conditional) mutual information terms, as mentioned in Section~\ref{sec:classicalcone}. Since the latter are always positive, all known non-Shannon inequalities are irrelevant (i.e., implied by existing constraints) for variable choices for which the causal restrictions imply that the Ingleton term is non-negative. This significantly reduces the choice of variables for which the known additional inequalities may be relevant.

\begin{example}
Consider Proposition~\ref{prop:zhangyeung} with $(X_1,~X_2,~X_3,~X_4)=(A,~B,~C,~X)$. The corresponding inequality can be written as \begin{equation} I(A:B|C)+I(A:B|X)+I(C:X)-I(A:B)+I(A:C|B)+I(B:C|A)+I(A:B|C) \geq 0 \ .\end{equation}
Whenever a causal structure $C^{\cC}$ implies $I(A:B)=0$, i.e., independence of $A$ and $B$, the above inequality is implied by the Shannon inequalities and the independence constraint $I(A:B)=0$. Hence, it will not improve on the outer approximation we already have.
\end{example}

The following proposition restricts the inequalities that may be relevant for the derivation of our improved approximations to $\marcone{\tri^\cC}$.\footnote{Notice that the proposition restricts the subsets of four out of the six random variables for which the known non-Shannon inequalities may be relevant. It is also possible to apply non-Shannon inequalities when interpreting the joint distribution of two or three random variables in the causal structure as one. We have, however, not looked into these here.} 

\begin{proposition} \label{prop:ingletonperm} Consider an entropy inequality
  that states the non-negativity of a positive linear combination of
  the Ingleton quantity~\eqref{eq:ingleton} and (conditional)
  mutual information terms. This inequality is implied by the Shannon
  inequalities  and the conditional independencies of
 $\tri^{\cC}$  (i.e., $I(A: XBC)=0$, $I(X: YZA|BC)=0$ and appropriate permutations) for all possible choices of four out of the six involved random variables, except
\begin{align}
\left( X_1,~X_2,~X_3,~X_4 \right) &= \left( X,~Y,~Z,~C \right)\\
&= \left( X,~Z,~Y,~B \right) \\
&= \left( Y,~Z,~X,~A \right),
\end{align}
up to exchange of $X_1$ and $X_2$ as well as $X_3$ and $X_4$. 
\end{proposition}

\begin{proof} 
For four random variables $X_1$, $X_2$, $X_3$ and $X_4$, the Ingleton inequality
\begin{equation}\label{eq:ingletproof}
I(X_1:X_2|X_3) + I(X_1:X_2|X_4) + I(X_3:X_4) - I(X_1:X_2)\geq 0\\
\end{equation}
can be equivalently rewritten in four more ways with the following equalities:
\begin{alignat}{2}
I(X_1:X_2|X_3)-I(X_1:X_2)&=I(X_1:X_3|X_2)&&-I(X_1:X_3)  \label{eq:rewritings1}\\
&=I(X_2:X_3|X_1)&&-I(X_2:X_3), \label{eq:rewritings2} \\
I(X_1:X_2|X_4)-I(X_1:X_2)&=I(X_1:X_4|X_2)&&-I(X_1:X_4) \label{eq:rewritings3} \\
&=I(X_2:X_4|X_1)&&-I(X_2:X_4) \ . \label{eq:rewritings4}
\end{alignat}
For the inequality \eqref{eq:ingletproof} not to be implied by the Shannon inequalities and the conditional independencies we need $X_1$, $X_2$, $X_3$ and $X_4$ to be such that
\begin{alignat}{3}
&I(X_1:X_2)& &> 0, \label{eq:infoterms1} \\ 
&I(X_1:X_3)& &> 0,\\
&I(X_1:X_4)& &> 0, \\
&I(X_2:X_3)& &> 0,  \\
&I(X_2:X_4)& &> 0, \label{eq:infoterms5}
\end{alignat}
hold simultaneously. If the conditional independencies of $\tri^{\cC}$
imply that one of these mutual informations is zero then the Ingleton
inequality can be expressed as a positive linear combination of
(conditional) mutual information terms in one of its five equivalent
forms and the corresponding non-Shannon inequality is redundant.

For the five constraints \eqref{eq:infoterms1}--\eqref{eq:infoterms5} to hold simultaneously,
$X_1$ and $X_2$ have to be correlated with one another as well as with two
further variables. This excludes the independent sources $A$, $B$ and
$C$ as candidates for $X_1$ and $X_2$; therefore $X_1,~X_2 \in \left\{X,~Y,~Z
\right\}$. Furthermore, the variables $X_3$ and
$X_4$ have to be correlated with both, $X_1$ and $X_2$. This excludes the two variables in
$\left\{ A,~B,~C \right\}$ that do not lie between $X_1$ and $X_2$ in
$\tri^{\cC}$. Hence, for each choice of $X_1$ and $X_2$, the variables $X_3$
and $X_4$ have to be chosen as the remaining element of $\left\{X,~Y,~Z
\right\}$ and the variable positioned opposite it in $\tri^{\cC}$.  
In summary, $\left(X_1,~X_2 \right) \left(X_3,~X_4 \right)$ can only be
$\left(X,~Y \right) \left(Z,~C \right)$, $\left(X,~Z \right)
\left(Y,~B \right)$ and $\left(Y,~Z \right) \left(X,~A \right)$ up to
permutations of the variables within a tuple.
\end{proof}

Of the 360 possible combinations of four variables (including permutations) only twelve may yield relevant non-Shannon inequalities to improve the approximation to $\marcone{\tri^{\cC}}$. We also remark that for most of the known non-Shannon inequalities, several of the remaining twelve choices can be shown to be redundant as well. Nonetheless, even after accounting for this redundancy, we cannot reduce to a finite number of inequalities. The outer approximation to $\marcone{\tri^{\cC}}$ could (potentially) be tightened by including all remaining inequalities. Hence, ideally, all of them should be added in step (2) of the computational procedure. However, in practice only a small number of inequalities can be added at a time until the task of marginalising becomes infeasible. The main reason is that the Fourier-Motzkin elimination algorithm used in step (4) is often too slow (see Section~\ref{sec:fm_elim} for information on the scaling of the algorithm). In our computations, the worst case scaling is usually not exhibited, since most of the inequalities we perform the elimination on contain few variables each and thus lead to far fewer than the maximal number of inequalities. However, we are still limited to adding a relatively small number of different inequalities at a time to avoid computational difficulties.

We have computed tighter approximations to $\marcone{\tri^{\cC}}$, by including the following
manageable sets of non-Shannon inequalities at a time:\smallskip

\noindent \emph{Case 1}: We include all permutations of the inequality of Proposition~\ref{prop:zhangyeung} as well as all six inequalities from~\cite{Dougherty2006} for the $12$ four variable combinations that are not shown to be redundant by Proposition~\ref{prop:ingletonperm}.\footnote{Note that these are all four variable non-Shannon inequalities that can be generated with at most two instances of the so-called copy lemma~\cite{Dougherty2011} (see also Section~\ref{sec:inner_general}).}
\smallskip

\noindent \emph{Case 2}: We include all inequalities of the form given
in~\eqref{eq:matus1} and~\eqref{eq:matus2} for $s=1,2,3$, applied to all four variable subsets of the six random variables of $\tri^\cC$.\bigskip

\noindent The resulting sets of constraints for the outer approximation to $\marcone{\tri^{\cC}}$ are available at~\cite{URL}. We have furthermore derived infinite families of valid inequalities.
\begin{proposition}\label{prop:matus1}
All entropy vectors $\bH \in \marcone{\tri^{\cC}}$ obey
\begin{multline}\label{eq:marginalmatus1}
\left(-\frac{1}{2} s^2 - \frac{3}{2} s \right) \left( H(X) +H(Z) \right)+ \left(-s-1\right) H(Y)\\
+\left( \frac{1}{2} s^{2} + \frac{3}{2} s + 1 \right) \left( H(XY) + H(YZ) \right)
+\left( s^{2} + 2s \right) H(XZ) + \left( -s^2-2s-1 \right) H(XYZ) \geq 0,
\end{multline}
for all $s \in \mathbb{N}$. The same holds for all permutations of $X$, $Y$ and $Z$.
\end{proposition}

\begin{proof} 
The instance of inequality \eqref{eq:matus1} with $(X_1,~X_2,~X_3,~X_4)=(X,~Y,~Z,~C)$ (which is known to hold for any entropy vector and hence for any $\bH \in \encone{\tri^{\cC}}$) can be rewritten as 
\begin{multline*}
\left(-\frac{1}{2} s^2 - \frac{3}{2} s \right) H(X)
+ \left(-s-1\right) H(Y) 
+ \left(-\frac{1}{2} s^2 - \frac{1}{2} s \right) H(Z)
+ s H(CX) 
+ s H(CY)
- s H(CZ)
- s H(CXY) \\
+\left( \frac{1}{2} s^{2} + \frac{3}{2} s + 1 \right) H(XY) 
+ \left( s^{2} + 2s \right) H(XZ)
+ \left( \frac{1}{2} s^{2} + \frac{3}{2} s + 1 \right) H(YZ) 
+ \left( -s^2-2s-1 \right) H(XYZ) \geq 0.
\end{multline*}
Applying $I(X:Y|C)=0$ and $I(Z:C)=0$, all terms containing the variable $C$ cancel, which leads to inequality~\eqref{eq:marginalmatus1}. Thus any $\bH \in \marcone{\tri^{\cC}}$ obeys~\eqref{eq:marginalmatus1}.
\end{proof}

The proof of Proposition~\ref{prop:matus1} relies on the validity of the family of inequalities~\eqref{eq:matus1}, which is the basis for the derivation of further infinite families of valid inequalities. However, extracting an explicit derivation for such inequalities is in general more involved, these being a consequence of several Shannon inequalities and permutations of~\eqref{eq:matus1}.

\begin{proposition}\label{prop:matus1fam2}
All entropy vectors $\bH \in \marcone{\tri^{\cC}}$ obey
\begin{multline}
\left(-\frac{1}{2} s^2 - \frac{3}{2} s -2\right) \left( H(X) + H(Y) + H(Z) - H(XY) \right) \\
+\left(\frac{1}{2} s^2 + \frac{3}{2} s +1\right)H(XZ)
+\left(  s + 2 \right) H(YZ) 
+ \left( -s-1 \right) H(XYZ) \geq 0, \label{eq:mres}
\end{multline}
for all $s \in \mathbb{N}$. The same holds for all permutations of $X$, $Y$ and $Z$.
\end{proposition}

Proposition~\ref{prop:matus1fam2} is derived from~\eqref{eq:matus1} by combining the inequalities for one value $s \in \mathbbm{N}$ at a time with Shannon and conditional independence constraints. Other than Proposition~\ref{prop:matus1}, it is the only family resulting from \eqref{eq:matus1} in this way, for which none of its inequalities are rendered redundant by those found in the calculations for Case~2, i.e., by inequalities that arise when inequalities with different $s$-values are simultaneously considered. (It is not known whether our families would survive if we were able to consider further $s$-values in Case~2.) The proposition is proven in Section~\ref{sec:pr1}.

Similar considerations can be applied to \eqref{eq:matus2} and could be extended to further families of inequalities~\cite{Dougherty2011}, Proposition~\ref{prop:matus2} is such an example, the proof of which is presented in Section~\ref{sec:pr2}.

\begin{proposition}\label{prop:matus2}
All entropy vectors $\bH \in \marcone{\tri^{\cC}}$ obey
\begin{multline}
\left(-\frac{1}{2} s^2 - \frac{3}{2} s -2 \right) \left( H(X) +H(Z)-H(XY) \right)
+ \left(-2s-2\right) H(Y) \\
+\left( s^{2} + 2 \right) H(XZ) 
+\left( \frac{1}{2} s^{2} + \frac{3}{2} s + 1 \right) \left(  H(YZ) \right)
+ \left( -s^2-1 \right) H(XYZ) \geq 0, \label{eq:nres}
\end{multline}
for all $s \in \mathbb{N}$. The same holds for all permutations of $X$, $Y$ and $Z$.
\end{proposition}

One might expect that adding five and six variable inequalities to $\outcone{\tri^{\cC}}$ could lead to further restrictions. However, this is observed not to be the case.
\begin{remark}
The known genuine five and six variable inequalities from~\cite{Zhang1998, Makarychev2002} are not relevant for the entropic outer approximation to $\marcone{\tri^{\cC}}$.
\end{remark}
 
This can be shown by expanding the inequalities into a linear combination of mutual information terms and a similar reasoning as is applied in the proof of Proposition~\ref{prop:ingletonperm}. As they are not particularly instructive, the details of these arguments are omitted here.

\begin{conjecture}\label{conj:matusconj}
$\marcone{\tri^{\cC}}$ is not polyhedral, i.e., infinitely many linear inequalities are needed for its characterisation.
\end{conjecture}

Our main evidence for this is that the family of inequalities~\eqref{eq:matus1}, used by Mat\'{u}\u{s} to prove that the analogue of this conjecture holds for $\enconek{4}$, leads to infinite families of inequalities for $\tri^{\cC}$ after marginalising (cf.\ Proposition~\ref{prop:matus1}). Hence, it may be that the structure of this region of entropy space is to some extent preserved under the causal constraints and the projection to the marginal scenario and thus retains a non-polyhedral segment of the cone's boundary. It is, however, also conceivable that none of the non-polyhedral boundary regions survive this mapping.

\section{Non-Shannon inequalities in quantum and hybrid causal structures} \label{sec:quantum_appl}
In the following, we analyse the quantum triangle scenario $\tri^{\qQ}$ as well as hybrid versions, $\tri^{\cCCQ}$ and $\tri^{\cCQQ}$, with non-Shannon inequalities.

\subsection{Quantum triangle scenario}\label{sec:quantcorr} 
\label{sec:quantumtriangle}
To motivate the separate entropic analysis of $\tri^{\qQ}$, it is important to establish that some  probability distributions of the observed variables $X$, $Y$ and $Z$ in $\tri^{\qQ}$ cannot be obtained classically, hence $\tri^{\qQ}$ might lead to a larger set of compatible entropy vectors than $\tri^{\cC}$. That there is a separation at the level of correlations was proven in~\cite{Fritz2012} by identifying a quantum distribution that is not classically reproducible, where each of the random variables $X$, $Y$ and $Z$ outputs two bits. In the following, we generalise this to a scenario where one party outputs one bit only.

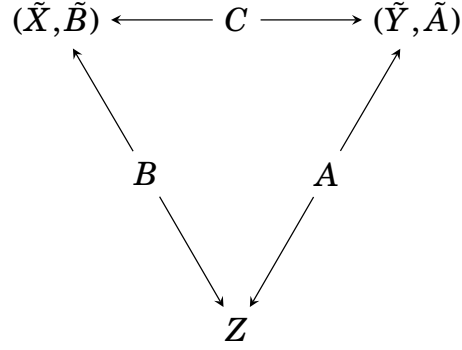
\begin{figure}
\centering
\resizebox{0.4 \columnwidth}{!}{%
\begin{tikzpicture}
\node (X) at (-2,2) {$(\tilde{X},\tilde{B})$};
\node (Y) at (2,2) {$(\tilde{Y},\tilde{A})$};
\node (Z) at (0,-1.46) {$Z$};
\node (A) at (1,0.28) {$A$};
\node (B) at (-1,0.28) {$B$};
\node (C) at (0,2) {$C$};
\draw [->,>=stealth] (A)--(Y);
\draw [->,>=stealth] (A)--(Z);
\draw [->,>=stealth] (B)--(X);
\draw [->,>=stealth] (B)--(Z);
\draw [->,>=stealth] (C)--(X);
\draw [->,>=stealth] (C)--(Y);
\end{tikzpicture}
}%
\caption[Strategies to derive quantum correlations that are not classically reproducible in the triangle causal structure]{Scenario involving unobserved quantum systems, leading to a
  distribution which is not reproducible with classical $A$, $B$ and
  $C$~\cite{Fritz2012}. The observed variables $X=(\tilde{X},\tilde{B})$ and $Y=(\tilde{Y},\tilde{A})$ are chosen such that
$P_{\mathrm{\tilde{X}\tilde{Y}|AB}}$ maximally violates the CHSH
inequality~\cite{Clauser1969}.
 $Z=(A',B')$ is such that $B'=\tilde{B}=B$ and
$A'=\tilde{A}=A$.  
In Proposition~\ref{thm:andextension}, we prove that a strategy where $Z=\mathrm{AND}(A',B')$ also leads to correlations that cannot be mocked up classically.}
\label{fig:fritzproof}
\end{figure}

\begin{proposition} \label{thm:andextension}
There are non-classical quantum correlations in $\tri$ where $X$ and $Y$ output two bits each while $Z$ outputs only one.
\end{proposition}

Before proceeding with the proof, we analyse the strategy that leads to a quantum distribution that is shown not to be classically reproducible in~\cite{Fritz2012}. We later adapt this strategy to prove our own result. In $\tri^{\qQ}$ one can take $X$ and $Y$ to correspond to two bits, which we call $(\tilde{X},\tilde{B})$ and $(\tilde{Y},\tilde{A})$ respectively. The quantum state corresponding to node
$C$ is a maximally entangled state $\ket{\Psi_\mathrm{C}}=\frac{1}{\sqrt{2}} ( \left| 01 \right\rangle - \left| 10 \right\rangle)$, the first half of which is the subsystem to $C_X$ and the second half is $C_Y$. $A$ and $B$ can be taken to be uniform classical bits. We introduce $\Pi_\theta=\ketbra{\theta}{\theta}$, where
$\ket{\theta}=\cos(\frac{\theta}{2})\ket{0}+\sin(\frac{\theta}{2})\ket{1}$, and the four POVMs,
\begin{align}
&E_0=\left\{\Pi_0,\Pi_{\pi}\right\} \ , &&E_1=\left\{\Pi_{\pi/2},\Pi_{3\pi/2}\right\} \ ,\\
&F_0=\left\{\Pi_{\pi/4},\Pi_{5\pi/4}\right\} \ , &&F_1=\left\{\Pi_{3\pi/4},\Pi_{7\pi/4}\right\} \ .
\end{align}
Consider a measurement on the $C_X$ subsystem with POVM $E_B$ (i.e., if $B=0$ then $E_0$ is measured and otherwise $E_1$), and likewise a measurement on $C_Y$ with POVM $F_A$.  Let us denote the corresponding
outcomes $\tilde{X}$ and $\tilde{Y}$.  With this choice $P_{\mathrm{\tilde{X}\tilde{Y}|AB}}$ violates the CHSH inequality~\cite{Clauser1969}.  The observed variables are then $X=(\tilde{X},\tilde{B})$, $Y=(\tilde{Y},\tilde{A})$ and $Z=(A',B')$, with the correlations set up such that $B'=\tilde{B}=B$ and
$A'=\tilde{A}=A$. In essence the reason that this cannot be realised in the causal
structure $\tri^{\cC}$ is the CHSH violation. Note though that it is also crucial that information about $A$ is present in both $Y$ and $Z$ (and analogously for $B$).  If for example, we consider the same
scenario but with $Y=\tilde{Y}$ then we could mock-up the correlations classically.  This can be done by removing $A$, replacing $B$ with $(B_1,B_2)$ and taking $B_1,B_2$ and $C$ to each be a uniform random bit.  We can then take $Y=C$, $Z=(B_1,B_2)$ and $X=(f(C,B_1,B_2),B_1)$, where $f$ is chosen appropriately. Since $f$ can depend on all of the other observed variables it can generate any correlations between them\footnote{This is like playing the CHSH game but where Alice knows Bob's input and output in addition to her own.}. In the causal structure $\tri$, taking $\tilde{A}=A'$ ensures that these are shared through $A$ and hence information about them cannot be used to generate $X$.

From this, one might expect that all information about the measurement choices in this Bell setup has to be exposed at the observed node $Z$. Our Proposition shows that this is not the case.
We prove that a strategy where $Z=\operatorname{AND}(A,B)$ also leads to correlations that cannot be reproduced classically. We consider the following observed distribution,
\begin{equation} \label{eq:violatingdistr}
P_\mathrm{\tilde{A}\tilde{B}\tilde{X}\tilde{Y}Z}
=\frac{1}{4}P_\mathrm{\tilde{X}\tilde{Y}|\tilde{A}
  \tilde{B}}\delta_{Z,\operatorname{AND}(\tilde{A},\tilde{B})}\, ,
\end{equation}
where $\tilde{A}=A$, $\tilde{B}=B$ and $P_\mathrm{\tilde{X}\tilde{Y}|\tilde{A}\tilde{B}}$ maximally violates the CHSH inequality, according to the strategy detailed above, but $Z=\operatorname{AND}(A,B)$. In a slight abuse of notation we refer to this as $P_\mathrm{XYZ}$. Our proof relies on the following lemma in which we ignore $\tilde{X}$ and $\tilde{Y}$ and which is proven in Section~\ref{sec:lemma_proof}.

\begin{lemma}\label{lemma:andlemma}
If $P_\mathrm{XYZ} \in \mathcal{P}_\mathcal{M}\left(C_3^{\cC}\right)$, then
$P_\mathrm{\tilde{A}|A C}=P_\mathrm{\tilde{A}|A}$ and $P_\mathrm{\tilde{B}|B C}=P_\mathrm{\tilde{B}|B}$.
\end{lemma}

\begin{proof}[Proof of Proposition~\ref{thm:andextension}]
To prove the proposition, we will suppose that $P_\mathrm{XYZ} \in \mardist{\tri^{\cC}}$ and derive a contradiction. First note that Lemma~\ref{lemma:andlemma} together with the form of $\tri^{\cC}$ implies
\begin{equation}\label{eq:anoc}
P_\mathrm{\tilde{A}|C}=P_{\mathrm{\tilde{A}}},\text{ and }
P_\mathrm{\tilde{B}|C}=P_{\mathrm{ \tilde{B}}}\, .
\end{equation}
Furthermore, from $P_\mathrm{XYZ} \in \mardist{\tri^{\cC}}$ we have
\begin{eqnarray}
P_\mathrm{\tilde{A}\tilde{B}\tilde{X}\tilde{Y}}=\sum_cP_\mathrm{C}(c)P_\mathrm{\tilde{A}|c}P_\mathrm{\tilde{B}|c}P_\mathrm{\tilde{X}\tilde{Y}|\tilde{A}\tilde{B}c}
\end{eqnarray}
which, using~\eqref{eq:anoc}, and the form of $\tri^{\cC}$ can be rewritten
\begin{equation}
P_\mathrm{\tilde{A}\tilde{B}\tilde{X}\tilde{Y}}=\sum_cP_\mathrm{C}(c)P_\mathrm{\tilde{A}}P_\mathrm{\tilde{B}}P_\mathrm{\tilde{X}|\tilde{B}c}P_\mathrm{\tilde{Y}|\tilde{A}c}
\, .
\end{equation}
However, that $P_\mathrm{\tilde{X}\tilde{Y}|\tilde{A}\tilde{B}}$ violates a Bell inequality means that this equation cannot hold, establishing a contradiction.
\end{proof}

Proposition~\ref{thm:andextension} shows that a strategy that involves local processing of the unobserved variables at \emph{each} observed node leads to correlations that are not classically reproducible, which sets this strategy apart from one achievable in a bipartite Bell scenario $\bellsc^{\qQ}$. This insight is useful for the comparison of $\marcone{\tri^{\cC}}$ and $\marcone{\tri^{\qQ}}$ below, where the local post-processing at node $Z$ shall turn out to be crucial.

\bigskip

Entropically, an outer approximation to $\marcone{\tri^{\qQ}}$ has been constructed in~\cite{Chaves2015}, which was presented as Example~\ref{example:quantumtriangle}. It is natural to ask whether tighter approximations to $\marcone{\tri^{\qQ}}$ can be realised with a similar procedure to the one that led to tighter approximations in the classical case.  Unfortunately, we do not know of any similar inequalities for the von Neumann entropy of multi-party quantum states. Furthermore, even if the known non-Shannon inequalities were to hold for von Neumann entropy we would not be able to use them to add constraints to $\tri^{\qQ}$ due to the lack of large enough sets of coexisting interdependent variables.

\begin{problem}\label{op:1}
Do the closures of the sets of compatible entropy vectors coincide in classical and quantum triangle scenario, i.e., does $\marcone{\tri^{\cC}}=\marcone{\tri^{\qQ}}$ hold?
\end{problem}
A way to solve the open problem would be by identifying an entropy vector compatible with $\tri^{\qQ}$ that lies outside an outer approximation of $\marcone{\tri^{\cC}}$. We have attempted to find such a vector with a few random searches as well as by considering games involving
entangled states, for which we know that there is distinctive quantum
behaviour (i.e., a separation at the level of
correlations). Intuitively such scenarios are good candidates to
exhibit violations of the classical entropic bounds.  However,
somewhat surprisingly, we have not been able to find any. Even though these searches were unsuccessful and are far from exhaustive, we present them in the following as a collection of (initial) attempts. We are also able to point out types of distributions that provably cannot lead to quantum violations of any classical constraints.  

\bigskip

\noindent
\textbf{Attempts to resolve Open Problem~\ref{op:1}}
\smallskip

\begin{itemize}
\item Random searches: We let the sources $A$, $B$ and $C$
distribute up to four randomly chosen qubits each and did not detect any violations of the classical inequalities.  However, the evidence from these random searches against a separation of the classical and the quantum
marginal cones is relatively weak: vectors, $v$ that lie in
$\marcone{\tri^{\qQ}}$ but not in $\marcone{\tri^{\cC}}$ necessarily violate \eqref{eq:intinfo}, i.e., $v \notin \inmarcone{\tri^{\cC}}$.  Random searches were unable to find such
vectors (although we know they exist), which clearly illustrates the weakness of this method, and shows that the region we are looking for (if it even exists) is small.

\item Entropy vectors related to the Bell scenario: The entropy vector corresponding to the CHSH correlations from~\cite{Fritz2012}, detailed in Figure~\ref{fig:fritzproof} and in Section~\ref{sec:quantcorr} (with $Z=(A',B')$), is a natural candidate, but this lies inside $\inmarcone{\tri^{\cC}}$ so is classically reproducible.  This particular distribution is also achievable in the bipartite Bell scenario $\bellsc$ (or equivalently the line-like causal structure $\pfour$ that shall be discussed in Chapter~\ref{chap:classquant}).  Any distribution compatible
with $\bellsc$ may be realised in $\tri$ by choosing one of the variables,
e.g.\ $Z$, to have two separate outputs, one depending only on the input from
node $A$ and the other one depending on the input from $B$.
Distributions realisable in $\bellsc^{\qQ}$ or $\bellsc^{\cC}$ are thus always realisable in $\tri^{\qQ}$ or $\tri^{\cC}$ respectively.
According to the results of Section~\ref{sec:linelike}, all entropy
vectors realised with distributions in $\bellsc^{\qQ}$ are also classically achievable, i.e., realisable in $\bellsc^{\cC}$ (at least asymptotically). 
Hence, no distribution in $\bellsc^{\qQ}$ may ever violate any of the classical entropy inequalities valid for $\tri^{\cC}$. Therefore, the only way of violating the entropy inequalities approximating $\marcone{\tri^{\cC}}$ is by processing the inputs to all three nodes $X$, $Y$ and $Z$.\footnote{Two distributions that share the same entropy vector can be very different and hence may be separated by local processing.}

\item Strategies with local processing at all observed nodes: We have considered the previous example where $Z$ was taken to be a function of $A$
and $B$. We have additionally considered further local processing of $X$ and $Y$, for instance by applying all possible functions from two bits to one bit.
Note that vectors outside $\inmarcone{\tri^{\cC}}$ can be constructed with appropriate post-processing of the (quantum) distribution. A possible way to achieve this is applying $\operatorname{AND}$ or $\operatorname{OR}$ functions appropriately. One may for instance consider the quantum scenario above, and take $X=\operatorname{AND}(\tilde{X}, \tilde{B})$, $Y=\operatorname{AND}(\tilde{Y}, \tilde{A})$ and $Z=\operatorname{OR}(A',B')$. This renders the interaction information of the entropy vector of the joint distribution of $X$, $Y$ and $Z$ positive, i.e., violates inequality \eqref{eq:intinfo}, which is not the case without applying the functions.

\item Further games and quantum strategies: We have considered further distributions in $\mardist{\tri^\qQ}$, ensuring
that they do not lie in $\mardist{\bellsc^{\qQ}}$. These include input states and measurements known to lead to violations of the chained Bell
inequalities~\cite{Braunstein1988} or the Mermin-Peres magic square
game~\cite{Mermin1990, Peres1990}.  We have also considered scenarios
where all three parties measure entangled states and use the
measurement outputs as inputs for further measurements. We have
furthermore attempted to incorporate functions known to lead to a
positive interaction information in the classical case as well as
functions from two bits to one bit in general into these scenarios. 
\end{itemize}

In a number of scenarios we have also considered
shared PR-boxes instead of entangled states, again without detecting
any violations of the inequalities. In most cases the corresponding
entropy vectors have a negative interaction information, and hence lie
in $\inmarcone{\tri^{\cC}}$, so they can also be realised with a classical distribution, like the correlations~\eqref{eq:w_cor}. 

\bigskip

If Open Problem~\ref{op:1} were to be answered in the affirmative, this would point towards deficiencies of the current entropic techniques for approximating $\marcone{\tri^{\qQ}}$, which are neither able to recover all constraints of $\outmarcone{\tri^{\cC}}$ nor to derive any additional inequalities similar to the non-Shannon inequalities found in the classical case.

\subsection{Hybrid triangle scenarios}
In a \emph{hybrid causal structure} some of the unobserved nodes are allowed to be quantum, whereas others are restricted to be classical. This is of interest, in cases where we have reason to assume that some involved unobserved systems are classical. For
instance, since sharing entanglement over large distances is challenging due to noise, we might assume that two observations that are taken far apart do not have joint quantum causes. 
In the case of the causal structure $\tri$, there are two hybrid scenarios: either one or two of the three unobserved variables can be restricted to be classical, while the others are quantum. We call these two causal structures $\tri^{\cCQQ}$ and $\tri^{\cCCQ}$ respectively. In the following, we will approximate the entropy cones for both scenarios. We show that in hybrid scenarios of the triangle causal structure non-Shannon inequalities are relevant.

\subsubsection{$\tri^{\cCQQ}$ scenario}
In this scenario one of the unobserved variables is classical, which we
take to be $A$.   
We find that the outer approximation to $\marcone{\tri^{\cCQQ}}$ obtained without taking non-Shannon inequalities into account coincides with that to $\marcone{\tri^{\qQ}}$, i.e.,
\begin{equation}
\outmarcone{\tri^{\cCQQ}}=\outmarcone{\tri^{\qQ}} \ .
\end{equation}
However, unlike in the fully quantum case $\tri^{\qQ}$, non-Shannon
constraints can be legitimately included for $\tri^{\cCQQ}$, for instance the inequality from Proposition~\ref{prop:zhangyeung} with variable choices
\begin{align}
\Diamond_{YZAX} \geq 0 \ , \\ 
\Diamond_{YZXA} \geq 0 \ . 
\end{align}
This results in a tighter approximation to $\marcone{\tri^{\cCQQ}}$, which comprises the Shannon inequalities for three variables,
the three constraints~\eqref{eq:fritzhenson} and\footnote{The second of these inequalities can be easily derived from $\Diamond_{YZXA} \geq 0$ and the conditional independencies, analogously to  Proposition~\ref{prop:matus1}. To derive the first inequality, on the other hand, several inequalities have to be combined.}
\begin{align}
-3 H(X)-3 H(Y)- 3 H(Z)+ 2H(XY)+ 2H(XZ)+ 3H(YZ)- H(XYZ) &\geq 0 \ , \label{eq:firstofthem} \\
-2 H(X)-2 H(Y)- 2 H(Z)+ 3H(XY)+ 3H(XZ)+ 3H(YZ)- 4H(XYZ) &\geq 0 \ .
\end{align}
Note that permutations of~\eqref{eq:firstofthem} are not included.
Further non-Shannon inequalities could be considered in order to improve on these approximations. Our inequalities show that some of the extremal rays of $\outmarcone{\tri^{\qQ}}$ are provably not achievable if $A$, $B$ and $C$ do not \emph{all} share entangled states.

\subsubsection{$\tri^{\cCCQ}$ scenario}
In this scenario we take $A$ and $B$ to be classical. This scenario can be understood as a Bell scenario, where the measurement choices of the two parties are unobserved and processed to one single observed output, $Z$.\footnote{Note that even though classical and quantum entropy cones  coincide in the bipartite Bell scenario this may not be the case here as very different distributions may lead to the same entropy vector in the classical and quantum case. These may be entropically separated through local processing.} The distributions from Section~\ref{sec:quantcorr}, that are provably not reproducible in $\tri^{\cC}$ can be generated in this causal structure. 
To approximate the marginal entropy cone of this scenario, $\marcone{\tri^{\cCCQ}}$, we proceed analogously to the $\tri^{\qQ}$ and $\tri^{\cCQQ}$ scenarios before. However, the result differs and leads to a tighter cone, even without considering non-Shannon inequalities,
\begin{equation}
\outmarcone{\tri^{\cCCQ}} \subsetneq \outmarcone{\tri^{\cCQQ}} \ .
\end{equation}
$\outmarcone{\tri^{\cCCQ}}$ is given by the three variable Shannon inequalities, the three constraints~\eqref{eq:fritzhenson} and
\begin{equation} \label{eq:ccq}
-3 H(X)-3 H(Y)- 3 H(Z)+ 2H(XY) + 3H(XZ)+ 2H(YZ)- H(XYZ)\geq 0 \ ,
\end{equation}
up to permutations of $Y$ and $Z$. These five inequalities are a
subset of the seven inequalities of Example~\ref{example:TRI_class_obs} that delimit
$\outmarcone{\tri^{\cC}}$ , hence
\begin{equation}
\outmarcone{\tri^{\cC}} \subsetneq \outmarcone{\tri^{\cCCQ}} \ .
\end{equation} 

When including the inequalities,
\begin{align}
\Diamond_{XZBY} \geq 0 \ , \\
 \Diamond_{YZAX} \geq 0 \ , \\
 \Diamond_{YZXA} \geq 0 \ , 
\end{align}
we obtain the additional inequalities
\begin{align}
-2H(X)-2H(Y)-2H(Z)+3H(XY)+3H(XZ)+3H(YZ)-4H(XYZ) &\geq 0 \ , \label{eq:ccqadditional1}\\
-6H(X)-6H(Y)-6H(Z)+5H(XY)+5H(XZ)+5H(YZ)-3H(XYZ) &\geq 0 \ , \label{eq:ccqadditional2} \\
-4H(X)-4H(Y)-4H(Z)+3H(XY)+4H(XZ)+3H(YZ)-2H(XYZ) &\geq 0 \ , \label{eq:ccqadditional3}
\end{align}
up to permutations of $X$ and $Y$ in the last inequality.  They render the two permutations of~\eqref{eq:ccq} redundant, while \eqref{eq:fritzhenson} remains irredundant (for all of its permutations). Note that \eqref{eq:ccqadditional1} is also part of $\outmarcone{\tri^{\cCQQ}}$. 
Just as for $\tri^{\cCQQ}$, further constraints could be derived here by considering additional non-Shannon inequalities.

\section{Non-Shannon inequalities and post-selection} \label{sec:nonshan_post}

Non-Shannon inequalities can be considered for post-selected causal structures and may lead to tighter constraints on the allowed entropy vectors in the post-selected causal structure. We will illustrate the success of this technique with the example of $\tilde{\inst}$ of Figure~\ref{fig:IC_inner_ch4}, the causal structure relevant for the information causality game. The entropic restriction imposed by $\tilde{\inst}$ has been analysed in~\cite{Chaves2015}, an analysis that was provided as Example~\ref{example:information_causality}. Here, we improve on corresponding results by using non-Shannon inequalities.~\footnote{As previously remarked, non-Shannon inequalities are also relevant for $\tilde{\inst}$ without post-selection.}

We post-select on the values of the binary variable $S$ and analyse $\tilde{\inst}^\cC_\mathrm{S}$ with the entropy vector method. Considering just the additional non-Shannon inequality from Proposition~\ref{prop:zhangyeung} (and permutations, but no applications to more than four variables) leads to a total of $1017$ entropic inequalities including permutations on the variables with corresponding observed distributions in $\tilde{\inst}^\cC$, i.e., on $H( X_1 X_2 Z  Y_\mathrm{ | S=1})$, $H( X_1 X_2 Z  Y_\mathrm{ | S=2})$ and their marginals. These include all $176$ irredundant inequalities that were obtained without non-Shannon constraints in~\cite{Chaves2015} but also $841$ new ones (a list of all inequalities is available at~\cite{URL}). Moreover, we expect further non-Shannon inequalities to lead to numerous additional constraints potentially rendering our inequalities redundant. Infinite families of inequalities similar to those found for the triangle scenario in Section~\ref{sec:newouter} could also be derived here.

In the quantum case, on the other hand, we can only apply the non-Shannon inequalities to the two coexisting sets of observed (classical) variables $\left\{ X_1, X_2, Z , Y_\mathrm{ | S=1} \right\}$ and $\left\{ X_1, X_2, Z , Y_\mathrm{ | S=2} \right\}$, which means that we can impose a set of $24$ additional constraints (including permutations) by adding all permutations of the inequality from Proposition~\ref{prop:zhangyeung} to the entropic description that is obtained without these.
 
It is worth pointing out that although our results imply that previous entropic approximations of $\marcone{\tilde{\inst}^\cC_\mathrm{S}}$ (particularly the approximations from~\cite{Chaves2015}) were not tight and even though non-Shannon inequalities improve on the entropic characterisation of $\tilde{\inst}_\mathrm{S}$ in both, classical and quantum case, the inequality from Example~\ref{example:information_causality} is not rendered redundant in our computations.

\section{Appendix}

\subsection{Proof of Proposition~\ref{prop:matus1fam2}} \label{sec:pr1}

The basic proof idea, which is in principle always feasible, is to consider the Shannon and independence constraints for $\tri^\cC$ and all (four) different permutations of the inequality~\eqref{eq:matus1} and then to perform the Fourier-Motzkin elimination of the unobserved variables manually, hence deriving inequalities that retain $s$ as a variable. The following proof proceeds along these lines but is more slender.

Inequality~\eqref{eq:matus1}, with variable choices
$(X_1,~X_2,~X_3,~X_4)=(Y,~X,~C,~Z)$ and using the independencies $I(X:Y|C)=0$ and $I(Z:C)=0$, can be rewritten as 
\begin{multline}
\left(\frac{1}{2} s^2  \frac{1}{2} s +1 \right) H(C)
+ \left(-s-1\right) H(X) 
+ \left(-\frac{1}{2} s^2 - \frac{3}{2} s  \right) H(Y)
- s H(Z)
+\left( -\frac{1}{2} s^{2} - \frac{1}{2} s \right) H(CX) \\
- H(CY) 
+\left( \frac{1}{2} s^{2} + \frac{3}{2} s + 1 \right) H(XY) 
+ s H(XZ) 
+ s H(YZ)
- s H(XYZ)\geq 0 \ .  \label{eq:m0}
\end{multline}
We also marginalise $\outcone{\tri^{\cC}}$ to obtain constraints on the vectors $$\left( H(C), H(X), H(Y), H(Z), H(CX), H(CY), H(XY), H(XZ), H(YZ), H(XYZ) \right) \ $$
that arise from Shannon and independence inequalities. Two inequalities this leaves us with are,
\begin{align}
-2H(C)-2H(X)-2H(Y)-3H(Z)+H(CX)+H(CY)+H(XY)+2H(XZ)+2H(YZ)-H(XYZ) \geq 0 \ , \label{eq:m1} \\
-H(C)-H(X)-H(Z)+H(CX)+H(XZ) \geq 0 \ . \label{eq:m2} 
\end{align}
We now use~\eqref{eq:m1} to remove $H(CY)$ from \eqref{eq:m0}, which yields
\begin{multline}
\left(\frac{1}{2} s^2 + \frac{1}{2} s - 1 \right) H(C)
+ \left(-s-3 \right) H(X) 
+ \left(-\frac{1}{2} s^2 - \frac{3}{2} s - 2 \right) H(Y)
+ \left( - s -3 \right) H(Z)  \\
+\left( -\frac{1}{2} s^{2} - \frac{1}{2} s +1 \right) H(CX)
+\left( \frac{1}{2} s^{2} + \frac{3}{2} s + 2\right) H(XY) 
+ \left(s +2 \right) H(XZ) 
+ \left(s +2 \right) H(YZ)
+ \left(- s +1 \right) H(XYZ) \geq 0 \ .  \label{eq:m3}
\end{multline}
With \eqref{eq:m2}, $H(CX)$ and $H(C)$ are eliminated from \eqref{eq:m3}, resulting in~\eqref{eq:mres}.

\subsection{Proof of Proposition~\ref{prop:matus2}} \label{sec:pr2}

In a similar manner as for the proof of Proposition~\ref{prop:matus1fam2}, we consider inequality~\eqref{eq:matus2}
with variable choices
$(X_1,~X_2,~X_3,~X_4)=(X,~Y,~C,~Z)$ and the independencies $I(X:Y|C)=0$ and $I(Z:C)=0$ to obtain
\begin{multline}
\left( s +1 \right) H(C)
+ \left(-\frac{1}{2} s^2 - \frac{3}{2} s \right) H(X) 
+ \left(- s -1  \right) H(Y)
+\left(-\frac{1}{2} s^2 - \frac{1}{2} s \right) H(Z)
- H(CX) \\
-s H(CY) 
+\left( \frac{1}{2} s^{2} + \frac{3}{2} s + 1 \right) H(XY) 
+ s^2 H(XZ) 
+\left( \frac{1}{2} s^{2} + \frac{1}{2} s  \right) H(YZ)
- s^2 H(XYZ)\geq 0 \ .  \label{eq:n0}
\end{multline}
We also consider two inequalities that are obtained from marginalising $\outcone{\tri^{\cC}}$ to vectors $$\left( H(C), H(X), H(Y), H(Z), H(CX), H(CY), H(XY), H(XZ), H(YZ), H(XYZ) \right) \ , $$ namely \eqref{eq:m1} as well as
\begin{equation}
-H(C)-H(Y)-H(Z)+H(CY)+H(YZ) \geq 0 \ . \label{eq:n1}
\end{equation}
Inequality \eqref{eq:m1} allows us to eliminate $H(CX)$ from \eqref{eq:n0} and \eqref{eq:n1} allows us to eliminate $H(C)$ and $H(CY)$, yielding \eqref{eq:nres}.

\subsection{Proof of Lemma~\ref{lemma:andlemma}} \label{sec:lemma_proof}
Due to the causal constraints we can write 
\begin{equation}
P_\mathrm{Z \tilde{A} \tilde{B}}=\sum_{abc} P_\mathrm{Z|ab}P_\mathrm{\tilde{A}|ac} P_\mathrm{\tilde{B}|bc} P_\mathrm{A}(a) P_\mathrm{B}(b) P_\mathrm{C}(c).
\end{equation}
Because $Z=\operatorname{AND}(\tilde{A}, \tilde{B})$, we can derive the following two conditions:
\begin{enumerate}[(1)]
\item Using $P_\mathrm{Z \tilde{A} \tilde{B}}(0,1,1)=0$, it follows
  that for each triple $(a,b,c)$ either $P_\mathrm{Z|ab}(0)=0$ or
  $P_\mathrm{\tilde{A}|ac}(1)=0$ or $P_\mathrm{\tilde{B}|bc}(1)=0$.
\item Using $P_\mathrm{Z\tilde{A}\tilde{B}}(1,0,0)=P_\mathrm{Z
    \tilde{A} \tilde{B}}(1,0,1)=P_\mathrm{Z \tilde{A}
    \tilde{B}}(1,1,0)=0$, it follows that for each triple
  $(a,b,c)$ either $P_\mathrm{Z|ab}(1)=0$ or
  $P_\mathrm{\tilde{A}|ac}(1)=P_\mathrm{\tilde{B}|bc}(1)=1$.
\end{enumerate}
We first argue that $Z$ is a deterministic function of $A$ and $B$,
i.e., $P_{Z|ab}\in\{0,1\}$ for all pairs $(a,b)$. From condition (2) we
know that either $P_\mathrm{Z|ab}(1)=0$ (and thus
$P_\mathrm{Z|ab}(0)=1$) deterministically, or that
$P_\mathrm{\tilde{A}|ac}(1)=P_\mathrm{\tilde{B}|bc}(1)=1$. But in the
latter case condition (1) implies that $P_\mathrm{Z|ab}(0)=0$ (and
thus $P_\mathrm{Z|ab}(1)=1$).

Now let us consider the two cases (a)~$P_\mathrm{Z|ab}(1)=1$; and
(b)~$P_\mathrm{Z|ab}(1)=0$ separately. 
\begin{enumerate}[(a)]
\item Let $(a,b)$ be such that
$P_\mathrm{Z|ab}(1)=1$. According to condition (2),
$P_\mathrm{\tilde{A}|ac}(1)=P_\mathrm{\tilde{B}|bc}(1)=1$ for all $c$,
and thus we have $P_\mathrm{\tilde{A}|ac}=P_\mathrm{\tilde{A}|a}$ as
well as $P_\mathrm{\tilde{B}|bc}=P_\mathrm{\tilde{B}|b}$. 
\item Let $(a,b)$ be such that $P_\mathrm{Z|ab}(1)=0$. Then $P_\mathrm{Z|ab}(0)=1$ and thus by condition (1) for
all $c$ either $P_\mathrm{\tilde{A}|ac}(1)=0$ or
$P_\mathrm{\tilde{B}|bc}(1)=0$.  We further divide into two cases:
either (i) $(a,b)$ are such that $P_\mathrm{Z|ab'}(1)=0$ for all
$b'$ and $P_\mathrm{Z|a'b}(1)=0$ for all $a'$; or (ii) they are
not. \smallskip

\noindent (ii) Suppose $\exists b'$ such that $P_\mathrm{Z|ab'}(1)=1$.
In this case $P_\mathrm{\tilde{A}|ac}(1)=1$ for all $c$ due to
condition (2) and thus from condition (1) we have $P_\mathrm{\tilde{B}|bc}(1)=0$. Thus for such pairs $(a,b)$, the relations
$P_\mathrm{\tilde{A}|ac}=P_\mathrm{\tilde{A}|a}$ as well as
$P_\mathrm{\tilde{B}|bc}=P_\mathrm{\tilde{B}|b}$ hold. Symmetric
considerations can be made in the case where $\exists a'$ such that
$P_\mathrm{Z|a'b}(1)=1$ instead.\smallskip

\noindent (i) It cannot be the case that all pairs $(a,b)$
have $P_\mathrm{Z|ab'}(1)=0$ for all $b'$ and $P_\mathrm{Z|a'b}(1)=0$ for all $a'$ (otherwise $P_Z(1)=0$). Hence there exists
$(a'',b'')$ for which $P_\mathrm{Z|a''b''}(1)=1$. By
condition (2), this implies that $P_\mathrm{\tilde{A}|a''c}(1)=P_\mathrm{\tilde{B}|b''c}(1)=1$ for all $c$. Thus, as $P_\mathrm{Z|ab''}(1)=0$ and $P_\mathrm{\tilde{B}|b''c}(1)=1$, it follows from
condition (1) that $P_\mathrm{\tilde{A}|ac}(1)= 0$ for any $c$;
$P_\mathrm{\tilde{B}|bc}(1)=0$ follows analogously, which concludes
the proof.
\end{enumerate}
\qed

%=========================================================

%% file: chapter06/chap_quantclass.tex
\let\textcircled=\pgftextcircled
\chapter{Entropic distinction between classical and quantum causal structures}
\label{chap:classquant}

\initial{I}n this chapter we analyse whether the entropy vector method is able to distinguish classical and quantum causal structures. We start by analysing the so-called bilocal causal structure in Section~\ref{sec:bilocality}, for which we show that the entropy vector method is unable to distinguish between the classical and the quantum scenario despite there being a well-known gap at the level of correlations. In Section~\ref{sec:linelike}, we then present the main result of this chapter, that in any line-like causal structure the sets of compatible entropy vectors coincide for classical, quantum and general non-signalling resources. In Section~\ref{sec:smallstructures} we show that the same is true for other small causal structures and discuss possible avenues towards a general statement. We also explain why such a general statement has, thus far, remained elusive.
In Section~\ref{sec:inabilitypost} we show how the post-selection technique can in some cases salvage the entropic approach but also point out its limitations.

\section{Inability of the entropy vector approach to distinguish classical and quantum causes in the bilocal causal structure}\label{sec:bilocality}
\begin{figure}
\centering
\resizebox{0.6\columnwidth}{!}{%
\begin{tikzpicture} [scale=1.3]
\node[draw=black,circle,scale=0.6]  (0) at (-1,0) {$A$};
\node[draw=black,circle,scale=0.6]  (1) at (0,1) {$X$};
\node (2) at (1,0) {$C_2$};
\node[draw=black,circle,scale=0.6]  (3) at (2,0) {$W$};
\node[draw=black,circle,scale=0.6]  (4) at (2,1) {$Y$};
\node (5) at (3,0) {$C_3$};
\node[draw=black,circle,scale=0.6]  (6) at (4,1) {$Z$};
\node[draw=black,circle,scale=0.6]  (7) at (5,0) {$B$};

\draw [->,>=stealth] (0)--(1);
\draw [->,>=stealth] (2)--(1);
\draw [->,>=stealth] (3)--(4);
\draw [->,>=stealth] (2)--(4);
\draw [->,>=stealth] (5)--(4);
\draw [->,>=stealth] (5)--(6);
\draw [->,>=stealth] (7)--(6);
\end{tikzpicture}
}
\caption[Bilocal causal structure]{Bilocal causal structure. $A$, $B$, $W$, $X$, $Y$ and $Z$ are observed variables, $C_2$ and $C_3$ are unobserved. The direct links from $A$ to $X$, from $B$ to $Z$ and from $W$ to $Y$ could be equivalently replaced with an unobserved variable shared between each pair.}
\label{fig:bilocality}
\end{figure}
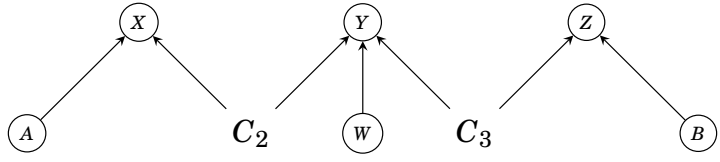

Entanglement swapping~\cite{Bennett1993} is a quantum operation that is applied in quantum repeaters to achieve long-distance quantum communication~\cite{Briegel1998, Sangouard2011} as well as in the event-ready detectors scheme~\cite{Zukowski1993} and has for instance been used in one of the recent loophole-free Bell tests~\cite{Hensen2015}.
The causal structure representing the corresponding setup is the \emph{bilocal causal structure}, $\bil$, displayed in Figure~\ref{fig:bilocality}. The correlations compatible with the scenario have been analysed in~\cite{Branciard2010, Branciard2012}, where it was shown that there is a gap between the sets of compatible distributions in classical and quantum case, i.e., $\mardist{\bil^{\cC}} \subsetneq \mardist{\bil^{\qQ}}$. 
The technological relevance of the scenario and its prominence in the community makes it a suitable example for illustrating our techniques.

\begin{proposition}\label{prop:bilocality}
The entropy vector method is unable to distinguish, classical, quantum and more general non-signalling resources in the bilocal causal structure,
\begin{equation}
\marcone{\bil^{\cC}}=\marcone{\bil^{\qQ}} \ . 
\end{equation}
\end{proposition} 

We prove this in the following by constructing an outer approximation to the set of achievable entropy vectors in $\bil^{\qQ}$ and by showing that both, $\marcone{\bil^{\cC}}$ and $\marcone{\bil^{\qQ}}$, coincide with this approximation, which also equals $\outmarcone{\bil^{\cC}}$.

\begin{proof} 
We first give an outer approximation to the marginal cone of the bilocal causal structure in terms of the \emph{observed} variables, which is thus valid for classical, quantum and general non-signalling unobserved variables. Let us consider the Shannon inequalities of the six observed variables $A$, $B$, $W$, $X$, $Y$ and $Z$ and the following conditional independence relations,\footnote{In classical and quantum case there are further valid (in)equalities implied by the causal structure, as long as the unobserved subsystems are included in our considerations. The following argument shows, however, that these do not impart any additional constraints on the entropy vectors of the observed marginal scenario.}
\begin{align}
I(A:BWYZ)&=0 \ , \\
I(B:AWXY)&=0 \ , \\
I(W:ABXZ)&=0 \ , \\
I(X:BWZ|A)&=0 \ , \\
I(Z:AWX|B)&=0 \ , \\
I(Y:AB|X)&=0 \ . 
\end{align}

Converting the $\hrep$-representation of this cone to its $\vrep$-representation yields $35$ extremal rays meeting at the zero vector, which is the tip of the pointed polyhedral cone. In the following, one entropy vector on each extremal ray is listed with components ordered according to 
$$\left( H(A),~H(B),~H(W),~H(X),~H(Y),~H(Z),~H(AB),~H(AW), ~\ldots~,~H(ABWXYZ) \right).$$  
{\small
$$\arraycolsep=0.8pt 
\begin{array}{cccccccccccccccccccccccccccccccccccccccccccccccccccccccccccccccc}
(  1) & 1 & 1 & 1 & 2 & 3 & 2 & 2 & 2 & 3 & 4 & 3 & 2 & 3 & 4 & 3 & 3 & 4 & 3 & 5 & 4 & 5 & 3 & 4 & 5 & 4 & 4 & 5 & 4 & 6 & 5 & 6 & 4 & 5 & 4 & 6 & 5 & 6 & 6 & 5 & 6 & 7 & 5 & 6 & 5 & 7 & 6 & 7 & 6 & 6 & 7 & 7 & 7 & 6 & 6 & 7 & 7 & 7 & 7 & 7 & 7 & 7 & 7 & 7 \\
(  2) & 1 & 1 & 1 & 2 & 2 & 2 & 2 & 2 & 3 & 3 & 3 & 2 & 3 & 3 & 3 & 3 & 3 & 3 & 4 & 4 & 4 & 3 & 4 & 4 & 4 & 4 & 4 & 4 & 5 & 5 & 5 & 4 & 4 & 4 & 5 & 5 & 5 & 5 & 5 & 5 & 6 & 5 & 5 & 5 & 6 & 6 & 6 & 6 & 6 & 6 & 7 & 6 & 6 & 6 & 7 & 7 & 7 & 7 & 7 & 7 & 7 & 7 & 7 \\
(  3) & 1 & 1 & 1 & 1 & 2 & 2 & 2 & 2 & 2 & 3 & 3 & 2 & 2 & 3 & 3 & 2 & 3 & 3 & 3 & 3 & 4 & 3 & 3 & 4 & 4 & 3 & 4 & 4 & 4 & 4 & 5 & 3 & 4 & 4 & 4 & 4 & 5 & 4 & 4 & 5 & 5 & 4 & 5 & 5 & 5 & 5 & 6 & 5 & 5 & 6 & 6 & 5 & 5 & 5 & 6 & 6 & 6 & 6 & 6 & 6 & 6 & 6 & 6 \\
(  4) & 1 & 1 & 1 & 2 & 2 & 1 & 2 & 2 & 3 & 3 & 2 & 2 & 3 & 3 & 2 & 3 & 3 & 2 & 4 & 3 & 3 & 3 & 4 & 4 & 3 & 4 & 4 & 3 & 5 & 4 & 4 & 4 & 4 & 3 & 5 & 4 & 4 & 5 & 4 & 4 & 5 & 5 & 5 & 4 & 6 & 5 & 5 & 5 & 5 & 5 & 6 & 6 & 5 & 5 & 6 & 6 & 6 & 6 & 6 & 6 & 6 & 6 & 6 \\
(  5) & 1 & 1 & 1 & 1 & 2 & 1 & 2 & 2 & 2 & 3 & 2 & 2 & 2 & 3 & 2 & 2 & 3 & 2 & 3 & 2 & 3 & 3 & 3 & 4 & 3 & 3 & 4 & 3 & 4 & 3 & 4 & 3 & 4 & 3 & 4 & 3 & 4 & 4 & 3 & 4 & 4 & 4 & 5 & 4 & 5 & 4 & 5 & 4 & 4 & 5 & 5 & 5 & 4 & 4 & 5 & 5 & 5 & 5 & 5 & 5 & 5 & 5 & 5 \\
(  6) & 0 & 1 & 1 & 1 & 2 & 2 & 1 & 1 & 1 & 2 & 2 & 2 & 2 & 3 & 3 & 2 & 3 & 3 & 3 & 3 & 4 & 2 & 2 & 3 & 3 & 2 & 3 & 3 & 3 & 3 & 4 & 3 & 4 & 4 & 4 & 4 & 5 & 4 & 4 & 5 & 5 & 3 & 4 & 4 & 4 & 4 & 5 & 4 & 4 & 5 & 5 & 5 & 5 & 5 & 5 & 5 & 5 & 5 & 5 & 5 & 5 & 5 & 5 \\
(  7) & 1 & 0 & 1 & 2 & 2 & 1 & 1 & 2 & 3 & 3 & 2 & 1 & 2 & 2 & 1 & 3 & 3 & 2 & 4 & 3 & 3 & 2 & 3 & 3 & 2 & 4 & 4 & 3 & 5 & 4 & 4 & 3 & 3 & 2 & 4 & 3 & 3 & 5 & 4 & 4 & 5 & 4 & 4 & 3 & 5 & 4 & 4 & 5 & 5 & 5 & 5 & 5 & 4 & 4 & 5 & 5 & 5 & 5 & 5 & 5 & 5 & 5 & 5 \\
(  8) & 1 & 1 & 1 & 1 & 1 & 1 & 2 & 2 & 2 & 2 & 2 & 2 & 2 & 2 & 2 & 2 & 2 & 2 & 2 & 2 & 2 & 3 & 3 & 3 & 3 & 3 & 3 & 3 & 3 & 3 & 3 & 3 & 3 & 3 & 3 & 3 & 3 & 3 & 3 & 3 & 3 & 4 & 4 & 4 & 4 & 4 & 4 & 4 & 4 & 4 & 4 & 4 & 4 & 4 & 4 & 4 & 5 & 5 & 5 & 5 & 5 & 5 & 5 \\
(  9) & 0 & 1 & 1 & 1 & 2 & 1 & 1 & 1 & 1 & 2 & 1 & 2 & 2 & 3 & 2 & 2 & 3 & 2 & 3 & 2 & 3 & 2 & 2 & 3 & 2 & 2 & 3 & 2 & 3 & 2 & 3 & 3 & 4 & 3 & 4 & 3 & 4 & 3 & 3 & 4 & 4 & 3 & 4 & 3 & 4 & 3 & 4 & 3 & 3 & 4 & 4 & 4 & 4 & 4 & 4 & 4 & 4 & 4 & 4 & 4 & 4 & 4 & 4 \\
( 10) & 1 & 0 & 1 & 1 & 2 & 1 & 1 & 2 & 2 & 3 & 2 & 1 & 1 & 2 & 1 & 2 & 3 & 2 & 3 & 2 & 3 & 2 & 2 & 3 & 2 & 3 & 4 & 3 & 4 & 3 & 4 & 2 & 3 & 2 & 3 & 2 & 3 & 4 & 3 & 3 & 4 & 3 & 4 & 3 & 4 & 3 & 4 & 4 & 4 & 4 & 4 & 4 & 3 & 3 & 4 & 4 & 4 & 4 & 4 & 4 & 4 & 4 & 4 \\
( 11) & 0 & 1 & 1 & 1 & 1 & 1 & 1 & 1 & 1 & 1 & 1 & 2 & 2 & 2 & 2 & 2 & 2 & 2 & 2 & 2 & 2 & 2 & 2 & 2 & 2 & 2 & 2 & 2 & 2 & 2 & 2 & 3 & 3 & 3 & 3 & 3 & 3 & 3 & 3 & 3 & 3 & 3 & 3 & 3 & 3 & 3 & 3 & 3 & 3 & 3 & 3 & 4 & 4 & 4 & 4 & 4 & 4 & 4 & 4 & 4 & 4 & 4 & 4 \\
( 12) & 1 & 0 & 1 & 1 & 1 & 1 & 1 & 2 & 2 & 2 & 2 & 1 & 1 & 1 & 1 & 2 & 2 & 2 & 2 & 2 & 2 & 2 & 2 & 2 & 2 & 3 & 3 & 3 & 3 & 3 & 3 & 2 & 2 & 2 & 2 & 2 & 2 & 3 & 3 & 3 & 3 & 3 & 3 & 3 & 3 & 3 & 3 & 4 & 4 & 4 & 4 & 3 & 3 & 3 & 3 & 4 & 4 & 4 & 4 & 4 & 4 & 4 & 4 \\
( 13) & 1 & 1 & 0 & 1 & 1 & 1 & 2 & 1 & 2 & 2 & 2 & 1 & 2 & 2 & 2 & 1 & 1 & 1 & 2 & 2 & 2 & 2 & 3 & 3 & 3 & 2 & 2 & 2 & 3 & 3 & 3 & 2 & 2 & 2 & 3 & 3 & 3 & 2 & 2 & 2 & 3 & 3 & 3 & 3 & 4 & 4 & 4 & 3 & 3 & 3 & 4 & 3 & 3 & 3 & 4 & 3 & 4 & 4 & 4 & 4 & 4 & 4 & 4 \\
( 14) & 0 & 0 & 1 & 1 & 2 & 1 & 0 & 1 & 1 & 2 & 1 & 1 & 1 & 2 & 1 & 2 & 3 & 2 & 3 & 2 & 3 & 1 & 1 & 2 & 1 & 2 & 3 & 2 & 3 & 2 & 3 & 2 & 3 & 2 & 3 & 2 & 3 & 3 & 3 & 3 & 3 & 2 & 3 & 2 & 3 & 2 & 3 & 3 & 3 & 3 & 3 & 3 & 3 & 3 & 3 & 3 & 3 & 3 & 3 & 3 & 3 & 3 & 3 \\
( 15) & 0 & 0 & 1 & 1 & 1 & 1 & 0 & 1 & 1 & 1 & 1 & 1 & 1 & 1 & 1 & 2 & 2 & 2 & 2 & 2 & 2 & 1 & 1 & 1 & 1 & 2 & 2 & 2 & 2 & 2 & 2 & 2 & 2 & 2 & 2 & 2 & 2 & 3 & 3 & 3 & 3 & 2 & 2 & 2 & 2 & 2 & 2 & 3 & 3 & 3 & 3 & 3 & 3 & 3 & 3 & 3 & 3 & 3 & 3 & 3 & 3 & 3 & 3 \\
( 16) & 0 & 1 & 0 & 1 & 1 & 1 & 1 & 0 & 1 & 1 & 1 & 1 & 2 & 2 & 2 & 1 & 1 & 1 & 2 & 2 & 2 & 1 & 2 & 2 & 2 & 1 & 1 & 1 & 2 & 2 & 2 & 2 & 2 & 2 & 3 & 3 & 3 & 2 & 2 & 2 & 3 & 2 & 2 & 2 & 3 & 3 & 3 & 2 & 2 & 2 & 3 & 3 & 3 & 3 & 3 & 3 & 3 & 3 & 3 & 3 & 3 & 3 & 3 \\
( 17) & 0 & 1 & 1 & 0 & 1 & 1 & 1 & 1 & 0 & 1 & 1 & 2 & 1 & 2 & 2 & 1 & 2 & 2 & 1 & 1 & 2 & 2 & 1 & 2 & 2 & 1 & 2 & 2 & 1 & 1 & 2 & 2 & 3 & 3 & 2 & 2 & 3 & 2 & 2 & 3 & 2 & 2 & 3 & 3 & 2 & 2 & 3 & 2 & 2 & 3 & 2 & 3 & 3 & 3 & 3 & 3 & 3 & 3 & 3 & 3 & 3 & 3 & 3 \\
( 18) & 1 & 0 & 0 & 1 & 1 & 1 & 1 & 1 & 2 & 2 & 2 & 0 & 1 & 1 & 1 & 1 & 1 & 1 & 2 & 2 & 2 & 1 & 2 & 2 & 2 & 2 & 2 & 2 & 3 & 3 & 3 & 1 & 1 & 1 & 2 & 2 & 2 & 2 & 2 & 2 & 3 & 2 & 2 & 2 & 3 & 3 & 3 & 3 & 3 & 3 & 3 & 2 & 2 & 2 & 3 & 3 & 3 & 3 & 3 & 3 & 3 & 3 & 3 \\
( 19) & 1 & 0 & 1 & 1 & 1 & 0 & 1 & 2 & 2 & 2 & 1 & 1 & 1 & 1 & 0 & 2 & 2 & 1 & 2 & 1 & 1 & 2 & 2 & 2 & 1 & 3 & 3 & 2 & 3 & 2 & 2 & 2 & 2 & 1 & 2 & 1 & 1 & 3 & 2 & 2 & 2 & 3 & 3 & 2 & 3 & 2 & 2 & 3 & 3 & 3 & 3 & 3 & 2 & 2 & 2 & 3 & 3 & 3 & 3 & 3 & 3 & 3 & 3 \\
( 20) & 0 & 0 & 0 & 1 & 1 & 1 & 0 & 0 & 1 & 1 & 1 & 0 & 1 & 1 & 1 & 1 & 1 & 1 & 2 & 2 & 2 & 0 & 1 & 1 & 1 & 1 & 1 & 1 & 2 & 2 & 2 & 1 & 1 & 1 & 2 & 2 & 2 & 2 & 2 & 2 & 2 & 1 & 1 & 1 & 2 & 2 & 2 & 2 & 2 & 2 & 2 & 2 & 2 & 2 & 2 & 2 & 2 & 2 & 2 & 2 & 2 & 2 & 2 \\
( 21) & 0 & 0 & 1 & 0 & 1 & 1 & 0 & 1 & 0 & 1 & 1 & 1 & 0 & 1 & 1 & 1 & 2 & 2 & 1 & 1 & 2 & 1 & 0 & 1 & 1 & 1 & 2 & 2 & 1 & 1 & 2 & 1 & 2 & 2 & 1 & 1 & 2 & 2 & 2 & 2 & 2 & 1 & 2 & 2 & 1 & 1 & 2 & 2 & 2 & 2 & 2 & 2 & 2 & 2 & 2 & 2 & 2 & 2 & 2 & 2 & 2 & 2 & 2 \\
( 22) & 0 & 0 & 1 & 1 & 1 & 0 & 0 & 1 & 1 & 1 & 0 & 1 & 1 & 1 & 0 & 2 & 2 & 1 & 2 & 1 & 1 & 1 & 1 & 1 & 0 & 2 & 2 & 1 & 2 & 1 & 1 & 2 & 2 & 1 & 2 & 1 & 1 & 2 & 2 & 2 & 2 & 2 & 2 & 1 & 2 & 1 & 1 & 2 & 2 & 2 & 2 & 2 & 2 & 2 & 2 & 2 & 2 & 2 & 2 & 2 & 2 & 2 & 2 \\
( 23) & 0 & 0 & 0 & 0 & 0 & 1 & 0 & 0 & 0 & 0 & 1 & 0 & 0 & 0 & 1 & 0 & 0 & 1 & 0 & 1 & 1 & 0 & 0 & 0 & 1 & 0 & 0 & 1 & 0 & 1 & 1 & 0 & 0 & 1 & 0 & 1 & 1 & 0 & 1 & 1 & 1 & 0 & 0 & 1 & 0 & 1 & 1 & 0 & 1 & 1 & 1 & 0 & 1 & 1 & 1 & 1 & 0 & 1 & 1 & 1 & 1 & 1 & 1 \\
( 24) & 0 & 0 & 0 & 0 & 1 & 0 & 0 & 0 & 0 & 1 & 0 & 0 & 0 & 1 & 0 & 0 & 1 & 0 & 1 & 0 & 1 & 0 & 0 & 1 & 0 & 0 & 1 & 0 & 1 & 0 & 1 & 0 & 1 & 0 & 1 & 0 & 1 & 1 & 0 & 1 & 1 & 0 & 1 & 0 & 1 & 0 & 1 & 1 & 0 & 1 & 1 & 1 & 0 & 1 & 1 & 1 & 1 & 0 & 1 & 1 & 1 & 1 & 1 \\
( 25) & 0 & 0 & 0 & 1 & 0 & 0 & 0 & 0 & 1 & 0 & 0 & 0 & 1 & 0 & 0 & 1 & 0 & 0 & 1 & 1 & 0 & 0 & 1 & 0 & 0 & 1 & 0 & 0 & 1 & 1 & 0 & 1 & 0 & 0 & 1 & 1 & 0 & 1 & 1 & 0 & 1 & 1 & 0 & 0 & 1 & 1 & 0 & 1 & 1 & 0 & 1 & 1 & 1 & 0 & 1 & 1 & 1 & 1 & 0 & 1 & 1 & 1 & 1 \\
( 26) & 0 & 0 & 1 & 0 & 0 & 0 & 0 & 1 & 0 & 0 & 0 & 1 & 0 & 0 & 0 & 1 & 1 & 1 & 0 & 0 & 0 & 1 & 0 & 0 & 0 & 1 & 1 & 1 & 0 & 0 & 0 & 1 & 1 & 1 & 0 & 0 & 0 & 1 & 1 & 1 & 0 & 1 & 1 & 1 & 0 & 0 & 0 & 1 & 1 & 1 & 0 & 1 & 1 & 1 & 0 & 1 & 1 & 1 & 1 & 0 & 1 & 1 & 1 \\
( 27) & 0 & 1 & 0 & 0 & 1 & 1 & 1 & 0 & 0 & 1 & 1 & 1 & 1 & 2 & 2 & 0 & 1 & 1 & 1 & 1 & 2 & 1 & 1 & 2 & 2 & 0 & 1 & 1 & 1 & 1 & 2 & 1 & 2 & 2 & 2 & 2 & 2 & 1 & 1 & 2 & 2 & 1 & 2 & 2 & 2 & 2 & 2 & 1 & 1 & 2 & 2 & 2 & 2 & 2 & 2 & 2 & 2 & 2 & 2 & 2 & 2 & 2 & 2 \\
( 28) & 0 & 1 & 0 & 0 & 0 & 0 & 1 & 0 & 0 & 0 & 0 & 1 & 1 & 1 & 1 & 0 & 0 & 0 & 0 & 0 & 0 & 1 & 1 & 1 & 1 & 0 & 0 & 0 & 0 & 0 & 0 & 1 & 1 & 1 & 1 & 1 & 1 & 0 & 0 & 0 & 0 & 1 & 1 & 1 & 1 & 1 & 1 & 0 & 0 & 0 & 0 & 1 & 1 & 1 & 1 & 0 & 1 & 1 & 1 & 1 & 0 & 1 & 1 \\
( 29) & 1 & 0 & 0 & 1 & 1 & 0 & 1 & 1 & 2 & 2 & 1 & 0 & 1 & 1 & 0 & 1 & 1 & 0 & 2 & 1 & 1 & 1 & 2 & 2 & 1 & 2 & 2 & 1 & 2 & 2 & 2 & 1 & 1 & 0 & 2 & 1 & 1 & 2 & 1 & 1 & 2 & 2 & 2 & 1 & 2 & 2 & 2 & 2 & 2 & 2 & 2 & 2 & 1 & 1 & 2 & 2 & 2 & 2 & 2 & 2 & 2 & 2 & 2 \\
( 30) & 1 & 0 & 0 & 0 & 0 & 0 & 1 & 1 & 1 & 1 & 1 & 0 & 0 & 0 & 0 & 0 & 0 & 0 & 0 & 0 & 0 & 1 & 1 & 1 & 1 & 1 & 1 & 1 & 1 & 1 & 1 & 0 & 0 & 0 & 0 & 0 & 0 & 0 & 0 & 0 & 0 & 1 & 1 & 1 & 1 & 1 & 1 & 1 & 1 & 1 & 1 & 0 & 0 & 0 & 0 & 0 & 1 & 1 & 1 & 1 & 1 & 0 & 1 \\
( 31) & 0 & 0 & 0 & 0 & 1 & 1 & 0 & 0 & 0 & 1 & 1 & 0 & 0 & 1 & 1 & 0 & 1 & 1 & 1 & 1 & 1 & 0 & 0 & 1 & 1 & 0 & 1 & 1 & 1 & 1 & 1 & 0 & 1 & 1 & 1 & 1 & 1 & 1 & 1 & 1 & 1 & 0 & 1 & 1 & 1 & 1 & 1 & 1 & 1 & 1 & 1 & 1 & 1 & 1 & 1 & 1 & 1 & 1 & 1 & 1 & 1 & 1 & 1 \\
( 32) & 0 & 0 & 0 & 1 & 1 & 0 & 0 & 0 & 1 & 1 & 0 & 0 & 1 & 1 & 0 & 1 & 1 & 0 & 1 & 1 & 1 & 0 & 1 & 1 & 0 & 1 & 1 & 0 & 1 & 1 & 1 & 1 & 1 & 0 & 1 & 1 & 1 & 1 & 1 & 1 & 1 & 1 & 1 & 0 & 1 & 1 & 1 & 1 & 1 & 1 & 1 & 1 & 1 & 1 & 1 & 1 & 1 & 1 & 1 & 1 & 1 & 1 & 1 \\
( 33) & 0 & 0 & 1 & 0 & 1 & 0 & 0 & 1 & 0 & 1 & 0 & 1 & 0 & 1 & 0 & 1 & 1 & 1 & 1 & 0 & 1 & 1 & 0 & 1 & 0 & 1 & 1 & 1 & 1 & 0 & 1 & 1 & 1 & 1 & 1 & 0 & 1 & 1 & 1 & 1 & 1 & 1 & 1 & 1 & 1 & 0 & 1 & 1 & 1 & 1 & 1 & 1 & 1 & 1 & 1 & 1 & 1 & 1 & 1 & 1 & 1 & 1 & 1 \\
( 34) & 0 & 1 & 0 & 0 & 0 & 1 & 1 & 0 & 0 & 0 & 1 & 1 & 1 & 1 & 1 & 0 & 0 & 1 & 0 & 1 & 1 & 1 & 1 & 1 & 1 & 0 & 0 & 1 & 0 & 1 & 1 & 1 & 1 & 1 & 1 & 1 & 1 & 0 & 1 & 1 & 1 & 1 & 1 & 1 & 1 & 1 & 1 & 0 & 1 & 1 & 1 & 1 & 1 & 1 & 1 & 1 & 1 & 1 & 1 & 1 & 1 & 1 & 1 \\
( 35) & 1 & 0 & 0 & 1 & 0 & 0 & 1 & 1 & 1 & 1 & 1 & 0 & 1 & 0 & 0 & 1 & 0 & 0 & 1 & 1 & 0 & 1 & 1 & 1 & 1 & 1 & 1 & 1 & 1 & 1 & 1 & 1 & 0 & 0 & 1 & 1 & 0 & 1 & 1 & 0 & 1 & 1 & 1 & 1 & 1 & 1 & 1 & 1 & 1 & 1 & 1 & 1 & 1 & 0 & 1 & 1 & 1 & 1 & 1 & 1 & 1 & 1 & 1 \\
\end{array}
$$
}

We now show that these vectors are all achievable with classical probability distributions that are compatible with the bilocal causal structure, i.e., $P \in \mardist{\bil^\cC}$. Hence, the conic hull of these vertices is $\marcone{\bil^{\cC}}$ and, therefore, the closures of the sets of entropy vectors compatible with $\bil^{\cC}$, $\bil^{\qQ}$ and $\bil^{\gG}$ all coincide. The strategies that lead to the  above $35$ entropy vectors are as follows.
\begin{itemize} 
\item $A$, $B$, $W$ are uniformly random bits. The nodes $C_2$ and $C_3$ distribute two uniformly random bits $C_2^{1}$ and $C_2^{2}$ as well as $C_3^{1}$ and $C_3^{2}$ respectively. $X$ consists of two bits, the first of which is $X_1= A \oplus C_2^{1}$, the second is $X_2= C_2^{2}$. Similarly, $Z_1=B \oplus C_3^{1}$ and $Z_2=C_3^{2}$. Furthermore, $Y$ outputs three bits $Y_1= W \oplus C_2^{2} \oplus C_3^{2}$, $Y_2=W \oplus C_2^{1}$ and $Y_3=W \oplus C_3^{1}$. Reproduces (1).

\item $A$, $B$, $W$ are uniformly random bits. The nodes $C_2$ and $C_3$ distribute two uniformly random bits $C_2^{1}$ and $C_2^{2}$ as well as $C_3^{1}$ and $C_3^{2}$ respectively. $X$ consists of two bits, the first of which is $X_1= A \oplus C_2^{1}$, the second is $X_2= C_2^{2}$. Similarly, $Z_1=B \oplus C_3^{1}$ and $Z_2=C_3^{2}$. Furthermore, $Y$ outputs two bits $Y_1= W \oplus C_2^{1} \oplus C_3^{2}$ and $Y_2=W \oplus C_2^{2} \oplus C_3^{1}$. Reproduces (2).

\item $A$, $B$, $W$ are uniformly random bits. The nodes $C_2$ distributes one uniformly random bit, the node $C_3$ distributes two uniformly random bits $C_3^{1}$ and $C_3^{2}$. $X$ outputs one bit $X= A \oplus C_2$. Furthermore, $Y$ outputs two bits $Y_1= W \oplus C_3^{2}$, $Y_2=C_2 \oplus C_3^{1}$. Similarly, $Z_1=B \oplus C_3^{1}$ and $Z_2=C_3^{1} \oplus C_3^{2}$. Reproduces (3), (4) is a permutation of this.

\item $A$, $B$, $W$ are uniformly random bits. The nodes $C_2$ and $C_3$ distribute one uniformly random bit each. $X$ outputs one bit, $X_1= A \oplus C_2$. Similarly, $Z=B \oplus C_3$. Furthermore, $Y$ outputs two bits $Y_1= W \oplus C_2$ and $Y_2=W \oplus C_3$. Reproduces (5).

\item $A=0$, $B$ and $W$ are uniformly random bits. The node $C_2$ distributes one uniformly random bit, the node $C_3$ distributes two, $C_3^{1}$ and $C_3^{2}$. $X$ outputs one bit, $X= C_2$. Furthermore, $Y$ outputs two bits $Y_1= W \oplus C_3^{1}$ and $Y_2=W \oplus C_2 \oplus C_3^{2}$. Also, $Z_1=B \oplus C_3^{1}$, $Z_2=C_3^{2}$.  Reproduces (6) and the ray (7) is obtained with a permutation of this strategy.

\item $A$, $B$, $W$ are uniformly random bits. The nodes $C_2$ and $C_3$ distribute one uniformly random bit each. $X$ outputs one bit, $X= A \oplus C_2$. Similarly, $Z=B \oplus C_3$. Furthermore, $Y= W \oplus C_2 \oplus C_3$. Reproduces (8).  

\item $A=0$, $B$ and $W$ are uniformly random bits. The nodes $C_2$, $C_3$ distribute one uniformly random bit each. $X$ outputs $X= C_2$. Furthermore, $Y$ outputs two bits $Y_1= W \oplus C_2$ and $Y_2=W \oplus C_3$. Also, $Z=B \oplus C_3$.  Reproduces (9). (10) is obtained with a permuted strategy.

\item $A=0$, $B$ and $W$ are uniformly random bits. The nodes $C_2$, $C_3$ distribute one uniformly random bit each. $X$ outputs $X= C_2$. Furthermore, $Y= W \oplus C_2 \oplus C_3$. Also, $Z=B \oplus C_3$.  Reproduces (11). (12) is obtained with a permuted strategy.

\item $W=0$, $A$ and $B$ are uniformly random bits. The nodes $C_2$, $C_3$ distribute one uniformly random bit each. $X$ outputs $X=W \oplus C_2$. Furthermore, $Y= C_2 \oplus C_3$. Also, $Z=B \oplus C_3$.  Reproduces (13).

\item $A=B=0$, $W$ is a uniformly random bit. The nodes $C_2$, $C_3$ distribute one uniformly random bit each. $X$ outputs $X=C_2$. Furthermore,$Y$ outputs two bits $Y_1=W \oplus C_2$ and $Y_2=W \oplus C_3$. Also, $Z=C_3$.  Reproduces (14).

\item $A=B=0$, $W$ is a uniformly random bit. The nodes $C_2$, $C_3$ distribute one uniformly random bit each. $X$ outputs $X=C_2$. Furthermore, $Y=W \oplus C_2\oplus C_3$. Also, $Z=C_3$.  Reproduces (15).

\item $A=W=0$, $B$ is a uniformly random bit. The nodes $C_2$, $C_3$ distribute one uniformly random bit each. $X$ outputs $X=C_2$. Furthermore, $Y=C_2\oplus C_3$. Also, $Z=B \oplus C_3$.  Reproduces (16). (18) is recovered similarly.

\item $A=X=0$, $B$ and $W$ are a uniformly random bits. The node $C_3$ distributes one uniformly random bit.  $Y=W\oplus C_3$. Also, $Z=B \oplus C_3$.  Reproduces (17). (19) is recovered similarly.

\item $A=B=W=0$. The node $C_3$ distributes one uniformly random bit. $X=C_2$ and $Y=C_2\oplus C_3$. Also, $Z= C_3$.  Reproduces (20).

\item $A=B=X=0$. $W$ is a uniformly random bit. The node $C_2$ and $C_3$ distribute one uniformly random bit each. $Y= W \oplus C_3$, $Z= C_3$.  Reproduces (21). (22) is recovered similarly

\item $A=B=W=X=Y=0$, $Z$ is a uniformly random bit and permutations of this recover (23)-(26), (28) and (30).

\item $A=W=X=0$, $B$ is a uniformly random bit. The node $C_3$ distributes one uniformly random bit. $Y=C_3$. Also, $Z=B \oplus C_3$.  Reproduces (27). (29) is recovered similarly.

\item $A=B=W=X=0$, $B$ is a uniformly random bit. The node $C_3$ distributes one uniformly random bit. $Y=C_3$. Also, $Z=C_3$.  Reproduces (31). (32) is recovered similarly.

\item $A=B=X=Z=0$, $W$ is a uniformly random bit. $Y=W$.  Recovers (33). (34) and (35) are recovered similarly.
\end{itemize}
\end{proof}

The above also implies that non-Shannon inequalities do not play any role for the characterisation of $\bil^\cC$ with the entropy vector method. 

Entropic considerations relying on the post-selection technique have been carried out in Refs.~\cite{Chaves2012, Chaves2016}, where it was shown that the post-selection technique leads to additional inequalities for this scenario.\footnote{Note that Refs.~\cite{Chaves2012, Chaves2016} actually analysed $\pfive$ introduced in the next section.}  These can be violated with PR-boxes~\cite{Tsirelson1993,Popescu1994FP}, whether they can certify a separation between classical and quantum correlations is unknown.

\section{Inability of the entropy vector method to distinguish classical and quantum causes in line-like causal structures}\label{sec:linelike}
\begin{figure}
\centering
\resizebox{0.8 \columnwidth}{!}{%
\begin{tikzpicture}[scale=1.5]
\node[draw=black,circle,scale=0.75] (1) at (0,1) {$X_1$};
\node (2) at (1,0) {$C_1$};
\node[draw=black,circle,scale=0.75] (3) at (2,1) {$X_2$};
\node (4) at (3,0) {$C_2$};
\node[draw=black,circle,scale=0.75] (5) at (4,1) {$X_3$};
\node (6) at (5,0) {$C_3$};
\node (7) at (6.5,0) {$C_{n-1}$};
\node[draw=black,circle,scale=0.75] (8) at (7.5,1) {$X_n$};

\draw [->,>=stealth] (2)--(1);
\draw [->,>=stealth] (2)--(3);
\draw [->,>=stealth] (4)--(3);
\draw [->,>=stealth] (4)--(5);
\draw [->,>=stealth] (6)--(5);
\draw [->,>=stealth] (7)--(8);
\draw [loosely dotted, line cap=round] (6)--(7);
\end{tikzpicture}
}%
\caption[Line-like causal structures]{The causal structure $\pn$. The nodes $X_i$ represent observed variables, whereas the 
$C_i$ denote the unobserved classical, quantum or more general non-signalling systems.}
\label{fig:Pn}
\end{figure}
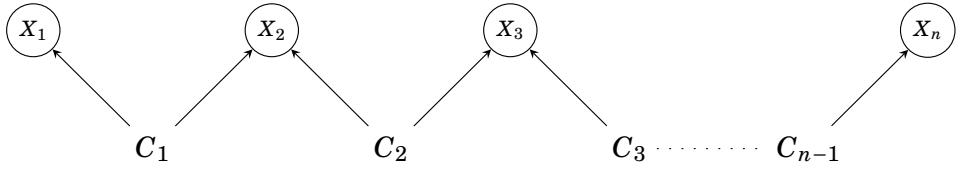

In this section, we consider the family of line-like causal structures, $\pn$, displayed in Figure~\ref{fig:Pn}. The causal structure $\pn$ has observed nodes $X_1,X_2,\ldots,X_n$.  Each pair of consecutive observed nodes $X_i$ and $X_{i+1}$ has an unobserved parent $C_i$.
As briefly mentioned before, the case $n=4$ is in one-to-one correspondence with the bipartite Bell
causal structure shown in Figure~\ref{fig:HensonStructures}(b)~\cite{Fritz2012}.  To make the identification, take $X_1=A$, $X_2=X$, $X_3=Y$, $X_4=B$ and
$C_2=C$.  We can assume without loss of generality that $C_1=A$ and
$C_3=B$: the same set of observed correlations can be generated in
either case (cf.\ Figure~\ref{fig:inputs}).

In the classical case the node $C$ corresponds to a local hidden variable.  Free choice of settings, crucial to the derivation of a Bell inequality, is naturally encoded in the causal structure (e.g., $P_{A|BYC}=P_A$ follows as $A$ has no parents but $BYC$ as its non-descendants), as are the
conditions of \emph{local causality}, that $P_{XY|ABC}=P_{X|AC}P_{Y|BC}$.  The only difference between $\pfour^\cC$ and the quantum case, $\pfour^\qQ$, is the nature of the node $C$.
Bell's original argument then implies that there are non-classical
correlations, i.e., there are distributions in $\mardist{\pfour^{\qQ}}$
that are not in $\mardist{\pfour^{\cC}}$.

In spite of this well-known separation, by looking at the entropy vectors of the observed variables no distinction can be made.  This is stated more formally as follows.
\begin{theorem}\label{thm:1}
  $\marcone{\pn^\qQ}=\marcone{\pn^\cC}$ for all $n\in\mathbb{N}$.
\end{theorem}
Note that for $n\leq 3$, $\mardist{\pn^\cC}=\mardist{\pn^\qQ}$ and hence in these cases the theorem immediately follows~\cite{Fritz2012}. The full proof of this theorem is conducted in Section~\ref{sec:linethm}. 

The theorem proves a considerable limitation of the entropy vector method, which is neither able to distinguish the sets of achievable classical and quantum (and more general non-signalling) entropy vectors in the Bell scenario, nor in its line-like generalisations. The post-selection technique may make up for this issue to some extent, as we shall discuss in Section~\ref{sec:inabilitypost}.

\section{Towards a general statement}\label{sec:smallstructures}
Theorem~\ref{thm:1} reveals major limitations of the entropy vector approach for line-like causal structures, we shall now analyse its performance for several other relevant examples.   
We first focus on causal structures with up to five nodes, for which there is a separation between classical and quantum (and potentially more general resources) at the level of correlations, i.e., for which $\mardist{\smcaus^\cC} \subsetneq \mardist{\smcaus^\qQ}$. In Ref.~\cite{Henson2014}, these causal structures were identified. There is only one with four nodes, the instrumental scenario, and $96$ with five nodes (which can be reduced to the three causal structures displayed in Figure~\ref{fig:HensonStructures} with so called reduction rules)~\cite{Henson2014}. In the following, we show that for the three examples from Figure~\ref{fig:HensonStructures}  the entropy vector method is also unable to distinguish classical and quantum. 
\begin{proposition} \label{prop:uptofive}
For the three small causal structures, $\smone$, $\smtwo$ and $\smthree$ respectively, the entropy vector method is unable to distinguish classical and quantum version.
\end{proposition}

For the Bell scenario, $\smtwo$, this has already been established through Theorem~\ref{thm:1} in the previous section, it is mentioned here for completeness. Whilst for this example the set of compatible entropy vectors can be identified by considering observed variables only, and hence is valid also in theories with unobserved non-signalling resources, the same cannot be said for structures $\smone$ and $\smthree$. However, we do not know of any violations of the entropy inequalities that are valid for $\smone$ and $\smthree$ in the classical and quantum case by means of more general non-signalling systems. The proof of Proposition~\ref{prop:uptofive} is deferred to Section~\ref{sec:smallprop}. 

The only causal structures with up to five nodes, where the entropy vector method may unveil a separation between classical and quantum correlations are the $94$ remaining examples from Ref.~\cite{Henson2014}, for all others the sets of compatible distributions generated with classical and quantum resources are identical, $\mardist{\smcaus^{\cC}}=\mardist{\smcaus^{\qQ}}$, and so are the sets of compatible entropy vectors. 
The causal structure $\hat{\inst}$ from Figure~\ref{fig:IC_inner_ch4} is an example of a five variable causal structure for which we were not able to establish that $\marcone{\hat{\inst}^{\cC}}=\marcone{\hat{\inst}^{\qQ}}$. For $\hat{\inst}$ we found a gap between inner and outer approximations to the classical entropy cone, which could be explored with non-Shannon inequalities, as outlined in Chapter~\ref{chap:nonshan}. Even including such inequalities, however, the linear characterisation of the set of compatible entropy vectors remains elusive in both the classical and the quantum case, leaving the question as to whether their boundaries coincide open. 

Out of all causal structures with six random variables, there are $10186$ for which a classical to quantum separation might exist~\cite{Henson2014} and the number of such examples increases rapidly with the number of variables. As we have observed in Chapter~\ref{chap:nonshan}, for the $18$ examples to which the former can be reduced, non-Shannon inequalities play a role in their entropic characterisation and we are lacking a complete linear characterisation in all cases. Our insufficient understanding of the mapping from probability distributions to entropies and of the structure of entropy space itself hinder us from proving equivalence (or inequivalence) in these comparably small examples.

\bigskip

Overall, there are several reasons why a general statement about the classical to quantum separation in entropy space is elusive. First of all, for $n \geq 4$ we cannot sufficiently characterise the entropy cone for $n$ random variables nor that for an $n$-party quantum state. In the classical case infinitely many inequalities are required for this endeavour in general, a practically infeasible task that is complicated further as not all of these inequalities are known. The quantum case is even less understood. No inequalities for the von Neumann entropy are known except for positivity of entropy, submodularity and weak-monotonicity. Hence, the entropy cone of a coexisting set of quantum nodes in a causal structure may or may not be sufficiently characterised in terms of linear entropy inequalities~\cite{Pippenger2003}. In addition, the transition from quantum states to classical observed variables through measurements, that in the entropic approximations is encoded in DPIs, allows CPTP maps with classical output states only. Whether and how this may restrict the compatible entropy vectors, and whether it may retain the structure of a convex cone is, to our knowledge, not understood. This leaves us with the following open problem.

\begin{problem}
Does there exist a causal structure $C$ for which $\marcone{C^\cC} \neq \marcone{C^\qQ}$?
\end{problem}

It is unknown whether the entropy vector method merely accounts for (conditional) independence relations and hence encodes theory-independent information about causal structure or whether it is able to distinguish unobserved resources from different underlying theories.

\section{Distinguishing classical and quantum causes with post-selection}\label{sec:inabilitypost}
The post-selection technique is applicable whenever at least one observed parentless node is part of the causal structure under consideration or when appropriate adaptations can be made (cf.\ Section~\ref{sec:post-selection}). This means that it is neither directly applicable in the triangle causal structure nor in line-like causal structures.

\begin{figure}
\centering
\resizebox{0.45 \columnwidth}{!}{%
\begin{tikzpicture}[scale=1]
\node (a)  at (-2,2.5) {$(a)$};
\node[draw=black,circle,scale=0.75] (X1) at (-1,2) {$X_2$};
\node[draw=black,circle,scale=0.75] (A1) at (-1,0) {$X_1$};
\node (Z1) at (-0.0,0.5) {};

\node (b) at (2,2.5) {$(b)$};
\node[draw=black,circle,scale=0.75] (X) at (3,2) {$X_2$};
\node[draw=black,circle,scale=0.75] (A) at (3,0) {$X_1$};
\node (ZZ) at (3,1) {$A$};
\node (Z) at (4,0.5) {};
\node (Y) at (5,0.5) {};

\draw [->,>=stealth] (A1)--(X1);
\draw [->,>=stealth] (ZZ)--(X);
\draw [->,>=stealth] (ZZ)--(A);
\draw [dotted,->,>=stealth] (Z1)--(X1);
\draw [dotted,->,>=stealth] (Z)--(X);
\end{tikzpicture}
}%
\caption[Causal structures producing the same correlations]{For a quantum causal structure with an observed input node, $X_1$--meaning a parentless node from which there is only one arrow to another observed node, $X_2$--there always exists another (quantum) causal structure that allows for exactly the same correlations and where the observed input is replaced by a shared quantum parent of $X_1$ and $X_2$. To simulate any correlations in (a) within scenario (b) we can
use a quantum system that sends perfectly correlated classical
states to both nodes $X_1$ and $X_2$, which are distributed as $X_1$. On the other hand, any correlations created in scenario (b) can be created in scenario (a) by having a random variable $X_1$ sent to node $X_2$, the relevant quantum states (the reduced states that would be present in (b) conditioned on the value
of $X_1$) are locally generated.}
\label{fig:inputs}
\end{figure}
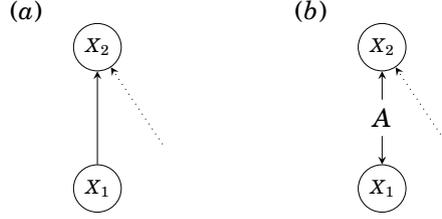

Whereas in the triangle scenario this issue cannot be easily overcome, we can apply the technique to the line-like causal structures. There, the two nodes $X_1$ and $X_n$ can be converted into observed parents of $X_2$ and $X_{n-1}$ respectively, in the sense that $\pn$ is in one-to-one correspondence with such an adapted scenario. This is illustrated in Figure~\ref{fig:inputs}.
Therefore, small line-like causal structures can be analysed with the post-selection technique. $\pfour$ and $\pfive$ have been previously analysed in this way in Refs.~\cite{Braunstein1988, Chaves2013} and~\cite{Chaves2012} respectively. While for $\pfour$ all quantum correlations that are not classically reproducible can be detected in this way~\cite{Chaves2014}, no quantum violations of the corresponding entropic inequalities for $\pfive$ are known, there are, however, violations generated with PR-boxes~\cite{Chaves2012}.

For $\psix$, we find an outer approximation with $28$ extremal rays, alternatively characterised by $16$ equalities and $153$ inequalities (including permutations).\footnote{We do not list these vertices and inequalities here, as they are not particularly instructive. They may, however, be provided upon request.} For larger line-like causal structures the number of variables soon becomes computationally unmanageable, rendering the method impractical.

\section{Appendix}

\subsection{Proof of Theorem~\ref{thm:1}} \label{sec:linethm} 
For our proof, we rely on the following lemma by Yeung~\cite{Yeung1997}, the proof of which is omitted here. 
\begin{lemma}[Yeung~\cite{Yeung1997}]
  Consider $n$ variables $X_1,X_2,\ldots,X_n$ and define
  $\Omega:=\left\{X_1,X_2,\ldots,X_n\right\}$.  Taking positivity of
  each entropy to be implicit, the Shannon inequalities for these are
  generated from a minimal set of $n+n(n-1)2^{n-3}$ inequalities:
\begin{eqnarray}
H(\Omega | \Omega \setminus \{X_i\} )  &\geq &  0,\label{eq:monot}\\
I(X_i:X_j|X_S)  &\geq & 0,\label{eq:submod}
\end{eqnarray}
where the first is needed for all $X_i\in\Omega$ and the second for
all $X_S\subsetneq\Omega$, $X_i,X_j\in\Omega$, $X_i,X_j\notin X_S$,
$i<j$.
\end{lemma}

Now take $X_1,X_2,\ldots,X_n$ to be the observed nodes in $\pn$ and
for $i+1<j$ define $M_{i,j}:=\left\{ X_k \right\}_{k=i+1}^{j-1}$ as
the set of nodes between $X_i$ and $X_j$. The first part of the proof
of Theorem~\ref{thm:1} is to show that from these $n+n(n-1)2^{n-3}$
Shannon inequalities at most $\frac{n(n+1)}{2}$ are not implied by the
conditional independence constraints and the remaining Shannon
inequalities.

\begin{lemma}\label{lemma:middlenodes}
Within the causal structure $P^\cC_n$, all of the submodularity inequalities~\eqref{eq:submod} with $M_{i,j}\not\subseteq X_S$ are implied by the causal constraints.
\end{lemma}
\begin{proof}
Let $M_{i,j} \not\subseteq X_S$, then there is at least one node $X_{k} \not\in X_S$ with $i<k<j$. For each such node we can partition $X_S=\left\{ X_S^{k-},~X_S^{k+} \right\}$, where $X_S^{k-}$ contains all $X_l \in X_S$ with $l<k$ and $X_S^{k+}$ contains the elements with $X_l \in X_S$ with $l>k$ (note that both sets may be empty). Since $\{X_i\}\cup X_S^{k-}$ is d-separated
from $\{X_j\}\cup X_S^{k+}$ we have
\begin{eqnarray}
H(\left\{X_i,~X_j \right\} \cup X_S)&=&H(\left\{X_i \right\} \cup X_S^{k-})+H(\left\{X_j \right\} \cup X_S^{k+}),\\
H(\left\{X_i \right\} \cup X_S)&=&H(\left\{X_i \right\} \cup X_S^{k-})+H(X_S^{k+}),\\
H(\left\{X_j \right\} \cup X_S)&=&H( X_S^{k-})+H(\left\{X_j \right\} \cup X_S^{k+}), \\
H(X_S)&=&H( X_S^{k-})+H(X_S^{k+}),
\end{eqnarray}
and thus~\eqref{eq:submod} is obeyed with equality.
\end{proof}

\begin{lemma}\label{lemma:lines}
  Within the causal structure $\pn^\cC$, the $\frac{n(n-1)}{2}$
  submodularity constraints of the form $I(X_i:X_j|M_{i,j})\geq 0$ for
  all $X_i,X_j$ with $i<j$ imply all submodularity
  constraints~\eqref{eq:submod}.
\end{lemma}
\begin{proof} 
  Lemma~\ref{lemma:middlenodes} shows this to hold in the
  case $M_{i,j}\not\subseteq X_S$.  Thus, we restrict to the case
  $M_{i,j}\subseteq X_S$. Let us write $X_S=M_{i,j}\cup X_T$, where
  $X_T=X_S\setminus M_{i,j}$.
  
  First consider the case where $X_{i-1},X_{j+1}\notin X_T$. Here
  $M_{i-1,j+1}$ and $X_T$ are d-separated and hence
  \begin{eqnarray}
    H(\{X_i,X_j\}\cup M_{i,j}\cup X_T)&=&H(\{X_i,X_j\}\cup M_{i,j})+H(X_T)\\
    H(\{X_i\}\cup X_T)&=&H(X_i)+H(X_T)\\H(\{X_j\}\cup X_T)&=&H(X_j)+H(X_T)\\H(M_{i,j}\cup X_T)&=&H(M_{i,j})+H(X_T)
  \end{eqnarray}
  so that $I(X_i:X_j|M_{i,j}\cup X_T)=I(X_i:X_j|M_{i,j})$.
Next, consider the case where $X_k\in X_T$ for $k=j+1,j+2,\ldots,j+L$, but $X_{i-1},X_{j+L+1}\notin X_T$. By
d-separation, we have $I(X_i:X_j|M_{i,j}\cup
X_T)=I(X_i:X_j|M_{i,j}\cup\{X_{j+1},\ldots,X_{j+L}\})$,
and the latter expression can be more concisely written as
$I(X_i:X_j|M_{i,j}\cup M_{j,j+L+1})$.  Then,
\begin{align}
I(X_i:X_j|M_{i,j}\cup
M_{j,j+L+1})
&=I(X_i:M_{j,j+L+1}\cup\{X_j\}|M_{i,j})-I(X_i:M_{j,j+L+1}|M_{i,j})\\
&=I(X_i:M_{j,j+L+1}\cup\{X_j\}|M_{i,j})\\
&=I(X_i:X_j|M_{i,j})+I(X_i:M_{j,j+L+1}|M_{i,j+1})\\
&=I(X_i:X_j|M_{i,j})+I(X_i:M_{j+1,j+L+1}\cup\{X_{j+1}\}|M_{i,j+1}),
\end{align}
where we have used $I(X_i:M_{j,j+L+1}|M_{i,j})=0$, which follows from
d-separation.  Noting the relation between the last term in the final line and the
third line, we can proceed to recursively decompose the
expression into
\begin{align}
I(X_i:X_j|&M_{i,j}\cup M_{j,j+L+1})
=\sum_{l=0}^L I(X_i:X_{j+l}|M_{i,j+l})\, .
\label{eq:removej}
\end{align}

Now suppose $X_k\in X_T$ for $k=i-1,i-2,\ldots i-K$ and
$k=j+1,j+2,\ldots,j+L$, but $X_{i-K-1},X_{j+L+1}\notin X_T$. By d-separation, we have $I(X_i:X_j|M_{i,j}\cup
X_T)=I(X_i:X_j|M_{i,j}\cup\{X_{i-K},\ldots,X_{i-1}\}\cup\{X_{j+1},\ldots,X_{j+L}\})$, and the latter expression can be more concisely written as $I(X_i:X_j|M_{i,j}\cup M_{i-K-1,i}\cup M_{j,j+L+1})$. Then,
\begin{align}
I(X_i:X_j & |  M_{i,j}\cup
M_{i-K-1,i}\cup
M_{j,j+L+1}) \nonumber \\
&=I(M_{i-K-1,i}\cup\{X_i\}:X_j|M_{i,j}\cup
M_{j,j+L+1})
-I(M_{i-K-1,i}:X_j|M_{i,j}\cup
M_{j,j+L+1})\\
&=I(M_{i-K-1,i}\cup\{X_i\}:X_j|M_{i,j}\cup
M_{j,j+L+1})\\
&=I(X_i:X_j|M_{i,j}\cup
M_{j,j+L+1})
+I(M_{i-K-1,i}:X_j|M_{i-1,j}\cup
M_{j,j+L+1})\\
&=I(X_i:X_j|M_{i,j}\cup M_{j,j+L+1})+I(M_{i-K-1,i-1}\cup\{X_{i-1}\}:X_j|M_{i-1,j}\cup M_{j,j+L+1}),
\end{align}
where we have used $I(M_{i-K-1,i}:X_j|M_{i,j}\cup
M_{j,j+L+1})=0$, which follows from d-separation.
Noting the relation between the last term in the final line and the
third line, we can hence proceed to recursively decompose the
expression into
\begin{equation}
I(X_i:X_j|M_{i,j}\cup
M_{i-K-1,i}\cup
M_{j,j+L+1})
=\sum_{k=0}^K I(X_{i-k}:X_j|M_{i-k,j}\cup
M_{j,j+L+1})\, .
\end{equation}
The latter can then be decomposed using~\eqref{eq:removej}.
\end{proof}

Including the $n$ monotonicity constraints, there are at most
$\frac{n(n+1)}{2}$ Shannon inequalities that are not implied by the
conditional independence relations of $\pn^\cC$. These inequalities
constrain a pointed polyhedral cone with the zero vector as its
vertex. They hold for all entropy vectors in $\pn^\cC$ and thus
approximate the entropy cone $\marcone{\pn^\cC}$ from the outside. They are
also valid for $\marcone{\pn^\qQ}$ (recall that two subsets of a coexisting
set are independent if they have no shared ancestors).  Note that the
causal constraints reduce the effective dimensionality of the problem
to $\frac{n(n+1)}{2}$, since the entropies of contiguous sequences are
sufficient to determine all entropies\footnote{There are $n$
  contiguous sequences of length $1$, $\{H(X_i)\}_{i=1}^n$, $n-1$ of
  length $2$, $\{H(X_iX_{i+1})\}_{i=1}^{n-1}$, and so on, leading to
  $\sum_{i=1}^ni=\frac{n(n+1)}{2}$ in total.}.
The $\frac{n(n+1)}{2}$ inequalities can lead to at most
$\frac{n(n+1)}{2}$ extremal rays, which corresponds to the number of
ways of choosing $\frac{n(n+1)}{2}-1$ inequalities to be
simultaneously obeyed with equality. In the following we show that
this bound is saturated by constructing $\frac{n(n+1)}{2}$ entropy
vectors from probability distributions in $\pn^\cC$, each of which lies
on a different extremal ray.

Consider the following set of distributions in $\pn^\cC$ (leading to
corresponding entropy vectors). Let $\{C_i\}_{i=1}^{n-1}$ be uniform
random bits, and $1\leq i\leq j\leq n$.  For each $i,~j$ we define a distribution $D_{i,j}$.
\begin{itemize}
\item For $i\leq n-1$, $D_{i,i}$ is formed by taking
  $X_i=C_i$ and $X_k=1$ for all $k\neq i$, while $D_{n,n}$ has
  $X_i=C_{i-1}$ and $X_k=1$ for all $k\neq i$.
\item For $i<j$, $D_{i,j}$ is constructed in the following. Note that depending on $i$ and $j$, each of the parts indexed by $k$ below may also be empty.
\begin{itemize}
\item $X_k=1$ for $1\leq k\leq i-1$,
\item $X_i=C_i$,
\item $X_k=C_{k-1} \oplus C_k$ for $i+1\leq k\leq j-1$, 
\item $X_j=C_{j-1}$,
\item $X_k=1$ for $j+1\leq k\leq n$.
\end{itemize}
\end{itemize}
Note that the set of distributions $\{D_{i,j}\}_{i,j}$ for $1 \leq i \leq j \leq n$ is in one-to-one correspondence with the contiguous sequences from $\Omega$.

\begin{lemma}\label{lem:5}
The $\frac{n(n+1)}{2}$ entropy vectors of the probability distributions $\{D_{i,j}\}_{i,j}$ with $1 \leq i \leq j \leq n$ are extremal rays of $\overline{\Gamma^*_\cM}(\pn^\cC)$.
\end{lemma}

\begin{proof} 
It is sufficient to prove the following:
\begin{itemize}
\item For each $i$, $D_{i,i}$ obeys all of the Shannon equalities with equality except the monotonicity relation $H(\Omega)-H(\Omega\setminus\{X_i\})\geq 0$, which is a strict inequality.
\item For $i<j$, $D_{i,j}$ obeys all of the Shannon inequalities with equality except $I(X_i:X_j|M_{i,j})\geq 0$, which is a strict inequality.
\end{itemize}

For the $n$ distributions $D_{i,i}$ all variables are independent and thus their entropy vectors automatically satisfy all submodularity inequalities with equality. Furthermore, for any $X_S \subsetneq \Omega$
with $X_i\notin X_S$ we have $H(\{X_i\}\cup X_S)=H(X_i)$. Thus, for $j\neq i$ we have
\begin{equation}
H(\Omega)-H(\Omega\setminus\{X_j\})=0\, ,
\end{equation}
while for $j=i$
\begin{equation}
\begin{split}
H(\Omega)-H(\Omega\setminus\{X_j\})&=H(X_i) \\
&>0.
\end{split}
\end{equation} 
This establishes the first statement.

Consider now the $(n-1)!$ distributions $D_{i,j}$ with $i<j$.  We
first deal with the monotonicity constraints.  For $k<i$ and $k>j$, we
have $H(\Omega)=H(\Omega\setminus X_k)=j-i$. Similarly, since any
$j-i-1$ elements of $M_{i-1,j+1}$ are sufficient to determine the
remaining element, we also have $H(\Omega\setminus X_k)=j-i$ for $i\leq
k\leq j$.  Thus, all the monotonicity constraints hold with equality.
For the submodularity constraints, it is useful to note that for any $D_{i,j}$ with $i<j$ we have
\begin{equation}
H(X_k|M_{k,l})=\begin{cases}
1,&k=i\text{ and }k<l\leq j,\\
1,&i<k\leq j\text{ and }k<l,\\
0,&\text{otherwise}.
\end{cases} 
\end{equation} 
Thus, $I(X_k:X_l|M_{k,l})=H(X_k|M_{k,l})-H(X_k|M_{k,l+1})$ is zero
unless $k=i$ and $l=j$ (in which case it is $1$). This establishes the
second statement, and hence completes the proof of Lemma~\ref{lem:5}.
\end{proof}

Note that the entropy vector of each of the $\frac{n(n+1)}{2}$
distributions belongs to a different extremal ray. We have thus shown
that for each extremal ray of $\marcone{\pn^\cC}$ there
is a distribution in $\mardist{\pn^\cC}$ whose entropy vector lies on
that ray.  It follows by convexity that any vector that satisfies all
the Shannon constraints and the causal constraints of the marginal
scenario in $\pn^\cC$ is realisable in $\pn^\cC$ (at least
asymptotically). Since the same outer approximation is valid for
$\marcone{\pn^{\qQ}}$ and any classical
distribution can be realised quantum mechanically, we have
$\marcone{\pn^{\cC}} \subseteq
\marcone{\pn^{\qQ}} \subseteq
\marcone{\pn^\cC}$ and 
therefore $\marcone{\pn^{\cC}} =
\marcone{\pn^{\qQ}}$. \qed

\subsection{Proof of Proposition~\ref{prop:uptofive}} \label{sec:smallprop}
For $\smone$, $\smtwo$ and $\smthree$, we have found their classical entropy cones, $\marcone{\smone^{\cC}}$, $\marcone{\smtwo^{\cC}}$ and $\marcone{\smthree^{\cC}}$ in Chapter~\ref{chap:inner} (Example~\ref{example:innerI} and Section~\ref{sec:inner_examples}). It thus remains to show that these coincide with the quantum sets $\marcone{\smone^{\qQ}}$, $\marcone{\smtwo^{\qQ}}$ and $\marcone{\smthree^{\qQ}}$ respectively.
For this purpose, we rely on the $\hrep$-representation of  $\marcone{\smone^{\cC}}$, $\marcone{\smtwo^{\cC}}$ and $\marcone{\smthree^{\cC}}$ and prove for each of them that the respective inequalities also hold in the quantum case.
\begin{enumerate}[(a)]
\item For $\smone$ the only inequality in addition to the Shannon constraints for three observed variables is $I(X : ZY )\leq H(Z)$ (cf.\ Example~\ref{example:IC_class_obs} or Ref.~\cite{Henson2014,Chaves2014b}). It is also valid in the quantum case because 
\begin{align}
I(X:ZY) &\leq I(X:ZA_Y) \\
&\leq  H(X)+ H(Z) + H(A_Y) - H(XZA_Y) \\
&= H(Z) + H(X A_Y) - H(XZA_Y) \\
&\leq H(Z) \ , 
\end{align}
where the first inequality is a DPI, then we use submodularity, the independence of $X$ and $A_Y$ and finally monotonicity for the cq-state $\rho_{XZA_Y}$.

\item For $\smtwo$ the only additional constraints are the independencies $I(A:BY)=0$ and $I(B:AX)=0$, (cf.\ Section~\ref{sec:inner_examples}) which hold in the quantum case by the d-separation rules of  Section~\ref{sec:quantum_ns_causal}.

\item For $\smthree$ the only additional inequality is $I(Y:Z|X) \leq H(X)$ (cf.\ Section~\ref{sec:inner_examples}). This holds in the quantum case as,
\begin{align}
I(Y:Z|X) &\leq I(Y:B_Z|X) \\
&\leq I(A_Y:B_Z|X) \\
&\leq H(A_Y X) + H(B_Z X) - H(X) - H(A_Y B_Z) \\
&= H(A_Y X) + H(B_Z X) - H(X) - H(A_Y) - H(B_Z) \\
&\leq H(X) \, ,
\end{align}
where the first two are DPI and then we use monotonicity. The equality holds as $A_Y$ and $B_Z$ are independent and the last inequality is implied by two submodularity inequalities. 
\end{enumerate}
\qed

%% file: chapter07/thermoaxioms.tex
\let\textcircled=\pgftextcircled
\part{Error-tolerant approach to microscopic thermodynamics}

\chapter{Axiomatic framework for error-tolerant resource theories}\label{chap:microthermo}

\initial{W}e introduce an axiomatic framework for resource theories that models processes with finite precision and errors, the significance of which we motivate in Section~\ref{sec:error_motivation}.
As a first original contribution, we introduce error-tolerant processes into the resource theoretic approach in Section~\ref{sec:introduce_errors}. In Section~\ref{sec:smoothentropies} we define quantities that give necessary conditions as well as sufficient conditions for error-tolerant transformations  and we analyse their properties, greatly simplifying the working with these resource theories. Our elaborations are illustrated with the example of adiabatic processes, for which we consider different types of errors in Sections~\ref{sec:introduce_errors} and~\ref{sec:alvaro_smoothing} respectively.

\section{Motivation}\label{sec:error_motivation}

In microscopic systems, errors play a much more significant role than for large, thermodynamic systems in equilibrium. If we consider the thermodynamic system par excellence, a large box filled with a gas in an equilibrium state, small deviations, for instance if the gas is contaminated with a single atom of a different gas, do not (notably) change the thermodynamic properties of the gas. On the other hand, replacing a single atom in a system of \emph{few} particles may result in a considerable change to the system's properties. Resource theoretic work has so far mostly considered idealised situations where there is perfect control over (usually microscopic) systems~\cite{Goold2016}, impeding experimental implementations. The consideration of finite size effects in terms of restrictions on available reservoirs~\cite{Gemmer2006, Sparaciari2016} and the consideration of probabilistic transformations~\cite{Alhambra2015}, have initiated a development towards more realistic resource theories. 
Modelling processes with limited precision, which are innate in any experimental setup, is a natural continuation of this trend that we pursue here. Recently -- concurrently with our considerations presented here -- this has been achieved for specific resource theories in Refs.~\cite{Hanson2017, VanderMeer2017, Horodecki2017}.

In general, different types of errors are quantified in different ways. For instance, when implementing a process, an experimenter will usually only be able to prepare a system in a specified state up to a certain level of precision, which is quantified depending on the experiment at hand. Furthermore, the implementation of a process on that system and the state of the system after preparation may be further affected by noise. Additionally, the complete failure of a process or experiment may also be accounted for as another type of error. The variety of situations we may want to describe motivates us to take an axiomatic approach that is capable of describing different types of errors. 

Axiomatic approaches have a tradition in thermodynamics dating back to the work of Carath\'{e}odory~\cite{Caratheodory1909} and have recently received new attention with further advancements being made~\cite{Giles1964,Lieb1998,Lieb1999,Lieb2001,Lieb2013,Lieb2014}. Even though resource theories that apply to microscopic systems have been shown to be compatible with such an axiomatic framework~\cite{Lieb1998,Lieb1999,Lieb2001,Lieb2013,Lieb2014,Weilenmann2015}, the frameworks at hand were not devised for such an application. The approach of Refs.~\cite{Lieb1998,Lieb1999,Lieb2001,Lieb2013,Lieb2014} relies on the existence of systems that can be scaled continuously, as considered in classical thermodynamics, a concept we aim to avoid here. In addition, previous axiomatisations are unable to model errors, since this would require a different axiomatic structure. One of the reasons for this is that sequential composition of processes is no longer transitive when errors are accounted for and transitivity is a prerequisite in~\cite{Lieb1998,Lieb1999,Lieb2001,Lieb2013,Lieb2014}. In the following, we provide an axiomatic framework that explicitly takes errors into account and that can do without a continuous scaling operation. Our framework is inspired by Lieb and Yngvason's axiomatisation of (macroscopic and mesoscopic) thermodynamics~\cite{Lieb1998,Lieb1999,Lieb2001,Lieb2013,Lieb2014}.

\section{Thermodynamic processes with error-tolerance}\label{sec:introduce_errors}

Like in the traditional resource theory approach introduced in Section~\ref{sec:resource}, we consider a state space, here denoted $\Gamma$ (as inspired by~\cite{Lieb1998, Lieb1999}), on which the possible state transformations introduce an ordering. This ordering is, however, changed when considering processes with potential errors. 
That is, instead of a preorder, $\prec$, an error-tolerant resource theory introduces a family of order relations, $ \prec^\eps $, one for each error-tolerance $\eps$:
for $\rho$, $\sigma \in \Gamma$ the relation $\rho \prec^\eps \sigma$ expresses that a transformation from $\rho$ to $\sigma$ with error at most $\eps$ is possible. For $\eps > 0 $, the relations $\prec^\eps$ are no longer transitive and thus no longer form a preorder on $\Gamma$. Nonetheless, we require $\prec^\eps$ to satisfy a few natural axioms. 

\begin{axiom}[Reflexivity]
\label{axiom:reflexivity}
For any state $\rho \in \Gamma$ and any $\eps \geq 0$, 
\begin{equation}
\rho \prec^{\eps} \rho \ .
\end{equation}
\end{axiom}

This captures the intuition that there should be an operation that leaves the system alone, even when allowing for errors. The second axiom captures the intuition that a larger error-tolerance should never render a transformation impossible. If a process can occur with a small error, it should also be possible if we allow errors to be larger.

\begin{axiom}[Ordering of error-tolerances]
\label{axiom:smoothing_order}
For any $\rho$, $\sigma \in \Gamma$ and any $\eps'\geq \eps \geq 0$, 
\begin{equation}
\rho\prec^\eps\sigma \implies \rho \prec^{\eps'} \sigma \ . 
\end{equation}
\end{axiom}

Furthermore, when sequentially composing processes, the errors of consecutive processes may add up. For instance, when errors express a probability of failure of a process, the worst case is encountered when an error in the first process renders the second process impossible and hence leads to a failure of the overall operation. Alternatively, when errors specify a deviation from a target, there are situations where the second process may correct for the deviations in the first but also situations where it may not. The latter situations also lead to the addition of errors.
This behaviour renders $\prec^{\eps}$ for $\eps >0$ intransitive, while the error free case $\prec^0$, which from here on in we denote by $\prec$, remains a preorder on $\Gamma$.

\begin{axiom}[Additive transitivity]
\label{axiom:additivetransitivity}
For any $\rho$, $\sigma$, $\omega \in \Gamma$ and any $\eps$, $\delta \geq 0$ it holds that 
\begin{equation}
\rho\prec^\eps\sigma, \sigma\prec^\delta \omega
\implies \rho\prec^{\eps+\delta} \omega \ . 
\end{equation}
\end{axiom}

In the following, we shall quantify errors in terms of probabilities, i.e., we shall require $0 \leq \eps, \delta \leq 1$. Whenever we add two errors we understand this as  $\eps+\delta = \min \left\{ \eps+\delta, \  1 \right\}$. However, errors may be quantified in other ways, in which case they may become arbitrarily large.

The last axiom concerns joint transformations of multiple systems. Composition of two systems $\rho \in \Gamma$, $\rho' \in \Gamma'$  is denoted in terms of a Cartesian product as $(\rho, \rho') \in \Gamma \times \Gamma'$, which is assumed to be commutative and associative. It describes the consideration of two independent systems as part of a larger one, where joint operations may then be applied to the two parts. If we consider an $n$-system or $n$-particle state space we shall denote this $\Gamma^{(n)}$, where in general $\Gamma^{(1)}\times \cdots \times \Gamma^{(1)} \subseteq \Gamma^{(n)}$, i.e., the $n$-system state space may contain states that cannot be obtained by composing $n$ systems. 
The following axiom captures the intuition that the presence of additional resources should not inhibit any transformations. It is an adapted version of the composition axiom from~\cite{Lieb1998,Lieb1999,Lieb2001,Lieb2013}.

\begin{axiom}[Consistent composition]
\label{axiom:consistency}
For any $\rho$, $\sigma \in \Gamma$, $\omega \in \Gamma'$ and any $\eps \geq 0$, it holds that
\begin{equation}
\rho\prec^{\eps}\sigma \implies (\rho,\omega)\prec^{\eps}(\sigma,\omega) \ .
\end{equation}
\end{axiom}

\subsection{The noise-tolerant resource theory of adiabatic processes} \label{sec:smoothingNO}

We explore different ways to define an error-tolerant version of the resource theory of adiabatic operations from Section~\ref{sec:noisy}. The state space is the set of density operators on a finite dimensional Hilbert space, $\Gamma= \cS(\cH)$\footnote{For simplicity, we take all states to live on the same Hilbert space. Our considerations could be generalised (see Section~\ref{sec:noisy}).}
and the composition of systems is defined as their tensor product, i.e., for $\rho \in \cS(\cH)$, $\rho' \in \cS(\cH')$, their composed state is $( \rho,~\rho' ) = \rho \otimes \rho' \in \cS(\cH \otimes \cH')$.

Adiabatic operations can be affected by different types of errors. In an experimental setup, it is usually impossible to have full control over a system and any state and any process can only be prepared and operated up to some finite precision. It may therefore not be possible to experimentally distinguish whether an exact transition between two states has occurred or whether a process between similar states has been achieved instead. Such errors can be quantified in terms of the deviations of input and output states from their respective targets. For instance, we may allow for imprecise input states, accounting for preparation errors, so that an error-tolerant transition is possible if an exact transition from an input state that is close, or even experimentally indistinguishable from the desired state to the target output state is possible. Alternatively, we can allow for imprecise output states, accounting for errors in the process: in this case, a transition is possible if an exact transformation could reach an output state close to the target. These two types of errors can moreover be combined to an overall error that accounts for imprecision in input and output. More specifically, we may consider any combination of deviations of the first two types such that they add up to some maximal error $\eps$. We shall call these three ways of quantifying errors \emph{smoothings}. Different types of errors shall be considered in Section~\ref{sec:alvaro_smoothing}.

In the following, we show that when quantifying the smoothings with the trace distance, all of them are mathematically equivalent.  Moreover, they can be nicely expressed by an adapted majorisation relation. We also remark here that for $\eps=0$ the usual majorisation relation is recovered. We shall call the corresponding noise-tolerant resource theory the resource theory of \emph{smooth adiabatic operations}.

\begin{proposition} 
\label{prop:smoothing_equivalence}
For the resource theory of adiabatic operations with $\Gamma=\cS(\cH)$, the following four definitions of $\maj^\epsilon$ are equivalent:
\begin{enumerate}[(1)]
\item $\rho\maj^\epsilon \sigma \iff
\exists \ \rho' \text{ s.t. } \rho'\maj \sigma\text{ and } \rho'\in \mathcal{B}^\epsilon(\rho)$, 
\item $\rho\maj^\epsilon \sigma \iff 
\exists \ \sigma' \text{ s.t. } \rho\maj \sigma'\text{ and } \sigma'\in \mathcal{B}^\epsilon(\sigma)$, 
\item $\rho\maj^\epsilon \sigma \iff 
\exists \ \rho',\sigma' \text{ and } \epsilon', \epsilon'' \text{ s.t. } \rho'\maj \sigma'\text{ and } \rho'\in \mathcal{B}^{\epsilon'}(\rho),\  \sigma'\in\mathcal{B}^{\epsilon''}(\sigma)$ with $\epsilon'+\epsilon''\leq\epsilon$, 
\item $\rho\maj^\epsilon \sigma \iff 
\int_0^x f_\rho(x') \ dx' \geq 
\int_0^x f_\sigma(x') \ dx' - {\epsilon} \ \ \forall x\geq 0$, where $f_\rho$ and $f_\sigma$ are the step functions of the spectrum of $\rho$ and $\sigma$ respectively.
\end{enumerate}
\end{proposition}

The proposition is proven in Section~\ref{sec:smoothing_equivalence}. We note here that the equivalence of different smoothings is particular to adiabatic operations, if we were to consider a different resource theory, for instance the resource theory of thermal operations~\cite{Janzing2000_cost, Horodecki2011, Renes2014}, a similar proposition would not hold~\cite{prep}.

The resource theory of smooth adiabatic operations allows us to analyse situations that could not be described within the resource theory of adiabatic processes, as is illustrated with the following example. 
\begin{example}
Let us consider a maximally mixed qubit, 
\begin{equation}
\rho=\frac{1}{2} \ket{0}\bra{0} + \frac{1}{2} \ket{1}\bra{1}\ .
\end{equation}
Measuring this qubit in the $\left\{ \ket{0}, \ket{1} \right\}$-basis brings the system into a pure state\footnote{At the same time it decreases its entropy by $\log_2(2)$.}. Such a measurement cannot be implemented as an adiabatic process. Nonetheless, it can occur if we allow for a large enough error and it can be modelled in the resource theory of  smooth adiabatic operations. The state  $\ket{1}$ is obtained from $\rho$ by means of the identity operation, which produces the state $\ket{0}$ with probability $\frac{1}{2}$ and $\ket{1}$ also with probability $\frac{1}{2}$. In this process, the transformation from $\rho$ to $\ket{1}$ has an error probability $\frac{1}{2}$, thus 
\begin{equation}
\rho \maj^{\frac{1}{2}} \ket{1}\bra{1} \ .
\end{equation}
\end{example} 

In the following, we show that the resource theory of smooth adiabatic operations satisfies the axioms of an error-tolerant resource theory. 

\begin{proposition}
\label{prop:NO_axioms}
The resource theory of smooth adiabatic operations with state space $\Gamma=\cS(\cH)$ and with composition of states defined as their tensor product obeys Axioms~\ref{axiom:reflexivity} to~\ref{axiom:consistency}.
\end{proposition}

\begin{proof}
We prove the proposition by showing that all four axioms hold. 
\begin{enumerate}[(1)]
\item Reflexivity: The identity operation is an adiabatic operation, thus $\rho \maj^\eps \rho$ for all $\eps \geq 0$.

\item Ordering of error-tolerances: let $\rho \maj^{\eps} \sigma$, i.e., let there exist $\rho' \sth \rho'\maj \sigma$ with $\rho'\in \epsball{\eps}{\rho}$. Now for any $\eps' \geq \eps$ also $\rho'\in \epsball{\eps'}{\rho}$, hence $\rho \maj^{\eps'} \sigma$.

\item Additive transitivity follows as 
$\rho\maj^\eps\sigma$ implies that there exists $\rho'\in\epsball{\eps}{\rho}$ such that $\rho'\maj\sigma$. Similarly, $\sigma\maj^\delta\tau$ implies that
$\exists \ \tau' \in \epsball{\epsilon}{\tau}$ with $\sigma \maj \tau'$. 
Since $\maj$ is transitive, this implies that there is also a state transformation $\rho' \maj \tau'$.
As $\rho' \in \epsball{\epsilon}{\rho}$ and $\tau' \in \epsball{\delta}{\tau}$,  also $\rho' \maj^{\epsilon+\delta} \tau'$.

\item The consistent composition axiom follows, since $\rho' \maj \sigma$ with $\rho'\in \epsball{\eps}{\rho}$ implies that also $\rho' \otimes \omega \maj \sigma \otimes \omega$ and $\rho'\otimes \omega \in \epsball{\eps}{\rho\otimes\omega}$. \footnote{This can for instance be seen as the trace distance is subadditive with respect to the tensor product.}

\end{enumerate}
\end{proof}

\section{Quantifying resources} \label{sec:smoothentropies}

In general, thermodynamic properties of a system are measured through an interaction with a measuring apparatus. The change that is observed in the state of the measuring device provides information about the state of the measured system. For instance, gas pressure can be measured with a manometer, where the change in height of the liquid specifies the pressure of the gas to be  measured compared to that of a reference gas. The measuring device is in that case a tube filled with a liquid whose state change (in terms of the relative height) specifies the pressure of different gases with respect to the reference. 

While in the above example it is implicitly assumed that the gas and its pressure remain essentially unaffected by the measuring process and are thus compared to a reference gas in a passive manner, similar assumptions may be inapplicable for small systems whose thermodynamic state is notably changed in the interaction with a similar apparatus. In microscopic quantum systems, for instance, it is even possible that quantum correlations with a measuring apparatus build up.
In the following, we hence consider interactions between systems and measuring devices that simultaneously change the state of the systems and of the devices. The state change of the measuring apparatus then provides information about the state change of the system.

Different properties of a system are explored with different devices, which we call \emph{meter systems}. The property we are interested in here is the value of the different states of a system as resources in an error-tolerant resource theory, which is encoded in the position of a state in the orderings with respect to the family of relations $\prec^\eps$. It can be specified with a meter system that is characterised by a single parameter~\cite{Lieb2014, Kraemer2016}. 
Later in this section, we shall introduce quantities that identify necessary conditions for state transformation and that allow for the derivation of sufficient conditions for each $\eps \geq 0$, in terms of such a meter.

We consider a meter system that can be in a family of states $\left\{ \chi_\lambda \right\}_{\lambda \in \Lambda} = \Gamma_{\lambda}$, which can be specified with a function
\begin{align*}
\chi:\ \Lambda&\to \Gamma_\lambda\\
\lambda&\mapsto\chi_\lambda
\end{align*}
with $\Lambda\subseteq\mathbb{R}_{\geq 0}$, where the parameter $\lambda$ labels the different states. The change in $\lambda$ produced during an interaction with a system shall specify the resource value of the system of interest.

The notion of a meter system was first introduced in Ref.~\cite{Lieb2014}, where an \emph{entropy meter} is a thermodynamic system in equilibrium, for which all states are characterised by a unique entropy function (to which our parameter $\lambda$ acts similarly). Instead of adopting the fully-fledged framework of systems in thermodynamic equilibrium~\cite{Lieb1998, Lieb1999, Lieb2001, Lieb2013, Lieb2014}, we specify a few properties of meter systems axiomatically. The main notion we aim to avoid is the requirement of a continuous parameter that labels the states of the meter system, which we judge undesirable when working with microscopic systems.

A meter system is supposed to \emph{measure} the value of different resources. It should therefore act passively in the sense that it should not enable otherwise impossible state transformations catalytically, i.e., without being consumed.

\begin{axiom}[Reduction] \label{axiom:reduction} For any two states of a system, $\rho$, $\sigma \in \Gamma$, and for any meter state $\chi_\lambda \in \Gamma_\lambda$ and any $\eps \geq 0$,
\begin{equation}
(\rho,\chi_\lambda) \prec^\eps (\sigma,\chi_{\lambda})
\implies \rho\prec^\eps \sigma \ .
\end{equation}
\end{axiom}

The following axiom specifies how to construct larger meter systems from smaller ones and how to label meter states on systems of different sizes consistently. More specifically, for two meter states $\chi_{\lambda_1}$, $\chi_{\lambda_2} \in \Gamma_\lambda$, the state $(\chi_{\lambda_1}, \chi_{\lambda_2}) \in \Gamma_\lambda \times \Gamma_\lambda$ is a state of a composed system. The next axiom ensures that it has the properties of a meter state on a larger system. 

\begin{axiom}[Additivity under Composition]
\label{axiom:additivity}
For any $\chi_{\lambda_1}$, $\chi_{\lambda_2} \in \Gamma_\lambda$,
\begin{equation}
\chi_\lambda \sim (\chi_{\lambda_1}, \chi_{\lambda_2}) \iff \lambda=\lambda_1+ \lambda_2 \ ,
\end{equation}
where $\chi_\lambda \sim (\chi_{\lambda_1}, \chi_{\lambda_2})$ means that $\chi_\lambda \prec (\chi_{\lambda_1}, \chi_{\lambda_2})$ as well as $(\chi_{\lambda_1}, \chi_{\lambda_2}) \prec  \chi_\lambda$.
\end{axiom}

We assume two more properties of the meter systems, that primarily concern the labelling of the meter's states.

\begin{axiom}[Order]\label{axiom:order}
A meter system has at least two inequivalent states and its states are labelled monotonically in $\lambda$, such that
for any $\chi_{\lambda_1}$, $\chi_{\lambda_2} \in \Gamma_\lambda$, 
\begin{equation}
\lambda_1 \leq \lambda_2 \iff \chi_{\lambda_1} \prec \chi_{\lambda_2} \ .
\end{equation}
\end{axiom}
Note that with this we assume that all states $\chi_\lambda \in \Gamma_\lambda$ can be compared with $\prec$, i.e., there are no $\chi_{\lambda_1}$, $\chi_{\lambda_2} \in \Gamma_\Lambda$ such that at the same time $\chi_{\lambda_1} \not\prec \chi_{\lambda_2}$ and $\chi_{\lambda_2} \not\prec \chi_{\lambda_1}$.

So far, we have not addressed processes with errors on the meter system, in particular, we have not specified which state transformations on $\Gamma_\lambda$ are enabled by the allowed errors. 
Axiom~\ref{axiom:smoothing_order} requires that a higher error-tolerance cannot inhibit any transformations but may enable more, which could be phrased as follows.  
\begin{axiom_modified}[Entropy scaling]
For any $\lambda_1 > \lambda_2 \in \Lambda$ and any $\eps \geq 0$,
\begin{equation}
\chi_{\lambda_1}\prec^\eps \chi_{\lambda_2} \implies \lambda_1 \leq \lambda_2 + f(\eps) \ .
\end{equation}
\end{axiom_modified}

The function $f(\eps)$ is non-decreasing in $\eps$ and $f(0)=0$ according to Axiom~\ref{axiom:order}.  
The exact form of $f$ is determined by the type of error that is considered. Whenever $\eps$ is understood as an error probability such that $0 \leq \eps \leq 1$, any state transformation should be possible in the extreme case of $\eps=1$.
Thus, in addition to $f(0)=0$, we require $\lim_{\eps \rightarrow 1}f(\eps)= \infty$ (as the allowed values for $\lambda$ are unbounded if we consider composed meter systems).
Furthermore, it should always be possible to run $n$ independent instances of the same process in parallel, in which case the success probabilities should multiply. In the case of a meter system we take it that there is also no alternative process with a lower error probability, i.e., one process is possible if and only if the $n$ parallel instances of the process are possible. 
We thus require that both of the following relations hold
\begin{align}\lambda_1 &\leq  \lambda_2 +f(1-(1-\eps)) \\ 
n \cdot \lambda_1 &\leq n \cdot \lambda_2 +f(1-(1-\eps)^n) \ .
\end{align}
This can be ensured with $f(1-(1-\eps)^n)= n \cdot f(1-(1-\eps))$, which is obeyed by $f(\eps)= - c \cdot \log_2(1-\eps)$. For our applications, which all quantify errors in terms of probabilities, we therefore replace Axiom~\ref{axiom:wpt}' by the following, where we fix $c=1$ for convenience.
\begin{axiom}[Work-error trade-off]\label{axiom:wpt} 
For any $\lambda_1 > \lambda_2 \in \Lambda$ and any $0 \leq \eps \leq 1$,
\begin{equation}
\chi_{\lambda_1}\prec^\eps \chi_{\lambda_2} \implies \lambda_1 \leq \lambda_2 -  \log_2(1-\eps) \ . 
\end{equation}
\end{axiom} 
Increasing the error tolerance and, hence, increasing $-\log_2(1-\eps)$, allows for a decrease in the difference $\lambda_2 -\lambda_1$. Since in thermodynamics the difference in $\lambda$ may be physically overcome by providing work, we call this axiom a work-error trade-off. This is illustrated with the following example.

\begin{example}\label{example:work}
Consider the resource theory of smooth adiabatic operations introduced in Section~\ref{sec:smoothingNO}. Let the meter system have four states, 
$\chi_0= \ket{00} \bra{00}$, $\chi_{\log_2(2)}= \frac{\ket{00} \bra{00}+\ket{01} \bra{01}}{2}$, $ \chi_{ \log_2(3)}= \frac{\ket{00} \bra{00}+\ket{01} \bra{01}+\ket{10} \bra{10}}{3}$ and $\chi_{\log_2(4)}= \frac{\mathbbm{1}_4}{4}$.
If we aim to erase the first qubit of the maximally mixed two qubit state, $\chi_{\log_2(4)}$, to obtain $\chi_{\log_2(2)}$, this transformation cannot be realised with an adiabatic operation, since $\chi_{\log_2(4)} \not\maj \chi_{\log_2(2)}$. According to Landauer's principle this operation has a work cost of $k_\mathrm{B} T \ln(2)$~\cite{Landauer1961, Bennett1982, LeffRex02, delRio2011, Faist2015} (which corresponds to $k_\mathrm{B} T \ln(2) \left( \log_2(4) -\log_2(2) \right)$). 

Accepting an error of $\eps=\frac{1}{3}$ (instead of $\eps=0$) can reduce this cost, since ${\chi_{\log_2(3)} \in \epsball{\frac{1}{3}}{\chi_{\log_2(2)}}}$. For the transformation from $\chi_{\log_2(4)}$ to $\chi_{\log_2(3)}$ a work expenditure of $k_\mathrm{B} T \ln(2) \left(\log_2(4)-\log_2(3)\right)$ is sufficient. 
If we increase the error tolerance to $\eps=\frac{1}{2}$, then $\chi_{\log_2(4)} \in \epsball{\frac{1}{2}}{\chi_{\log_2(2)}}$ and it holds that ${\chi_{\log_2(4)} \maj^{\frac{1}{2}} \chi_{\log_2(2)}}$, i.e., no work has to be invested.
Hence, the work expenditure is reduced by increasing the error-tolerance.\footnote{Note that Landauer's principle is resource-theoretically formulated in the realm of thermal operations, where a heat bath at temperature $T$ is considered and the work is quantified in terms of free energy, hence the factor $k_\mathrm{B} T$ above. We remark here, that when considering a trivial Hamiltonian on the system where the erasure takes place this is mathematically equivalent to our adiabatic operations. In the context of adiabatic processes, however, the terminology entropy-probability trade-off may be more apt.} 
\end{example}

\begin{definition}
A meter system with state space $\Gamma_\lambda$ is \emph{suitable}
for measuring a system with state space $\Gamma$, if it obeys Axioms~\ref{axiom:reflexivity} to~\ref{axiom:consistency} and the axioms for meter systems (Axioms~\ref{axiom:reduction} to~\ref{axiom:wpt}), and, if there exists a reference state $\sigma_1 \in \Gamma$ such that for any  state $\rho \in \Gamma$  there exist meter states  $\chi_{\lambda_1(\rho)}$, $\chi_{\lambda_2(\rho)}$, $\chi_{\lambda_3(\rho)}$ and $\chi_{\lambda_4(\rho)}$ such that $(\sigma_1, \chi_{\lambda_1(\rho)}) \prec (\rho, \chi_{\lambda_2(\rho)})$ and $(\rho, \chi_{\lambda_3(\rho)}) \prec (\sigma_1, \chi_{\lambda_4(\rho)})$ hold. 
\end{definition}

Relying on the notion of a suitable meter system, we define the following quantities. 

\begin{definition} \label{definition:smoothentropies}
For an error-tolerant resource theory with state space $\Gamma$ and a suitable meter system $\Gamma_\lambda$ we define for each $\eps \geq 0$ and for each $\rho \in \Gamma$,
\begin{alignat}{2}
S^\epsilon_-(\rho)&\defeq \sup &&\left\{\lambda_1-\lambda_2 \ \middle| \ (\sigma_1,\chi_{\lambda_1})\prec^\epsilon (\rho,\chi_{\lambda_2})\right\} \ , \\
S^\epsilon_+(\rho)&\defeq \inf &&\left\{\lambda_2-\lambda_1 \ \middle| \ (\rho,\chi_{\lambda_1})\prec^\epsilon (\sigma_1,\chi_{\lambda_2})\right\} \ ,
\end{alignat}
where $\sigma_1 \in \Gamma$ is a fixed reference state and $\chi_{\lambda_1}$, $\chi_{\lambda_2} \in \Gamma_\lambda$. 
\end{definition}

Due to the suitability of the entropy meter, $S^\eps_-$ and $S^\eps_+$ are defined for any $\rho \in \Gamma$.   
In terms of these quantities, we can derive necessary conditions and sufficient conditions for state transformations. Because $\prec^{\eps}$ is intransitive for $\eps>0$, these quantities are not monotonic with respect to $\prec^{\eps}$ (except for $\eps=0$).

\begin{proposition}[Adapted monotonicity of smooth entropic quantities]
\label{proposition:monotonicitylike}
Consider an error-tolerant resource theory with state space $\Gamma$ that obeys Axioms~\ref{axiom:reflexivity}--\ref{axiom:consistency} and a suitable meter system. Then for any $\rho$, $\sigma \in \Gamma$, if $\rho \prec^{\eps} \sigma$ for some $\eps \geq 0$, then for any $\delta \geq 0$, 
\begin{align}
    S_-^{\delta}(\rho) &\leq S_-^{\delta+\eps}(\sigma) \\
    S_+^{\delta+\eps}(\rho) &\leq S_+^{\delta}(\sigma) \ .
\end{align}
\end{proposition}
\begin{proof} 
Let $\rho$, $\sigma \in \Gamma$ such that $\rho \prec^{\eps} \sigma$ and let $\Gamma_\lambda$ be the state space of a suitable meter system. Then for $\chi_{\lambda_1}$, $\chi_{\lambda_2} \in \Gamma_\lambda$ that obey
\begin{equation}
(\sigma_1,\chi_{\lambda_1})\prec^\delta (\rho,\chi_{\lambda_2}) \ ,
\end{equation}
additive transitivity and the consistent composition axiom imply that also
\begin{equation}
(\sigma_1,\chi_{\lambda_1})\prec^{\delta+\eps} (\sigma,\chi_{\lambda_2})
\end{equation}
and therefore
$S_-^{\delta}(\rho) \leq S_-^{\delta+\eps}(\sigma)$. The statement for $S_+$ follows analogously and is omitted here.
\end{proof}

\begin{proposition}[Sufficient condition for state transformations]
\label{proposition:sufficient_condition_axiom_implication}
Consider an error-tolerant resource theory with state space $\Gamma$ that obeys Axioms~\ref{axiom:reflexivity}--\ref{axiom:consistency} and a suitable meter system. Then, for $\rho$, $\sigma \in \Gamma$, the inequality $S_+^\eps(\rho) < S_-^{\eps'}(\sigma)$ implies that $\rho \prec^{\eps+\eps'} \sigma $.
\end{proposition}

\begin{proof}
According to Definitions~\ref{definition:smoothentropies} the condition 
$S_+^\eps(\rho) < S_-^{\eps'}(\sigma)$ means that there exist $\chi_{\lambda_1}$, $\chi_{\lambda_2}$, $\chi_{\lambda_3}$ and $\chi_{\lambda_4}$ such that $ (\rho,\chi_{\lambda_1})
\prec^\eps (\sigma_1,\chi_{\lambda_2})$ and $ (\sigma_1,\chi_{\lambda_3})\prec^{\eps'} (\sigma,\chi_{\lambda_4})$ with $\lambda_2-\lambda_1 < \lambda_3-\lambda_4$, i.e., $\lambda_2+\lambda_4 < \lambda_1+\lambda_3$. 
By Axiom~\ref{axiom:consistency}, it holds that
\begin{equation}
 (\rho,\chi_{\lambda_1},\chi_{\lambda_3})
\prec^\eps (\sigma_1,\chi_{\lambda_2},\chi_{\lambda_3})
\prec^{\eps'} (\sigma,\chi_{\lambda_2},\chi_{\lambda_4}) \ . 
\end{equation}
Now due to additive transitivity and the additivity of meter states, it follows that
\begin{equation}
(\rho,\chi_{\lambda_1+\lambda_3}) \prec^{\eps+\eps'} (\sigma,\chi_{\lambda_2+\lambda_4}) 
\end{equation}
and therefore  
\begin{equation}\label{eq:laststep}
 \rho \prec^{\eps+\eps'} \sigma \ .
\end{equation}
This can be seen since, by the additivity of meter states, $\chi_{\lambda_1+\lambda_3}$ and $\chi_{\lambda_2+\lambda_4}$ are meter states on $\Gamma_\lambda \times \Gamma_\lambda$.
By Axiom~\ref{axiom:order}, the relation $\chi_{\lambda_2+\lambda_4} \prec \chi_{\lambda_1+\lambda_3}$ holds and, hence, due to Axiom~\ref{axiom:additivetransitivity} also $(\rho,\chi_{\lambda_1+\lambda_3}) \prec^{\eps+\eps'} (\sigma,\chi_{\lambda_1+\lambda_3})$. Equation~\eqref{eq:laststep} then follows by Axiom~\ref{axiom:reduction}.
\end{proof}

To consistently define $S_-^{\eps}$ and $S_+^{\eps}$ for systems of different sizes, we define reference states on composed systems such that for $\sigma_1 \in \Gamma$, $\sigma'_1 \in \Gamma'$, $( \sigma_1, \sigma'_1 )$ is the reference state on $\Gamma \times \Gamma'$. This captures the intuition that, when they are interpreted as part of a composed system, the scale on each individual system should not change.   
In the following, we show that $S_-^\eps$ and $S_+^\eps$ obey further natural properties such as super and sub-additivity. The proofs are all deferred to Section~\ref{sec:properties}.

\begin{proposition} 
\label{prop:additivityproperties}
Consider an error-tolerant resource theory with state space $\Gamma$ that obeys Axioms~\ref{axiom:reflexivity} to \ref{axiom:consistency} and a suitable meter system. Then,
 for $\eps, \delta \geq 0$ and for $\rho$, $\sigma \in \Gamma$, 
\begin{align}
    S_-^{\eps+\delta}((\rho,\sigma)) 
    &\geq 
    S_-^\eps(\rho) + S_-^\delta(\sigma) \ ,\\
    S_+^{\eps+\delta}((\rho,\sigma))
    &\leq 
    S_+^\eps(\rho) + S_+^\delta(\sigma) \ .
\end{align}
\end{proposition} 
In particular, the proposition implies that for $\eps,\delta=0$ usual super and sub-additivity are recovered,
\begin{align}
    S_-((\rho,\sigma)) 
    &\geq 
    S_-(\rho) + S_-(\sigma) \ , \\
    S_+((\rho,\sigma))
    &\leq 
    S_+(\rho) + S_+(\sigma) \ .
\end{align}

For meter states $\chi_\lambda \in \Gamma_\lambda$, one can furthermore relate the two entropic quantities $S_-^\eps (\chi_\lambda) $ and $S_+^\eps (\chi_\lambda) $ as follows. 

\begin{lemma} 
\label{lemma:equilibriumbound} 
Consider an error-tolerant resource theory with state space $\Gamma_\lambda$ that obeys Axioms~\ref{axiom:reflexivity} to~\ref{axiom:consistency} and Axioms~\ref{axiom:reduction} to~\ref{axiom:wpt} and let $\Gamma_\lambda$ also be the state space of the meter system under consideration. Then, for any $\chi_\lambda \in \Gamma_\lambda$, 
\begin{equation}
0 \leq S_-^\eps(\chi_\lambda) -S_+^\eps(\chi_\lambda) \leq -2 \log_2(1-\eps)\ .
\end{equation} 
\end{lemma}

Note that this lemma implies that meter states in the corresponding error-free resource theory obey $S_-(\chi_\lambda)=S_+(\chi_\lambda)$.
For meters for which the work-error trade-off (Axiom~\ref{axiom:wpt}) is achieved  with equality, that is, for which the scale on the meter system is sufficiently fine-grained, further properties may be derived (see Section~\ref{sec:fine-grained} for details). The meter system introduced in the next section is such a fine-grained example.

\subsection{Smooth entropies for the resource theory of smooth adiabatic operations}\label{sec:entropies_majorization}

We now introduce a meter system for the resource theory of smooth adiabatic operations, which, 
in accordance with~\cite{Gour2013,Weilenmann2015,Kraemer2016}, is chosen such that the meter states have flat step functions on the range from $0$ to $2^\lambda$, 
\begin{equation}
 f_{ \chi_\lambda}(x) = 
\begin{cases} 
2^{-\lambda} & x \leq 2^\lambda \ ,\\
0 & 2^\lambda < x \leq \dim (S) \ .
\end{cases}
\end{equation}
We let the parameter $\lambda$ take values $\lambda=\log_2(n)$ for $n=1, \ldots, \dim{(S)}$, where $\dim{(S)}>1$  and we assume that a meter system with arbitrarily large dimension, $\dim (S)$, may be chosen. In the definition of $S_-^{\eps}$ and $S_+^{\eps}$ we shall optimise over these meter systems of different sizes, i.e., effectively consider an infinite meter system $\Gamma_\Lambda$. 
The following lemma ensures that with respect to smooth majorisation $\Gamma_\Lambda$ has all properties required for a meter system. It is proven in Section~\ref{sec:meter_NO}. 
\begin{lemma}\label{lemma:meter_axioms}
The meter system $\Gamma_\Lambda$ obeys Axioms~\ref{axiom:reflexivity} to Axiom~\ref{axiom:consistency} as well as Axioms~\ref{axiom:reduction} to~\ref{axiom:wpt} with respect to the resource theory of smooth adiabatic operations and is suitable for measuring systems with a  state space $\Gamma=\cS(\cH)$. 
\end{lemma}

In the following we derive exact expressions for the smooth entropic quantities $S_-^\eps$ and $S_+^\eps$, which provide necessary and sufficient conditions for the existence of smooth adiabatic operations between different states according to Propositions~\ref{proposition:monotonicitylike} and~\ref{proposition:sufficient_condition_axiom_implication}.

\begin{proposition}\label{prop:smoothentr}
For the resource theory of smooth adiabatic operations acting on a state space $\Gamma=\cS(\cH)$ with a reference state $\sigma_1=\ket{0}\bra{0} \in \cS(\cH)$ and for the meter system $\Gamma_\Lambda$, 
\begin{align}
S^\epsilon_-(\rho)&=\Hmineps{\eps}{\rho} \ , \label{eq:sma1} \\
S^\epsilon_+(\rho)&=\HHeps{1-\epsilon}{\rho} + \log_2(1-\eps) \ . \label{eq:sma2}
\end{align}
\end{proposition}
Note that for $\eps=0$ we recover $\Hmin{\rho}$ and $\Hzero{\rho}$. The proposition is proven in Section~\ref{sec:smoothentr}. Hence, in the context of smooth adiabatic operations $\lambda$ is a measure of entropy.

We can quantify the decrease in $\lambda$, i.e., the entropy reduction, that a given quantum state $\rho \in \Gamma$ can catalytically enable on the meter system in a smooth adiabatic process. More specifically, we consider smooth adiabatic operations with error-tolerance $\eps$ that achieve the transition 
\begin{equation}
( \rho, \chi_{\lambda_1}) \maj^\eps ( \rho, \chi_{\lambda_2}) \ .
\end{equation}
We find that there are states and operations such that $\lambda_1-\lambda_2 > -\log_2{(1-\eps)}$. Hence, certain states $\rho$ enable transformations on the meter system that do not obey the work-error or in this case entropy-probability trade-off, as is illustrated with the following example.

\begin{example}
Consider the state
$ \rho = \frac{2}{3} \ket{0}\bra{0} + \frac{1}{3} \ket{1}\bra{1} $. Then for $\eps=\frac{1}{3}$, it holds that
\begin{equation} 
\rho \otimes \frac{\id_2}{2} \maj^\eps \rho \otimes \ket{0}\bra{0}  \ ,
\end{equation}
since the eigenvalues of
$\rho \otimes \frac{\id_2}{2}$ are
$\left(\frac{1}{3},\frac{1}{3},\frac{1}{6},\frac{1}{6}\right)$ and the eigenvalues of 
$\rho\otimes\ket{0}\bra{0} $
are
$\left(\frac{2}{3},\frac{1}{3},0,0\right)$,
 and thus, for all $x\geq 0$,
\begin{equation}
 \int_0^x f_{\rho \otimes \frac{\id_2}{2}}(x')dx'\geq\int_0^x f_{\rho \otimes\ket{0}\bra{0}}(x')dx'-\frac{1}{3} \ .
 \end{equation}
At the same time the work-error trade-off is not obeyed by $\chi_{\lambda_1}=\frac{\id_2}{2}$ and $\chi_{\lambda_2}=\ket{0}\bra{0}$, since $\lambda_1 -\lambda_2= 1 > \log_2 \left(\frac{3}{2} \right)$.
\end{example}

In principle, by suitably engineering $\rho$ (allowing it to have arbitrary rank) we can have an arbitrary small failure probability $\eps$ while reducing $\lambda$ by an arbitrarily large amount.  
 This may sound counterintuitive, we can see it as an example of an embezzling-type effect~\cite{VanDam2003, Ng2014, Brandao2015b}.

\begin{lemma} \label{lemma:embezzling} 
For any meter state $\chi_\lambda \in \Gamma_\Lambda$ with $\lambda = \log_2(m)$ for some $m \in \mathbb{N}$ and for any $\eps > 0$ we can find a quantum state $\rho$ such that
\begin{equation}
\rho \otimes \chi_{\lambda} \maj^\eps \rho \otimes \ket{0}\bra{0} \ . 
\end{equation}
\end{lemma}
This is proven in Section~\ref{sec:embezzling}.

\section{Alternative types of errors in adiabatic operations} \label{sec:alvaro_smoothing}
There are types of errors that cannot be quantified in terms of the trace distance. For instance, in Ref.~\cite{Alhambra2015} probabilistic transformations, $\cptp$, from a state $\rho \in \cS(\cH)$ to a state $\sigma \in \cS(\cH)$ with error probability $\eps$,
\begin{equation}
\rho \rightarrow \cptp(\rho)= (1- \eps) \sigma + \eps \xi
\end{equation}
were considered, where $\xi$ is an arbitrary state. This expresses that the process, which aims to transform $\rho$ to $\sigma$ succeeds with probability $1-\eps$, whereas with probability $\eps$ any output can be produced. For adiabatic operations such \emph{adiabatic probabilistic transformations} define an order relation
\begin{equation}
\rho \pmaj^{\eps} \sigma  \iff \exists \ \xi \in \cS(\cH) \ \sth \rho \maj (1- \eps) \sigma + \eps \xi \ .
\label{eq:alvaro_smoothing}
\end{equation} 
Composition of states is defined as their tensor product. Sequential composition of two processes, 
\begin{align}
\rho &\rightarrow \cptp_1(\rho)=(1-\eps)\sigma + \eps \xi_1 \ , \\
\sigma &\rightarrow \cptp_2(\sigma)=(1-\delta)\omega + \delta \xi_2 \ , 
\end{align}
yields, by linearity,
\begin{align}
\cptp_2 \circ \cptp_1(\rho) &=\cptp_2((1-\eps)\sigma+ \eps \xi_1) \\
&=(1-\eps)\cptp_2(\sigma)+ \eps \cptp_2(\xi_1) \\
&=(1-\eps)(1-\delta)\omega+ (1-\eps)\delta \xi_2 + \eps \cptp_2(\xi_1) \ .
\end{align} 
\begin{lemma}
Adiabatic probabilistic transformations on a state space $\Gamma=\cS(\cH)$, where composition of states is defined as their tensor product, obey Axioms~\ref{axiom:reflexivity} to~\ref{axiom:consistency}. Furthermore, the meter system $\Gamma_\Lambda$, defined in Section~\ref{sec:entropies_majorization}, is a suitable meter system.
\end{lemma}

\begin{proof}
In the following we show that the probabilistic adiabatic operations obey all axioms. 
\begin{enumerate}[(1)]
\item Reflexivity is obeyed, which can be seen by choosing $\cptp$ to be the identity map on $\cS(\cH)$.

\item The ordering of error-tolerances is obeyed, as for $\rho \maj^\eps \sigma$ and $\eps' \geq \eps$ we can rewrite
\begin{align}
\cptp(\rho)&=(1-\eps)\sigma+\eps \xi_1 \\
&=(1-\eps')\sigma+ \eps' \left((1-\frac{\eps}{\eps'}) \sigma + \frac{\eps}{\eps'}\xi_1 \right) \ , 
 \end{align}
where $\xi_2=(1-\frac{\eps}{\eps'}) \sigma + \frac{\eps}{\eps'}\xi_1 \in \cS(\cH)$, as it is  a convex combination of two states.

\item Additive transitivity is obeyed, as 
$\rho \pmaj^{\eps} \sigma$ and $\sigma \pmaj^{\delta} \omega$ imply that there are $\xi_1$ and $\xi_2$ and adiabatic operations $\cptp_1$ and $\cptp_2$ such that
\begin{align}
\cptp_1(\rho) &=(1-\eps) \sigma + \eps \xi_1 \ , \\
\cptp_2(\sigma) &=(1-\delta) \omega + \delta \xi_2 \ .
\end{align}
Sequentially composing the processes yields
\begin{align}
\cptp_2 \circ \cptp_1(\rho) &= (1-\eps)(1-\delta)\omega+ (1-\eps)\delta \xi_2 + \eps \cptp_2(\xi_1) \ , \\
&= (1-(\eps+\delta))\omega+ (\eps + \delta) \left( \frac{\eps\delta}{\eps+\delta} \omega +\frac{(1-\eps)\delta}{\eps+\delta}  \xi_2 + \frac{\eps}{\eps+\delta}  \cptp_2(\xi_1) \right) \ ,
\end{align}
where $\frac{\eps\delta}{\eps+\delta} \omega +\frac{(1-\eps)\delta}{\eps+\delta}  \xi_2 + \frac{\eps}{\eps+\delta}  \cptp_2(\xi_1)$ is a state as it is a convex combination of states. Therefore, it holds that 
$\rho \pmaj^{\eps+ \delta} \omega$.

\item The consistent composition axiom is obeyed, as 
$\rho \prec^{\eps} \sigma$ means that there is a $\xi_1$ and an adiabatic operation, $\cptp_1$, such that
\begin{equation}
\cptp_1(\rho) = (1-\eps) \sigma+ \eps \xi_1 \ ,
\end{equation}
hence, $\rho \maj  (1-\eps) \sigma+ \eps \xi_1$.
Then, it also holds that
\begin{equation}
\rho \otimes  \omega \maj (1-\eps) \sigma \otimes \omega+ \eps \xi_1 \otimes \omega \ ,
\end{equation}
and thus $\rho \otimes  \omega \pmaj^{\eps} \sigma \otimes \omega$.

\item The reduction axiom for meter states follows, as whenever there exists a $\xi$ such that 
\begin{equation}
\rho \otimes  \chi_\lambda \maj (1-\eps) \sigma \otimes \chi_\lambda+ \eps \xi \ , 
\end{equation}
then there also exists a state $\tilde{\xi}$ that is constant on equal eigenvalues of $\sigma \otimes \chi_\lambda$ and such that 
\begin{equation}
\rho \otimes \chi_\lambda \maj (1-\eps) \sigma \otimes \chi_\lambda+ \eps \tilde{\xi} \ . 
\end{equation}
$\tilde{\xi}$ can hence be written as $\tilde{\xi}=\xi' \otimes \chi_\lambda$ for some state $\xi'$. 
Hence, it is equivalent to $\rho \maj (1-\eps) \sigma + \eps \xi'$ and there exists an adiabatic operation achieving this transformation.

\item Additivity under composition, the ordering of meter states and the suitability of the meter system do not rely on the definition of errors. Their validity has been confirmed for adiabatic operations and  the meter system in question in Section~\ref{sec:entropies_majorization}.

\item The work-error trade-off is obeyed. Requiring that there is an adiabatic operation $\chi_{\lambda_1} \rightarrow {(1-\eps)} \chi_{\lambda_2} + \eps \xi$ is equivalent to the requirement that $(1-\eps) 2^{-\lambda_2} \leq 2^{-\lambda_1}$.\footnote{ Note that $\xi$ can always be chosen appropriately. If $\lambda_1 \leq \lambda_2$, then $\xi=\chi_{\lambda_2}$ is a possible choice, if $\lambda_1 > \lambda_2$, then $\xi$ can be chosen as $\xi=\chi_{\lambda_1}-(1-\eps)\chi_{\lambda_2}$ for instance, which is a valid state.} Hence, $\lambda_1 \leq \lambda_2 - {\log_2(1-\eps)}$.
\end{enumerate}
\end{proof}

We remark here that the above proof also shows that 
$\rho \pmaj^\eps \sigma$ and $\sigma \pmaj^\delta \omega$ imply $\rho \pmaj^{\eps+\delta-\eps \delta} \omega$, which for $\eps, \delta \neq 0$ is a strictly smaller error than the axiomatically required $\eps+\delta$.
The entropic quantities $S_-^\eps$ and $S_+^\eps$ that give us necessary conditions and sufficient conditions for state transformations with adiabatic probabilistic transformations are derived in the following proposition.

\begin{proposition} For adiabatic probabilistic transformations on a state space $\Gamma=\cS(\cH)$ with reference state $\sigma_1=\ket{0}\bra{0} \in \cS(\cH)$ and the meter system $\Gamma_\Lambda$, 
\begin{align}
S_-^\eps(\rho)&=\Hmin{\rho}+ \log_2(1-\eps) \ , \label{eq:sme1}\\
S_+^\eps(\rho)&=\Hzero{\rho} \ , \label{eq:sme2}
\end{align}
hold for any $\rho \in \cS(\cH)$.
\end{proposition}

\begin{proof}
In order to prove Equation~\eqref{eq:sme1} for any $\rho \in \cS(\cH)$, we consider the majorisation condition 
$\sigma_0 \otimes \chi_{\lambda_1} \pmaj^\eps \rho \otimes \chi_{\lambda_1}$, 
which is satisfied if and only if
\begin{equation}
2^{-\lambda_1} \geq (1-\eps) 2^{-\lambda_2} \maxEV{(\rho)} \ ,
\end{equation}
where $\maxEV{(\rho)}$ is the maximal eigenvalue of $\rho$. This inequality can be rewritten as 
\begin{equation}
\lambda_1-\lambda_2 \leq \log_2 \left(\frac{1}{\maxEV{(\rho)}} \right)-\log_2(1-\eps) \ ,
\end{equation}
i.e., $\lambda_1-\lambda_2 \leq \Hmin{\rho}-\log_2(1-\eps)$. Since $\lambda_1=\log_2(m)$ and $\lambda_2=\log_2(n)$ for some $m,n \in \mathbb{N}$, we can rewrite $\lambda_1-\lambda_2=\log_2(\frac{m}{n})$. As the meter system has states for any natural $m$ and $n$, the supremum over $\lambda_1-\lambda_2$ always achieves the value $\Hmin{\rho}-\log_2(1-\eps)$, which thus equals $S_-^\eps(\rho)$.

To prove Equation~\eqref{eq:sme2}, let us consider the majorisation condition 
$\rho \otimes \chi_{\lambda_1} \pmaj^\eps \sigma_1 \otimes \chi_{\lambda_2}$, 
which is satisfied if and only if $\rank(\rho)\cdot \lambda_1 \geq \lambda_2$. Hence, $\lambda_2-\lambda_1 \geq \rank(\rho)$, which implies that $S_+^\eps(\rho)=\Hzero{\rho}$.
\end{proof}

We remark here that the existence of an adiabatic probabilistic transformation from a state $\rho$ to a state $\sigma$ implies that there is also a smooth adiabatic transformation from $\rho$ to $\sigma$. We can see this by considering the output state of \eqref{eq:alvaro_smoothing}, which obeys
\begin{equation}
\trdist{\sigma}{(1-\eps)\sigma+\eps \xi}= \eps \trdist{\sigma}{ \xi} 
\leq \eps \ ,
\end{equation}
since the trace distance of two states $\sigma$ and $\xi$ is bounded by $1$. The converse statement is not true. This can be realised by considering the states $\rho=\frac{1}{2}\ket{0} \bra{0} + \frac{1}{2}\ket{1} \bra{1}$ and $\sigma=\frac{3}{4} \ket{0}\bra{0}+ \frac{1}{4} \ket{1}\bra{1}$.  For an error tolerance of $\eps=\frac{1}{4}$, there is a smooth adiabatic operation from $\rho$ to $\sigma$, since $\rho \maj^{\frac{1}{4}} \sigma$, whereas with probabilistic adiabatic processes the transformation is not possible, as $\rho \not\pmaj^{\frac{1}{4}} \sigma$. Instead, an error tolerance of (at least) $\frac{1}{3}$ would be needed in the latter case.

\section{Appendix}
\subsection{Proof of Proposition~\ref{prop:smoothing_equivalence}} \label{sec:smoothing_equivalence}
We shall prove this proposition in four steps, showing that (1) $\implies$ (2) $\implies$ (3) $\implies$ (4) $\implies$ (1).
\begin{enumerate}
\item[(1) $\implies$ (2)] Note that $\rho'\maj\sigma$ for $\rho'\in\epsball{\eps}{\rho}$  implies that there exists a unital map $\mathcal{E}$ such that $\mathcal{E}(\rho')=\sigma$~\cite{Gour2013}. Since the trace distance is monotonically decreasing under general CPTP maps~\cite{BookNielsenChuang2000}, we automatically obtain $\sigma'=\mathcal{E}(\rho)\in\epsball{\eps}{\sigma}$ and thus  $\rho \maj \sigma'\in\epsball{\eps}{\sigma}$.
    
\item[(2) $\implies$ (3)] Since in (3) we can set $\epsilon'=0$ and $\epsilon''=\epsilon$ this immediately follows from (2).
    
\item[(3) $\implies$ (4)] 
For  $\sigma\in\epsball{\epsilon''}{\sigma'}$ and $\rho\in\epsball{\eps'}{\rho'}$,
\begin{align} 
\int_0^x \vert f_\sigma(x')-f_{\sigma'}(x')\vert_+ \ dx' &\leq{\eps''} \ , \label{eq:fun1} \\
\int_0^x \vert f_{\rho'}(x')-f_{\rho}(x')\vert_+ \ dx'  &\leq{\eps'} \label{eq:fun2} 
\end{align}
 for all $x\geq 0$, where $\vert \cdot \vert_+$ denotes the positive part, i.e., for $x \in \mathbb{R}$, $\vert x \vert_+ = \max \left\{x,0 \right\}$. 
Since $\rho'\maj\sigma'$ means that
    \begin{equation}  
    \int_0^x f_{\sigma'}(x') \ dx' 
    \leq \int_0^x f_{\rho'}(x') \ dx'
    \end{equation}
    for all $x$ we have
    \begin{align}
    \int_0^x (f_\sigma(x')-f_{\rho'}(x')) \ dx'
    &\leq\int_0^x (f_\sigma(x')-f_{\sigma'}(x')) \ dx' \\
    &\leq \int_0^x \vert f_\sigma(x')-f_{\sigma'}(x')\vert_+ \ dx'\\
    &\leq{\epsilon''} \ ,
    \end{align}
where the last inequality follows  by \eqref{eq:fun1}. 
According to \eqref{eq:fun2},
    \begin{align}
    \int_0^x (f_\sigma(x')-f_{\rho}(x')) \ dx'
    &= \int_0^x (f_\sigma(x')-f_{\rho'}(x'))+(f_{\rho'}(x')-f_\rho(x')) \ dx'\\
    &\leq{\epsilon''} + \int_0^x \vert f_{\rho'}(x')-f_{\rho}(x')\vert_+ \ dx'\\
    &\leq {\epsilon'+\epsilon''}\\
    &\leq{\epsilon} \ .
    \end{align} 
\item[(4) $\implies$ (1)] Finally, let $\rho$, $\sigma \in \Gamma$, such that
\begin{equation} \label{eq:start}
 \int_0^x f_\rho(x') \ dx' \geq \int_0^x f_\sigma(x') \ dx'-{\epsilon}
\end{equation} holds for all $x\geq 0$. 
Let $\left\{\ithEV{\rho} \right\}_{i=1}^{d}$ be the decreasingly ordered eigenvalues of $\rho$. Now, choose $\rho'\in\epsball{\eps}{\rho}$ such that 
\begin{equation} 
\ithEVk{\rho'}{1} = \min \left\{ 1, \ \ithEVk{\rho}{1} + \eps \right\} \ ,
\end{equation}
and, to preserve normalisation, 
\begin{equation}
\ithEV{\rho'} = \ithEV{\rho} - \eps_i
\end{equation}
for $1 \leq i \leq d$, where
\begin{equation} 
\eps_i = \begin{cases} 
    \min \left\{ \ithEVk{\rho}{d}, \ \tilde{\eps} \right\} &\text{for } i = d , \\
    \min \left\{ \ithEV{\rho} , \ \tilde{\eps} - \sum_{j=i+1}^d \eps_j \right\} &\text{otherwise},
\end{cases} 
\end{equation}
with   $\tilde{\eps}=\ithEVk{\rho'}{1}-\ithEVk{\rho}{1}$. Now if $\ithEVk{\rho'}{1}=1$, $\rho' \prec \sigma$ for any $\sigma \in \Gamma$. If, on the other hand, $\ithEVk{\rho'}{1}\neq1$, \eqref{eq:start} implies that $ f_\rho(x)\geq f_\sigma(x)-{\eps}$ for $0\leq x< 1$, and therefore $f_{\rho'}(x) \geq f_\sigma(x)$ for $0\leq x < 1$ . Moreover, for $x \geq 1$ we have
\begin{equation}
\int_0^x f_{\rho'}(x') \ dx'  \geq \int_0^x f_\rho(x') \ dx'  + \eps  
\end{equation}
and therefore
\begin{equation}
\int_0^x f_{\rho'}(x') \ dx'  \geq \int_0^x f_\sigma(x') \ dx'    
\end{equation}
for all $x \geq 0$.
Hence $\rho'\prec\sigma$ for $\rho'\in\epsball{\eps}{\rho}$, which concludes the proof. 
\end{enumerate}
\qed

\subsection{Proof of Proposition~\ref{prop:additivityproperties}} \label{sec:properties}

Let $\left( \lambda_1^\mu-\lambda_2^\mu \right)_\mu$ and $\left( \tilde{\lambda}_1^{\nu} -\tilde{\lambda}_2^{\nu} \right)_\nu$ be sequences that converge to $S_-^\eps(\rho)$ and $S_-^\delta(\sigma)$ respectively, such that for all $\mu$, $\nu$,
\begin{align}
    (\sigma_1,\chi_{\lambda_1^\mu})
    &\prec^\epsilon (\rho,\chi_{\lambda_2^\mu}) \ , \\
    (\sigma_1,\chi_{\tilde{\lambda}_1^{\nu}})
    &\prec^\delta (\sigma,\chi_{\tilde{\lambda}_2^{\nu}}) \ .
\end{align}
Due to the consistent composition axiom 
\begin{equation} \left((\sigma_1,\chi_{\lambda_1^\mu}),(\sigma_1,\chi_{\tilde{\lambda}_1^{\nu}}) \right) \prec^{\eps} \left((\rho,\chi_{\lambda_2^\mu}),(\sigma_1,\chi_{\tilde{\lambda}_1^{\nu}}) \right)  \prec^{\delta} \left((\rho,\chi_{\lambda_2^\mu}),(\sigma,\chi_{\tilde{\lambda}_2^{\nu}}) \right) \ , 
\end{equation}
and by additive transitivity,
\begin{equation} 
\left((\sigma_1,\chi_{\lambda_1^\mu}),(\sigma_1,\chi_{\tilde{\lambda}_1^{\nu}}) \right) \prec^{\eps+\delta} \left((\rho,\chi_{\lambda_2^\mu}),(\sigma,\chi_{\tilde{\lambda}_2^{\nu}}) \right) \ . 
\end{equation}
Due to the associativity and commutativity of the composition operation, it follows that 
\begin{equation} \left((\sigma_1,\sigma_1),(\chi_{\lambda_1^\mu},\chi_{\tilde{\lambda}_1^{\nu}}) \right) \prec^{\eps+\delta} \left((\rho,\sigma),(\chi_{\lambda_2^\mu},\chi_{\tilde{\lambda}_2^{\nu}}) \right)  
\end{equation}
and due to the additivity of the meter states 
\begin{equation} 
\left((\sigma_1,\sigma_1),\chi_{\lambda_1^\mu+\tilde{\lambda}_1^{\nu}} \right) \prec^{\eps+\delta} \left((\rho,\sigma),\chi_{\lambda_2^\mu+\tilde{\lambda}_2^{\nu}} \right) \ . 
\end{equation}
Since $(\sigma_1,\sigma_1)$ is the reference state on the larger system,
\begin{equation}
 S_-^{\eps+\delta}((\rho,\sigma)) 
\geq \lambda_1^\mu +\tilde{\lambda}_1^{\nu}-\lambda_2^\mu -\tilde{\lambda}_2^{\nu}
\end{equation}
and it follows that this also holds in the limit, i.e.,
$ S_-^{\eps+\delta}((\rho,\sigma)) 
    \geq S_-^\eps(\rho) + S_-^\delta(\sigma)$. 

\bigskip

Similarly, for any $\left( \lambda_2^\mu-\lambda_1^\mu \right)_\mu$ and $\left( \tilde{\lambda}_2^{\nu} - \tilde{\lambda}_1^{\nu} \right)_\nu$ converging to $S_+^\eps(\rho)$ and $S_+^\delta(\sigma)$ respectively, such that for all $\mu$, $\nu$,
\begin{align}
    (\rho,\chi_{\lambda_2^\mu}) 
    &\prec^\epsilon
    (\sigma_1,\chi_{\lambda_1^\mu}), \\
    (\sigma,\chi_{\tilde{\lambda}_2^{\nu}})
    &\prec^\delta
    (\sigma_1,\chi_{\tilde{\lambda}_1^{\nu}}) \ ,
\end{align}
consistent composition and additive transitivity imply that
\begin{equation}
\left((\rho,\chi_{\lambda_2^\mu}),(\sigma,\chi_{\tilde{\lambda}_2^{\nu}}) \right) 
\prec^{\eps+\delta}
\left((\sigma_1,\chi_{\lambda_1^\mu}),(\sigma_1,\chi_{\tilde{\lambda}_1^{\nu}}) \right) 
\end{equation}
and so from the associativity and commutativity of the composition operation and from the definition of $S_+^{\eps+\delta}((\rho,\sigma))$ it follows that  
$ S_+^{\eps+\delta}((\rho,\sigma)) \leq S_+^\eps(\rho) + S_+^\delta(\sigma)$.

\subsection{Proof of Lemma~\ref{lemma:equilibriumbound}} 
First, to prove positivity, let the reference state on the meter system be an arbitrary meter state $\sigma_1= \chi_{\lambda'}$. Now, 
\begin{equation} 
(\sigma_1,\chi_{\lambda_1})\prec (\chi_\lambda,\chi_{\lambda_2})
\end{equation}
implies by the order of meter states and their additivity that $\lambda_1 -\lambda_2 \leq \lambda- \lambda'$.
As for any meter state $\chi_{\lambda} \in \Gamma_\lambda$ one can take $\chi_{\lambda_1}=\chi_{\lambda}$, $\chi_{\lambda_2}=\sigma_1$ to satisfy $(\sigma_1,\chi_{\lambda_1}) \sim (\chi_{\lambda},\chi_{\lambda_2})$ and hence $\lambda_1 -\lambda_2=\lambda- \lambda'$, i.e.,
\begin{align}
  S_-(\chi_\lambda) &= \sup\{\lambda_1-\lambda_2:\ (\sigma_1,\chi_{\lambda_1})\prec (\chi_\lambda,\chi_{\lambda_2})\} \\
  &=\lambda- \lambda'.
\end{align} 
Analogously, it follows that with $\chi_{\lambda_1}=\sigma_1$ and $\chi_{\lambda_2}= \chi_\lambda$ the following infimum is attained, i.e., 
\begin{align}
    S_+(\chi_\lambda) &= \inf\{\lambda_2-\lambda_1:\ (\chi_\lambda,\chi_{\lambda_1})\prec (\sigma_1,\chi_{\lambda_2})\} \\
     &=\lambda- \lambda'.
\end{align}
Hence, meter states obey $S_-(\chi_\lambda)=S_+(\chi_\lambda)$ and Axiom~\ref{axiom:smoothing_order} implies that $S_-^\eps (\chi_\lambda) \geq  S_+^\eps (\chi_\lambda)$ for $\eps \geq 0$.

To prove the upper bound, note that for any $\lambda_1$, $\lambda_2$ such that 
\begin{equation}
( \chi_\lambda,~\chi_{\lambda_1}) \prec^\eps ( \sigma_1,~\chi_{\lambda_2} ) \ ,
\end{equation}
 the work-error trade-off and the additivity of meter states imply that 
$ \lambda+ \lambda_1 \leq \lambda' + \lambda_2 - \log_2(1-\eps)$. 
 Hence, it directly follows that  
$ S_+^{\eps} (\chi_\lambda) \geq \lambda -\lambda' + \log_2(1-\eps)$. 
Similarly,
\begin{equation}
(  \sigma_1,~\chi_{\lambda_3}) 
  \prec^\eps ( 
  \chi_\lambda,~\chi_{\lambda_4} ) 
\end{equation}
  implies that 
 $\lambda'+\lambda_3 \leq \lambda+ \lambda_4 - \log_2(1-\eps)$  and so
$ S_-^\eps(\chi_\lambda) \leq \lambda - \lambda'- \log_2(1-\eps)$. 
 Thus, 
$S_-^\eps(\chi_\lambda) - S_+^\eps(\chi_\lambda) \leq -2 \log_2(1-\eps)$ directly follows.

\subsection{Properties of smooth entropies measured with meter systems that achieve the work-error trade-off} \label{sec:fine-grained}

\begin{definition}[Fine-grained meter system]
$\Gamma_\lambda$ is the state space of a fine-grained meter system  with respect to an error tolerant resource theory and error $\eps$ if there exist $\chi_{\lambda_1}$, $\chi_{\lambda_2} \in \Gamma_\lambda$ such that
\begin{equation}
\sup \left\{ \lambda_1 - \lambda_2 \ \middle| \ \chi_{\lambda_1} \prec^\eps \chi_{\lambda_2} \right\} = -\log_2(1-\eps)\ , 
\end{equation}
and $\Gamma_\lambda$ obeys the axioms of a meter system (Axioms~\ref{axiom:reflexivity} to~\ref{axiom:consistency} and Axioms~\ref{axiom:reduction} to~\ref{axiom:wpt}).
\end{definition}
In particular, the infinite dimensional meter system introduced in  Section~\ref{sec:entropies_majorization} has this property for any $0\leq \eps \leq 1$.
For a fine-grained meter system we obtain a direct relation between the entropic quantities $S_+^\eps$ and $S_-^\eps$ with different smoothing parameters.\footnote{The following lemma tightens the relations $S_-^{\eps+\delta}(\rho) 
\geq S_-^\eps (\rho)$ and $S_+^{\eps+\delta}(\rho) \leq S_+^\eps (\rho)$ that follow from Proposition~\ref{proposition:monotonicitylike}.}

\begin{lemma}\label{lemma:lemmasmoothingparameter}
Consider an error-tolerant resource theory with state space $\Gamma$ that obeys Axioms~\ref{axiom:reflexivity}--\ref{axiom:consistency} and a suitable meter system and a fine-grained meter system with respect to error $\delta$ with state space $\Gamma_\lambda$. Then for all states $\rho \in \Gamma$,
\begin{align}
S_-^{\eps+\delta}(\rho) 
&\geq S_-^\eps (\rho) - \log_2(1-\delta) \ , \\
S_+^{\eps+\delta}(\rho) 
&\leq S_+^\eps (\rho) + \log_2(1-\delta)  \ .
\end{align}
\end{lemma}

\begin{proof}
Let $\left(\lambda_1^\mu -\lambda_2^\mu \right)_\mu$ be a sequence of parameters of meter states $\chi_{\lambda_1^\mu}$, $\chi_{\lambda_2^\mu}$ with $ (\sigma_1,\chi_{\lambda_1^\mu} )\prec^\epsilon (\rho,\chi_{\lambda_2^\mu }) $ that converges to $S^\epsilon_-(\rho)$ 
for $\mu \rightarrow \infty$. For a fine-grained meter system with respect to $\prec^\delta$ there is furthermore a sequence of $\left(\lambda_3^\nu -\lambda_4^\nu \right)_\nu$ with $\chi_{\lambda_3^\nu} \prec^\delta \chi_{\lambda_4^\nu}$ that converges to $-\log_2(1-\delta)$ for $\nu \rightarrow \infty$. Now for any such $\lambda_1^\mu$, $\lambda_2^\mu$, $\lambda_3^\nu$, $\lambda_4^\nu$, by the consistent composition axiom (Axiom~\ref{axiom:consistency}) and by additivity of the meter states (Axiom~\ref{axiom:additivity}),
\begin{equation}
    (\sigma_1,\chi_{\lambda_1^\mu+ \lambda_3^\nu})\prec^\epsilon (\rho,\chi_{\lambda_2^\mu+ \lambda_3^\nu}) \prec^\delta (\rho,\chi_{\lambda_2^\mu+ \lambda_4^\nu}). 
\end{equation}
Hence for all $\mu$ and $\nu$,
\begin{equation}
S^{\epsilon+ \delta}_-(\rho) \geq \lambda_1^\mu-\lambda_2^\mu+ \lambda_3^\nu - \lambda_4^\nu \ ,
\end{equation}
and in the limit $\mu \rightarrow \infty$, $\nu \rightarrow \infty$, 
$S_-^{\eps+\delta}(\rho) 
\geq S_-^\eps (\rho) - \log_2(1-\delta)$. 
The proof for $S_+^\eps$ proceeds analogously:
Let $\left(\lambda_2^{\mu}-\lambda_1^{\mu} \right)_\mu$ be a sequence of states obeying
$(\rho,\chi_{\lambda_1^\mu})\prec^\epsilon (\sigma_1,\chi_{\lambda_2}^\mu)$ that converges to $S^\epsilon_+(\rho)$ 
and let $\left(\lambda_3^\nu -\lambda_4^\nu \right)_\nu$ be a sequence with $\chi_{\lambda_3^\nu} \prec^\delta \chi_{\lambda_4^\nu}$ that converges to $-\log_2(1-\delta)$. Now it follows that 
\begin{align}
    (\rho,\chi_{\lambda_1^\mu+\lambda_3^\nu})\prec^\epsilon (\sigma_1,\chi_{\lambda_2^\mu+\lambda_3^\nu}) \prec^\delta
    (\sigma_1,\chi_{\lambda_2^\mu+\lambda_4^\nu}) \ .
\end{align}
Then, for all $\mu$ and $\nu$,
\begin{equation}
S_+^{\eps+\delta}(\rho) \leq \lambda_2^\mu-\lambda_1^\mu +\lambda_4^\nu-\lambda_3^\nu
\end{equation}
and in the limit $\mu \rightarrow \infty$ and $\nu \rightarrow \infty$, 
$S_+^{\eps+\delta}(\rho) 
\leq S_+^\eps (\rho) + \log_2(1-\delta)$. 
\end{proof}

There is furthermore a duality relation for meter states in the sense that $S_-^\eps$ and $S_+^{1-\eps}$ only differ by logarithmic terms in $\eps$ for such states.
\begin{corollary}[Duality relation for meter states]
Consider an error-tolerant resource theory whose state space is that of a meter system, $\Gamma_\lambda$, and that is fine-grained with respect to $\eps$ and $1-\eps$. 
\begin{equation}
S_-^{\eps}(\chi_\lambda) 
= S_+^{1-\eps}(\chi_\lambda)+ \log_2{\left(\frac{1}{\eps (1-\eps)}\right)}\ .
\end{equation}
\end{corollary}

\begin{proof}
For a meter state $\chi_\lambda \in \Gamma_\lambda$,
$S_+^{\eps}(\chi_\lambda)=S_+(\chi_\lambda)+\log_2(1-\eps)$ and $S_+^{(1-\eps)}(\chi_\lambda)=S_+(\chi_\lambda)+\log_2(\eps)$ as well as $S_-^{\eps}(\chi_\lambda)=S_-(\chi_\lambda)-\log_2(1-\eps)$. This implies that $S_+^{\eps}(\chi_\lambda)=S_+^{1-\eps}(\chi_\lambda)-\log_2(\eps)+\log_2(1-\eps)$ and, using Lemma~\ref{lemma:equilibriumbound} (for zero-error tolerance),
\begin{align}
S_-^{\eps}(\chi_\lambda)
&= S_+^{\eps}(\chi_\lambda)-2\log_2(1-\eps)\\
&=S_+^{1-\eps}(\chi_\lambda)-\log_2(\eps)+\log_2(1-\eps)-2\log_2(1-\eps) \ .
\end{align}
\end{proof}

\subsection{Proof of Lemma~\ref{lemma:meter_axioms}}\label{sec:meter_NO}

To prove the Lemma, we remark that since $\Gamma_\Lambda \subseteq \cS(\cH)$ Axioms~\ref{axiom:reflexivity} to~\ref{axiom:consistency} can be seen to hold analogously to the proof of Lemma~\ref{prop:NO_axioms}. We proceed with the proof of the axioms that are specific to meter states.
\begin{enumerate}[(1)]
\item To prove the reduction axiom, realise that since the $\left\{\chi_\lambda\right\}_{\lambda}$ are states with a flat spectrum, for any state $\rho$ and meter state $\chi_{\lambda}$,
\begin{equation}
f_{\rho \otimes \chi_\lambda}(x)=\frac{1}{2^{\lambda}} f_\rho \left(\frac{x}{2^{\lambda}}\right)\ ,
\end{equation}
for all $x \in \mathbbm{R}_{\geq 0}$.
Hence, $\rho \otimes \chi_\lambda \maj^\eps \sigma \otimes \chi_\lambda$, which can be written
\begin{equation}
\int_0^x f_{\rho\otimes\chi_\lambda}(x')dx' \geq
\int_0^x f_{\sigma\otimes\chi_\lambda}(x')dx'-{\epsilon} \quad \forall x\geq 0 
\end{equation}
is equivalent to 
\begin{equation}
 \int_0^x f_{\rho}(y')dy' \geq
\int_0^x f_{\sigma}(y')dy'-{\epsilon} \quad \forall y\geq 0,
\end{equation}
which is established with the substitution $y'=\frac{x'}{2^{\lambda}}$. Therefore also $\rho \maj^\eps \sigma$.
\item To establish additivity under composition, note that for $\chi_{\lambda_1}$, $\chi_{\lambda_2} \in \Gamma_\Lambda$ the composed state $\chi_{\lambda_1} \otimes \chi_{\lambda_2}$ has a step function 
\begin{equation}
 f_{ \chi_{\lambda_1} \otimes \chi_{\lambda_2}}(x) = 
\begin{cases} 
2^{-(\lambda_1+\lambda_2)} & 0 \leq x \leq 2^{\lambda_1+\lambda_2} \\
0 & 2^{\lambda_1+\lambda_2} < x \ .
\end{cases} 
\end{equation}
This corresponds to the step function of a state $\chi_\lambda \in \Gamma_\Lambda \times \Gamma_\Lambda$ with $\lambda=\lambda_1+\lambda_2$.
\item To see that the order axiom holds, realise that for states $\chi_{\lambda_1}$,  $\chi_{\lambda_2} \in \Gamma_\Lambda$ with $\lambda_1 \leq \lambda_2 $, the majorisation condition $\chi_{\lambda_1} \maj \chi_{\lambda_2}$ always holds, since all non-zero eigenvalues of $\chi_{\lambda_1}$ are larger than those of $\chi_{\lambda_2}$. 
\item Work-error trade-off: For $\chi_{\lambda_1}$, $\chi_{\lambda_2} \in \Gamma_\Lambda$, 
$\chi_{\lambda_1}\prec^\eps \chi_{\lambda_2}$ holds, if and only if the maximal eigenvalue of $\chi_{\lambda_2}$ can be reduced below $2^{-\lambda_1}$ with an error of at most $\eps$. This means that the smoothing has to occur uniformly over the rank of $\chi_{\lambda_2}$, so that the new maximal eigenvalue of $\chi_{\lambda_2'} \in \epsball{\eps}{\chi_{\lambda_2}}$ satisfies
$2^{-\lambda_2'}=2^{-\lambda_2}(1-\eps)\leq 2^{-\lambda_1}$, that is,
\begin{equation}
\lambda_1\leq \lambda_2 - \log_2(1-\eps) \ .
\end{equation}
\end{enumerate}

Furthermore, we show that the meter system is suitable for $\Gamma=\cS(\cH)$ and adiabatic operations. For any $\rho \in  \cS(\cH)$ with maximal eigenvalue $\maxEV{\left(\rho \right)}$, $\sigma_1=\ket{0} \bra{0} \maj^\eps \rho$. Hence for $\lambda_1=\lambda_2$, as $\sigma_1 \otimes \chi_{\lambda_1} \maj \rho \otimes \chi_{\lambda_2}$. Moreover, for $\lambda_3=0$ and $\lambda_4 \geq \log_2(\rank{\rho})$, $\sigma_1 \otimes \chi_{\lambda_3} \maj \rho \otimes \chi_{\lambda_4}$. This is possible since $\Gamma_\Lambda$ contains states with arbitrarily large rank.

\subsection{Proof of Proposition~\ref{prop:smoothentr}} \label{sec:smoothentr}
To prove Equation~\eqref{eq:sma1}, we first consider 
\begin{equation}
S^\epsilon_-(\rho) = \sup\left\{\lambda_1-\lambda_2 \ \middle| \ \sigma_1 \otimes \chi_{\lambda_1} \maj^\eps \rho \otimes \chi_{\lambda_2} \right\} \ .
\end{equation}
Since $\sigma_1 \otimes \chi_{\lambda_1}$ has a flat step function, we find that 
\begin{equation}
\sigma_1 \otimes \chi_{\lambda_1} \maj^\eps \rho \otimes \chi_{\lambda_2} \iff f_{\sigma_1 \otimes \chi_{\lambda_1}}(0) \geq f_{\tilde{\rho}}(0)\  , \label{eq:aaa}
 \end{equation}
where $\tilde{\rho}\in \epsball{\eps}{\rho \otimes \chi_{\lambda_2}}$. Moreover, the optimal smoothing strategy to obtain $\tilde{\rho}$ is to reduce the largest values of $f_{\rho \otimes \chi_{\lambda_2}}$ so that they equal those of $ f_{\sigma_1 \otimes \chi_{\lambda_1}}$. We therefore choose $\eps(x)$ as small as possible and such that for all $x$ for which $f_{\rho \otimes \chi_{\lambda_2}} (x) > f_{\sigma_1 \otimes \chi_{\lambda_1}} (x)$, 
\begin{equation}
f_{\tilde{\rho}} (x)=  f_{\rho \otimes \chi_{\lambda_2}} (x) - \eps (x) \leq f_{\sigma_1 \otimes \chi_{\lambda_1}} (x)  \ ,
\end{equation}
and such that $\eps(x)=0$ otherwise. By definition of $\maj^\eps$, for any $\sigma_1 \otimes \chi_{\lambda_1}$ and $ \rho \otimes \chi_{\lambda_2}$ such that $\sigma_1 \otimes \chi_{\lambda_1} \maj^\eps \rho \otimes \chi_{\lambda_2}$ there exists such an $\eps(x)$ with $ \int_0^x \eps(x') \ dx' \leq \eps$ for all $x$ (this corresponds to bounding the trace distance by $\eps$, where normalisation is achieved by increasing the smallest eigenvalues appropriately). 
Then,
\begin{align}
f_{\tilde{\rho}}(0) &= \maxEV \left( \tilde{\rho} \right) \\
&= \maxEV \left( \rho \right) 2^{-\lambda_2}-\eps(0) \\
&= 2^{-\lambda_2} \left( \maxEV \left( \rho \right) - 2^{\lambda_2} \eps(0) \right) \ .
\end{align}
Now $f_{\tilde{\rho}}(0)$ is equivalent to the maximal eigenvalue of a sub-normalised state $\rho' \otimes \chi_{\lambda_2}$ with eigenvalues $\ithEV{ \rho' } =  \ithEV{\rho } - 2^{\lambda_2} \eps(i-1)$. This obeys $\rho' \in \epsballsub{\eps}{\rho}$ because the multiplicity of eigenvalues of $\rho'$ is the one of $\rho \otimes \chi_{\lambda_2}$ divided by $2^{\lambda_2}$ and since $ \int_0^x \eps(x') \ dx' \leq \eps $ for all $x$ \footnote{Note that there is a contribution to the trace distance due to the sub-normalisation of $\rho'$ (see Definition~\ref{def:trdist}).} and we can rewrite 
\begin{equation}
f_{\tilde{\rho}}(0)= 2^{-\lambda_2} 2^{-\Hmin{\rho'}}.
\end{equation}
As $ f_{\sigma_1 \otimes \chi_{\lambda_1}}(0) = 2^{-\lambda_1}$, \eqref{eq:aaa} can be rewritten as 
\begin{equation}
\sigma_1 \otimes\chi_{\lambda_1} \maj \tilde{\rho} \iff 2^{\lambda_2 -\lambda_1} \geq 2^{- \Hmin{\rho'}} \ . 
\end{equation}
Since any real number $2^{ - \Hmin{\rho'}}$ can be approximated arbitrarily closely by means of a rational $\frac{2^{\lambda_2}}{2^{\lambda_1}}$ (both $2^{\lambda_1}$ and $2^{\lambda_2}$ are assumed to be natural numbers), we obtain 
\begin{equation}
 S^\eps_-(\rho)=\sup_{\rho'\in\epsballsub{\eps}{\rho}} \Hmin{\rho'}= \Hmineps{\eps}{\rho} \ . 
\end{equation} 

\bigskip

To prove Equation~\eqref{eq:sma2}, we first analyse how the infimum comes about in $S_+^\eps$,
\begin{align}
    S_+^\eps(\rho) &= \inf \left\{\lambda_2 - \lambda_1 \ \middle| \ \rho \otimes \chi_{\lambda_1} \maj^\eps \sigma_1 \otimes \chi_{\lambda_2} \right\}  \\
&= \inf_{\lambda_1, \lambda_2; \rho'\in \mathcal{B}^\eps(\rho\otimes \chi_{\lambda_1})}
    \left\{ \lambda_2-\lambda_1 \ \middle| \ \rho' \maj \sigma_1 \otimes\chi_{\lambda_2} \right\} \ . 
\end{align}
Since $\sigma_1 \otimes \chi_{\lambda_2}$ has a flat step function of rank $2^{\lambda_2}$, $\rho' \maj \sigma_1 \otimes \chi_{\lambda_2}$ if and only if  $2^{\lambda_2} \geq \rank(\rho')$.

\begin{figure}
\centering
\includegraphics[width=1.0\textwidth]{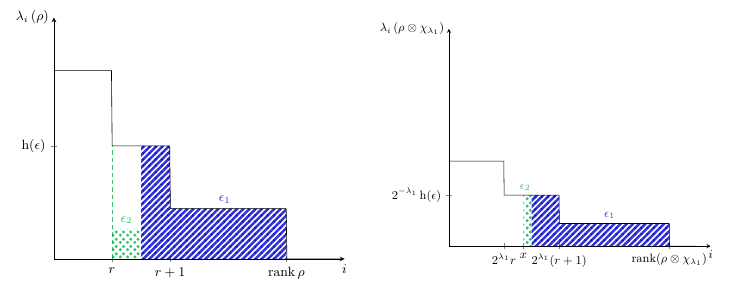}
\caption[Optimal smoothing strategy]{Illustration of the optimal smoothing strategy that leads to $S_+^\eps$. While for $\rho$ its rank can be reduced to a value $r+1$ (which is sometimes achieved for a value $\eps_1 < \eps$), displayed on the left, the tensor product with a state $\chi_{\lambda_1}$ can allow for a reduction of the rank of $\rho \otimes \chi_{\lambda_1}$ below $2^{\lambda_1}(r+1)$, displayed on the right. In the limit of large $\lambda_1$ this reduction tends to a rank of $2^{\lambda_1}\left( r+1-\frac{\eps_2}{h(\eps)} \right)$ with $\eps_2=\eps - \eps_1$.}
\label{fig:optimal_S_+}
\end{figure}

Figure~\ref{fig:optimal_S_+} illustrates that the above optimisation is optimal for  $\lambda_1$ and $\lambda_2$ such that the smoothing can be entirely used to reduce the rank of $\rho \otimes \chi_{\lambda_1}$\footnote{Note that to preserve normalisation larger eigenvalues of $\rho \otimes \chi_{\lambda_1}$ are thereby increased.}  to obtain $\rho'\in \mathcal B^\eps(\rho\otimes \chi_{\lambda_1})$, such that
\begin{equation} 2^{\lambda_2} \geq (r+1) 2^{\lambda_1} - \frac{\eps_2}{h(\eps)} 2^{\lambda_1} 
\end{equation}
  with $r$, $\eps_2$ and $h(\eps)$ as in the figure, and so
\begin{equation}
\lambda_2 - \lambda_1 \geq \log_2 \left(r+1-\frac{\eps_2}{h(\eps)}\right) \  . 
\end{equation}
For meter states $\chi_{\lambda_1}$ and $\chi_{\lambda_2}$,
$\lambda_1-\lambda_2=  \log_2(\frac{m}{n})$ for $m$, $n \in \mathbb{N}$, hence we can approximate $\log_2 \left(r+1-\frac{\eps_2}{h(\eps)}\right)$ to arbitrary precision by $\lambda_1-\lambda_2$ and we obtain the infimum 
\begin{equation} S_+^\eps(\rho) = \inf_{\lambda_1, \lambda_2; \rho'\in \mathcal{B}^\eps(\rho\otimes \chi_{\lambda_1})} \left\{ \lambda_2 - \lambda_1 \ \middle| \ \rho'\prec \sigma_1 \otimes \chi_{\lambda_2}  \right\} = \log_2 \left(r+1-\frac{\eps_2}{h(\eps)}\right)\ , 
\end{equation}
which we have shown to equal $\HHeps{1-\eps}{\rho}+\log_2(1-\eps)$ in the proof of Lemma~\ref{lemma:max_entropies}.

\subsection{Proof of Lemma~\ref{lemma:embezzling}} \label{sec:embezzling}
Let $\chi_\lambda$ have $\lambda=\log_2(m)$ as specified in the lemma. Consider the following construction for the state $\rho$. Let $\lambda_1\left(\rho\right), \ldots, \lambda_n \left(\rho \right)$ denote the $n$ non-zero eigenvalues of $\rho$ in decreasing order. Take $c=\frac{m\eps}{m-1}$ and set
\begin{equation}
    \lambda_i\left(\rho\right) = 
    \begin{cases}
     c &i = 1, \\
    \frac{c}{m}   &1 < i \leq m \\
    \frac{c}{m^2} &m < i \leq m^2 \\
    \frac{c}{m^3} &m^2 < i \leq m^3 \\
    \text{etc.\ }
    \end{cases}
\end{equation} 
until normalisation is reached, i.e., until
\begin{equation} 
\sum_{i=1}^n \lambda_i\left(\rho \right) \geq 1 \ . 
\end{equation}
If the above procedure leads to $ \sum_{i=1}^n \lambda_i\left(\rho \right) > 1 $, decrease $ \lambda_n\left( \rho \right) $ to be such that the state is normalised. We remark here that the $\lambda_i\left(\rho \right)$ are also the non-zero eigenvalues of $\rho \otimes \ket{0} \bra{0}$.

Now, consider the eigenvalues of $\rho \otimes \chi_\lambda$,  $ \lambda_1\left(\rho \otimes \chi_{\lambda} \right), \dots, \lambda_k \left(\rho \otimes \chi_{\lambda} \right) $.  Then,
\begin{equation}
    \lambda_i \left(\rho \otimes \chi_{\lambda} \right) = 
    \begin{cases}
    \frac{c}{m}    &1 \leq i \leq m \ , \\
    \frac{c}{m^2}  &m < i \leq m^2 \ , \\
    \frac{c}{m^3}  &m^2 < i \leq m^3 \ ,\\
    \text{etc.\ }
    \end{cases}
\end{equation}
until normalisation is reached, where $\lambda_k \left(\rho \otimes \chi_\lambda \right)$ is reduced to the value that guarantees normalisation as above. 
Since all the eigenvalues up to $n$ apart from the first one are equal for $\rho \otimes \ket{0}\bra{0}$ and $\rho \otimes \chi_\lambda$ (note that $k>n$) and since $\lambda_1 \left(\rho \otimes \ket{0} \bra{0} \right) - \lambda_1\left( \rho \otimes \chi_{\lambda} \right) = \frac{(m-1)\eps}{m - 1} = \eps $ , indeed,
\begin{equation} \rho \otimes \chi_{\lambda} \maj^\eps \rho \otimes \ket{0}\bra{0} \ . 
\end{equation}

%=========================================================

%% file: chapter08/thermomacro.tex
\let\textcircled=\pgftextcircled
\chapter{Macroscopic thermodynamics from microscopic considerations}\label{chap:thermomacro}

\initial{R}esource theories with error-tolerance give insights into processes in microscopic thermodynamics, as was discussed in the previous chapter. Here, we show that the behaviour of macroscopic systems that we know from thermodynamics emerges naturally from the microscopic framework. We first introduce macroscopic state spaces and thermodynamic equilibrium states in Section~\ref{sec:aep_typ}. We furthermore show that the microscopic considerations of the previous chapter imply a macroscopic behaviour governed by an effective order relation that is characterised by a single entropic quantity for thermodynamic equilibrium states, in agreement with standard thermodynamics. We go on to give several examples in Section~\ref{sec:examples} and recover the von Neumann entropy and the Boltzmann entropy in this context.

\section{States with macroscopic behaviour} \label{sec:aep_typ}

In thermodynamics, the behaviour of systems in equilibrium is governed and described by a few large-scale properties of the system that do not rely on the properties of each individual, microscopic constituent and that scale linearly in the system size. Hence, one can define equivalence classes of equilibrium states that all have essentially the same properties and differ in their amount of substance only. Keeping this in mind, we define the elements of a  macroscopic state space to be sequences of states corresponding to systems of different sizes, $\rho_n \in \Gamma^{(n)}$,
\begin{equation}
\Gamma_\mathrm{N} \defeq \left\{ \rho_\mathrm{N}  \  \middle|  \ \rho_\mathrm{N}  =  \left\{ \rho_n \right\}_{n=1}^{\infty} \sth \rho_n \in \Gamma^{(n)} \right\} \ .
\end{equation}
So far, this definition has not imposed any restrictions on the  states in a sequence that make up a macroscopic state. However,  physically, further restrictions should be introduced. For systems of indistinguishable particles, for instance -- which thermodynamics is  (usually) concerned with -- permutation symmetry on $n$-particle states should be required~\cite{Renner2007, Finetti, Caves2002, Konig2005}. (We leave the implications of such restrictions for future work.) Moreover, arbitrary sequences of states of different sizes are not physically meaningful but there should be a relation of the individual states of the systems $\rho_n \in \Gamma^{(n)}$ within a meaningful sequence. 
Below, we introduce thermodynamic equilibrium states, a subset of the possible macroscopic states that are subject to specific restrictions. In order to introduce them we need a bit more terminology and another axiom.  

\begin{axiom}[Existence of a non-catalytic state] \label{axiom:typicality}
An error-tolerant resource theory on a state space $\Gamma$ has at least one state that cannot be used catalytically with respect to the resource theory, i.e., there is a state $\omega \in \Gamma$ that for any $\rho$, $\sigma \in \Gamma$ obeys
\begin{equation}
 \rho \prec^\eps \sigma  \ \iff \   (\rho, \omega) \prec^\eps (\sigma, \omega) 
\end{equation} 
and for any meter states, $\chi_{\lambda_1}$, $\chi_{\lambda_2} \in \Gamma_\lambda$, obeys
\begin{equation}
\chi_{\lambda_1} \prec^\eps \chi_{\lambda_2}  \ \iff \   (\chi_{\lambda_1}, \omega) \prec^\eps (\chi_{\lambda_2}, \omega) \ . 
\end{equation} 
\end{axiom}

The popular quantum resource theories have such states, for instance the maximally mixed state for adiabatic processes (and the previously considered meter $\Gamma_\Lambda$) and the Gibbs states for thermal operations. There are further examples in both cases. 
From now on, we fix the reference state on a system, $\sigma_1 \in \Gamma$, to have this property. We then define the following subset of $\Gamma$, 
\begin{equation}
\Gamma_\mathrm{T}= \left\{ \tau_\lambda \in \Gamma \ \middle| \ S_-(\tau_\lambda)= S_+(\tau_\lambda)=\lambda \right\} \subseteq \Gamma \ ,
\end{equation} 
which contains at least one state $\tau_0=\sigma_1$. The states $\tau_\lambda \in \Gamma_\mathrm{T}$ have properties similar to meter states, for instance, they obey the work-error trade-off.\footnote{The details of such considerations are left to~\cite{prep} and to future work, the work-error trade-off is proven in Lemma~\ref{lemma:wptlemma}.}
Relying on such states, we define thermodynamic equilibrium states as macroscopic states that in the limit of large $n$ behave in a similar manner. In that sense our definition bears some resemblance with~\cite{Lieb2013, Lieb2014}, where thermodynamic equilibrium states are used as a meter system. 

\begin{definition} \label{def:equil}
$\Gamma_\mathrm{N}$ is a set of \emph{thermodynamic equilibrium states} if  
for all $\rho_\mathrm{N} \in \Gamma_\mathrm{N}$, there exists $\lambda_{\rho_\mathrm{N}}$ such that for any $0 < \eps < 1$ and for any $\delta > 0$, there exists an $n_0$ such that for all $ n \geq n_0$, there exist $\tau_{\lambda_-}, \tau_{\lambda_+} \in \Gamma_\mathrm{T}^{(n)}$ with $\lambda_- \geq n(\lambda_{\rho_\mathrm{N}}-\delta)$, $\lambda_+ \leq n(\lambda_{\rho_\mathrm{N}}+\delta)$ and
\begin{equation}
\tau_{\lambda_-} \prec^\eps \rho_n \prec^\eps \tau_{\lambda_+} \ .
\end{equation}
\end{definition}

In the following we derive properties for thermodynamic equilibrium states. They are essentially characterised by a single entropic quantity that gives necessary and sufficient conditions for state transformations and hence leads to the emergence of a macroscopic preorder relation. We proceed by deriving an asymptotic equipartition property and a corollary, the proofs of which are deferred to Section~\ref{sec:aep_proof}.

\begin{proposition}[Asymptotic Equipartition Property]
\label{lemma:axiomaticaep1}
Let there be an error-tolerant resource theory on state spaces $\Gamma^{(n)}$ that obeys Axioms~\ref{axiom:reflexivity} to~\ref{axiom:consistency} and Axiom~\ref{axiom:typicality} and let there be a suitable meter system. Furthermore, let $\rho_\mathrm{N} \in \Gamma_\mathrm{N}$ be a thermodynamic equilibrium state. 
Then, for any $1 > \tilde{\eps} >0$,  
\begin{equation}
S_{\infty}(\rho_\mathrm{N})\defeq\lim_{n \rightarrow \infty} \frac{S_-^{\tilde{\eps}}(\rho_n)}{n}= \lim_{n \rightarrow \infty} \frac{S_+^{\tilde{\eps}}(\rho_n)}{n}= \lambda_{\rho_\mathrm{N}} \ .
\end{equation} 
\end{proposition}
Note that as opposed to the asymptotic equipartition property from~\cite{TCR}, we need not consider the limit $\tilde{\eps} \rightarrow 0$ here.

\begin{corollary} \label{cor:aep}
Let the composition of $n$ copies of a state $\rho \in \Gamma$ obey the requirements of the Proposition~\ref{lemma:axiomaticaep1}, i.e., let all axioms be obeyed and let  $\rho_N=\left\{(\rho, \ldots, \rho) \in \Gamma^{(n)}\right\}_n$ be a thermodynamic equilibrium state, then
\begin{equation}
S_-(\rho) \leq S_{\infty}(\rho_\mathrm{N}) \leq S_+(\rho) \ .
\end{equation}
\end{corollary}
We remark that this implies that if all states $\rho_\mathrm{N}=\left\{(\rho, \dots, \rho) \in \Gamma^{(n)} \right\}_n$ are thermodynamic equilibrium states according to Definition~\ref{def:equil}, then for all (microscopic) states, $\rho \in \Gamma$, the relation $S_-(\rho) \leq S_+(\rho)$ holds.

Besides, Proposition~\ref{lemma:axiomaticaep1} establishes that the behaviour of thermodynamic equilibrium states with respect to an error-tolerant resource theory is characterised by the single quantity $S$. We now derive the corresponding necessary conditions and sufficient conditions for state transformations, which are proven in Section~\ref{sec:typicality_proof}. 
\begin{proposition}[Necessary and sufficent conditions for state transformations]\label{lemma:typicality}
Let there be an error-tolerant resource theory on state spaces $\Gamma^{(n)}$ that obeys Axioms~\ref{axiom:reflexivity} to~\ref{axiom:consistency} and Axiom~\ref{axiom:typicality} and let there be a suitable meter system. Furthermore, let $\rho_\mathrm{N}$, $ \sigma_\mathrm{N} \in \Gamma_\mathrm{N}$ be thermodynamic equilibrium states.  
Then, for any $\delta > 0$ and any $0 < \eps < 1$, there exists an $n_0$ such that for all $n \geq n_0$,
\begin{equation}
S_{\infty}(\rho_\mathrm{N}) + \delta \leq S_{\infty}(\sigma_\mathrm{N}) \implies \rho_n \prec^{\eps} \sigma_n \ .
\end{equation} 
Furthermore, if for some $0 < \eps < 1$, there exists an $n_0$ such that for all $n \geq n_0$, the relation $\rho_{n} \prec^{\eps} \sigma_{n}$ holds, then 
\begin{equation}
S_{\infty}(\rho_\mathrm{N}) \leq S_{\infty}(\sigma_\mathrm{N}) \ .
\end{equation}
\end{proposition}

\bigskip

We are now in a position to introduce an effective macroscopic order relation $\prec_\mathrm{N}$: we write $\rho_{N} \prec_\mathrm{N} \sigma_{N}$ if for any $\eps >0$ there exists $n_0$ such that for all $n \geq n_0$ we have $\rho_{n} \prec^{\eps} \sigma_{n}$. Then $S$ is monotonic with respect to $\prec_\mathrm{N}$ and $S_{\infty}(\rho_\mathrm{N}) < S_{\infty}(\sigma_\mathrm{N})$ implies $\rho_{N} \prec_\mathrm{N} \sigma_{N}$ by Proposition~\ref{lemma:typicality}. The macroscopic transformations according to $\prec_\mathrm{N}$ are fully characterised by a necessary \emph{and} sufficient condition in terms of $S$ (except for pairs of states where $S_{\infty}(\rho_\mathrm{N}) = S_{\infty}(\sigma_\mathrm{N})$). The relation $\prec_\mathrm{N}$ is furthermore transitive and, hence, a preorder relation which establishes the connection to the structure of traditional resource theories~\cite{Janzing2000_cost, Horodecki2003, Horodecki2003b, Horodecki2011, Renes2014} and to resource theories for macroscopic thermodynamics~\cite{Lieb1998, Lieb1999, Lieb2001}.

Furthermore, we may introduce composition and scaling operations for thermodynamic equilibrium states, as considered in the axiomatic work~\cite{Lieb1998, Lieb1999, Lieb2001, Lieb2013}, which may allow for the recovery of corresponding axioms. A natural way to define a space of composed systems is $\Gamma_\mathrm{N} \times \Gamma'_\mathrm{N} \defeq \left\{  (\rho_\mathrm{N},~\sigma_\mathrm{N} )  \ \middle| \  ( \rho_\mathrm{N},~\sigma_\mathrm{N} )=\left\{ (\rho_n,~\sigma_n ) \right\}_{n=1}^{\infty} \sth \rho_\mathrm{N} \in \Gamma_\mathrm{N},~\sigma_\mathrm{N} \in \Gamma'_\mathrm{N}  \right\}$. For this composition operation the macroscopic states and order relation obey the consistent composition axiom. 
A scaling operation that produces a state space $\alpha \Gamma_\mathrm{N}$ of states that are scaled by $\alpha \in \mathbb{R}_{\geq 0}$ could be introduced by changing the spreading of the elements in the sequences $\rho_\mathrm{N}=\left\{ \rho_n \right\}_n \in \Gamma_\mathrm{N}$. For instance, for $\alpha=2$ the state $2\rho_\mathrm{N} \in 2\Gamma_\mathrm{N}$ should only contain the elements of the original sequence with even index, i.e., $2\rho_\mathrm{N}=\left\{ \rho'_n \right\}_n$, where $\rho'_n=\rho_{2n}$. We leave the generalisation to arbitrary real scaling factors and the exploration of whether such definitions obey all axioms from Refs.~\cite{Lieb1998, Lieb1999, Lieb2001, Lieb2013} for future work.

\section{Thermodynamic equilibrium states under adiabatic processes} \label{sec:examples}
To find examples of thermodynamic equilibrium states with respect to adiabatic processes, let us first consider the set $\Gamma_\mathrm{T}$, which is the set of all states $\rho \in \cS(\cH)$ that obey $\Hmin{\rho}=\Hzero{\rho}$, since these entropies correspond to $S_-$ and $S_+$ respectively (see also~\cite{Lieb2013, Weilenmann2015}). Hence, $\Gamma_\mathrm{T}$ is the set of all states that obey $\maxEV \left( \rho \right)=\frac{1}{\rank (\rho)}$, where $\maxEV \left( \rho \right)$ denotes the maximal eigenvalue of $\rho$. These are all states for which the step function is flat in the sense that all non-zero eigenvalues are equal.

In the following, we show that the states $\rho_\mathrm{N} \in \Gamma_\mathrm{N}$ with $\rho_n=\rho^{\otimes n}$ are thermodynamic equilibrium states with respect to smooth adiabatic operations (as introduced in Section~\ref{sec:smoothingNO}). 

\begin{lemma}
\label{lemma:typicalityNO}
For smooth adiabatic operations on quantum states with state space $\Gamma=\cS(\cH)$ and the meter system $\Gamma_\Lambda$ introduced in Section~\ref{sec:entropies_majorization}, 
\begin{equation}
\Gamma_\mathrm{N}=\left\{ \rho_\mathrm{N}  \ \middle| \  \rho_\mathrm{N}=\left\{\rho_n \right\}_{n=1}^{\infty} \sth \rho_n=\rho^{\otimes n}  \right\}
\end{equation}
is a set of equilibrium states. Furthermore, $\lambda_\rho$ is the von Neumann entropy, i.e., $\lambda_\rho = H(\rho)$.
\end{lemma}

\begin{proof} 
First, note that for $\rho=\tau_\lambda \in \Gamma_\mathrm{T}$ the lemma follows immediately with $\lambda_-=\lambda_+=\lambda_{\rho_\mathrm{N}}$. Now let $\rho \not\in \Gamma_\mathrm{T}$, 
let $0 < \eps < 1$ and let $\delta  > 0$ be arbitrary. Now choose $\tilde{\eps}$ such that  $\eps= \sqrt{\tilde{\eps}}$ and define
\begin{equation}
\tilde{\rho}^{n}\defeq \frac{\Pi_n^{\tilde{\eps}} \rho^{\otimes n} \Pi_n^{\tilde{\eps}}}{\tr{\Pi_n^{\tilde{\eps}}\rho^{\otimes n}\Pi_n^{\tilde{\eps}}}} \ ,
\end{equation} 
where $\Pi_n^{\tilde{\eps}}$ is a projector onto the $\tilde{\eps}$-typical subspace of $\rho^{\otimes n}$. Now choose $n_0'$ large enough such that for $n \geq n_0'$,  
\begin{equation}
\trdist{ \tilde{\rho}^{n}}{\rho^{\otimes n} } \leq \eps \ , 
\end{equation}
a choice which is always possible~\cite{Wilde2013}. 
Now for any eigenvalue $p$ of $\tilde{\rho}^{n}$,
\begin{equation}
2^{-n H(\rho) -n \beta} \leq p \leq 2^{-n H(\rho) +n \beta} \ ,
\end{equation}
where $\beta \propto \frac{1}{\sqrt{n}}$~\cite{MacKay2003,Wilde2013}. 
Now let $n_0''$ be the minimal value such that the following two conditions are obeyed\footnote{ For an $n_0''$ such that $\delta >\beta$ and $\min \left\{ H(\rho), \delta \right\}-\beta+1 >0$ both conditions are satisfied. Such an $n_0''$ can always be found as $\beta \propto \frac{1}{\sqrt{n}}$.}
\begin{align}
2^{n_0''(H(\rho)+\delta)}-2^{n_0''(H(\rho)+\beta)} &> 1 \ , \\
2^{n_0''(H(\rho)-\beta)}-2^{n_0''(H(\rho)-\delta)} &> 1 \ ,
\end{align}
and set $n_0'''=\max \left\{n_0', n_0'' \right\}$.
Then, for any $n \geq n_0'''$ there exist $k$, $m \in \mathbb{N}$ such that
\begin{align}
2^{n(H(\rho)+\delta)} \geq k \geq 2^{n(H(\rho)+\beta)} \ , \\
2^{n(H(\rho)-\beta)} \geq m \geq 2^{n(H(\rho)-\delta)} \ .
\end{align}
Since $H(\rho)< \log_2 \left(\rank (\rho) \right)$ and $\beta \propto \frac{1}{\sqrt{n}}$, we can choose $n_0 \geq n_0'''$ such that for all $n \geq n_0$, it holds that $k,~m < \rank(\rho)^n$, hence, we can choose $\lambda_-=\log_2(k)$ and $\lambda_+=\log_2(m)$. 
Thus, Definition~\ref{def:equil} is obeyed with $\lambda_\rho= H(\rho)$. 
\end{proof}

For adiabatic operations on quantum states $S$ is the von Neumann entropy, in agreement with the results from Ref.~\cite{TCR}.~\footnote{Note that in Ref.~\cite{TCR} the max entropy $H_0^{\eps}$ instead of $H_\mathrm{H}^{1-\eps}$ is considered, however, the relation has also been proven for the latter~\cite{Dupuis2013_DH}.} 
With respect to adiabatic \emph{probabilistic} transformations, the set of all sequences of states $\rho_n=\rho^{\otimes n}$ is not a set of thermodynamic equilibrium states according to Definition~\ref{def:equil}. To see this, let $\eps \leq \frac{1}{4}$, let $\delta=\frac{1}{100}$ and take a state $\rho=\frac{3}{4} \ket{0} \bra{0} + \frac{1}{4} \ket{1}\bra{1}$. Then $\rho^{\otimes n}$ has a maximal eigenvalue $\left(\frac{3}{4}\right)^{n}$ and rank $2^{n}$.
Now consider $\tau_{\lambda_-} \pmaj^\eps \rho^{\otimes n}$ and $\rho^{\otimes n} \pmaj^\eps \tau_{\lambda_+}$, i.e.,
\begin{align}
\tau_{\lambda_-} &\rightarrow (1-\eps) \rho^{\otimes n} + \eps \xi_1 \\
\rho^{\otimes n} &\rightarrow (1-\eps) \tau_{\lambda_+} + \eps \xi_2.
\end{align}
For there to be such transformations, the following necessary conditions have to be met~\footnote{This follows as the state $\tau_{\lambda_-}$ has to be chosen such that it majorises  $(1-\eps) \rho^{\otimes n}$. On the other hand, $\rho^{\otimes n}$ also has to majorise the $(1-\eps) \tau_{\lambda_+}$, hence its rank should be smaller than that of $\tau_{\lambda_+}$. }
\begin{align}
\lambda_- &\leq n \log_2{\left( \frac{4}{3} \right)} - \log_2{(1- \eps)} \\
\lambda_+ &\geq n \log_2{(2)}, 
\end{align} 
which imply that $\lambda_+ - \lambda_- \geq n \log_2{\left(\frac{3}{2} \right)} +\log_2{( 1- \eps)} > 2 n \delta$.

\bigskip

A second class of thermodynamic equilibrium states with respect to the smooth adiabatic operations are the microcanonical states, 
\begin{equation}
\Gamma_\mathrm{N}= \left\{ \rho_\mathrm{N} \ \middle| \ \rho_\mathrm{N}=\left\{\frac{\Pi_{\Omega_\mathrm{micro}(nE,nV,n)}}{\Omega_\mathrm{micro}(nE,nV,n)} \right\}_n \right\} \ ,
\end{equation}
where $\Omega_\mathrm{micro}$ is the microcanonical partition function and $\Pi_{\Omega_\mathrm{micro}}$ is the projector onto its subspace. The state $\rho_\mathrm{N}$ is the sequence of microcanonical states of different particle number that all have the same energy $E$ and volume $V$ per particle, for which we obtain the entropy per particle known from statistical mechanics, $S_{\infty}(\rho_\mathrm{N})=\lambda_\rho= \log_2(\Omega_\mathrm{micro}(E,V,1))$.

\bigskip

We remark here, that necessary and sufficient conditions for state transformations can also be directly derived for specific examples, without going through the axiomatisation. For instance, for smooth majorisation on quantum states with i.i.d.\ states $\rho_n=\rho^{\otimes n}$, we could work with $H_\mathrm{min}^\eps$ and $H_\mathrm{0}^\eps$ and arrive at statements analogous to Proposition~\ref{lemma:typicality}, where S is the von Neumann entropy. Such a derivation is, however, more cumbersome than verifying the axioms and it has to be considered separately for each example.  
The statement here is more general, as it applies to any resource theory that obeys the axioms, such as the resource theory of thermal operations~\cite{prep}. Besides, the axioms reveal natural features that lead to thermodynamic behaviour, which may also be applicable to systems that are not represented as density operators or probability distributions.

\section{Appendix}

\subsection{Proof of Proposition~\ref{lemma:axiomaticaep1} and Corollary~\ref{cor:aep}}\label{sec:aep_proof}

We first prove the following lemma.
\begin{lemma} \label{lemma:tau}
For any state $\tau_\lambda \in \Gamma_\mathrm{T}$ the following inequalities hold,
\begin{align}
\lambda \leq S_-^\eps(\tau_\lambda) &\leq \lambda -\log_2(1-\eps) \ , \label{eq:e1} \\
\lambda + \log_2(1-\eps) \leq S_+^\eps(\tau_\lambda) &\leq \lambda \ . \label{eq:e2}  
\end{align}
\end{lemma}

\begin{proof}
To prove the inequalities~\eqref{eq:e1}, note first that $S_-^\eps(\tau_\lambda) \geq S_-(\tau_\lambda)=\lambda$.
Furthermore, let $\left(\lambda_1^\mu , \lambda_2^\mu \right)_\mu$ label a sequence of pairs of states $\chi_{\lambda_1^\mu}$, $\chi_{\lambda_2^\mu} \in \Gamma_\Lambda$ such that
\begin{equation}
(\sigma_1,~\chi_{\lambda_1^\mu}) \prec^\eps (\tau_\lambda,~\chi_{\lambda_2^\mu}) \ ,
\end{equation}
with $S_-^\eps(\tau_\lambda)=\lim_{\mu \rightarrow \infty} \lambda_1^\mu -\lambda_2^\mu$. Furthermore, let $\left(\tilde{\lambda}_1^\nu , \tilde{\lambda}_2^\nu \right)_\nu$ label another sequence of pairs of states obeying
\begin{equation}
(\tau_\lambda,~\chi_{\tilde{\lambda}_1^\nu}) \prec (\sigma_1,~\chi_{\tilde{\lambda}_2^\nu}) \ 
\end{equation}
with $S_-(\tau_\lambda)=\lambda=\lim_{\nu \rightarrow \infty} \tilde{\lambda}_2^\nu -\tilde{\lambda}_1^\nu$.
Then, by consistent composition
\begin{equation}
(\sigma_1,~\chi_{\lambda_1^\mu},~\chi_{\tilde{\lambda}_1^\nu}) \prec^\eps(\tau_\lambda,~\chi_{\lambda_2^\mu}~\chi_{\tilde{\lambda}_1^\nu}) \prec (\sigma_1,~\chi_{\lambda_2^\mu},~\chi_{\tilde{\lambda}_2^\nu}) \ ,
\end{equation}
and by additive transitivity, Axiom~\ref{axiom:typicality} and the additivity of meter states,
\begin{equation}
\chi_{\lambda_1^\mu+\tilde{\lambda}_1^\nu} \prec^\eps \chi_{\lambda_2^\mu+\tilde{\lambda}_2^\nu} \ .
\end{equation}
The work-error trade-off then implies $\lambda_1^\mu-\lambda_2^\mu \leq \tilde{\lambda}_2^\nu-\tilde{\lambda}_1^\nu + \log_2(1-\eps)$. 
Since this holds for all $\mu$, $\nu$, it remains true in the limit and proves the inequality. The proof of inequalities~\eqref{eq:e2} proceeds analogously and is omitted here. 
\end{proof}

To prove the proposition, let $\eps = \frac{1}{2} \min \left\{\tilde{\eps}, (1-\tilde{\eps}) \right\}$, and let $\delta > 0$ be arbitrary, now let  $\lambda_{\rho_\mathrm{N}}$ and $n$ be such that the conditions of Definition~\ref{def:equil} are met for $\eps$ and $\delta$.
Then, by Proposition~\ref{proposition:monotonicitylike},
\begin{align}
S_-^{\tilde{\eps}-\eps}(\tau_{\lambda_-}) &\leq S_-^{\tilde{\eps}}(\rho_n) \leq S_-^{\tilde{\eps}+\eps}(\tau_{\lambda_+}) \ , \\
S_+^{\tilde{\eps}+\eps}(\tau_{\lambda_-}) &\leq
S_+^{\tilde{\eps}}(\rho_n) \leq S_+^{\tilde{\eps}-\eps}(\tau_{\lambda_+}) \ ,
\end{align} 
and by Lemma~\ref{lemma:tau},
\begin{align}
\lambda_- &\leq S_-^{\tilde{\eps}}(\rho_n) \leq \lambda_+ -\log_2(1-(\tilde{\eps}+\eps)) \ ,\\
\lambda_- +\log_2(1-(\tilde{\eps}+\eps)) &\leq S_+^{\tilde{\eps}}(\rho_n) \leq \lambda_+ \ ,
\end{align}
and as $\lambda_- \geq n(\lambda_{\rho_\mathrm{N}}-\delta)$ and $\lambda_+ \leq n(\lambda_{\rho_\mathrm{N}}+\delta)$,
\begin{align}
n(\lambda_{\rho_\mathrm{N}}-\delta) &\leq S_-^{\tilde{\eps}}(\rho_n) \leq n(\lambda_{\rho_\mathrm{N}}+\delta)-\log_2(1-(\tilde{\eps}+\eps)) \ ,\\
n(\lambda_{\rho_\mathrm{N}}-\delta)+\log_2(1-(\tilde{\eps}+\eps)) &\leq S_+^{\tilde{\eps}}(\rho_n) \leq n(\lambda_{\rho_\mathrm{N}}+\delta) \ ,
\end{align}
as well as 
\begin{equation}-\frac{2n  \delta}{n} \leq \frac{S_-^{\tilde{\eps}}(\rho_n)- S_+^{\tilde{\eps}}(\rho_n)}{n} \leq  \frac{2n  \delta -2\log_2(1-(\tilde{\eps}+ \eps))}{n}.\label{eq:bound}
\end{equation}
As this holds for any $\delta >0$, we have that 
\begin{equation}\label{eq:lambdaequ} \lim_{n \rightarrow \infty} \frac{S_-^{\tilde{\eps}}(\rho_n)}{n}= \lim_{n \rightarrow \infty} \frac{S_+^{\tilde{\eps}}(\rho_n)}{n}= \lambda_{\rho_\mathrm{N}} \ .
\end{equation} 
\qed

\bigskip

To prove the corollary we rely on Proposition~\ref{prop:additivityproperties}, according to which, for $\rho_n=(\rho, \ldots, \rho) \in \Gamma^{(n)}$,
\begin{align}
S_-^{\tilde{\eps}}(\rho_n) &\geq S_-^{\tilde{\eps}}(\rho)+ (n-1) S_-(\rho), \\
S_+^{\tilde{\eps}}(\rho_n) &\leq S_+^{\tilde{\eps}}(\rho)+ (n-1) S_+(\rho). 
\end{align}
Then, by Proposition~\ref{proposition:monotonicitylike},
\begin{align}
S_-^{\tilde{\eps}}(\rho_n) &\geq n S_-(\rho), \\
S_+^{\tilde{\eps}}(\rho_n) &\leq n S_+(\rho). 
\end{align}
From \eqref{eq:bound} it follows that
\begin{align}
n S_-(\rho) &\leq S_-^{\tilde{\eps}}(\rho_n) \\
&\leq S_+^{\tilde{\eps}}(\rho_n)+ 2 n \delta -2\log_2(1-(\tilde{\eps}+ \eps))
\\
&\leq n S_+(\rho)+ 2 n  \delta-2\log_2(1-(\tilde{\eps}+ \eps)) \
. 
\end{align}
Now since $S_{\infty}(\rho_\mathrm{N})\defeq \lim_{n \rightarrow \infty} \frac{S_-^{\tilde{\eps}}(\rho_n)}{n}$, dividing the above by $n$ and taking the limit ${n \rightarrow \infty}$ leads to
\begin{equation}
 S_-(\rho)  \leq S_{\infty}(\rho_\mathrm{N}) \leq S_+(\rho)+ 2 \delta \ .
\end{equation} 
As the above applies for any $\delta >0$ this concludes the proof.
\qed

\subsection{Proof of Proposition~\ref{lemma:typicality}} \label{sec:typicality_proof}
We first prove the following lemma. 
\begin{lemma} \label{lemma:wptlemma}
For states $\tau_\lambda$, $\tau_{\tilde{\lambda}} \in \Gamma_\mathrm{T}$ the work-error trade-off is obeyed, i.e.,
\begin{equation}
\tau_\lambda \prec^\eps \tau_{\tilde{\lambda}} \implies \lambda \leq \tilde{\lambda} - \log_2(1-\eps) \ .
\end{equation}
\end{lemma}
\begin{proof}
Let $\tau_\lambda$, $\tau_{\tilde{\lambda}} \in \Gamma_\mathrm{T}$ such that $\tau_\lambda \prec^\eps \tau_{\tilde{\lambda}}$. Furthermore, let $\left\{ \lambda_1^\mu, \lambda_2^\mu \right\}_\mu$ and $\left\{ \tilde{\lambda}_2^\nu, \tilde{\lambda}_1^\nu \right\}_\nu$ label two sequences of pairs of states $\chi_{\lambda_1^\mu}$, $\chi_{\lambda_2^\mu}$ and $\chi_{\tilde{\lambda}^\nu}$, $\chi_{\tilde{\lambda}_2^\nu}$ respectively, such that
\begin{align} 
(\sigma_1,~\chi_{\lambda_1^\mu}) \prec (\tau_\lambda,~\chi_{\lambda_2^\mu}) \ , \\
(\tau_{\tilde{\lambda}},~\chi_{\tilde{\lambda}_1^\nu}) \prec (\sigma_1,~\chi_{\tilde{\lambda}_2^\nu}) \ ,
\end{align}
with $\lim_{\mu \rightarrow \infty}\lambda_1^\mu-\lambda_2^\mu= \lambda$ and $\lim_{\nu \rightarrow \infty}\tilde{\lambda}_2^\nu-\tilde{\lambda}_1^\nu=\tilde{\lambda}$.
Then, by consistent composition and additive transitivity,
\begin{equation}
(\sigma_1,~\chi_{\lambda_1^\mu},~\chi_{\tilde{\lambda}_1^\nu}) \prec^\eps ( \sigma_1,~\chi_{\lambda_2^\mu},~\chi_{\tilde{\lambda}_2^\nu}) \ ,
\end{equation}
which implies that, $\chi_{\lambda_1^\mu+\tilde{\lambda}_1^\nu} \prec^\eps \chi_{\lambda_2^\mu+\tilde{\lambda}_2^\nu}$ and thus $\lambda_1^\mu-\lambda_2^\mu \leq \tilde{\lambda}_1^\nu - \tilde{\lambda}_2^\nu - \log_2(1-\eps)$, which remains valid in the limits.
\end{proof}

To prove the proposition, let $\mu=\frac{\eps}{2}$ and let $\tilde{\delta}=\frac{\delta}{2}$. It follows from Equation~\eqref{eq:lambdaequ} that
one can choose $n_0$ such that for $n \geq n_0$, the following two inequalities hold,
\begin{align}
S_{\infty}(\rho_\mathrm{N}) + \tilde{\delta} &\geq \frac{S_+^{\mu}(\rho_{n})}{n} \ , \\
S_{\infty}(\sigma_\mathrm{N}) - \tilde{\delta} &\leq \frac{S_-^{\mu}(\sigma_{n})}{n} \ .
\end{align}
This implies that under the assumption that $S_{\infty}(\rho_\mathrm{N})+\delta \leq S_{\infty}(\sigma_\mathrm{N})$ for $n \geq n_0$,
\begin{equation}
S_+^{\mu}(\rho_{n}) \leq S_-^{\mu}(\sigma_{n}) \ , 
\end{equation}
which by Proposition~\ref{proposition:sufficient_condition_axiom_implication} directly implies 
\begin{equation}
\rho_{ n} \prec^{\eps} \sigma_{n} \ .
\end{equation}

\bigskip

Furthermore, let $n_0$ be such that $\rho_{ n} \prec^{\eps} \sigma_{ n}$ for all $n \geq n_0$. Then, take $\eps'<\frac{1-\eps}{2}$, $\delta>0$ and let $\lambda_{\rho_\mathrm{N}}$, $\lambda_{\sigma_\mathrm{N}}$ and $n_0'$ be such that for all $n \geq n_0'$ Definition~\ref{def:equil} holds for both states $\rho_\mathrm{N}$ and $\sigma_\mathrm{N}$.
This implies that for $n \geq \max\left\{ n_0, n_0' \right\}$,
\begin{equation}
\chi_{\lambda_-(\rho)} \prec^{\eps'}\rho_{ n} \prec^{\eps} \sigma_{n}\prec^{\eps'}\chi_{\lambda_+(\sigma)} \ .
\end{equation}
Hence, due to Axiom~\ref{axiom:additivetransitivity} and  Lemma~\ref{lemma:wptlemma},
\begin{equation}
\lambda_-(\rho) \leq \lambda_+(\sigma) - \log_2(1-(\eps+2\eps'))
\end{equation}
and thus also
\begin{equation}
n(\lambda_{\rho_\mathrm{N}}-\delta) \leq n(\lambda_{\sigma_\mathrm{N}}+\delta) - \log_2(1-(\eps+2\eps')),
\end{equation}
which implies
\begin{equation}
\lambda_{\rho_\mathrm{N}}  \leq \lambda_{\sigma_\mathrm{N}} + 2 \delta - \frac{\log_2(1-(\eps+2\eps'))}{n} \ . 
\end{equation}
As this holds for every $\delta>0$ this concludes the proof as $S_{\infty}(\rho_\mathrm{N})=\lambda_{\rho_\mathrm{N}}$ and $S_{\infty}(\sigma_\mathrm{N})=\lambda_{\sigma_\mathrm{N}}$.
\qed

%=========================================================

%% file: chapter09/conclusion.tex
\let\textcircled=\pgftextcircled
\chapter{Conclusions and Outlook} \label{chap:conclusion}

In Part~I of this thesis, we improved on the current entropic techniques for analysing causal structures in several ways. We introduced the application of non-Shannon inequalities, which improve outer approximations to the entropy cones of (classical) causal structures, and derived inner approximations. Our techniques are applicable to both the entropy vector approach and the post-selection technique. Our new outer approximations improve on the distinction of the sets of correlations compatible with different causal structures and the inner approximations help us identify the actual boundary of these cones. They furthermore aid us in identifying situations where it is necessary to look beyond entropy vectors for certifying incompatibility with a causal structure.

We have also discovered important deficiencies of the entropy vector method when it comes to distinguishing between classical and quantum correlations. More specifically, we have shown that for line-like causal structures and a few more examples there is no gap between the sets of compatible entropy vectors in the classical and quantum case, despite there being a separation at the level of probabilities. Hence, no function of the entropies of the observed variables can make such a distinction. Whether the entropy vector method is ever able to distinguish classical and quantum causes or whether it merely accounts for independence relations (oblivious to the underlying non-signalling theory), remains an open problem. 

Our results fuel the discussion of whether entropies are suitable for analysing causal structures, or whether the mapping from probabilities to entropies oversimplifies the problem to such an extent that the search for other techniques should be prioritised. Criteria for certifying whether a set of entropy inequalities is able to detect non-classical correlations remains scarce. For the CHSH scenario, the known entropic constraints are sufficient for detecting any non-classical correlations~\cite{Braunstein1988, Chaves2013}, however, a tool to analyse other causal structures in a similar way is unavailable. This is partly because the proof for the CHSH scenario relies on the convexity of the set of compatible probability distributions and cannot be straightforwardly generalised. For many of the known entropic inequalities relevant to classical causal structures it is unknown whether they remain valid in the corresponding quantum scenario.

A worthwhile improvement on current entropic techniques could be made by considering the family of Tsallis entropies~\cite{Havrda,Tsallis}, that, contrary to the R\'{e}nyi entropies, seem to have the properties needed for deriving non-trivial entropy inequalities for causal structures. By constructing compatible entropy vectors for different Tsallis entropies, one could encode more information about the underlying probability distributions or quantum systems than with conventional entropy vectors, which may in turn facilitate the detection of quantum violations of classically valid inequalities.

In any case, searches for quantum violations of entropic inequalities, which have so far been based on random searches in low dimension (conducted here for the triangle causal structure and previously in~\cite{Cadney2012} without imposing causal constraints), should be made systematic. This could be achieved by performing local optimisation with random starting points, for instance relying on a randomised gradient descent algorithm.

An alternative relaxation of the problem of finding constraints on the distributions that are compatible with a classical causal structure is the inflation technique~\cite{Wolfe2016}. It is able to certify any observed distribution that is not realisable in a causal structure as incompatible if a large enough inflation is considered~\cite{Navascues2017}. In order to derive constraints that are independent of the system's dimensionality, it may be possible to combine this technique with entropic methods. Such methods could apply to causal structures that cannot, so far, be sufficiently analysed entropically, as for instance the triangle causal structure. Nonetheless, computational feasibility is not guaranteed, neither for the detection of incompatible distributions with the current inflation technique, nor with potential entropic extensions.

The problem of whether there is a causal structure that leads to a gap between the sets of compatible entropy vectors in classical and quantum case may also inform the open problem of whether there are (unconstrained) inequalities for the von Neumann entropy of a multi-party quantum state beyond strong-subadditivity~\cite{Linden2005, Cadney2012}, similar to the non-Shannon inequalities, which hold for the entropy of jointly distributed random variables.

Understanding the gap between classical and quantum correlations in causal structures beyond the Bell scenario is important regarding their application in cryptographic protocols. Implementations of protocols between distant agents that in principle rely on the bipartite Bell scenario often establish the required entanglement through entanglement swapping at intermediate stations. Hence, these protocols are not actually implemented in the Bell causal structure but rather in a larger, line-like network, which allows eavesdroppers to devise additional attacks~\cite{Lee2017}. Besides this, there are various cryptographic tasks such as secret sharing schemes~\cite{Shamir1979, Blakley1979} that rely on a causal structure other than the Bell scenario. The analysis of such causal structures is therefore practically relevant. Conversely, the characterisation and comparison of causal structures may be of use for adapting and improving current protocols.

Finally, the exploration of the difference between correlations in classical and quantum causal structures is only a small contribution on the route to an understanding of cause in quantum mechanics. This understanding may be a first step towards resolving the conceptual discrepancy between the classical notion of causality that underlies our understanding of the laws of physics and the probabilistic nature of quantum theory~\cite{Hardy2005, Hardy2007, Hardy2009}. A problem whose solution might fundamentally influence discussions about how to attempt a unification of quantum mechanics and the notion of space-time in general relativity.

\bigskip

In Part~II of this thesis, we presented an axiomatic framework for analysing quantum thermodynamics in the microscopic regime, designed to bring resource theoretic considerations closer to applications by modelling imprecisions and errors. The framework leads to necessary conditions and sufficient conditions for the existence of processes between states of a system in terms of real-valued functions. In the specific case of smooth adiabatic processes these are a smooth min-entropy and a smooth max-entropy from the generalised entropy framework~\cite{Dupuis2013_DH}, which in turn both obtain an operational meaning. For probabilistic transformations we recover slightly different entropy measures. This contributes to the discussion of which of the various smooth min and max-entropies from the recent literature are operationally sensible~\cite{Renyi1960_MeasOfEntrAndInf, Renner2004ISIT, Koenig2009IEEE_OpMeaning, Datta2009IEEE, TomamichelBook, Dupuis2013_DH, MDSFT, Wilde2014, Audenaert2015, Dupuis2016}.

Our considerations move the resource theoretic approach closer to experimental implementations, where finite precision and errors are pertinent topics.
This may also allow us to compare different types of microscopic processes with and without quantum effects in one resource theoretic framework. Furthermore, different sources of error, possibly including quantum fluctuations, can be accounted for, which might be useful for analysing questions related to whether quantum devices can outperform their classical analogues. 

Our framework is not designed for modelling situations where quantum side information about a system is held, which could facilitate and enable additional transformations between states. A corresponding framework may need a careful reconsideration of several components, such as of the sequential composition of operations. Such an axiomatisation might, however, lead to the re-derivation of smooth conditional entropy measures, which would hence obtain an operational meaning in respective resource theories.

Our axiomatic framework also allowed us to consider macroscopic systems and to define thermodynamic equilibrium states.  
This allowed us to recover thermodynamic behaviour, particularly a macroscopic preorder relation that specifies whether a transformation between two different equilibrium states is possible. The framework is also equipped with a single quantity that provides necessary and sufficient conditions for state transformations, in the spirit of the second law and in agreement with axiomatic frameworks for macroscopic thermodynamics~\cite{Lieb1998, Lieb1999, Lieb2001, Lieb2013}. For adiabatic operations, we recover the von Neumann entropy in this manner, as well the Boltzmann entropy for microcanonical states, in agreement with and extending Ref.~\cite{Weilenmann2015}.

Our derivation of necessary and sufficient conditions for state transformations among thermodynamic equilibrium states proceeds along relatively simple lines. Furthermore, checking that certain classes of states obey the prerequisites, i.e., that the corresponding axioms and definitions apply, is relatively straightforward compared to the derivation of these conditions in the individual examples, at least for adiabatic (and thermal) operations. Our framework may therefore be seen to simplify and clarify the derivation of these conditions.

States that qualify as thermodynamic equilibrium states in the case of smooth adiabatic operations are those that are close to their projection onto a typical subspace. Our definition of thermodynamic states can hence be seen as a specification of typical behaviour. This may be of interest beyond thermodynamics. Our results illustrate, for instance, that an asymptotic equipartition property for i.i.d.\ states can be obtained without letting the smoothing parameter tend to zero, contrary to the original proof~\cite{TCR}.

In addition, our exploration of macroscopic systems opens up many avenues for future research. While sequences of states are natural for describing systems with thermodynamic behaviour, where we can define equivalence classes of states that have essentially the same properties, how to generalise these ideas to arbitrary macroscopic systems is unclear. A first step may be to restrict the macroscopic state space to physically meaningful states by imposing that certain criteria be obeyed. One such restriction could be to require permutation symmetry of indistinguishable parts of a system~\cite{Finetti, Konig2005, Renner2007}. One may also define different ways of composing systems, which may lead to meaningful sequences of states. 
The definition of a full macroscopic resource theory that encompasses not just thermodynamic equilibrium states but any physical state and emerges from microscopic considerations remains an open problem.

In this thesis, we have not addressed resource theories other than adiabatic operations within our axiomatic framework. In~\cite{prep}, we show that a resource theory of smooth thermal operations also obeys our axioms. Our framework may also lead to new insights for other, less explored resource theories. Such investigations will be important for gauging the generality of our framework and its potential impact beyond thermodynamics.

\bigskip

The two parts of this thesis have been explored disjointly, except for their remote connection
through quantum technologies. Nonetheless, it may be possible to relate them,
in the sense that one could potentially phrase both of them in terms of resource theories. Causal structures
can be interpreted in terms of quantum channels between the involved systems, and, it may be
possible to understand these channels as resources that allow for the generation of a restricted
set of correlations. Hence, it may be that one day we see the theory of causal structures as a resource theory of causality.

%=========================================================